%
%
%
\def\unredoffs{} \def\redoffs{\voffset=-.31truein\hoffset=-.48truein}
\def\speclscape{}
%
%
%
%
%
\newbox\leftpage \newdimen\fullhsize \newdimen\hstitle \newdimen\hsbody
\tolerance=1000\hfuzz=2pt
\catcode`\@=11 
\ifx\hyperdef\UNd@FiNeD\def\hyperdef#1#2#3#4{#4}\def\hyperref#1#2#3#4{#4}\fi
\def\bigans{b }
\def\answ{b }
%
\ifx\answ\bigans\message{(This will come out unreduced.}
\magnification=1200\unredoffs\baselineskip=16pt plus 2pt minus 1pt
\hsbody=\hsize \hstitle=\hsize 
\else\message{(This will be reduced.} \let\l@r=L
\magnification=1000\baselineskip=16pt plus 2pt minus 1pt \vsize=7truein
\redoffs \hstitle=8truein\hsbody=4.75truein\fullhsize=10truein\hsize=\hsbody
\output={\ifnum\pageno=0 
  \shipout\vbox{\speclscape{\hsize\fullhsize\makeheadline}
    \hbox to \fullhsize{\hfill\pagebody\hfill}}\advancepageno
  \else
  \almostshipout{\leftline{\vbox{\pagebody\makefootline}}}\advancepageno
  \fi}
\def\almostshipout#1{\if L\l@r \count1=1 \message{[\the\count0.\the\count1]}
      \global\setbox\leftpage=#1 \global\let\l@r=R
 \else \count1=2
  \shipout\vbox{\speclscape{\hsize\fullhsize\makeheadline}
      \hbox to\fullhsize{\box\leftpage\hfil#1}}  \global\let\l@r=L\fi}
\fi
%
\newcount\yearltd\yearltd=\year\advance\yearltd by -2000

\def\Title#1#2{\nopagenumbers\abstractfont\hsize=\hstitle\rightline{#1}%
\vskip 1in\centerline{\titlefont #2}\abstractfont\vskip .5in\pageno=0}
\def\Date#1{\vfill\leftline{#1}\tenpoint\supereject\global\hsize=\hsbody%
\footline={\hss\tenrm\hyperdef\hypernoname{page}\folio\folio\hss}}%
%

\def\draftmode{\message{ DRAFTMODE }\def\draftdate{{\rm preliminary draft:
\number\month/\number\day/{0}\number\yearltd\ \ \hourmin}}%
\headline={\hfil\draftdate}\writelabels\baselineskip=20pt plus 2pt minus 2pt
 {\count255=\time\divide\count255 by 60 \xdef\hourmin{\number\count255}
  \multiply\count255 by-60\advance\count255 by\time
  \xdef\hourmin{\hourmin:\ifnum\count255<10 0\fi\the\count255}}}
\def\nolabels{\def\wrlabeL##1{}\def\eqlabeL##1{}\def\reflabeL##1{}}
\def\writelabels{\def\wrlabeL##1{\leavevmode\vadjust{\rlap{\smash%
{\line{{\escapechar=` \hfill\rlap{\sevenrm\hskip.03in\string##1}}}}}}}%
\def\eqlabeL##1{{\escapechar-1\rlap{\sevenrm\hskip.05in\string##1}}}%
\def\reflabeL##1{\noexpand\llap{\noexpand\sevenrm\string\string\string##1}}}
\nolabels
%
\global\newcount\secno \global\secno=0
\global\newcount\meqno \global\meqno=1
\def\s@csym{}
\def\newsec#1{\global\advance\secno by1%
{\toks0{#1}\message{(\the\secno. \the\toks0)}}%
\global\subsecno=0\eqnres@t\let\s@csym\secsym\xdef\secn@m{\the\secno}\noindent
{\bf\hyperdef\hypernoname{section}{\the\secno}{\the\secno.} #1}%
\writetoca{{\string\hyperref{}{section}{\the\secno}{\it\the\secno.}} {{\it #1} }}%
\par\nobreak\medskip\nobreak}
\def\eqnres@t{\xdef\secsym{\the\secno.}\global\meqno=1\bigbreak\bigskip}
\def\sequentialequations{\def\eqnres@t{\bigbreak}}\xdef\secsym{}
\global\newcount\subsecno \global\subsecno=0
\def\subsec#1{\global\advance\subsecno by1%
{\toks0{#1}\message{(\s@csym\the\subsecno. \the\toks0)}}%
\ifnum\lastpenalty>9000\else\bigbreak\fi       \global\subsubsecno=0
\noindent{\it\hyperdef\hypernoname{subsection}{\secn@m.\the\subsecno}%
{\secn@m.\the\subsecno.} #1}\writetoca{\string\quad
{\string\hyperref{}{subsection}{\secn@m.\the\subsecno}{\secn@m.\the\subsecno.}}
{#1}}\par\nobreak\medskip\nobreak}
\def\appendix#1#2{\global\meqno=1\global\subsecno=0\xdef\secsym{\hbox{#1.}}%
\bigbreak\bigskip\noindent{\bf Appendix \hyperdef\hypernoname{appendix}{#1}%
{#1.} #2}{\toks0{(#1. #2)}\message{\the\toks0}}%
\xdef\s@csym{#1.}\xdef\secn@m{#1}%
\writetoca{\string\hyperref{}{appendix}{#1}{{\it Appendix} {\it #1.}} {\it #2}}%
\par\nobreak\medskip\nobreak}
%
%
\def\checkm@de#1#2{\ifmmode{\def\f@rst##1{##1}\hyperdef\hypernoname{equation}%
{#1}{#2}}\else\hyperref{}{equation}{#1}{#2}\fi}
\def\eqnn#1{\DefWarn#1\xdef #1{(\noexpand\relax\noexpand\checkm@de%
{\s@csym\the\meqno}{\secsym\the\meqno})}%
\wrlabeL#1\writedef{#1\leftbracket#1}\global\advance\meqno by1}
\def\f@rst#1{\c@t#1a\em@ark}\def\c@t#1#2\em@ark{#1}
\def\eqna#1{\DefWarn#1\wrlabeL{#1$\{\}$}%
\xdef #1##1{(\noexpand\relax\noexpand\checkm@de%
{\s@csym\the\meqno\noexpand\f@rst{##1}}{\hbox{$\secsym\the\meqno##1$}})}
\writedef{#1\numbersign1\leftbracket#1{\numbersign1}}\global\advance\meqno by1}
\def\eqn#1#2{\DefWarn#1%
\xdef #1{(\noexpand\hyperref{}{equation}{\s@csym\the\meqno}%
{\secsym\the\meqno})}$$#2\eqno(\hyperdef\hypernoname{equation}%
{\s@csym\the\meqno}{\secsym\the\meqno})\eqlabeL#1$$%
\writedef{#1\leftbracket#1}\global\advance\meqno by1}
\def\xeqn{\expandafter\xe@n}\def\xe@n(#1){#1}
\def\xeqna#1{\expandafter\xe@n#1}
\def\eqns#1{(\e@ns #1{\hbox{}})}
\def\e@ns#1{\ifx\UNd@FiNeD#1\message{eqnlabel \string#1 is undefined.}%
\xdef#1{(?.?)}\fi{\let\hyperref=\relax\xdef\next{#1}}%
\ifx\next\em@rk\def\next{}\else%
\ifx\next#1\xeqn#1\else\def\n@xt{#1}\ifx\n@xt\next#1\else\xeqna#1\fi
\fi\let\next=\e@ns\fi\next}

\def\DefWarn#1{\ifx\UNd@FiNeD#1\else
\immediate\write16{*** WARNING: the label \string#1 is already defined ***}\fi}
%
\newskip\footskip\footskip14pt plus 1pt minus 1pt 
\def\footnotefont{\ninepoint}\def\f@t#1{\footnotefont #1\@foot}
\def\f@@t{\baselineskip\footskip\bgroup\footnotefont\aftergroup\@foot\let\next}
\setbox\strutbox=\hbox{\vrule height9.5pt depth4.5pt width0pt}
\global\newcount\ftno \global\ftno=0
\def\foot{\global\advance\ftno by1\def\foot@rg{\hyperref{}{footnote}%
{\the\ftno}{\the\ftno}\xdef\foot@rg{\noexpand\hyperdef\noexpand\hypernoname%
{footnote}{\the\ftno}{\the\ftno}}}\footnote{$^{\foot@rg}$}}
%
\newwrite\ftfile
\def\footend{\def\foot{\global\advance\ftno by1\chardef\wfile=\ftfile
\hyperref{}{footnote}{\the\ftno}{$^{\the\ftno}$}%
\ifnum\ftno=1\immediate\openout\ftfile=\jobname.fts\fi%
\immediate\write\ftfile{\noexpand\smallskip%
\noexpand\item{\noexpand\hyperdef\noexpand\hypernoname{footnote}
{\the\ftno}{f\the\ftno}:\ }\pctsign}\findarg}%
\def\footatend{\vfill\eject\immediate\closeout\ftfile{\parindent=20pt
\centerline{\bf Footnotes}\nobreak\bigskip\input \jobname.fts }}}
\def\footatend{}
%
%
\global\newcount\refno \global\refno=1
\newwrite\rfile
\def\ref{[\hyperref{}{reference}{\the\refno}{\the\refno}]\nref}
\def\nref#1{\DefWarn#1%
\xdef#1{[\noexpand\hyperref{}{reference}{\the\refno}{\the\refno}]}%
\writedef{#1\leftbracket#1}%
\ifnum\refno=1\immediate\openout\rfile=\jobname.refs\fi
\chardef\wfile=\rfile\immediate\write\rfile{\noexpand\item{[\noexpand\hyperdef%
\noexpand\hypernoname{reference}{\the\refno}{\the\refno}]\ }%
\reflabeL{#1\hskip.31in}\pctsign}\global\advance\refno by1\findarg}
\def\findarg#1#{\begingroup\obeylines\newlinechar=`\^^M\pass@rg}
{\obeylines\gdef\pass@rg#1{\writ@line\relax #1^^M\hbox{}^^M}%
\gdef\writ@line#1^^M{\expandafter\toks0\expandafter{\striprel@x #1}%
\edef\next{\the\toks0}\ifx\next\em@rk\let\next=\endgroup\else\ifx\next\empty%
\else\immediate\write\wfile{\the\toks0}\fi\let\next=\writ@line\fi\next\relax}}
\def\striprel@x#1{} \def\em@rk{\hbox{}}
\def\lref{\begingroup\obeylines\lr@f}
\def\lr@f#1#2{\DefWarn#1\gdef#1{\let#1=\UNd@FiNeD\ref#1{#2}}\endgroup\unskip}

\def\addref#1{\immediate\write\rfile{\noexpand\item{}#1}} 
\def\listrefs{\footatend\vfill\supereject\immediate\closeout\rfile\writestoppt
\baselineskip=\footskip\centerline{{\bf References}}\bigskip{\parindent=20pt%
\frenchspacing\escapechar=` \input \jobname.refs\vfill\eject}\nonfrenchspacing}
\def\startrefs#1{\immediate\openout\rfile=\jobname.refs\refno=#1}
\def\xref{\expandafter\xr@f}\def\xr@f[#1]{#1}
\def\refs#1{\count255=1[\r@fs #1{\hbox{}}]}
\def\r@fs#1{\ifx\UNd@FiNeD#1\message{reflabel \string#1 is undefined.}%
\nref#1{need to supply reference \string#1.}\fi%
\vphantom{\hphantom{#1}}{\let\hyperref=\relax\xdef\next{#1}}%
\ifx\next\em@rk\def\next{}%
\else\ifx\next#1\ifodd\count255\relax\xref#1\count255=0\fi%
\else#1\count255=1\fi\let\next=\r@fs\fi\next}
%

%
\newwrite\ffile\global\newcount\figno \global\figno=1
\def\fig{fig.~\hyperref{}{figure}{\the\figno}{\the\figno}\nfig}
\def\nfig#1{\DefWarn#1%
\xdef#1{fig.~\noexpand\hyperref{}{figure}{\the\figno}{\the\figno}}%
\writedef{#1\leftbracket fig.\noexpand~\xfig#1}%
\ifnum\figno=1\immediate\openout\ffile=\jobname.figs\fi\chardef\wfile=\ffile%
{\let\hyperref=\relax
\immediate\write\ffile{\noexpand\medskip\noexpand\item{Fig.\ %
\noexpand\hyperdef\noexpand\hypernoname{figure}{\the\figno}{\the\figno}. }
\reflabeL{#1\hskip.55in}\pctsign}}\global\advance\figno by1\findarg}
\def\listfigs{\vfill\eject\immediate\closeout\ffile{\parindent40pt
\baselineskip14pt\centerline{{\bf Figure Captions}}\nobreak\medskip
\escapechar=` \input \jobname.figs\vfill\eject}}
\def\xfig{\expandafter\xf@g}\def\xf@g fig.\penalty\@M\ {}
\def\figs#1{figs.~\f@gs #1{\hbox{}}}
\def\f@gs#1{{\let\hyperref=\relax\xdef\next{#1}}\ifx\next\em@rk\def\next{}\else
\ifx\next#1\xfig #1\else#1\fi\let\next=\f@gs\fi\next}
\def\figin{\epsfcheck\figin}\def\figins{\epsfcheck\figins}
\def\epsfcheck{\ifx\epsfbox\UNd@FiNeD
\message{(NO epsf.tex, FIGURES WILL BE IGNORED)}
\gdef\figin##1{\vskip2in}\gdef\figins##1{\hskip.5in}
\else\message{(FIGURES WILL BE INCLUDED)}%
\gdef\figin##1{##1}\gdef\figins##1{##1}\fi}
\def\DefWarn#1{}
\def\figinsert{\goodbreak\midinsert}
\def\ifig#1#2#3{\DefWarn#1\xdef#1{Fig.~\noexpand\hyperref{}{figure}%
{\the\figno}{\the\figno}}\writedef{#1\leftbracket fig.\noexpand~\xfig#1}%
\figinsert\figin{\centerline{#3}}\medskip\centerline{\vbox{\baselineskip12pt
\advance\hsize by -1truein\noindent\wrlabeL{#1=#1}\footnotefont%
{\bf Fig.~\hyperdef\hypernoname{figure}{\the\figno}{\the\figno}:} #2}}
\bigskip\endinsert\global\advance\figno by1}
\newwrite\lfile
{\escapechar-1\xdef\pctsign{\string\%}\xdef\leftbracket{\string\{}
\xdef\rightbracket{\string\}}\xdef\numbersign{\string\#}}
\def\writedefs{\immediate\openout\lfile=\jobname.defs \def\writedef##1{%
{\let\hyperref=\relax\let\hyperdef=\relax\let\hypernoname=\relax
 \immediate\write\lfile{\string\def\string##1\rightbracket}}}}%
\def\writestop{\def\writestoppt{\immediate\write\lfile{\string\pageno
 \the\pageno\string\startrefs\leftbracket\the\refno\rightbracket
 \string\def\string\secsym\leftbracket\secsym\rightbracket
 \string\secno\the\secno\string\meqno\the\meqno}\immediate\closeout\lfile}}
\def\writestoppt{}\def\writedef#1{}
\def\seclab#1{\DefWarn#1%
\xdef #1{\noexpand\hyperref{}{section}{\the\secno}{\the\secno}}%
\writedef{#1\leftbracket#1}\wrlabeL{#1=#1}}
\def\subseclab#1{\DefWarn#1%
\xdef #1{\noexpand\hyperref{}{subsection}{\secn@m.\the\subsecno}%
{\secn@m.\the\subsecno}}\writedef{#1\leftbracket#1}\wrlabeL{#1=#1}}
\def\applab#1{\DefWarn#1%
\xdef #1{\noexpand\hyperref{}{appendix}{\secn@m}{\secn@m}}%
\writedef{#1\leftbracket#1}\wrlabeL{#1=#1}}
\newwrite\tfile \def\writetoca#1{}
\def\leaderfill{\leaders\hbox to 1em{\hss.\hss}\hfill}
\def\writetoc{\immediate\openout\tfile=\jobname.toc
   \def\writetoca##1{{\edef\next{\write\tfile{\noindent ##1
   \string\leaderfill {\string\hyperref{}{page}{\noexpand\number\pageno}%
                       {\noexpand\number\pageno}} \par}}\next}}}
\newread\ch@ckfile
\def\listtoc{\immediate\closeout\tfile\immediate\openin\ch@ckfile=\jobname.toc
\ifeof\ch@ckfile\message{no file \jobname.toc, no table of contents this pass}%
\else\closein\ch@ckfile\centerline{\bf Contents}\nobreak\medskip%
{\baselineskip=18.5pt  \footnotefont
\parskip=2pt\catcode`\@=12\input\jobname.toc
\catcode`\@=12\bigbreak\bigskip}\fi}
\catcode`\@=12 
%
\edef\tfontsize{\ifx\answ\bigans scaled\magstep3\else scaled\magstep4\fi}
\font\titlerm=cmr10 \tfontsize \font\titlerms=cmr7 \tfontsize
\font\titlermss=cmr5 \tfontsize \font\titlei=cmmi10 \tfontsize
\font\titleis=cmmi7 \tfontsize \font\titleiss=cmmi5 \tfontsize
\font\titlesy=cmsy10 \tfontsize \font\titlesys=cmsy7 \tfontsize
\font\titlesyss=cmsy5 \tfontsize \font\titleit=cmti10 \tfontsize
\skewchar\titlei='177 \skewchar\titleis='177 \skewchar\titleiss='177
\skewchar\titlesy='60 \skewchar\titlesys='60 \skewchar\titlesyss='60
\def\titlefont{\def\rm{\fam0\titlerm}
\textfont0=\titlerm \scriptfont0=\titlerms \scriptscriptfont0=\titlermss
\textfont1=\titlei \scriptfont1=\titleis \scriptscriptfont1=\titleiss
\textfont2=\titlesy \scriptfont2=\titlesys \scriptscriptfont2=\titlesyss
\textfont\itfam=\titleit \def\it{\fam\itfam\titleit}\rm}
 \ifx\answ\bigans\else scaled\magstep1\fi
\ifx\answ\bigans\def\abstractfont{\tenpoint}\else
\font\absit=cmti10 scaled \magstep1
\font\abssl=cmsl10 scaled \magstep1
\font\absrm=cmr10 scaled\magstep1 \font\absrms=cmr7 scaled\magstep1
\font\absrmss=cmr5 scaled\magstep1 \font\absi=cmmi10 scaled\magstep1
\font\absis=cmmi7 scaled\magstep1 \font\absiss=cmmi5 scaled\magstep1
\font\abssy=cmsy10 scaled\magstep1 \font\abssys=cmsy7 scaled\magstep1
\font\abssyss=cmsy5 scaled\magstep1 \font\absbf=cmbx10 scaled\magstep1
\skewchar\absi='177 \skewchar\absis='177 \skewchar\absiss='177
\skewchar\abssy='60 \skewchar\abssys='60 \skewchar\abssyss='60
\def\abstractfont{\def\rm{\fam0\absrm}
\textfont0=\absrm \scriptfont0=\absrms \scriptscriptfont0=\absrmss
\textfont1=\absi \scriptfont1=\absis \scriptscriptfont1=\absiss
\textfont2=\abssy \scriptfont2=\abssys \scriptscriptfont2=\abssyss
\textfont\itfam=\absit \def\it{\fam\itfam\absit}\def\footnotefont{\tenpoint}%
\textfont\slfam=\abssl \def\sl{\fam\slfam\abssl}%
\textfont\bffam=\absbf \def\bf{\fam\bffam\absbf}\rm}\fi
\def\tenpoint{\def\rm{\fam0\tenrm}
\textfont0=\tenrm \scriptfont0=\sevenrm \scriptscriptfont0=\fiverm
\textfont1=\teni  \scriptfont1=\seveni  \scriptscriptfont1=\fivei
\textfont2=\tensy \scriptfont2=\sevensy \scriptscriptfont2=\fivesy
\textfont\itfam=\tenit \def\it{\fam\itfam\tenit}\def\footnotefont{\ninepoint}%
\textfont\bffam=\tenbf \def\bf{\fam\bffam\tenbf}\def\sl{\fam\slfam\tensl}\rm}
\font\ninerm=cmr9 \font\sixrm=cmr6 \font\ninei=cmmi9 \font\sixi=cmmi6
\font\ninesy=cmsy9 \font\sixsy=cmsy6 \font\ninebf=cmbx9
\font\nineit=cmti9 \font\ninesl=cmsl9 \skewchar\ninei='177
\skewchar\sixi='177 \skewchar\ninesy='60 \skewchar\sixsy='60
\def\ninepoint{\def\rm{\fam0\ninerm}
\textfont0=\ninerm \scriptfont0=\sixrm \scriptscriptfont0=\fiverm
\textfont1=\ninei \scriptfont1=\sixi \scriptscriptfont1=\fivei
\textfont2=\ninesy \scriptfont2=\sixsy \scriptscriptfont2=\fivesy
\textfont\itfam=\ninei \def\it{\fam\itfam\nineit}\def\sl{\fam\slfam\ninesl}%
\textfont\bffam=\ninebf \def\bf{\fam\bffam\ninebf}\rm}
%
%
\def\noblackbox{\overfullrule=0pt}
\hyphenation{anom-aly anom-alies coun-ter-term coun-ter-terms}
\def\inv{^{\raise.15ex\hbox{${\scriptscriptstyle -}$}\kern-.05em 1}}

\def\Dsl{\,\raise.15ex\hbox{/}\mkern-13.5mu D} 
\def\dsl{\raise.15ex\hbox{/}\kern-.57em\partial}

\def\tr{{\rm tr}} \def\Tr{{\rm Tr}}
\def\lspace{\ifx\answ\bigans{}\else\qquad\fi}
\def\lbspace{\ifx\answ\bigans{}\else\hskip-.2in\fi} 
\def\boxeqn#1{\vcenter{\vbox{\hrule\hbox{\vrule\kern3pt\vbox{\kern3pt
	\hbox{${\displaystyle #1}$}\kern3pt}\kern3pt\vrule}\hrule}}}
\def\mbox#1#2{\vcenter{\hrule \hbox{\vrule height#2in
		\kern#1in \vrule} \hrule}}  
%

\def\e#1{{\rm e}^{^{\textstyle#1}}}

\def\vev#1{\langle #1 \rangle}

\def\darr#1{\raise1.5ex\hbox{$\leftrightarrow$}\mkern-16.5mu #1}

\def\roughly#1{\raise.3ex\hbox{$#1$\kern-.75em\lower1ex\hbox{$\sim$}}}

\global\newcount\subsubsecno \global\subsubsecno=0
\def\subsubsec#1{\global\advance\subsubsecno by1%
{\toks0{#1}\message{(\the\secno\the\subsecno\the\subsubsecno. \the\toks0)}}%
\ifnum\lastpenalty>9000\else\bigbreak\fi
\noindent{\it\hyperdef\hypernoname{subsubsection}{\the\secno.\the\subsecno\the\subsubsecno}%
{\the\secno.\the\subsecno.\the\subsubsecno.} #1}
\par\nobreak\medskip\nobreak}
\def\boxit#1{\vbox{\hrule\hbox{\vrule\kern8pt
\vbox{\hbox{\kern8pt}\hbox{\vbox{#1}}\hbox{\kern8pt}}
\kern8pt\vrule}\hrule}}
\def\mathboxit#1{\vbox{\hrule\hbox{\vrule\kern8pt\vbox{\kern8pt
\hbox{$\displaystyle #1$}\kern8pt}\kern8pt\vrule}\hrule}}
\def\slashchar#1{\setbox0=\hbox{$#1$}           
   \dimen0=\wd0                                 
   \setbox1=\hbox{/} \dimen1=\wd1               
   \ifdim\dimen0>\dimen1                        
      \rlap{\hbox to \dimen0{\hfil/\hfil}}      
      #1                                        
   \else                                        
      \rlap{\hbox to \dimen1{\hfil$#1$\hfil}}   
      /                                         
   \fi}
\def\sqr#1#2{{\vcenter{\vbox{\hrule height.#2pt
         \hbox{\vrule width.#2pt height#1pt \kern#1pt
            \vrule width.#2pt}
         \hrule height.#2pt}}}}


\noblackbox
\ifx\answ\bigans
\magnification=1200\baselineskip=14pt plus 2pt minus 1pt
\else\baselineskip=16pt 
\fi

\def\crr{\noalign{\vskip5pt}}
\def\ds#1{{\displaystyle{#1}}}
\def\comment#1{{}}
\def\ss#1{{\scriptstyle{#1}}}

\def\ap{\alpha'}

\def\cf{{\it cf.\ }}
\def\ie{{\it i.e.\ }}
\def\eg{{\it e.g.\ }}
\def\eqq{{\it Eq.\ }}
\def\eqqs{{\it Eqs.\ }}
\def\th{\theta}
\def\eps{\epsilon}
\def\al{\alpha}

\def\si{\sigma}\def\Si{{\Sigma}}
\def\Om{\Omega}
\def\bet{\beta}

\def\Hc{{{\bf H}_+}}
\def\Dc{{\bf D}}

\def\FF#1#2{{_#1F_#2}}
\newif\ifnref

\def\doubref#1#2{\refs{{#1},{#2} }}
\def\threeref#1#2#3{\refs{{#1},{#2},{#3} }}

\nreffalse

\input epsf

\def\figin{\epsfcheck\figin}\def\figins{\epsfcheck\figins}
\def\epsfcheck{\ifx\epsfbox\UnDeFiNeD
\message{(NO epsf.tex, FIGURES WILL BE IGNORED)}
\gdef\figin##1{\vskip2in}\gdef\figins##1{\hskip.5in}
\else\message{(FIGURES WILL BE INCLUDED)}%
\gdef\figin##1{##1}\gdef\figins##1{##1}\fi}
\def\DefWarn#1{}
\def\figinsert{\goodbreak\midinsert}  
\def\ifig#1#2#3{\DefWarn#1\xdef#1{Fig.~\the\figno}
\writedef{#1\leftbracket fig.\noexpand~\the\figno}%
\figinsert\figin{\centerline{#3}}\medskip\centerline{\vbox{\baselineskip12pt
\advance\hsize by -1truein\noindent\footnotefont\centerline{{\bf
Fig.~\the\figno}\ \sl #2}}}
\bigskip\endinsert\global\advance\figno by1}

\def\iifig#1#2#3#4{\DefWarn#1\xdef#1{Fig.~\the\figno}
\writedef{#1\leftbracket fig.\noexpand~\the\figno}%
\figinsert\figin{\centerline{#4}}\medskip\centerline{\vbox{\baselineskip12pt
\advance\hsize by -1truein\noindent\footnotefont\centerline{{\bf
Fig.~\the\figno}\ \ \sl #2}}}\smallskip\centerline{\vbox{\baselineskip12pt
\advance\hsize by -1truein\noindent\footnotefont\centerline{\ \ \ \sl #3}}}
\bigskip\endinsert\global\advance\figno by1}


\def\appA{A}
\def\appB{B}
\def\appC{C}
\def\appD{D}\def\appDi{D.1.}\def\appDii{D.2.}

\def\tilde{\widetilde}
\def\hatt{\widehat}
\def\h {{1\over 2}}

\def\ov {\overline}
\def\o {\over}
\def\fc#1#2{{#1 \o #2}}

\def\IZ{ {\bf Z}}\def\IC{{\bf C}}\def\IR{ {\bf R}}


\def\br{\hfill\break}
\def\tr {{\rm tr}}
\def\det {{\rm det}}

\def\lf {\left}
\def\ri {\right}
\def\ra {\rightarrow}

\def\re {{\rm Re}}
\def\im {{\rm Im}}
\def\p {\partial}

\def\Cc {{\cal C}} 
 
 \def\Ac {{\cal A}}
\def\Pc {{\cal P}} 
 
\def\Ic {{\cal I}} 
\def\Kc {{\cal K}}


\lref\Gubser{
I.R.~Klebanov and L.~Thorlacius,
``The Size of p-Branes,''
  Phys.\ Lett.\  B {\bf 371}, 51 (1996)
  [arXiv:hep-th/9510200];\br
  S.S.~Gubser, A.~Hashimoto, I.R.~Klebanov and J.M.~Maldacena,
``Gravitational lensing by $p$-branes,''
  Nucl.\ Phys.\  B {\bf 472}, 231 (1996)
  [arXiv:hep-th/9601057].
}

\lref\Mangano{
  M.L.~Mangano and S.J.~Parke,
``Multi-Parton Amplitudes in Gauge Theories,''
  Phys.\ Rept.\  {\bf 200}, 301 (1991)
  [arXiv:hep-th/0509223].
}

\lref\Kleiss{
  R.~Kleiss and H.~Kuijf,
``Multi - Gluon Cross-Sections And Five Jet Production At Hadron Colliders,''
  Nucl.\ Phys.\  B {\bf 312}, 616 (1989).
}

\lref\Bern{
  Z.~Bern, J.J.M.~Carrasco and H.~Johansson,
``New Relations for Gauge-Theory Amplitudes,''
  Phys.\ Rev.\  D {\bf 78}, 085011 (2008)
  [arXiv:0805.3993 [hep-ph]].
}

\lref\DelDuca{
  V.~Del Duca, L.J.~Dixon and F.~Maltoni,
``New color decompositions for gauge amplitudes at tree and loop level,''
  Nucl.\ Phys.\  B {\bf 571} (2000) 51
  [arXiv:hep-ph/9910563].
}

\lref\oocc{S. Stieberger, 
``Disk Amplitudes of Open and Closed String Moduli in Calabi--Yau Orientifolds,''
MPP--2008--02, to appear.}

\lref\AM{L.F.~Alday and J.M.~Maldacena,
``Gluon scattering amplitudes at strong coupling,''
  JHEP {\bf 0706}, 064 (2007)
  [arXiv:0705.0303 [hep-th]].
}

\lref\McGreevy{J.~McGreevy and A.~Sever,
``Quark scattering amplitudes at strong coupling,''
  JHEP {\bf 0802}, 015 (2008)
  [arXiv:0710.0393 [hep-th]].
}

\lref\Sen{
A.~Sen, ``Open and closed strings from unstable D-branes,''
  Phys.\ Rev.\  D {\bf 68}, 106003 (2003)
  [arXiv:hep-th/0305011];
``Open-closed duality at tree level,''
  Phys.\ Rev.\ Lett.\  {\bf 91}, 181601 (2003)
  [arXiv:hep-th/0306137].
}

\lref\Plahte{E. Plahte,
``Symmetry properties of dual tree-graph N--point amplitudes,'' 
Nuovo Cimento {\bf 66 A} (1970) 713.}

\lref\Brazil{L.A.~Barreiro and R.~Medina,
``5-field terms in the open superstring effective action,''
  JHEP {\bf 0503}, 055 (2005)
  [arXiv:hep-th/0503182];\br
R.~Medina, F.T.~Brandt and F.R.~Machado,
``The open superstring 5-point amplitude revisited,''
  JHEP {\bf 0207}, 071 (2002)
  [arXiv:hep-th/0208121].
}

\lref\Herbst{
  M.~Herbst, C.I.~Lazaroiu and W.~Lerche,
``Superpotentials, $A_\infty$ relations and WDVV equations for open
topological strings,''
  JHEP {\bf 0502}, 071 (2005)
  [arXiv:hep-th/0402110].
}

\lref\HK{A.~Hashimoto and I.R.~Klebanov,
``Scattering of strings from D-branes,''
  Nucl.\ Phys.\ Proc.\ Suppl.\  {\bf 55B}, 118 (1997)
  [arXiv:hep-th/9611214].
}

\lref\HKi{
  A.~Hashimoto and I.R.~Klebanov,
``Decay of Excited D-branes,''
  Phys.\ Lett.\  B {\bf 381}, 437 (1996)
  [arXiv:hep-th/9604065].
}

\lref\GM{
  M.R.~Garousi and R.C.~Myers,
``Superstring Scattering from D-Branes,''
  Nucl.\ Phys.\  B {\bf 475}, 193 (1996)
  [arXiv:hep-th/9603194].
}

\lref\russo{
  M.~Bertolini, M.~Billo, A.~Lerda, J.F.~Morales and R.~Russo,
``Brane world effective actions for D-branes with fluxes,''
  Nucl.\ Phys.\ B {\bf 743}, 1 (2006)
  [arXiv:hep-th/0512067].
}
\lref\DF{V.S.~Dotsenko and V.A.~Fateev,
``Four Point Correlation Functions And The Operator Algebra In The
Two-Dimensional Conformal Invariant Theories With The Central Charge $C < 1$,''
  Nucl.\ Phys.\  B {\bf 251}, 691 (1985); 
``Conformal algebra and multipoint correlation functions in  2D statistical models,''
  Nucl.\ Phys.\  B {\bf 240}, 312 (1984);\br
P. di Francesco, P. Mathieu, and David Senechal, 
``Conformal Field Theory'', Springer  2nd  edition, 1999.}

\lref\Minahan{
  J.A.~Minahan,
``One Loop Amplitudes On Orbifolds And The Renormalization Of Coupling Constants,''
  Nucl.\ Phys.\  B {\bf 298}, 36 (1988).
}

\lref\foto{M.R.~Garousi and R.C.~Myers,
``World-volume potentials on D-branes,''
  JHEP {\bf 0011}, 032 (2000)
  [arXiv:hep-th/0010122];\br
A.~Fotopoulos,
``On $(\ap)^2$ corrections to the D-brane action for non-geodesic
 world-volume embeddings,''
  JHEP {\bf 0109}, 005 (2001)
  [arXiv:hep-th/0104146].
}

\lref\fotoii{A.~Fotopoulos and A.A.~Tseytlin,
``On gravitational couplings in D-brane action,''
  JHEP {\bf 0212}, 001 (2002)
  [arXiv:hep-th/0211101].
}

\lref\garousi{M.R.~Garousi and E.~Hatefi,
``On Wess-Zumino terms of Brane-Antibrane systems,''
  Nucl.\ Phys.\  B {\bf 800}, 502 (2008)
  [arXiv:0710.5875 [hep-th]];
``More on WZ action of non-BPS branes,''
  JHEP {\bf 0903}, 008 (2009)
  [arXiv:0812.4216 [hep-th]];\br
M.R.~Garousi,
``On the effective action of D-brane-anti-D-brane system,''
  JHEP {\bf 0712}, 089 (2007);
 ``Higher derivative corrections to Wess-Zumino action of Brane-Antibrane systems,''
  JHEP {\bf 0802}, 109 (2008)
  [arXiv:0712.1954 [hep-th]];
 ``On Wess-Zumino terms of non-BPS D-branes and their higher derivative corrections,''
  arXiv:0802.2784 [hep-th];\br
M.R.~Garousi and H.~Golchin,
 ``On higher derivative corrections of the tachyon action,''
  Nucl.\ Phys.\  B {\bf 800}, 547 (2008)
  [arXiv:0801.3358 [hep-th]].
}

\lref\hunter{D.~L\"ust, S.~Stieberger and T.R.~Taylor,
``The LHC String Hunter's Companion,''
  Nucl.\ Phys.\  B {\bf 808}, 1 (2009)
  [arXiv:0807.3333 [hep-th]].
}

\lref\August{S.~Stieberger and T.R.~Taylor,
``Supersymmetry Relations and MHV Amplitudes in Superstring Theory,''
  Nucl.\ Phys.\  B {\bf 793}, 83 (2008)
  [arXiv:0708.0574 [hep-th]].
}

\lref\Potsdam{S.~Stieberger and T.R.~Taylor,
``Complete Six-Gluon Disk Amplitude in Superstring Theory,''
  Nucl.\ Phys.\  B {\bf 801}, 128 (2008)
  [arXiv:0711.4354 [hep-th]].
}

\lref\ACNY{E.S.~Fradkin and A.A.~Tseytlin,
"Nonlinear Electrodynamics From Quantized Strings,''
Phys.\ Lett.\ B {\bf 163}, 123 (1985);\br
A.~Abouelsaood, C.G. Callan, C.R.~Nappi and S.A.~Yost,
"Open Strings In Background Gauge Fields,''
Nucl.\ Phys.\ B {\bf 280}, 599 (1987).
}
\lref\SW{N.~Seiberg and E.~Witten,
"String theory and noncommutative geometry,''
JHEP {\bf 9909}, 032 (1999)
[arXiv:hep-th/9908142].
}

\lref\Jaxo{D.~Binosi and L.~Theussl,
``JaxoDraw: A graphical user interface for drawing Feynman diagrams,''
  Comput.\ Phys.\ Commun.\  {\bf 161}, 76 (2004)
  [arXiv:hep-ph/0309015].
}

\lref\VH{N.E.J.~Bjerrum-Bohr, P.H.~Damgaard and P.~Vanhove,
``Minimal Basis for Gauge Theory Amplitudes,''
  arXiv:0907.1425 [hep-th].
}

\lref\FISI{S. Stieberger,``Disk scattering of open and closed strings,'' Workshop
{\it New Perspectives in String Theory,}
The Galileo Galilei Institute for Theoretical Physics (GGI), Firenze, May 20, 2009,
http://ggi-www.fi.infn.it/talks/talk1211.pdf;\br
{\it see also:} http://wwwth.mppmu.mpg.de/members/stieberg/index.html.}

\lref\GHMR{
  D.J.~Gross, J.A.~Harvey, E.J.~Martinec and R.~Rohm,
``Heterotic String Theory. 2. The Interacting Heterotic String,''
  Nucl.\ Phys.\ B {\bf 267}, 75 (1986).
}

\lref\JSCH{M.B.~Green and J.H.~Schwarz,
``Supersymmetrical Dual String Theory. 2. Vertices And Trees,''
Nucl.\ Phys.\ B {\bf 198}, 252 (1982);\br
J.H.~Schwarz,
``Superstring Theory,''
Phys.\ Rept.\  {\bf 89}, 223 (1982).
}

\lref\JOE{
J. Polchinski, "String Theory'', Sections 6 \& 12, Cambridge University Press 1998.}

\lref\LMRS{
D.~L\"ust, P.~Mayr, R.~Richter and S.~Stieberger,
``Scattering of gauge, matter, and moduli fields from intersecting  branes,''
  Nucl.\ Phys.\ B {\bf 696}, 205 (2004)
  [arXiv:hep-th/0404134].
}

\lref\POLCH{
  J.~Polchinski,
``Dirichlet-Branes and Ramond-Ramond Charges,''
  Phys.\ Rev.\ Lett.\  {\bf 75}, 4724 (1995)
  [arXiv:hep-th/9510017].
}

\lref\Gradst{
I.S. Gradshteyn and I.M. Ryzhik,
``Table of Integrals, Series and Products'', A. Jeffrey and D. Zwillinger (eds.),
Academic Press, London 2007.}

\lref\Erdel{A. Erd\'elyi; W. Magnus, F. Oberhettinger, and
F.G. Tricomi, 
{``Higher Transcendental Functions''}, Vol. 1, McGraw--Hill Book
Company, New York 1953.}

\lref\MSD{
P.~Mayr and S.~Stieberger,
``Dilaton, antisymmetric tensor and gauge fields in string effective theories
at the one loop level,''
Nucl.\ Phys.\ B {\bf 412}, 502 (1994)
[arXiv:hep-th/9304055].
}

\lref\Kawai{
  H.~Kawai, D.C.~Lewellen and S.H.H.~Tye,
``A Relation Between Tree Amplitudes Of Closed And Open Strings,''
  Nucl.\ Phys.\ B {\bf 269}, 1 (1986).
}

\lref\LM{D.~Lancaster and P.~Mansfield,
``Relations Between Disk Diagrams,''
  Phys.\ Lett.\  B {\bf 217}, 416 (1989).
}

\lref\China{Y.X.~Chen, Y.J.~Du and Q.~Ma,
``Relations Between Closed String Amplitudes at Higher-order Tree Level and
Open String Amplitudes,''
  arXiv:0901.1163 [hep-th].
}

\lref\Dan{D.~Oprisa and S.~Stieberger,
``Six gluon open superstring disk amplitude, multiple hypergeometric  series
and Euler-Zagier sums,''
  arXiv:hep-th/0509042.
}

\lref\Bain{P.~Bain and M.~Berg,
"Effective action of matter fields in four-dimensional string  orientifolds,''
JHEP {\bf 0004}, 013 (2000)
[arXiv:hep-th/0003185].
}

\lref\STi{S.~Stieberger and T.R.~Taylor,
``Amplitude for N-gluon superstring scattering,''
  Phys.\ Rev.\ Lett.\  {\bf 97}, 211601 (2006)
  [arXiv:hep-th/0607184].
}

\lref\STii{S.~Stieberger and T.R.~Taylor,
``Multi-gluon scattering in open superstring theory,''
  Phys.\ Rev.\  D {\bf 74}, 126007 (2006)
  [arXiv:hep-th/0609175].
}

\Title{\vbox{\rightline{MPP--2008--01}
}}
{\vbox{\centerline{Open $\&$ Closed vs. Pure Open String Disk Amplitudes}}}
\medskip
\centerline{S. Stieberger}
\bigskip
\centerline{\it Max--Planck--Institut f\"ur Physik}
\centerline{\it Werner--Heisenberg--Institut}
\centerline{\it 80805 M\"unchen, Germany}

\vskip15pt

\medskip
\bigskip\bigskip\bigskip
\centerline{\bf Abstract}
\vskip .2in
\noindent

We establish a relation between disk amplitudes involving $N_o$ open and $N_c$
closed strings and disk amplitudes with only $N_o+2N_c$ open strings.
This map, which represents a sort of generalized KLT relation on the disk, 
reveals important structures between
open $\&$ closed and pure open string disk amplitudes:
it relates couplings of brane and bulk string states to pure brane couplings.

On the string world--sheet this becomes a non--trivial monodromy problem,
which reduces the disk amplitude of $N_o$ open and $N_c$ closed strings to a sum of many color ordered partial subamplitudes of $N_o+2N_c$ open strings.
This sum can be further reduced to a sum over  $(N_o+2N_c-3)!$ subamplitudes 
of $N=N_o+2N_c$ open strings only. Hence, the computation of disk amplitudes involving open and closed strings is reduced to computing these subamplitudes in the open string sector.

In this sector we find a string theory generalization and proof of the Kleiss--Kuijf and Bern--Carrasco--Johansson relations: All order $\ap$ identities between open string subamplitudes are derived, which reproduce these field--theory relations in the  limit $\ap\ra0$. These identities  allow to reduce the number of independent subamplitudes of an 
open string $N$--point amplitude to $(N-3)!$. This number is identical to the dimension
of a minimal basis of generalized Gaussian 
hypergeometric functions describing the full $N$--point open string amplitude.

\Date{}
\noindent
\listtoc 
\writetoc
\break
\newsec{Introduction}

The famous Kawai, Lewellen, Tye (KLT) 
relation allows to express pure closed string tree--level 
(sphere) amplitudes as sum of squares of pure open string tree--level (disk)
amplitudes \Kawai. 
This way at tree--level \eg graviton scattering amplitudes are expressed by
gluon amplitudes. More  precisely, a graviton amplitude can be expressed as a sum of
squares of partial color ordered gluon amplitudes (multiplied by some $\sin$--factors).
In the low--energy effective action this duality leads to drastic simplifications 
of gravitational interactions.
In more technical terms the absence of interactions of left-- and
right--moving world--sheet fields of the closed string allows to factorize
any closed string tree--level amplitude into products of disk amplitudes
of left--moving fields and right--moving fields.

When considering tree--level amplitudes of both open and closed strings
the interacting string world--sheet is a disk. In that case the left-- and right--moving 
world--sheet fields of the closed string do interact and the full amplitude 
{\it cannot} be factorized into sums of squares of pure open string disk amplitudes.
This is the generic situation in string theories with D--branes with brane and 
bulk fields. In the presence of D--branes boundary conditions have to be imposed
resulting in an interaction of left-- and right--moving fields. 

On the other hand, in this work we show concretely how disk amplitudes of open and closed strings are computed and how these amplitudes can be expressed in terms of pure open string disk amplitudes. This way any amplitude involving open and closed strings is mapped to pure open string amplitudes.
By this \eg a disk amplitude involving both brane and bulk fields is related to an
amplitude of only brane fields.
Hence this map gives a dictionary between mixed couplings involving brane and bulk fields to pure brane couplings.
As a consequence disk amplitudes involving both members from gauge multiplets 
and members from the supergravity multiplet are related to pure amplitudes
involving only members from gauge multiplets.
By using these arguments the string expansion w.r.t. to flat background may be
reduced to fluctuations in the open string sector on the D--brane world--volume.
The corresponding fermionic couplings may simply be obtained by acting
with the space--time supersymmetry currents on the relevant correlators in the open string sector \August. At any rate, we believe that explicit map of (tree--level)
couplings of brane and bulk fields to pure brane couplings may have some
deeper insight into the dynamics of D--branes, \cf also \Sen. 

In practice the map between disk amplitudes of open and closed strings and pure open
string disk amplitudes is much more involved than in the KLT case due to the
additional mixing between left-- and right--moving fields.
On the string world--sheet this becomes a non--trivial monodromy problem,
which reduces the disk amplitude of $N_o$ open and $N_c$ closed strings to a sum of many color ordered partial subamplitudes of $N_o+2N_c$ open strings (supplemented
with phases and $\sin$--factors).
By explicitly deforming the underlying contours we considerably reduce the 
number of terms in the expression.
This step is equivalent to finding relations between partial subamplitudes of open string amplitudes.
Hence, the problem of computing disk amplitudes of open and closed strings
is reduced to finding relations between pure open string amplitudes such, that
the sum over color ordered partial subamplitudes of $N_o+2N_c$ open strings
can be reduced.
In this work we will find, that for open strings there a many more identities between color ordered subamplitudes beyond the usual cyclic, reflection and parity symmetries. These new string theory relations between open string subamplitudes, which are derived in this work, allow  to eventually reduce the full disk amplitude of $N_o$ open and $N_c$ closed strings to a minimal number of open string disk amplitudes, namely $(N_o+2N_c-3)!$.

In field--theory such relations are known as Kleiss--Kuijf, Bern--Carrasco--Johansson and dual Ward identities. However we find a string theory generalization and proof of these relations: All order $\ap$ identities between open string subamplitudes are derived, which reproduce these field--theory relations in the  limit $\ap\ra0$. These identities  allow to reduce the number of independent subamplitudes of an 
open string $N$--point amplitude to $(N-3)!$. This number is identical to the dimension of a minimal basis of generalized Gaussian 
hypergeometric functions describing the full $N$--point open string amplitude
\refs{\Dan,\STii,\August}

When relating disk amplitudes of open and closed strings to pure open string
disk amplitudes the open string subamplitudes generically do not yet appear
as world--sheet integrals in canonical form, \ie with integrations along
the segment $[0,1]$, which would give a direct translation to  (generalized)
Euler integrals. We explicitly map 
the corresponding world--sheet integrals to open string amplitudes given in
their canonical form  by (generalized) Euler integrals (along the segment $[0,1]$).

In the past the computation of disk amplitudes of open and closed strings
has been accomplished explicitly only for very simple and restricted cases, 
\ie low number of specific
external states, see {\it Refs.} \refs{\Gubser,\HK,\HKi,\GM}, 
\refs{\LMRS,\russo,\Bain}, and \refs{\foto,\fotoii,\garousi}.
In this work we generalize these results and present the formalism and tools to compute disk amplitudes for any number of open and closed strings.

The organization of this article is as follows.
In Section 2 after presenting the world--sheet and space--time tools to
compute disk amplitudes of $N_o$ open and $N_c$ closed strings we express the
latter as sum over $N_o+2N_c$ open string color ordered partial subamplitudes.
Furthermore, we explicitly present the generic complex 
world--sheet integrals for the cases $(N_o,N_c)=(2,1),\ (3,1),\ (2,2),\
(4,1),\ (0,3)$ and $(3,2)$. 
In Section 3 we demonstrate, how a disk amplitudes involving $N_o$ open and
$N_c$ closed strings is reduced to a sum over color ordered open string 
$N_o+2N_c$--point amplitudes. We present explicit results for the cases 
$(N_o,N_c)=(2,1),\ (3,1),\ (2,2),\ (4,1),\ (0,3)$ and $(3,2)$
by expressing the results in terms of generalized Euler integrals
describing $4,5,6$-- and $7$--point open string scattering, respectively.
In Section 4 we derive (string) relations between partial subamplitudes involving 
$N$ open strings. These relations allow to express all partial amplitudes
in terms of a minimal basis of $(N-3)!$ subamplitudes.
For $N=N_o+2N_c$ these relations are then used to simplify the sum over 
partial amplitudes as it appears for disk amplitudes involving $N_o$ 
open and $N_c$ closed strings.
In Section 5 we explicitly compute disk amplitudes of open and closed strings
and show in detail, how they are mapped to pure open string disk amplitudes
thus giving a dictionary of couplings of brane and bulk fields to pure brane
couplings.
In the appendix we compute complex world--sheet integrals by splitting them
into holomorphic and anti--holomorphic pieces.

\newsec{Disk scattering from D--branes}

In the following we shall discuss tree--level disk scattering from D--branes.
The world--sheet diagram of a string $S$--matrix describing the interaction of 
open and closed strings at (open string) tree--level can be conformally mapped
to a surface with one boundary. According to the Riemann mapping theorem 
the latter is equivalent to the unit disk $\Dc=\{z\in\IC\ |\ |z|\leq1\}$.
Eventually with the M\"obius transformation $z\ra i\fc{1+z}{1-z}$ the unit disk ${\bf D}$
can be conformally mapped to the upper (complex) half--plane 
$\Hc=\{z\in \IC\ |\ \im(z)\geq0\}$. 
The string states, which correspond to asymptotic states in the string
$S$--matrix formulation,
are created through vertex operators. 
In theories with D--branes massless (charged) fields 
as \eg gauge vectors or open string moduli originate from open
string excitations living on the D--brane world--volume.
Hence the boundary of the disk diagram is attached to the D--brane world--volume.
On the other hand, the massless closed string or bulk fields  
representing \eg the graviton, dilaton 
field  and closed (geometric) string moduli live in the bulk and are inserted 
in the bulk of the disk.

\subsec{Disk scattering of open and closed strings}

The generic expression for the superstring disk amplitude involving 
$N_o$ massless open and $N_c$ massless closed string states is:
\eqn\Startwith{
\Ac(N_o,N_c)=\sum_{\si\in S_{N_o-1}}V_{\rm CKG}^{-1}\ \lf(
\int_{\Ic_\si} \prod_{j=1}^{N_o}dx_j\ \prod_{i=1}^{N_c} \int_\Hc d^2z_i\ri)\ \vev{\ 
\prod_{j=1}^{N_o}:V_o(x_j):\ \ \prod_{i=1}^{N_c} :V_c(\ov z_i,z_i):\ }.}
The open string vertex operators $V_o(x_i)$ are inserted 
at the positions $x_i$ on the boundary of the disk. 
The latter are integrated along the boundary of $\Hc$, subject to the integration
region $\Ic_\si$.
On the other hand, the closed string vertex operators $V_c(\ov z_i,z_i)$ 
are inserted at points $z_i$ inside the disk, \cf the next Figure:
\ifig\redui{Open and closed string vertex positions on the disk $\Hc$.}
{\epsfxsize=0.35\hsize\epsfbox{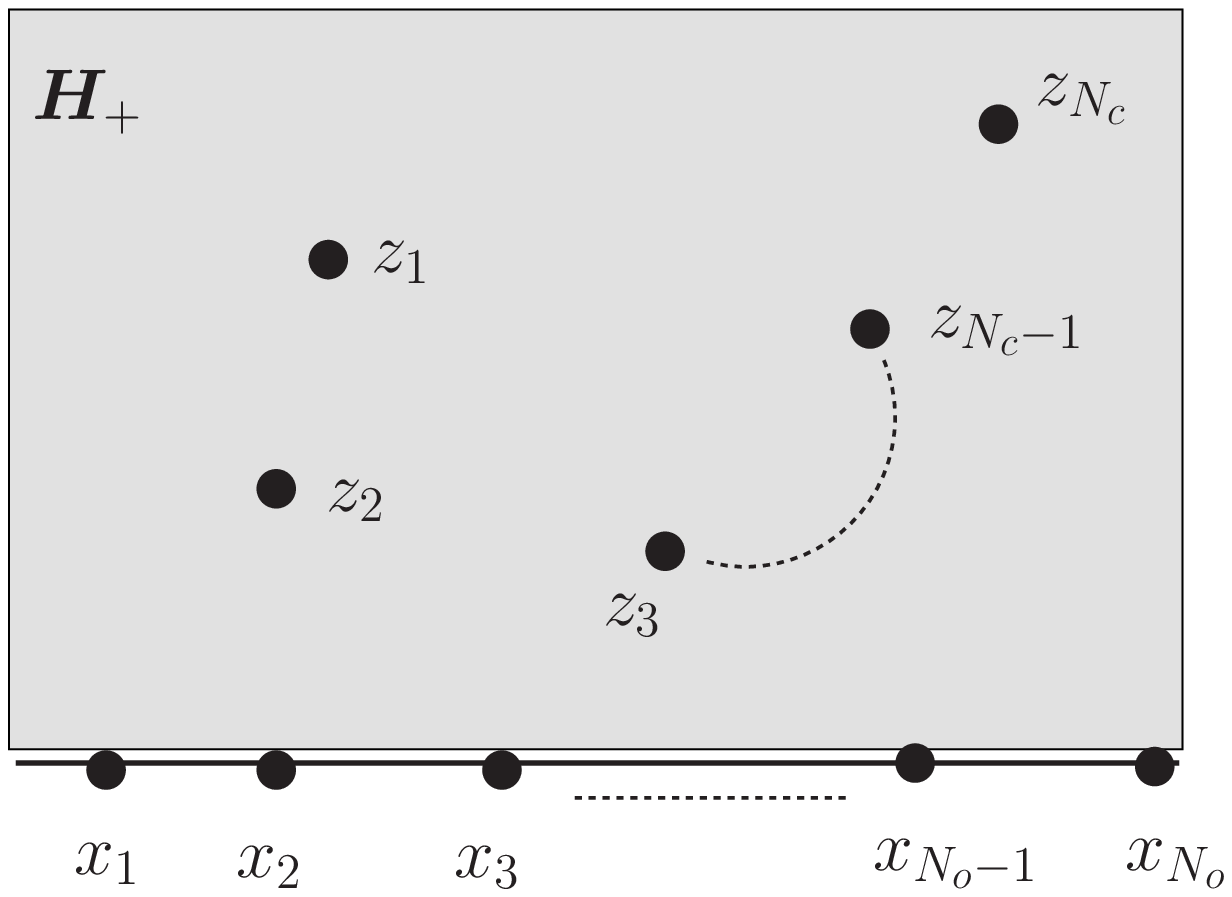}} 
\noindent
The sum runs over all $(N_o-1)!$ cyclic inequivalent orderings $\si\in S_{N_o-1}=S_{N_o}/\IZ_{N_o}$ of the $N_o$ 
open string vertex operators along the boundary of the disk. 
Each permutation $\pi$ gives rise to an integration region 
$\Ic_\si=\{x\in \IR\ |\ x_{1}<x_{\si(2)}< \ldots <x_{\si(N_o)}\}$.
The open string vertex operators $V(x_i)$ contain 
Chan--Paton factors $T^a$ carrying the gauge degrees of freedom of the open string
state. Depending on the ordering $\Ic_\si$ of the vertex
operator positions we obtain $(N_o-1)!$ partial amplitudes with the group factor 
$\Tr(T_{1}T_{\si(2)}\ldots T_{\si(N_o)})$.
Furthermore, in \eqq \Startwith\ the factor $V_{CKG}$ accounts for the volume of the
conformal Killing group of the disk after choosing the conformal gauge.
It will be canceled by fixing three vertex positions on the disk and introducing the
respective $c$--ghost correlator.

The upper half plane $\Hc$ may be obtained from the full complex plane $\IC$ 
representing the sphere through a $\IZ_2$ identification $z\simeq \ov z$. 
It is convenient to perform the computations in the double cover, \ie
in the  complex plane $\IC$, and extend the definition of these fields to the entire 
complex plane $\IC$ by taking into account the interaction
between left--moving and right--moving closed string fields \HK. 
Hence to evaluate the disk--integration in \Startwith\ over
the closed string vertex positions $z_i$ it is convenient to go to the 
double--cover and take into account the mixing of left-- and right--moving fields. 
This is described in more detail in the next Subsection.
An other way of arguing stems from the fact, that in \Startwith\ only vertex
operators, which are invariant under the world--sheet parity $\Om: z\leftrightarrow \ov
z$, enter. 
Hence, in the following we use vertex operators  invariant under the
world--sheet parity and integrate their positions $z_i$ over the 
whole complex plane $\IC$.

\subsec{Open and closed string vertex operators and their disk interactions}

The vertex operator for the gauge vector $A^a$, which comprises the massless
bosonic Neveu--Schwarz (NS) open string mode, is in the $(-1)$-ghost picture given by 
\eqn\fieldsiii{
V_{A^a}^{(-1)}(z,\xi,p) = g_A\ T^a\ e^{-\phi(z)}\ \xi^\mu\ \psi_\mu(z)\ e^{ip_\rho X^\rho(z)}\ ,}
while in the zero--ghost picture it takes the form:
\eqn\gaugevertexzero{
V_{A^a}^{(0)}(z,\xi,p)=\fc{g_A}{(2\ap)^{1/2}}\
T^a\xi_\mu\ [\ i\p X^\mu+2\ap\ (p\psi)\ \psi^\mu\ ]\ e^{ip_\rho X^\rho(z)}\ .}
On the other hand, the vertex operator for gaugino $\chi^a$, which gives rise to the massless Ramond (R) open string mode is given by
\eqn\fieldsii{
V_{\chi^{a}}^{(-1/2)}(z,u,p)=g_\chi\ T^a\ e^{-\h\phi(z)}\ u^\al\ S_\al(z)\ e^{ip_\rho X^\rho(z)}\ ,}
in the $(-1/2)$-ghost picture,
In the above definitions\foot{The open string vertex couplings are $g_\phi=(2\ap)^{1/2}\ g_{Y\! M},$\ 
$g_\chi=(2\ap)^{1/2}\ap^{1/4}\ g_{Y\! M}$ and $g_A=(2\ap)^{1/2}\ g_{Y\! M}$ for the scalar, gaugino and vector, respectively \JOE.
The gauge coupling $g_{Y\! M}$ can be expressed in terms of the ten--dimensional gauge coupling $g_{10}$
and the dilaton field $\phi_{10}$ through the relation $g_{Y\! M}=g_{10}e^{\phi_{10}/2}$. Finally, the closed string coupling $g_c$ is
given by $g_c=\fc{\kappa_{10}e^{\phi_{10}}}{2\pi}$ \JOE. In this work we consider open and closed strings scattering off one stack $a$ of
D--branes. Hence the overall normalization of all disk amplitudes has to
be supplemented by the factor $C_{D_2}=\fc{1}{2g_{Dp_a}^2\ap^2}$~\hunter.}
$T^a$ are the Chan--Paton factors accounting for the gauge degrees of freedom of the two open string ends.
Furthermore, the on--shell constraints $p^2=0,\ \slashchar{p}u=0$ are imposed.
The vertex operator for the massless  bosonic  NSNS closed string modes is given by
\eqn\dilaton{\eqalign{
&V_G^{(-1,-1)}(\ov z,z,\eps,q)=g_c\ \epsilon_{\mu\nu}\ e^{-\tilde\phi(\ov z)}\ 
e^{-\phi(z)}\ \tilde\psi^\mu(\ov z)\ \psi^\nu(z)\ e^{iq_\nu X^\nu(\ov z,z)}\ ,\cr
V_G^{(0,0)}(\ov z,z,\eps,q)&=-\fc{2g_c}{\ap}\ \epsilon_{\mu\nu}\ 
[i\ov\p X^\mu+\fc{\ap}{2}\ (q\tilde\psi)\ \tilde\psi^\mu(\ov z)]\ 
[i\p X^\nu+\fc{\ap}{2}\ (q\psi)\ \psi^\nu(z)]\ e^{iq_\nu X^\nu(\ov z,z)}}}
in the $(-1,-1)$ and $(0,0)$ ghost picture, respectively. 
The polarization tensor $\eps_{\mu\nu}$, which is subject to the on--shell conditions 
$\eps_{\mu\nu} q^\mu=0=\eps_{\mu\nu} q^\nu$ and $q^2=0$, is symmetric for the graviton and 
dilaton field and anti--symmetric for anti--symmetric tensor field. For further details, see also 
\MSD.
The vertex operator of the $n+1$--form RR field strengths $F_{n+1}$ is given by  \POLCH
\eqn\TEN{\eqalign{
V_{F_{n+1}}^{(-1/2,-1/2)}(\ov z,z,f,q)&=\fc{g_c}{(n+1)!}\ 
f_{\mu_0\ldots \mu_n}\ 
\lf(P^+\Gamma^{\mu_0}\ldots\Gamma^{\mu_n}\ri)^{\al\bet}\cr
&\times  e^{-\h\tilde\phi(\ov z)}\ e^{-\h\phi(z)}\ S_\al(z)\ 
\tilde S_\bet(\ov z)\  e^{iq_\rho X^\rho(\ov z,z)}\ ,}}
with the ten  $32\times32$ $\Gamma$--matrices $\Gamma^\mu$, the
spin fields $S_\al,\tilde S_\bet$ and the chiral projection operator 
$P^+=\h(I_{32}+\Gamma_{11})$ in $D=10$.
The $n+1$--tensor $f_{\mu_0\ldots \mu_n}$ is the Fourier transform of the Ramond $n$--form potential $c_n$, \ie $f_{\mu_0\ldots \mu_n}=i(n+1)\ q_{[\mu_0}\ c_{\mu_1\ldots \mu_n]}$, with
$c_{\mu\mu_2\ldots \mu_n}q^\mu=0$ and $q^2=0$.

The open string vertex operators $V_o(x_i)$ involve the holomorphic fields
$X^\mu,\psi^\nu,S_\al,\phi$ whose interactions are described by the usual correlators
\eqn\green{\eqalign{
&\vev{X^\mu(z_1)\ X^\nu(z_2)}=-2\ap\ g^{\mu\nu}\ \ln(z_1-z_2)\ \ \ ,\ \ \ 
\vev{\psi^\mu(z_1)\ \psi^\nu(z_2)}=\fc{g^{\mu\nu}}{z_1-z_2}\ ,\cr
&\vev{S_\al(z_1)S_\bet(z_2)}=\fc{C_{\al\bet}}{(z_1-z_2)^{5/4}}\ \ \ ,\
\ \ \vev{\phi(z_1)\ \phi(z_2)}=-\ln(z_1-z_2)}}
on the sphere $\IC$. Here $g^{\mu\nu}$ is the background metric
and $C_{\al\bet}$ the $D=10$ charge conjugation matrix with non--vanishing entries
only for spinor indices of opposite chirality.
The holomorphic fields $X^\mu,\psi^\nu,S_\al,\phi$ are 
defined on the upper half plane $\Hc$.
At the boundary of $\Hc$ boundary conditions are imposed for these fields:
\eqn\boundary{\eqalign{
(g_{\mu\nu}+2\pi\ap\ f_{\mu\nu})\ \p X^\nu(z)&=(g_{\mu\nu}-2\pi\ap\ f_{\mu\nu})\ 
\ov\p X^\nu(\ov z)\ ,\cr
(g_{\mu\nu}+2\pi\ap\ f_{\mu\nu})\ \psi^\nu(z)&=(g_{\mu\nu}-2\pi\ap\ f_{\mu\nu})\ 
\tilde\psi^\nu(\ov z)\ .}}
The matrix $f_{\mu\nu}$ depends on whether Dirichlet, Neumann or mixed
boundary conditions are imposed
at the open string end points attached to the  $Dp$--brane. Mixed
open string boundary conditions are specified by a non--trivial background
flux $f$ on the D--brane world--volume. 
If $N_c\neq 0$ in \Startwith\ also anti--holomorphic fields are
involved and the boundary 
conditions \boundary\ produce non--trivial interactions
between holomorphic and anti--holomorphic fields.
Then their interactions follow from the sphere correlators \green\ 
after taking  into account the matrix $D^{\mu\nu}$:
\eqn\greeni{\eqalign{
&\vev{X^\mu(z_1)\ \tilde X^\nu(\ov z_2)}=-2\ap\ D^{\mu\nu}\ \ln(z_1-\ov
z_2)\ \ \ ,\ \ \
\vev{\psi^\mu(z_1)\ \tilde \psi^\nu(\ov z_2)}=\fc{D^{\mu\nu}}{z_1-\ov z_2}\ ,\cr
&\vev{S_\al(z_1)\tilde S_\bet(\ov z_2)}=\fc{M_{\al\bet}}{(z_1-\ov z_2)^{5/4}}\ \ \ ,\
\ \ \vev{\phi(z_1)\ \tilde \phi(\ov z_2)}=-\ln(z_1-\ov z_2)\ .}}
The matrix $D$ is given by \threeref\ACNY\SW\LMRS:
\eqn\Dmatrix{
D=-g^{-1}+2\ (g+2\pi\ap\ f)^{-1}\ .}
The matrix $M$ may be obtained from the relations
$\Gamma^\mu_{\al\bet}=D^\mu_{\ \ \nu}\ (M^{-1}\Gamma^\nu M)_{\al\bet}$
and $M_{\al\gamma}\; M_{\beta\delta}\; C^{\gamma\delta}=C_{\al\bet}$,
which follow by considering the OPEs $\psi^\mu(z) S_\al(w)$ and $S_\al(z)S_\bet(w)$,
respectively, \cf {\it Ref.} \GM\ for more details.

In order to cancel the total background ghost charge of $-2$ on the disk, 
the vertices in the correlator \Startwith\ have to be chosen in the 
appropriate ghost picture.
The correlator of vertex operators in the integrand of \Startwith\ is evaluated 
by performing all possible Wick contractions and by using the Greens functions
as \green\ and \greeni.

The amplitude \Startwith\ involves an omnipresent correlator 
of a product of exponentials of space--time bosonic fields $X^\mu$
\eqn\Exp{\eqalign{
\vev{\prod_{j=1}^{N_o} :e^{ip_{j\mu} X^\mu(x_j)}:\ \prod_{i=1}^{N_c}& \ 
:e^{i q_{i\mu} X^\mu(\ov z_i, z_i)}:}=
(2\pi)^{10}\ \delta\lf(\sum\limits_{j=1}^{N_o}p_j+\sum\limits_{i=1}^{N_c}q_i\ri)\cr
&\times\prod_{j_1<j_2}^{N_o}    |x_{j_1}-x_{j_2}|^{2\ap
p_{j_1}p_{j_2}}\ \prod_{i=1}^{N_c}|z_i-\ov z_i|^{\ap q_{i\parallel}^2}\cr 
&\times\prod_{j=1}^{N_o}\prod_{i=1}^{N_c} |x_j-z_i|^{2\ap p_jq_i}
\ \prod_{i_1<i_2}^{N_c}   |z_{i_1}-z_{i_2}|^{\ap q_{i_1}q_{i_2}}\ 
|z_{i_1}-\ov z_{i_2}|^{\ap q_{i_1}Dq_{i_2}}\ ,}}
subject to momentum conservation 
\eqn\conservation{
\sum\limits_{j=1}^{N_o}p_j+\sum\limits_{i=1}^{N_c}q_{i\parallel }=0}
along the longitudinal brane directions, with $q_{i\parallel}=\h(q_i+Dq_i)$.

In the remainder of this article we shall work with the choice $\ap=\ap_{\rm closed}\equiv2$
for both type I and type II strings. In order to correctly accommodate this choice in the open sector with $\ap_{\rm open}\equiv\h$ 
the momentum $p_i$ in the open string vertex operator 
has to be doubled, \ie the operator $V_o(z,2p)$ is used in the amplitudes. 
This has the consequence, that in open string amplitudes
the momenta $p_i$ of the open strings appear with an additional factor of~$2$.

To summarize, after working out all Wick contractions in the double cover $\IC$
the amplitude \Startwith\ assumes the generic form
\eqn\STartwith{\eqalign{
\Ac(N_o,N_c)&=\sum_{\si\in S_{N_o}/\IZ_{N_o}} V_{\rm CKG}^{-1}\ \lf(
\int_{\Ic_\si} \prod_{j=1}^{N_o}dx_j\ \prod_{i=1}^{N_c} \int_\IC d^2z_i\ri)\ \sum_{I}\Kc_I\cr 
&\times\prod_{j_1<j_2}^{N_o}    |x_{j_1}-x_{j_2}|^{4p_{j_1}p_{j_2}}\ 
(x_{j_1}-x_{j_2})^{n^I_{j_1j_2}}\ \prod_{i=1}^{N_c}|z_i-\ov
z_i|^{2q_{i\parallel}^2}\ 
(z_i-\ov z_i)^{r^I_{ii}}\cr 
&\times\prod_{j=1}^{N_o}\prod_{i=1}^{N_c} |x_j-z_i|^{4 p_jq_i}(x_j-z_i)^{m^I_{ij}}
(x_j-\ov z_i)^{\ov m^I_{ij}}\ 
\prod_{i_1<i_2}^{N_c}   |z_{i_1}-z_{i_2}|^{2 q_{i_1}q_{i_2}}\ \cr
&\times |z_{i_1}-\ov z_{i_2}|^{2 q_{i_1}Dq_{i_2}}\ 
(z_{i_1}-z_{i_2})^{r^I_{i_1i_2}}\ 
\ (\ov z_{i_1}-\ov z_{i_2})^{\ov r^I_{i_1i_2}}\ (z_{i_1}-\ov z_{i_2})^{\tilde
r^I_{i_1i_2}}\ (\ov z_{i_1}-z_{i_2})^{\ov{\tilde r}^I_{i_1i_2}}\cr
&:=\sum_{\si\in S_{N_o}/\IZ_{N_o}}\sum_I \ \Kc_I\ 
\Ac^I(1,\si(2),\ldots,\si(N_o);N_o+1,\ldots,N_o+N_c)\ ,}}
with some integers $n_{ij}^I,m_{ij}^I,\ov m^I_{ij},r_{ij}^I,\tilde r_{ij}^I,\ov
r_{ij}^I,\ov{\tilde r}_{ij}^I \in\IZ$ referring to the kinematical factor
$\Kc_I$. For massless external states these numbers must obey:
\eqn\Kapstadt{\eqalign{
&\sum_{k<j}^{N_o}n_{kj}^I+\sum_{j<k}^{N_o}n_{jk}^I+
\sum_{i=1}^{N_c}(m^I_{ij}+\ov m^I_{ij})+2=0\ \ \ ,\ \ \ j=1,\ldots,N_o\ ,\cr
&r_{ii}^I+\sum_{j=1}^{N_o}m_{ij}^I+
\sum_{k<i}^{N_o}(r_{ki}^I+\ov{\tilde r}_{ki}^I)+
\sum_{i<k}^{N_o}(r_{ik}^I+\tilde r_{ik}^I)+2=0
\ \ \ ,\ \ \ i=1,\ldots,N_c\ ,\cr
&r_{ii}^I+\sum_{j=1}^{N_o}\ov m_{ij}^I+
\sum_{k<i}^{N_o}(\ov r_{ki}^I+\tilde r_{ki}^I)+
\sum_{i<k}^{N_o}(\ov r_{ik}^I+\ov{\tilde r}_{ik}^I)+2=0
\ \ \ ,\ \ \ i=1,\ldots,N_c\ .}}
Note, that these integers and the kinematical factor $\Kc_I$ do
not depend on the ordering~$\si$.

\subsec{Splitting the complex world--sheet integrals}

After performing all Wick contractions the amplitude \Startwith\ boils down to
a product of various polynomials in differences of 
the open and closed string positions $x_k$ and $z_j,\ov z_j$, \cf \STartwith.
To compute the integral over these positions  we write it as an integral 
over holomorphic and
anti--holomorphic coordinates following the method proposed in \Kawai.
After introducing the parameterization
$z_j=z_{1j}+i z_{2j}\ ,\ j=1,\ldots,N_c$ the integrand becomes an analytic
function in $z_{2j}$.
We then deform the $z_{2j}$--integral along the real axis $\im(z_{2j})=0$ to the
pure imaginary axis $\re(z_{2j})=0$, \ie $iz_{2j}\in\IR$.
This way, the variables 
\eqn\real{
\xi_j=z_{1j}+i\ z_{2j}\equiv z_j\ \ \ ,\ \ \ \eta_j=z_{1j}-i\ z_{2j}\equiv \ov
z_j\ \ \ ,\ \ \ j=1,\ldots,N_c}
become real quantities, \ie $\xi_i,\eta_j\in\IR$. 
We may concentrate on one term of the sum \STartwith, \ie one particular 
subamplitude, in the following denoted by 
$\Ac^I_\si(N_o,N_c):=\Ac^I(1,\si(2),\ldots,\si(N_o);N_o+1,\ldots,N_o+N_c)$ 
referring to one specific kinematics $\Kc_I$.
With the Jacobian  $\det\fc{\p(z_{1i},z_{2j})}{\p(\xi_i,\eta_j)}=\lf(\fc{i}{2}\ri)^{N_c}$, 
after fixing the position of the first open string vertex at $x_1=-\infty$ and fixing two other open string positions subject to the ordering $\Ic_\si$ 
for a given kinematics $\Kc_I$ one subamplitude of \Startwith\ or \STartwith\ 
may then be written
\eqn\AMPLITUDE{\eqalign{
\Ac^I_\si(N_o,N_c)&=\lf(\fc{i}{2}\ri)^{N_c}\ \lf(\ \int_{\Ic_\si} 
\prod_{l=2}^{N_o-2}dx_l\ \prod_{i,j=1}^{N_c} \int_{-\infty}^\infty d\xi_i\ 
\int_{-\infty}^\infty d\eta_j\ \ri)\ \Pi(x_l,\xi_i,\eta_j)\cr
&\times\prod^{N_o}_{2\leq l_1<l_2} |x_{l_1}-x_{l_2}|^{4p_{l_1}p_{l_2}}\
(x_{l_1}-x_{l_2})^{n^I_{l_1l_2}}\ 
\prod_{i=1}^{N_c} |\xi_i-\eta_i|^{2q_{i\parallel}^2} \ (\xi_i-\eta_i)^{r^I_{ii}}\cr 
&\times\prod^{N_o}_{l=2}\ \prod_{i=1}^{N_c}|x_l-\xi_i|^{2p_lq_i}\
|x_l-\eta_i|^{2p_lq_i}\ (x_j-\xi_i)^{m^I_{ij}}\ 
(x_j-\eta_i)^{\ov m^I_{ij}}\cr  
&\times \prod_{i,j=1}^{N_c}|\xi_i-\xi_j|^{q_iq_j}\ |\eta_i-\eta_j|^{q_iq_j}\ 
|\xi_i-\eta_j|^{q_iDq_j}\ |\eta_i-\xi_j|^{q_iDq_j}\cr
&\times (\xi_i-\xi_j)^{r_{ij}^I}\ 
\ (\eta_i-\eta_j)^{\ov r_{ij}^I}\ (\xi_i-\eta_j)^{\tilde
r_{ij}^I}\ \ (\eta_i-\xi_j)^{\ov{\tilde r}_{ij}^I}}}
as an integral over $N_o+2N_c-3$ real positions $x_l,\xi_i,\eta_j$ with the phase factor:
\eqn\PHASE{\eqalign{
\Pi(x_l,\xi_i,\eta_j)&=e^{2\pi i p_lq_i \th[-(x_l-\xi_i)(x_l-\eta_i)]}\ 
e^{i\pi q_iq_j \th[-(\xi_i-\xi_j)(\eta_i-\eta_j)]}\cr 
&\times e^{i\pi q_iDq_j \th[-(\xi_i-\eta_j)(\eta_i-\xi_j)]}\ 
e^{2\pi i q_{i\parallel}^2 \th(\eta_i-\xi_i)}\ .}}
In \AMPLITUDE\ the phase factor $\Pi(x_l,\xi_i,\eta_j)$ accounts for the
correct branch of the integrand. Note, that this phase is independent on the 
kinematical structure $\Kc_I$ and the integers as the branching is caused by
the kinematic invariants only.

The amplitude \AMPLITUDE\ may be interpreted as a disk amplitude involving $N_o+2N_c$
open strings with the following $N_o+2N_c$  open string vertex operator positions $z_r$:
\eqn\DISENT{\eqalign{
&z_1=-\infty\ ,\ \ \ \ \ \ \ \ \ z_l=x_l\ ,\ \ \ l=2,\ldots,N_o\ ,\cr
&z_{N_o+2i-1}=\xi_i\ ,\ \ \ z_{N_o+2i}=\eta_i\ ,\ \ \ i=1,\ldots,N_c\ .}}
For the sequel we introduce the $N_o+2N_c$ open string momenta $k_r$
\eqn\momenta{\eqalign{
k_i&=p_i\ \ \ ,\ \ \ i=1,\ldots,N_o\cr
k_{N_o+2j-1}&=\h\ Dq_j\ \ \ ,\ \ \ k_{N_o+2j}=\h\ q_j\ \ \ ,\ \ \ j=1,\ldots,N_c\ ,}}
which fulfill the massless condition $k_r^2=0$ for $D=1$.
In terms of these momenta, the momentum conservation \conservation\ reads:
\eqn\Conservation{
\sum\limits_{r=1}^{N_o+2N_c}k_r=0\ .}
The (open string) kinematic invariants $\hatt s_{ij}=\ap(k_i+k_j)^2$ become ($\ap=2$):
\eqn\INV{
\hatt s_{ij}=4\ k_ik_j\ .}

The expression \AMPLITUDE\ is by far not the final expression\foot{
A variant of \AMPLITUDE\ (for the simpler case $D=1, q_{i\parallel}^2=0$ 
and $n,m,\ov m,r,\ov r,\tilde r,\ov{\tilde r}=0$) 
has been recently  given in~\China. 
}. 
The position integrals
in $\xi_i$ and $\eta_j$ take into account all possible open string orderings along the real axis.
The set of these positions $\xi_i,\eta_j$ has also a relative ordering w.r.t. to the open
string positions $x_l$.  
In other words, the amplitude \AMPLITUDE\ decomposes into a sum of 
ordered partial amplitudes $A(1,\Si(2),\ldots\Si(N_o+2N_c))$ involving $N_o+2N_c$ open strings,
with $\Si\in S_{N_o+2N_c}/\IZ_{N_o+2N_c}\Big/S_{N_o}/\IZ_{N_o}$ subject to the given ordering $\si$ of the $N_o$ open strings. As we shall see, there exist many non--trivial relations between those partial amplitudes and the expression \AMPLITUDE\ may be written in terms of a basis of a minimal set of partial amplitudes. Indeed, the number of terms in
the expression \AMPLITUDE\ may be considerably reduced by deforming the
contours in the complex $\eta_j$--planes. This procedure is mathematically equivalent to
finding relations between various partial amplitudes and expressing \AMPLITUDE\ in terms of a minimal set. The method of deforming the contours in the complex $\eta_j$--planes
and reducing \AMPLITUDE\ to a minimal set is somewhat similar
as the method in \Kawai, however much more involved due to the mixing of the holomorphic 
and anti--holomorphic coordinates  $\xi_i$ and $\eta_j$   showing up in the phase \PHASE.
Eventually, the open string world--sheet integrals, which appear in \AMPLITUDE\ as partial amplitudes, 
should be mapped to 
canonical form given by (generalized) Euler integrals along the segment $[0,1]$.
This program will be pursued in Section 3.
Apart from \Kawai\ some earlier work on handling complex (sphere) integrals may be found in {\it Refs.} \DF. However, these integrals are much simpler than \STartwith\
due to the absence of the additional mixing of holomorphic and anti--holomorphic coordinates, which
causes additional branchings in the integrand. Hence these reference are not of any help.

\subsec{Higher--point disk amplitudes with open and closed strings}

After having presented the vertex operators, their interactions and the general form of a disk amplitude of $N_o$ open and $N_c$ closed strings, given in \STartwith\ and  \AMPLITUDE\
we are now prepared to compute  general disk amplitudes \Startwith.
In the remainder of this Subsection we shall discuss the six cases
$(N_o,N_c)=(2,1),\ (3,1),\ (2,2),\ (4,1),\ (0,3)$ and $(3,2)$.
We shall be concerned with the general case in Section 3.

\subsubsec{Three--point disk amplitudes with two open and one closed string}

For an amplitude of two open and one closed string in \Startwith\ we have 
one partial amplitude with group ordering $\Tr(T_1T_2)$.
Due to $PSL(2,\IR)$ invariance on the disk we may fix three vertex positions.
A convenient choice\foot{An alternative choice is:
$x_1=-x,\ x_2=x,\ \ov z=-i,\ z=i,$ with $x\in\IR_+$  \HKi.} is
\eqn\Fixi{
x_1=-\infty,\ \ \ x_2=1\ \ \ ,\ \ \ \ov z=-ix\ \ \ ,\ \ \ z=ix\ ,}
with $x\in\IR^+$, \cf also Subsection 2.4.3.
This choice implies the $c$--ghost contribution:
\eqn\cghost{
\vev{c(x_1)c(x_2)\tilde c(\ov z)}=(x_1-x_2)(x_1-\ov z)(x_2-\ov z)=x_\infty^2\ (1+i x)\ .}
Therefore, the partial amplitude of \Startwith\ becomes:
\eqn\startwithiv{
\Ac(1,2;3)=x_\infty^2\ \int_{0}^\infty dx\ (1+ix)\ \ \vev{:V_o(x_\infty):\ :V_o(1):\ :V_c(-ix,ix):}\ .}
For this three--point process to give a result, which does not vanish on--shell, the closed
string momentum $q$ has to have also a non--vanishing direction $q_\perp\neq 0$ 
transverse to the D--brane world--volume, with $q=q_\parallel+q_\perp$.
For the choice \Fixi\ the correlator \Exp\ assumes the form
\eqn\expi{
\vev{e^{2ip_{1\mu}X^\mu(-\infty)}e^{2ip_{2\mu}X^\mu(1)}e^{iq_{\mu}X^\mu(-ix,ix)}}=
(2x)^s\ (1+ix)^t\ (1-ix)^t\ ,}
with the kinematic invariants
\eqn\inviv{
s=4\ p_1p_2\ \ \ ,\ \ \ t=u=2\ p_1q=2\ p_2q=-2\ p_1p_2=-q_\parallel^2\ ,}
\ie $s=-2t$ \doubref\HK\LMRS.
After performing all Wick contractions, for each kinematics $\Kc_I$
the expression \startwithiv\ generically reduces to the form 
\eqn\firenze{\eqalign{
\Ac^I(1,2;3)&=2\ \int_{0}^\infty dx\  (2x)^{s-2-2n^I_1}\ (1+ix)^{t+n^I_1}\ (1-ix)^{t+n^I_1}\cr
&=2^{s-2-2n^I_1}\ B\lf(-\fc{s}{2}-t+\h\ ,\ \fc{s}{2}-\h-n^I_1\ri)\ ,}}
with some integer $n_1^I$.

The result \firenze\ originates from the more general class\foot{An often encountered case is 
$\al_0=\delta-1,\ \al_1=\al-\delta,\ \al_2=-\al-\delta$, for which 
we have \doubref\Bain\LMRS:
\eqn\bain{
I(\delta-1,\al-\delta,-\al-\delta)=\sqrt\pi\ 2^{-\delta}\ e^{-i\pi \fc{\alpha}{2}}\ 
\fc{\Gamma\lf(\fc{\delta}{2}\ri)\ 
\Gamma\lf(\h+\fc{\delta}{2}\ri)}{\Gamma\lf(\h+\fc{\delta}{2}-\fc{\al}{2}\ri) 
\Gamma\lf(\h+\fc{\delta}{2}+\fc{\al}{2}\ri)}\ .}} of complex integrals
\eqn\baini{\eqalign{
I(\al_0,\al_1,\al_2)&=\int_0^\infty dx\ x^{\alpha_0}\ (x-i)^{\alpha_1}\ 
(x+i)^{\alpha_2}=
i\ e^{\h i \pi (\al_0+\al_1+\al_2)}\cr 
&\times\lf\{\ B(1+\al_0+\al_1,-1-\al_0-\al_1-\al_2)\ 
\FF{2}{1}\lf[{-\al_1,-1-\al_0-\al_1-\al_2\atop-\al_0-\al_1};-1\ri]\ri.\cr
&\hskip1.5cm-\lf.e^{i\pi (\al_0+\al_1)}\ B(1+\al_0,-1-\al_0-\al_1)\ 
\FF{2}{1}\lf[{-\al_2,1+\al_0\atop 2+\al_0+\al_1};-1\ri]\ \ri\}\ ,}}
with  $\al_0,\al_1,\al_2\in\IR$, constrained by analyticity to $\re(\al_0)>-1,\ 
\re(\al_0+\al_1+\al_2)<-1$.
The integral \baini\ along the positive real axis may be derived from 
a closed contour $C_0+C_\infty$ in the complex $x$--plane. This contour, which in Fig. 2 is drawn in blue, consists of one large $C_\infty$ 
and one small circle $C_0$. The latter encircles the point $x=0$ clockwise. Except the point $x=0$ all 
singularities of the integrand are inside of the large circle $C_\infty$.
\ifig\erdel{Complex $x$--plane and contour integrations.}
{\epsfxsize=0.45\hsize\epsfbox{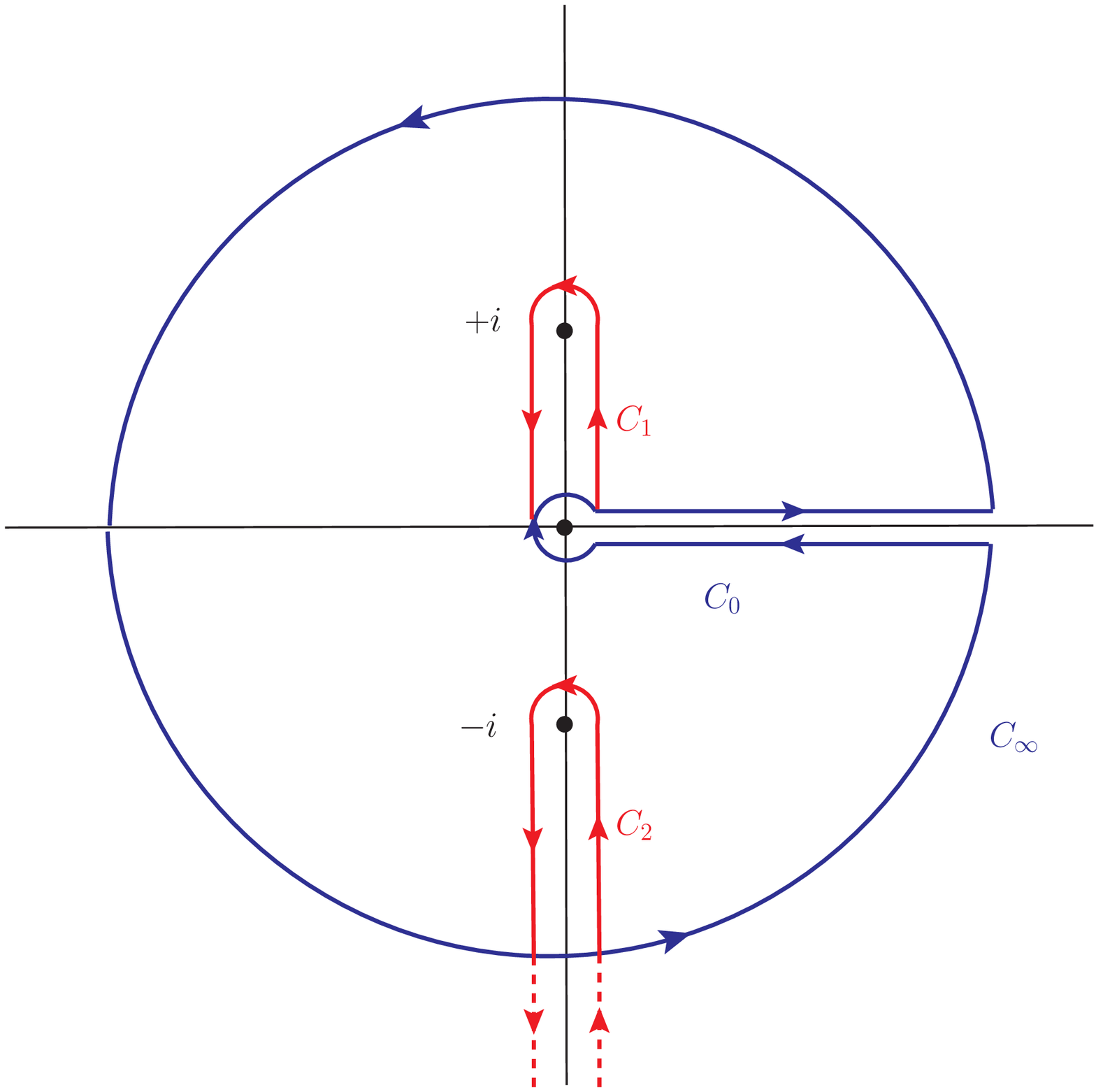}} 
\noindent
The circle $C_\infty$ may be deformed to infinity, where the integrand approaches zero, and hence gives a zero
contribution. On the other hand, the full contour $C_0+C_\infty$ may be deformed to the two loops $C_1$ and $C_2$ in Fig.~2 drawn in red: The loop\foot{The integral $\int_z^{(w+)} dx\ f(x)$ defines an integral taken along a contour $\Cc$, which starts at a point $z$ in the complex $x$--plane, encircles the point $w$ once counter--clockwise and returns to its starting point. All singularities of the integrand except $x=w$ are outside of $\Cc$.} $C_1$ $\int_0^{(i+)}$
starts at $x=0$, encircles once the point $x=i$ and returns to its starting point. The second loop $C_2$ $\int_{-i\infty}^{(-i+)}$ starts at $x=-i\infty$ encircles once the point $x=-i$ and returns to its starting point. After inspecting the other contours in the complex $x$--plane
we find\foot{Here $(-z)^{\alpha_0}:=e^{\alpha_0\ln(-z)}$ is
defined through the principal value of the logarithm.}
$$\eqalign{
\int_{+\infty}^{(0^-)} dx\ x^{\alpha_0}\ (x-i)^{\alpha_1} &  (x+i)^{\alpha_2}=2i\ \sin[\pi(\al_0+\al_1)]\ (-1)^{\al_0}\ I(\al_0,\al_1,\al_2)\cr
&=-2\ e^{i \pi (\al_0+\al_1)}\ \sin(\pi\al_1)\ \int_0^1 dx\ x^{\alpha_0}\ (1-x)^{\alpha_1}\ (1+x)^{\alpha_2}\cr
&+2\ \sin(\pi\al_2)\ \int_{-\infty}^{-1} dx\ (-x)^{\alpha_0}\ (1-x)^{\alpha_1}\ (-1-x)^{\alpha_2}\ ,}$$
which proves \baini. 
Alternatively, the result \baini\ may be written in the following form
\eqn\bainii{\eqalign{
&I(\al_0,\al_1,\al_2)=-i\ e^{\h i\pi (\al_0+\al_1+\al_2)} \ e^{i\pi (\al_0+\al_1+\al_2)}\cr
&\hskip1cm\times
\lf\{\ B(1+\al_0+\al_2,-1-\al_0-\al_1-\al_2)\ \FF{2}{1}\lf[{-\al_2,-1-\al_0
-\al_1-\al_2\atop-\al_0-\al_2};-1\ri]\ri.\cr
&\hskip3cm-\lf.e^{-i\pi (\al_0+\al_2)}\ B(1+\al_0,-1-\al_0-\al_2)\ 
\FF{2}{1}\lf[{-\al_1,1+\al_0\atop 2+\al_0+\al_2};-1\ri]\ \ri\}\ ,}}
which may be derived by deforming the full contour $C_0+C_\infty$ differently. In this case the contour $C_1$ starts at $x=i\infty$, encircles once the point $x=i$ and returns to its starting point, while the contour $C_2$ starts at $x=0$, encircles once the point $x=-i$ and returns to the origin. Hence we have:
$$\eqalign{
\int_{+\infty}^{(0^-)} dx\ x^{\alpha_0}\ (x-i)^{\alpha_1} &  (x+i)^{\alpha_2}=2i\sin[\pi(\al_0+\al_2)]\ (-1)^{\al_0}\ I(\al_0,\al_1,\al_2)\cr
&=-2\ e^{-i\pi (\al_0+\al_2)}\ \sin(\pi\al_2)\ \int_{-1}^0 dx\ (-x)^{\alpha_0}\ (1-x)^{\alpha_1}\
(1+x)^{\alpha_2}\cr
&+2\ \sin(\pi\al_1)\ \int_1^\infty dx\ x^{\alpha_0}\ (x-1)^{\alpha_1}\ (1+x)^{\alpha_2}\ .}$$
Furthermore, the result \baini\ may also be derived by using (3.197.3) of \Gradst:
\eqn\bainiii{\eqalign{
I(\al_0,\al_1,\al_2)&=-i\ e^{\h i\pi (\al_0+\al_1+\al_2)} \ e^{i\pi (\al_0+\al_1)}\cr
&\times B(1+\al_0,-1-\al_0-\al_1-\al_2)\ \FF{2}{1}\lf[{1+\al_0,-\al_2\atop-\al_1-\al_2};2\ri]\ .}}
With \eqq 2.10(4) of \Erdel\ this becomes \eqq \baini.

After having discussed the results \baini\ and \bainii\ for the 
general complex integral $I(\al_0,\al_1,\al_2)$, we are now prepared to
write the result \firenze\ in a different form.
For $\al_0=-2\al_2-2=-2t -2n_1-2$ and $\al_1=\al_2=t+n_1=u+n_1$
the two expressions \baini\ and \bainii\ may be combined to give \firenze:
\eqn\Firenze{\eqalign{
\Ac(1,2;3)=\fc{i}{4}\ &\lf\{e^{i\pi(t+n_1)}\ B(s-2n_1-1,u+n_1+1)+B(u+n_1+1,t+n_1+1)\ri.\cr
&\lf.+e^{i\pi(t+n_1)}\ B(t+n_1+1,s-2n_1-1)\ri\}\ .}}
Indeed, with the identity
\eqn\KITP{\eqalign{
e^{i\pi(s+n_1)}\ &  B(s+n_1+1,u-n_1-n_2-1)+B(t+n_2+1,s+n_1+1)\cr
&+e^{-i\pi(t+n_2)}\ B(t+n_2+1,u-n_1-n_2-1)=0\ \ \ ,\ \ \ s+t+u=0\ ,\ n_i\in\IZ}}
we may prove the equivalence of \firenze\ and \Firenze.
Above we have dropped the kinematical index $I$.

\subsubsec{Three open strings and one closed string}

For an amplitude of three open and one closed string in \Startwith\ we have two 
different partial amplitudes.
Due to $PSL(2,\IR)$ invariance on the disk we may fix three vertex positions.
For the partial amplitude with group ordering\foot{The partial amplitude with the group ordering $\Tr(T_1T_3T_2)$ may be simply obtained by interchanging of the second and third open string.}
 $\Tr(T_1T_2T_3)$ a convenient
choice is
\eqn\fixi{
x_1=x_\infty:=-\infty\ \ \ ,\ \ \ x_2=0\ \ \ ,\ \ \ x_3=1\ .}
This choice implies the $c$--ghost contribution:
\eqn\cghost{
\vev{c(x_1)c(x_2)c(x_3)}=(x_1-x_2)\ (x_1-x_3)\ (x_2-x_3)=x_\infty^2\ .}
For this choice in the amplitude \Startwith\ we are left over with one
integration of the closed string position $z$ over the complex plane
$\IC$. Therefore, the partial amplitude \Startwith\ becomes:
\eqn\startwithv{
\Ac(1,2,3;4)=\vev{c(-\infty)c(0)c(1)}\int_\IC d^2z\ 
\vev{\ :V_o(-\infty):\ :V_o(0):\ :V_o(1):\ \ :V_c(\ov z,z):\ }\ .}
For the choice \fixi\ the correlator \Exp\ assumes the form
\eqn\expi{
\vev{e^{2ip_{1\mu}X^\mu(-\infty)}e^{2ip_{2\mu}X^\mu(0)}e^{2ip_{3\mu}X^\mu(1)}
e^{iq_{\mu}X^\mu(\ov z,z)}}=z^t\ \ov z^t\ (1-z)^{s}\ (1-\ov z)^{s}\ |z-\ov z|^{2q_\parallel^2}\ ,}
with the kinematic invariants:
\eqn\invv{
s=2\ p_3q\ \ \ ,\ \ \ t=2\ p_2q\ \ \ ,\ \ \ u=2\ p_1q\ .}
From \conservation\ it follows: $s+t+u=-2\ q_\parallel^2$.
Furthermore, we have: $p_1p_2=\fc{s}{2}+\fc{q_\parallel^2}{2},\ p_2p_3=\fc{u}{2}+\fc{q_\parallel^2}{2}$ and $p_1p_3=\fc{t}{2}+\fc{q_\parallel^2}{2}$. 
After performing all Wick contractions the amplitude \startwithv\ boils down to
\eqn\boilv{
\Ac^I(1,2,3;4)=G^{(\al^I)}\lf[{t+n^I_1\ ,\ s+m^I_1\atop t+n^I_2\ ,\ s+m^I_2}\ri]\ ,}
for any kinematics $\Kc^I$ with four integers $n^I_i,m^I_i$ and
$\al^I\in\IR$  and the integral
\eqn\Provee{\eqalign{
G^{(\al)}\lf[{\lambda_1,\gamma_1\atop\lambda_2,\gamma_2}\ri]&:=
\int_\IC d^2z \ z^{\lambda_1}\ \ov z^{\lambda_2}\ 
(1-z)^{\gamma_1}\ (1-\ov z)^{\gamma_2}\ |z-\ov z|^{\hatt\al}\ (z-\ov z)^{\tilde\al}\cr
&=\pi\ \lambda_2\ \fc{\Gamma(1+\gamma_1)\ 
\Gamma(-1-\al-\lambda_2-\gamma_2)\
\Gamma(2+\al+\lambda_1+\lambda_2)}{\Gamma(1-\lambda_2)\ \Gamma(-\al-\gamma_2)\
\Gamma(3+\al+\lambda_1+\lambda_2+\gamma_1)}\cr
&\times\FF{3}{2}\lf[{-\gamma_2,1+\lambda_2,1+\gamma_1\atop 
-\al-\gamma_2,3+\al+\lambda_1+\lambda_2+\gamma_1};1\ri]+\pi\ e^{i\pi(m_1+m_2)}\cr 
&\times\al\ \fc{\Gamma(1+\gamma_1)\ 
\Gamma(-1-\al-\lambda_2-\gamma_2)\ \Gamma(-2-\al-\lambda_1-\lambda_2-\gamma_1-\gamma_2)}
{\Gamma(1-\al)\ 
\Gamma(-1-\al-\lambda_1-\lambda_2-\gamma_2)\ \Gamma(-\lambda_2-\gamma_2)}\cr
&\times \FF{3}{2}\lf[{-\gamma_2,-1-\al-\lambda_2-\gamma_2,-\al-2-\lambda_1-
\lambda_2-\gamma_1-\gamma_2\atop
-\lambda_2-\gamma_2,-1-\al-\lambda_1-\lambda_2-\gamma_2};1\ri]\ ,}}
with  
\eqn\withv{\lambda_i=t+n_i\ \ \ ,\ \ \ \gamma_i=s+m_i\ \ \
,\ \ \ \al=\hatt\al+\tilde\al\ ,}
and $n_i,m_i,\tilde\al\in\IZ$. Here $\hatt\al$ denotes the non--integer part 
of $\al$, \ie $\hatt\al=2q_\parallel^2$.
The integral \Provee\ is evaluated in Appendix \appA.
A special case arises for $\hatt\al=\tilde\al=0$, 
$\ss{G^{(0)}\lf[{\lambda_1,\gamma_1\atop\lambda_2,\gamma_2}\ri]:=
V\lf[{\lambda_1,\gamma_1\atop\lambda_2,\gamma_2}\ri]}$, with \refs{\GHMR,\Kawai}:
\eqn\Special{
V\lf[{\lambda_1,\gamma_1\atop\lambda_2,\gamma_2}\ri]:=
\int_\IC d^2z \ z^{\lambda_1}\ \ov z^{\lambda_2}\ 
(1-z)^{\gamma_1}\ (1-\ov z)^{\gamma_2}=\pi\ \fc{\Gamma(1+\lambda_1)\
\Gamma(1+\gamma_1)}{\Gamma(2+\lambda_1+\gamma_1)}\ 
\fc{\Gamma(-1-\lambda_2-\gamma_2)}{\Gamma(-\lambda_2)\ \Gamma(-\gamma_2)}\ .}

\subsubsec{Two open strings and two closed strings}

For an amplitude of two open and two closed strings in \Startwith\ we have 
one partial amplitude with group ordering $\Tr(T_1T_2)$.
The conformal killing group $PSL(2,\IR)$ of the disk 
allows to fix three vertex positions.
On the double cover $\IC$, with an appropriate $PSL(2,\IC)$
transformation we
may fix two positions on the boundary and the real part of a closed string modulus.
A convenient choice is ($z_1:=x_1,\ z_2:=x_2$)
\eqn\fix{
z_1=z_\infty:=
-\infty\ \ \ ,\ \ \ z_2=1\ \ \ ,\ \ \ \ov z_3=-ix\ \ \ ,\ \ \ z_3=ix\ \ \ ,
\ \ \ \ov z_4=\ov z\ \ \ ,\ \ \ z_4=z\ ,}
with $z\in \Hc$ and $x\in\IR^+$.
Three arbitrary points $w_1,w_2\in\IR$ and $w_3\in \IC$ may be mapped to the
points $z_1,z_2,z_3$ of \fix\ by the following $PSL(2,\IC)$ transformation
\eqn\psl{
P=\lf[\ w_{12}\ w_{13}\ w_{23}\ (1-ix)\ \ri]^{-1/2}\ 
\pmatrix{-w_{12}+ix\ w_{13}&w_3\ w_{12}-ix\ w_2\ w_{13}\cr
w_{23}&-w_1\ w_{23}}\in PSL(2,\IC)\ ,}
with $w_i\ra\fc{p_{11}w_i+p_{12}}{p_{21}w_i+p_{22}}$ and $w_{ij}=w_i-w_j$.
The transformation \psl\ must map boundary points $w\in\IR$ onto boundary points.
This requirement\foot{We wish to stress, that 
this condition rules out the more symmetric choice: $z_1=-\infty,\ z_2=0,\ 
\ov z_3=-ix,\ z_3=ix,\ \ov z_4=\ov z,\ z_4=z,$ which would give the empty solution for
$x$.}   yields the following relation between 
$w_1,w_2,w_3$ and~$x$
\eqn\psli{
x=\fc{\im(w_3)\ (w_1-w_2)}
{\lf|\ w_3-\h\ (w_1+w_2)\ \ri|^2-\fc{1}{4}\ (w_1-w_2)^2}\in\IR\ ,}
which constrains \psl\ to be an element of $PSL(2,\IR)$.
Indeed with \psli\ the matrix \psl\ becomes:
$$\eqalign{
&\hskip1cm P=|w_{13}|\ w_{12}^{1/2}\ \lf[\ \lf|w_3-\h\ (w_1+w_2)\ri|^2-\fc{1}{4}\
w_{12}^2\ \ri]^{1/2}\cr
&\times\pmatrix{[\ \h\ (w_3+\ov w_3)-w_1\ ]\ w_{12}&[\ \h\ (w_3+\ov w_3)\
w_1-|w_3|^2\ ]\ w_{12}\cr
&\cr
\lf|w_3-\h\ (w_1+w_2)\ri|^2-\fc{1}{4}\ w_{12}^2&-w_1\ \lf[
\lf|w_3-\h\ (w_1+w_2)\ri|^2-\fc{1}{4}\ w_{12}^2\ri]}\in PSL(2,\IR)\ .}$$
Depending on the ordering of the points $w_1,w_2$ along the boundary and the
values of $\im(w_3)$, the number $x$ may
take positive or negative real values, respectively, \cf the next Figure.
\iifig\redui{Complex $w_3$--plane for $w_1<w_2$}
{and its regions of positive and negative $x$.}
{\epsfxsize=0.3\hsize\epsfbox{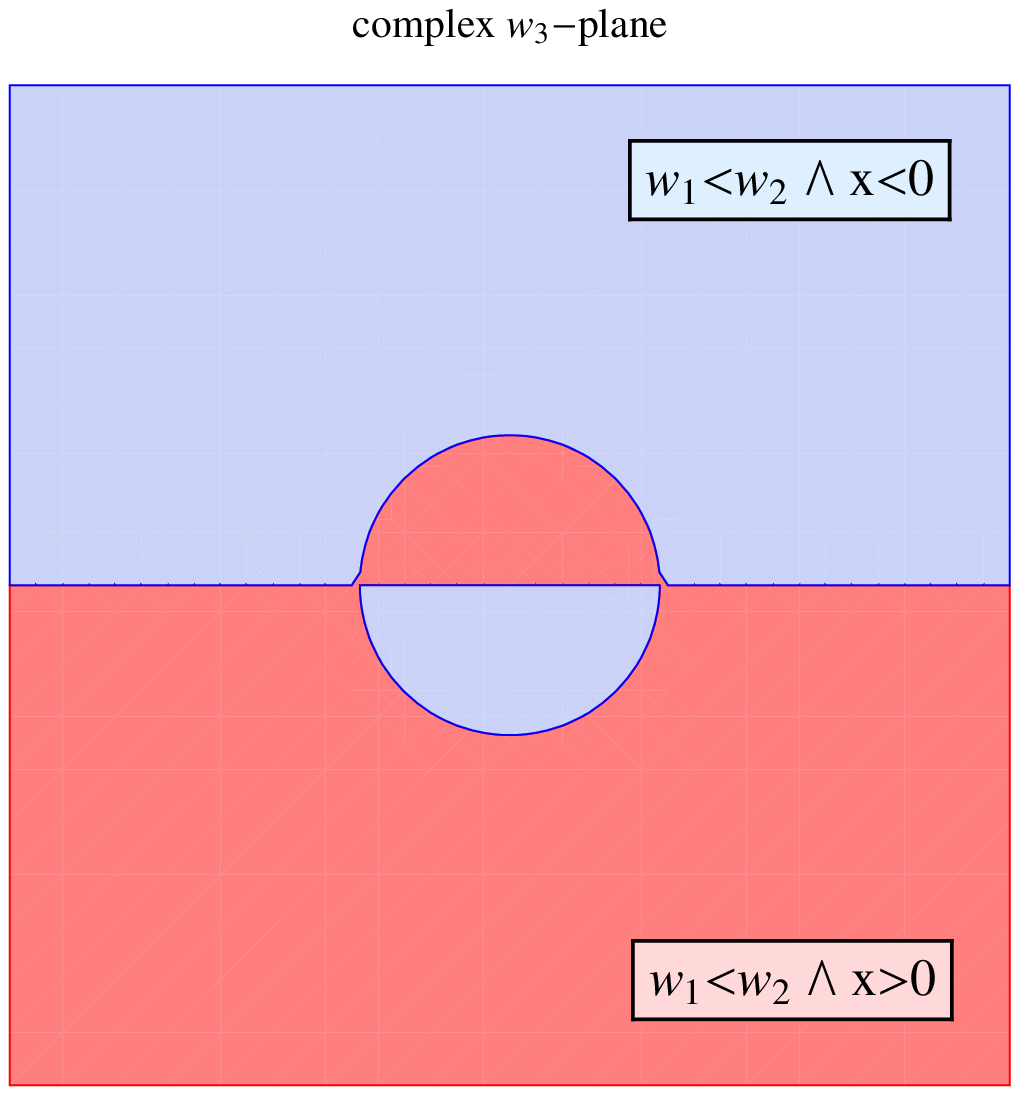}}
\noindent In the double cover the closed string position $z$  is 
integrated over the full complex plane $\IC$ and the $x$--integration 
goes from $-\infty$ to $\infty$. 
The choice \fix\ implies the $c$--ghost contribution:
\eqn\cghost{
\vev{c(z_1)c(z_2)\tilde c(\ov z_3)}=(z_1-z_2)\ (z_1-\ov z_3)\ (z_2-\ov z_3)
=(1+ix)\ z_\infty^2\ .}
Eventually, the partial amplitude \Startwith\ becomes:
\eqn\startwithvi{\eqalign{
\Ac(1,2;3,4)&=\int_{-\infty}^\infty  dx\ 
\vev{c(-\infty)c(1)c(ix)}\cr 
&\times\int_{\IC} d^2z\  
\vev{\ :V_o(-\infty):\ :V_o(1):\ :V_c(-ix,ix):\ :V_c(\ov z, z):\ }\ .}}
For the choice \fix\ the correlator \Exp\ becomes
\eqn\exp{\eqalign{
&\vev{e^{2ip_{1\mu}X^\mu(-\infty)}e^{2ip_{2\mu}X^\mu(1)}e^{iq_{1\mu}X^\mu(-ix,ix)}
e^{iq_{2\mu}X^\mu(\ov z,z)}}\cr
&\hskip0.75cm=(1+ix)^{u}\ (1-ix)^{u}\ (1-z)^{t}
\ (1-\ov z)^{t}\ (z+ix)^{s/2}\ (z-ix)^{s/2}\ (\ov z-ix)^{s/2}\ (\ov z+ix)^{s/2}\ ,}}
with the kinematic invariants:
\eqn\invvi{
u=2\ p_2q_1=2\ p_1q_2\ \ \ ,\ \ \ t=2\ p_2q_2=2\ p_1q_1\ \ \ ,\ \ \ s=2\ q_1q_2=2\ p_1p_2\ .}
Of course, with \conservation, we have: $s+t+u=0$.  
After performing all Wick contractions for each kinematics $\Kc_I$ the amplitude \startwithvi\ 
boils down to
\eqn\boilvi{
\Ac^I(1,2;3,4)=W^{(\kappa^I,\al^I_0)}\lf[{u+n^I_0\ ,\ t+n^I_1\ ,\ \h s+n^I_3\ ,\ \h s+n^I_5\atop
u+m^I_0\ ,\ t+n^I_2\ ,\ \h s+n^I_4\ ,\ \h s+n^I_6}\ri]\ ,}
with the ten integers $m^I_i,n^I_i,\kappa^I,\al^I_0\in \IZ$ 
and the integrals of the following type\foot{Alternatively, instead of the choice 
\fix\ we could have chosen
\eqn\alterfix{
z_1=z_\infty:=-y\ \ \ ,\ \ \ z_2=y\ \ \ ,\ \ \ \ov z_3=-i\ \ \ ,\ \ \ z_3=i\ \ \ ,
\ \ \ \ov z_4=\ov w\ \ \ ,\ \ \ z_4=w\ ,}
with $y\in\IR,\ w\in\IC$ in lines of Footnote 3. 
For this case we find:
\eqn\expalter{\eqalign{
\vev{&e^{2ip_{1\mu}X^\mu(-y)}e^{2ip_{2\mu}X^\mu(y)}e^{iq_{1\mu}X^\mu(-i,i)}
e^{iq_{2\mu}X^\mu(\ov w,w)}}=(2y)^{2s}\ (y+i)^{t+u}\ (y-i)^{t+u}\cr
&\hskip1.5cm \times(w-i)^{s/2}\ (w+i)^{s/2}\ (\ov w-i)^{s/2}\ (\ov w+i)^{s/2}\ 
(y-w)^t\ (y-\ov w)^t\ (y+w)^u\ (y+\ov w)^u\ .}}
However, with the transformation 
\eqn\transpsl{
z=\fc{2y}{1-y^2}\ \fc{1-yw}{y+w}\ \ \ ,\ \ \ x=-\fc{2y}{1-y^2}}
we may identify the two choices \fix\ and \alterfix, \ie \eqqs 
\exp\ and \expalter\ become identical. 
Furthermore, the transformation \transpsl\ gives the alternative 
representation for $W$:
$$\eqalign{
W^{(\kappa,\al_0)}\lf[{\alpha_1,\lambda_1,\gamma_1,\beta_1\atop
\alpha_2,\lambda_2,\gamma_2,\beta_2}\ri]&=2^{-4}
\int_{-\infty}^\infty dy\ \int_\IC d^2w\ (w-\ov w)^\kappa\ 
(2y)^{\al_0+\gamma_1+\gamma_2+\beta_1+\beta_2+\kappa}\cr
&\times  (y+i)^{2\alpha_1+\lambda_1+\lambda_2+\gamma_2+\beta_1+\kappa}
(y-i)^{2\alpha_2+\lambda_1+\lambda_2+\gamma_1+\beta_2+\kappa}\cr
&\times (w-i)^{\gamma_1}\ (w+i)^{\beta_1}\ (\ov w-i)^{\beta_2}\ 
(\ov w+i)^{\gamma_2}\cr
&\times (y-w)^{\lambda_1}\ (y-\ov w)^{\lambda_2}\ 
(y+w)^{-\lambda_1-\gamma_1-\beta_1-\kappa}\ (y+\ov w)^{-\lambda_2-\gamma_2-\beta_2-\kappa}\ ,}$$
subject to the condition (2.57).} 
\eqn\GEN{
W^{(\kappa,\al_0)}\lf[{\alpha_1,\lambda_1,\gamma_1,\beta_1\atop
\alpha_2,\lambda_2,\gamma_2,\beta_2}\ri]:=
\int_{-\infty}^\infty dx\ x^{\al_0}\ (1+ix)^{\alpha_1}\ 
(1-ix)^{\alpha_2}\ I^{(\kappa)}\lf[{\lambda_1,\gamma_1,\beta_1\atop
\lambda_2,\gamma_2,\beta_2}\ri](ix)\ ,}
with
\eqn\ANGEL{\eqalign{
I^{(\kappa)}\lf[{\lambda_1,\gamma_1,\beta_1\atop
\lambda_2,\gamma_2,\beta_2}\ri](ix)&=
\int_\IC d^2z \ (1-z)^{\lambda_1}\ (1-\ov z)^{\lambda_2}\ 
(z-ix)^{\gamma_1}\ (z+ix)^{\beta_1}\cr 
&\times (\ov z+ix)^{\gamma_2}\ (\ov z-ix)^{\beta_2}\ (z-\ov z)^\kappa\ ,}}
and the assignments
\eqn\parameter{
\matrix{&\al_1=u+n_0\ ,&\lambda_1=t+n_1,\
&\gamma_1=\h\ s+n_3,\ &\beta_1=\h\ s+n_5\ ,\cr\cr
&\al_2=u+m_0,\ &\lambda_2=t+n_2,\
&\gamma_2=\h\ s+n_4,\ &\beta_2=\h\ s+n_6\ ,}}
and the integers $m_i,n_i,\kappa,\al_0\in \IZ$ subject to the analyticity condition
\Kapstadt:
\eqn\conditionvi{
\al_0+\al_1+\al_2+\lambda_1+\lambda_2+\gamma_1+\gamma_2+\bet_1+\bet_2+\kappa+4=0\ .}
Clearly, the complex integral \ANGEL\ fulfills:
\eqn\fulfill{
I^{(\kappa)}\lf[{\lambda_1,\gamma_1,\beta_1\atop
\lambda_2,\gamma_2,\beta_2}\ri](-x)=I^{(\kappa)}\lf[{\lambda_1,\beta_1,\gamma_1\atop
\lambda_2,\beta_2,\gamma_2}\ri](x)\ .}

The integrand of \Startwith\ is invariant under the parity symmetries
$\ov z_3\leftrightarrow z_3$ and $\ov z_4\leftrightarrow z_4$ acting on the 
positions of the closed string vertex operators. 
After the fixing \fix\ in \eqq \startwithvi\ these symmetries become the operations
$x\rightarrow -x$ and $\ov z\leftrightarrow z$, respectively.
Moreover, the integral \GEN\ shares the following symmetry 
\eqn\SYMMETRY{\eqalign{
&W^{(\kappa,\al_0)}\lf[{\alpha_1,\lambda_1,\gamma_1,\beta_1\atop
\alpha_2,\lambda_2,\gamma_2,\beta_2}\ri]=(-1)^{\al_0+\al_1+\al_2+\lambda_1+\lambda_2}\cr
&\hskip0.75cm \times
W^{(\kappa,\al_0)}\lf[{2+\kappa+\alpha_1+\lambda_1+\lambda_2+\beta_1+\gamma_2,\ 
-2-\kappa-\lambda_1-\gamma_1-\beta_1,\ \beta_1,\ \gamma_1\atop
2+\kappa+\alpha_2+\lambda_1+\lambda_2+\gamma_1+\beta_2,\ 
-2-\kappa-\lambda_2-\gamma_2-\beta_2,\ \beta_2,\ \gamma_2}\ri]\ ,}}
which exchanges the invariants $t$ and $u$ in the exponents \parameter.
It may be proven by performing the following change of variables
$$z\ra\fc{w+y^2}{w-1}\ \ \ ,\ \ \ x\ra y$$
in the integrand of the l.h.s. of \SYMMETRY.
Furthermore, for \conditionvi\ the integral \GEN\ enjoys the following identity: 
\eqn\SYMMETRYi{\eqalign{
&W^{(\kappa,\al_0)}\lf[{\alpha_1,\lambda_1,\gamma_1,\beta_1\atop
\alpha_2,\lambda_2,\gamma_2,\beta_2}\ri]=2^{\kappa-\al_0}\ (-1)^{\kappa+\lambda_1+\lambda_2}\cr
&\hskip0.75cm \times
W^{(\al_0,\kappa)}\lf[{-2-\kappa-\lambda_1-\gamma_1-\beta_1,2+\kappa+\al_1+\lambda_1+\lambda_2+
\bet_1+\gamma_2,\ \beta_2,\ \gamma_1\atop
-2-\kappa-\lambda_2-\gamma_2-\beta_2,2+\kappa+\al_2+\lambda_1+\lambda_2+
\gamma_1+\beta_2,\ \beta_1,\ \gamma_2}\ri]\ .}}
This relation may be proven by first performing the change of
variables
$$x\ra \fc{z_2}{1-z_1}\ \ ,\ \ z_1\ra-\fc{z_1}{1-z_1}\ \ ,\ \ 
z_2\ra\fc{x}{1-z_1}\ ,$$
with $z_1=\h(z+\ov z)$ and $z_2=\fc{1}{2i}(z-\ov z)$ in the integrand of the 
l.h.s. of \SYMMETRYi\ and then applying \SYMMETRY.
The relation \SYMMETRYi\ proves to be useful for converting negative 
powers of $(z-\ov z)$ into negative powers of $x$, \eg:
\eqn\Golm{
W^{(-2,0)}\lf[{u-1,t,\fc{s}{2},\fc{s}{2}\atop u-1,t,\fc{s}{2},\fc{s}{2}}\ri]=\fc{1}{4}\ 
W^{(0,-2)}\lf[{u,t-1,\fc{s}{2},\fc{s}{2}\atop u,t-1,\fc{s}{2},\fc{s}{2}}\ri]\ .}

Let us now discuss the evaluation of the integral \GEN. 
To compute the integral \ANGEL\ over the complex $z$--plane we split it up into holomorphic and
anti--holomorphic contour integrals along the method proposed in \Kawai.
After introducing the parameterization $z=z_1+iz_2$ the integrand may be
considered as an analytic function in $z_2$.
We then deform the $z_2$--integral along the real axis $\im(z_2)=0$ to the
pure imaginary axis $\re(z_2)=0$, \ie $iz_2\in\IR$.
This way, the variables $\xi=z_1+i z_2\equiv z,\ \eta=z_1-i z_2\equiv \ov z$
become real quantities, \ie $\xi,\eta\in\IR$. 
Similarly, we deform the $x$--integration from the real axis to a contour
along the pure imaginary axis, \ie $\rho=ix$ becomes real.
With the Jacobian $\det\fc{\p(x,z_1,z_2)}{\p(\rho,\xi,\eta)}=\h$ we arrive at:
\eqn\GENN{\eqalign{
W^{(\kappa,\al_0)}\lf[{\alpha_1,\lambda_1,\gamma_1,\beta_1\atop
\alpha_2,\lambda_2,\gamma_2,\beta_2}\ri]&=\h\ \int_{-\infty}^{\infty} d\rho\
|1+\rho|^{\hatt\alpha_1}\ |1-\rho|^{\hatt\alpha_2}\ \rho^{\al_0}\ (1+\rho)^{n_0}\ (1-\rho)^{m_0}\ \cr 
&\times\int_{-\infty}^\infty d\xi\ \int_{-\infty}^\infty d\eta \ 
|1-\xi|^{\hatt\lambda_1}\ |\xi-\rho|^{\hatt\gamma_1}\ |\xi+\rho|^{\hatt\beta_1}\cr 
&\times |1-\eta|^{\hatt\lambda_2}\ |\eta+\rho|^{\hatt\gamma_2}\ |\eta-\rho|^{\hatt\beta_2}\ 
(\xi-\eta)^\kappa\ \Pi(\rho,\xi,\eta)\cr
&\times (1-\xi)^{n_1}\ (\xi-\rho)^{n_3}\ (\xi+\rho)^{n_5}\ 
(1-\eta)^{n_2}\ (\eta+\rho)^{n_4}\ (\eta-\rho)^{n_6}\ .}}
The quantities with a hat refer to their non--integer part.
In \GENN\ the phase factor $\Pi(\rho,\xi,\eta)$ following from \PHASE\ accounts for the correct 
branch of the integrand. 
The phases are analyzed in Appendix \appB. Eventually, in the integrand the
latter are accommodated by choosing the 
respective contours in the complex $\eta$-- and $\rho$--plane.
More precisely, for a given pair of $\rho,\xi\in\IR$ we may consider the
$\eta$--integral as an integration in the complex $\eta$--plane and the phases
$\Pi(\rho,\xi,\eta)$ give rise to the integration contours in the complex
$\eta$--plane as shown in Appendix \appB. 
The result for \boilvi\ is presented in Subsection~3.4.

\subsubsec{Four open strings and one closed string}

For an amplitude of four open and one closed string in \Startwith\ in total we have 
six partial amplitudes $\Ac(1,\si(2),\si(3),\si(4);5)$, with $\si\in S_3$ permuting 
the open strings.
Due to $PSL(2,\IR)$ invariance on the disk we may fix three vertex positions.
A convenient choice is ($z_i:=x_i,\ i=1,\ldots,4$)
\eqn\fixHU{
z_1=-\infty\ \ \ ,\ \ \ z_2=1\ \ \ ,\ \ \ z_3=-x,\ \ \ z_4=x \ \ ,\ \ \ \ov z_5=\ov z\ \ \ ,\ \ \ z_5=z\ ,}
with $z\in{\bf H}_+$ and $x\in\IR$. After analytic continuation in $x$ this setup becomes similar to the setup \fix, 
relevant for two open and two closed strings. 
In the double cover the closed string position $z$  is integrated over the full complex plane $\IC$.
The choice \fixHU\ implies the $c$--ghost contribution:
\eqn\cghostHU{
\vev{c(z_1)c(z_2)c(z_3)}=(z_1-z_2)\ (z_1-z_3)\ (z_2-z_3)=-(1+x)\ z_\infty^2\ .}
Eventually, the partial amplitude \Startwith\ becomes:
\eqn\startwithHU{\eqalign{
\Ac(1,\si(2),\si(3),\si(4);5)&=\int_{\Ic_\si}  dx\ \vev{c(-\infty)c(-x)c(1)}\cr 
&\times\int_{\IC} d^2z\  
\vev{\ :V_o(-\infty):\ :V_o(1):\  :V_o(-x):\  :V_o(x):\  :V_c(\ov z, z):\ }\ .}}
For the choice \fixHU\ the integration range $\Ic_\si$ in the real variable $x$ is 
related to the specific ordering $\si$ as follows:
\eqn\Orderings{\eqalign{
\Ac(1,3,4,2;5)\simeq\Ac(1,2,4,3;5):\ \ \ 
\Ic_{\si_1}&=\{x\in\IR\ |\ 0<x<1\}\ ,\cr
\Ac(1,4,3,2;5)\simeq\Ac(1,2,3,4;5):\ \ \ 
\Ic_{\si_2}&=\{x\in\IR\ |\ -1<x<0\}\ ,\cr
\Ac(1,4,2,3;5)\simeq\Ac(1,3,2,4;5):\ \ \ 
\Ic_{\si_3}&=\{x\in\IR\ |\ -\infty<x<-1\ \cup\ 1<x<\infty\}\ .}}
The correlator \Exp\ becomes for the choice \fixHU:
\eqn\expHU{\eqalign{
\vev{e^{2ip_{1\mu}X^\mu(-\infty)}&e^{2ip_{2\mu}X^\mu(1)}e^{2ip_{3\mu}X^\mu(-x)}
e^{2ip_{4\mu}X^\mu(x)}e^{iq_{\mu}X^\mu(\ov z,z)}}\cr
&= |2x|^{4p_3p_4}\ \ |1+x|^{4p_2p_3}\ |1-x|^{4p_2p_4}\ (1-z)^{2p_2q}\ (1-\ov z)^{2p_2q}\cr 
&\times   (x+z)^{2p_3q}\ (x+\ov z)^{2p_3q}\  (x-z)^{2p_4q}\ (x-\ov z)^{2p_4q}\ 
|z-\ov z|^{2q_\parallel^2}\ .}}
We have the five kinematic invariants:
\eqn\invvii{
s_1=4\ p_1p_2\ ,\ s_2=4\ p_2p_3\ ,\ s_3=4\ p_3p_4\ ,\ s_4=2\ p_4q\ ,\ s_5=2\ p_1q\ .}
Furthermore, we have 
$p_1p_3=-\fc{s_1}{4}-\fc{s_2}{4}+\fc{s_4}{2}+\fc{q_\parallel^2}{2},\ 
p_1p_4=\fc{s_2}{4}-\fc{s_4}{2}-\fc{s_5}{2}-\fc{q_\parallel^2}{2},\
p_2p_4=-\fc{s_2}{4}-\fc{s_3}{4}+\fc{s_5}{2}+\fc{q_\parallel^2}{2},$ and 
$p_2q=-\fc{s_1}{4}+\fc{s_3}{4}-\fc{s_5}{2}-\fc{q_\parallel^2}{2},\ 
p_3q=\fc{s_1}{4}-\fc{s_3}{4}-\fc{s_4}{2}-\fc{q_\parallel^2}{2}$.
After performing all Wick contractions for a given kinematics $\Kc_I$ 
the amplitude \startwithHU\ boils down to 
\eqn\boilvii{
\Ac_\si^I(1,\si(2),\si(3),\si(4);5)=
W^{(\kappa^I,\al^I_0)}_\si\lf[{\alpha^I_1,\lambda^I_1,\gamma^I_1,\beta^I_1\atop
\alpha^I_2,\lambda^I_2,\gamma^I_2,\beta^I_2}\ri]\ ,}
and the integrals of the following type
\eqn\GENHU{\eqalign{
W^{(\kappa,\al_0)}_\si\lf[{\alpha_1,\lambda_1,\gamma_1,\beta_1\atop
\alpha_2,\lambda_2,\gamma_2,\beta_2}\ri]&:=2^{\alpha_0}\ 
\int_{\Ic_\si} dx\ |x|^{\hatt\al_0}\ |1+x|^{\hatt\alpha_1}\ 
|1-x|^{\hatt\alpha_2}\cr
&\times x^{m_0}\ (1+x)^{m_1}\ (1-x)^{m_2}\  
I^{(\kappa)}\lf[{\lambda_1,\gamma_1,\beta_1\atop\lambda_2,\gamma_2,\beta_2}\ri](x)\ ,}}
with
\eqn\ANGELHU{\eqalign{
I^{(\kappa)}\lf[{\lambda_1,\gamma_1,\beta_1\atop
\lambda_2,\gamma_2,\beta_2}\ri](x)&=
\int_\IC d^2z \ (1-z)^{\lambda_1}\ (1-\ov z)^{\lambda_2}\ 
(z-x)^{\gamma_1}\ (\ov z-x)^{\gamma_2}\cr 
&\times (z+x)^{\beta_1}\ (\ov z+x)^{\beta_2}\ |z-\ov z|^\kappa\ (z-\ov z)^{\tilde\kappa}\ ,}}
the assignments
\eqn\parameterHU{
\matrix{&\al_0=s_3+m_0\ ,&\al_1=s_2+m_1\ ,&\al_2=-s_2-s_3+2s_5+m_2+2\ q_\parallel^2\ ,\cr\cr
&&\lambda_1=-\fc{s_1}{2}+\fc{s_3}{2}-s_5-q_\parallel^2+n_1\ ,\  
&\lambda_2=-\fc{s_1}{2}+\fc{s_3}{2}-s_5-q_\parallel^2+n_2\ ,\cr\cr
&&\gamma_1=s_4+n_3\ ,\ &\gamma_2=s_4+n_4\ ,\cr\cr
&&\beta_1=\fc{s_1}{2}-\fc{s_3}{2}-s_4-q_\parallel^2+n_5\ ,
&\beta_2=\fc{s_1}{2}-\fc{s_3}{2}-s_4-q_\parallel^2+n_6\ ,\cr\cr
&&& \kappa=2\ q_\parallel^2+\tilde\kappa}}
and the integers $m_i,n_i,\tilde\kappa\in \IZ$ subject to the analyticity condition
\Kapstadt:
\eqn\conditionvii{
\al_0+\al_1+\al_2+\lambda_1+\lambda_2+\gamma_1+\gamma_2+\beta_1+\beta_2+
\kappa+4=0\ .}
The complex integral \GENHU\ fulfills:
\eqn\fulfillHU{
I^{(\kappa)}\lf[{\lambda_1,\gamma_1,\beta_1\atop
\lambda_2,\gamma_2,\beta_2}\ri](-x)=I^{(\kappa)}\lf[{\lambda_1,\beta_1,\gamma_1\atop
\lambda_2,\beta_2,\gamma_2}\ri](x)\ .}
This relation is used in Appendix \appC.

Let us now discuss the evaluation of the integral \GENHU. 
To compute the integral \ANGELHU\ over the complex $z$--plane we split it up into 
holomorphic and
anti--holomorphic contour integrals along the method proposed in \Kawai.
After introducing the parameterization $z=z_1+iz_2$ the integrand may be
considered as an analytic function in $z_2$.
We then deform the $z_2$--integral along the real axis $\im(z_2)=0$ to the
pure imaginary axis $\re(z_2)=0$, \ie $iz_2\in\IR$.
This way, the variables $\xi=z_1+i z_2\equiv z,\ \eta=z_1-i z_2\equiv \ov z$
become real quantities, \ie $\xi,\eta\in\IR$. 
With the Jacobian $\det\fc{\p(z_1,z_2)}{\p(\xi,\eta)}=\fc{i}{2}$ we arrive at:
\eqn\GENNHU{\eqalign{
W^{(\kappa,\al_0)}_\si\lf[{\alpha_1,\lambda_1,\gamma_1,\beta_1\atop
\alpha_2,\lambda_2,\gamma_2,\beta_2}\ri]&=\fc{i}{2}\ \int_{\Ic_\si} dx\
|2x|^{\hatt\al_0}\ |1+x|^{\hatt\alpha_1}\ |1-x|^{\hatt\alpha_2}\ (2x)^{m_0}\ (1+x)^{m_1}\ (1-x)^{m_2}\cr 
&\times\int_{-\infty}^\infty d\xi\ \int_{-\infty}^\infty d\eta \ 
|1-\xi|^{\hatt\lambda_1}\ |\xi-x|^{\hatt\gamma_1}\ |\xi+x|^{\hatt\beta_1}\cr 
&\times |1-\eta|^{\hatt\lambda_2}\ |\eta-x|^{\hatt\gamma_2}\ |\eta+x|^{\hatt\beta_2}\ 
|\xi-\eta|^{\hatt\kappa}\ (\xi-\eta)^{\tilde\kappa}\ \Pi(x,\xi,\eta)\cr
&\times (1-\xi)^{n_1}\ (\xi-x)^{n_3}\ (\xi+x)^{n_5}\ (1-\eta)^{n_2}\
(\eta-x)^{n_4}\ (\eta+x)^{n_6}\ .}}
The quantities with a hat refer to their non--integer part.
In \GENNHU\ the phase factor $\Pi(x,\xi,\eta)$ following from \PHASE\ 
accounts for the correct branch of the integrand. In this phase factor the variable $x$ enters as a parameter.
The phases are analyzed in Appendix \appC. Eventually, in the integrand the
latter are accommodated by choosing the 
respective contours in the complex $\eta$--plane for a given range $x\in\Ic_\si$.
More precisely, for a given pair of $x,\xi\in\IR$ we may consider the
$\eta$--integral as an integration in the complex $\eta$--plane and the phases
$\Pi(x,\xi,\eta)$ give rise to the integration contours in the complex
$\eta$--plane as shown in Appendix \appC. 
The result for \boilvii\ is presented in Subsection~3.5.

\subsubsec{Three  closed strings}

In this Subsection we discuss the disk amplitude \Startwith\ of three closed strings 
$\Ac(1,2,3)$.
With a $PSL(2,\IR)$ transformation two arbitrary points 
$w_1,w_2\in {\bf H}_+$ can be mapped to the two special points 
$z_1=i,z_2=ix_\pm$ along the positive imaginary axis, with 
\eqn\withchoce{
x_\pm=\fc{\re(w_1-w_2)^2+\im(w_1)^2+\im(w_2)^2
\mp  |w_1-w_2|\ |w_1-\ov w_2|}{2\ \im(w_1)\ \im(w_2)}\in\IR^+} 
and $0<x_+<1$ and $x_->1$.
Therefore, this map  allows for the following choice of two closed string vertex positions
\eqn\Fixto{
\ov z_1=-i,\ \ \ z_1=i\ \ \ ,\ \ \ \ov z_2=-ix\ \ \ ,\ \ \ z_2=ix\ \ \ ,}
with $0<x<1$, see also \HK.
This choice implies the $c$--ghost contribution:
\eqn\cghostto{
\vev{\tilde c(\ov z_1)c(z_1)\tilde c(\ov z_2)}=2i\ (1-x^2)\ .}
Hence, in the double cover of ${\bf H}_+$ the disk amplitude \Startwith\ of three closed strings becomes
\eqn\startwithto{
\Ac(1,2,3)=2\ \int_{-1}^1 dx\ (1-x^2)\ \int_\IC d^2z\ \vev{:V_c(-i,i):\ :V_c(-ix,ix):\ :V_c(\ov z,z):}+{\it h.c.}\ .}
For this three--point process to give a result, which does not vanish onÐ-shell, the closed string momenta $q_i$ have to have also a non-Ðvanishing directions $q_{\perp i}\neq0$ transverse to the D-Ðbrane worldÐvolume, with $q_i = q_{i\parallel}+ q_{i\perp},\ i=1,2,3$ and momentum conservation \conservation:
\eqn\consto{
q_{1\parallel}+q_{2\parallel}+q_{3\parallel}=0\ .}
For the choice \Fixto\ the correlator \Exp\ becomes ($z_3\equiv z$):
\eqn\expto{\eqalign{
&\vev{e^{iq_{1\mu}X^\mu(\ov z_1,z_1)}e^{iq_{2\mu}X^\mu(\ov z_2,z_2)}
e^{iq_{3\mu}X^\mu(\ov z_3,z_3)}}=2^{2q_{1\parallel}^2+2q_{2\parallel}^2}\ |x|^{2q_{2\parallel}^2}\ |z-\ov z|^{2q_{3\parallel}^2}\cr
&\hskip1.0cm\times |1-x|^{2q_1q_2}\ |1+x|^{2q_1Dq_2}\  |i-z|^{2q_1q_3}\ |i-\ov z|^{2q_1Dq_3}\ |z-ix|^{2q_2q_3}\ |z+ix|^{2q_2Dq_3}\ .}}
We have the following six kinematic invariants
\eqn\invto{\matrix{
&s_1=2\ q_{1\parallel}^2=q_1Dq_1\ , & s_2=q_1Dq_2\ , &s_3=2\ q_{2\parallel}^2=q_2Dq_2\ ,\cr\cr
&s_4=q_2Dq_3\ ,& s_5=2\ q_{3\parallel}^2=q_3Dq_3\ ,&s_6=q_1Dq_3}}
to describe the scattering process.
From \consto\ and \invto\ we find:
\eqn\momto{\eqalign{
q_1q_2&=\h\ (-s_1-s_3+s_5)-s_2\ ,\cr
q_2q_3&=\h\ (s_1-s_3-s_5)-s_4\ ,\cr
q_1q_3&=\h\ (-s_1+s_3-s_5)-s_6\ .}}
After performing all Wick contractions and including the ghost correlator \cghostto\ 
for any kinematics $\Kc_I$ the amplitude \startwithto\ reduces to
\eqn\boilto{\eqalign{
\lf(1+e^{i\pi s_5}\ri)\ \Ac^I(1,2,3)&=(-1)^{m^I_3+n^I_6+n^I_8}\cr 
&\hskip-2cm\times
\lf\{\ W^{(\kappa^I,\al_0^I,\al_3^I)}\lf[{\al^I_1,\lambda_1^I,\gamma_1^I,\beta_1^I,\eps_1^I\atop\al^I_2,\lambda_2^I,\gamma_2^I,\beta_2^I,\eps_2^I}\ri]+
W^{(\kappa^I,\al_0^I,\al_3^I)}\lf[{\al^I_2,\gamma_1^I,\lambda_1^I,\beta_1^I,\eps_1^I
\atop\al^I_1,\gamma_2^I,\lambda_2^I,\beta_2^I,\eps_2^I}\ri]\ri\}\ ,}}
and the complex integral of the following type
\eqn\GENto{\eqalign{
W^{(\kappa,\al_0,\al_3)}\lf[{\al_1,\lambda_1,\gamma_1,\beta_1,\eps_1\atop\al_2,\lambda_2,
\gamma_2,\beta_2,\eps_2}\ri]&:=
2^{1+\al_0+\al_3}\ \int_{-1}^1 dx\ |x|^{\hatt \al_3}\ |1+x|^{\hatt\al_1}\ |1-x|^{\hatt\al_2}\cr
&\times x^{m_3}\ (1+x)^{1+m_1}\ (1-x)^{1+m_2}\ 
I^{(\kappa)}\lf[{\lambda_1,\gamma_1,\beta_1,\eps_1\atop\lambda_2,\gamma_2,\beta_2,\eps_2}\ri](x)\ ,}}
with
\eqn\GENTO{\eqalign{
I^{(\kappa)}\lf[{\lambda_1,\gamma_1,\beta_1,\eps_1\atop\lambda_2,\gamma_2,\beta_2,\eps_2}\ri](x)&=
\int_\IC d^2z\ (1-z)^{\lambda_1}\ (1-\ov z)^{\lambda_2}\ (1+z)^{\gamma_1}\ 
(1+\ov z)^{\gamma_2}\cr 
&\times (z-x)^{\beta_1}\ (\ov z-x)^{\beta_2}\ (z+x)^{\eps_1}\ 
(\ov z+x)^{\eps_2}\ |z+\ov z|^{\hatt\kappa}\ (z+\ov z)^{\tilde\kappa}\ ,}}
the assignments
\eqn\parameterto{
\matrix{&&\al_0=s_1+m_0\ ,&\al_3=s_3+m_3\ ,\ \ \ \kappa=s_5+\tilde\kappa\ ,\cr\cr
&&\al_1=2s_2+m_1\ ,&\al_2=-s_1-s_3+s_5-2s_2+m_2\ ,\cr\cr
&&\lambda_1=s_6+n_1\ ,\  &\lambda_2=s_6+n_2\ ,\cr\cr
&&\gamma_1=\h\ (-s_1+s_3-s_5)-s_6+n_3\ ,\ 
&\gamma_2=\h\ (-s_1+s_3-s_5)-s_6+n_4\ ,\cr\cr
&&\beta_1=s_4+n_5\ ,\ &\beta_2=s_4+n_6\ ,\cr\cr
&&\eps_1=\h\ (s_1-s_3-s_5)-s_4+n_7\ ,\ &\eps_2=\h\ (s_1-s_3-s_5)-s_4+n_8\ ,}}
and the integers $m_i,n_i,\tilde\kappa\in \IZ$ subject to the analyticity conditions
\Kapstadt:
\eqn\conditionto{\eqalign{
2\ \al_0+\al_1&+\al_2+\lambda_1+\lambda_2+\gamma_1+\gamma_2+4=0\ ,\cr
2\ \al_3+\al_1&+\al_2+\beta_1+\beta_2+\eps_1+\eps_2+4=0\ ,\cr
&\lambda_1+\gamma_1+\beta_1+\eps_1+\kappa+2=0\ ,\cr
&\lambda_2+\gamma_2+\beta_2+\eps_2+\kappa+2=0\ .}}
Note, that we have the following relation:
\eqn\follRel{
I^{(\kappa)}\lf[{\lambda_1,\gamma_1,\beta_1,\eps_1\atop\lambda_2,\gamma_2,\beta_2,\eps_2}\ri](-x)=I^{(\kappa)}\lf[{\lambda_1,\gamma_1,\eps_1,\beta_1
\atop\lambda_2,\gamma_2,\eps_2,\beta_2}\ri](x)\ .}

Let us now discuss the evaluation of the integral \GENto. 
To compute the integral \GENTO\ over the complex $z$--plane as before we split it up into  holomorphic and
anti--holomorphic contour integrals.
With the Jacobian $\det\fc{\p(z_1,z_2)}{\p(\xi,\eta)}=\fc{i}{2}$ we arrive at:
\eqn\GENNTO{\eqalign{
W^{(\kappa,\al_0,\al_3)}\lf[{\alpha_1,\lambda_1,\gamma_1,\beta_1,\eps_1\atop
\alpha_2,\lambda_2,\gamma_2,\beta_2,\eps_2}\ri]&=2^{\al_0}\ i\  
\int_{-1}^1 dx\ |2x|^{\hatt\al_3}\ |1+x|^{\hatt\alpha_1}\ 
|1-x|^{\hatt\alpha_2}\cr 
&\times (2x)^{m_3}\ (1+x)^{1+m_1}\ (1-x)^{1+m_2}\cr 
&\times\int_{-\infty}^\infty d\xi\ \int_{-\infty}^\infty d\eta \ 
|1-\xi|^{\hatt\lambda_1}\ |1+\xi|^{\hatt\gamma_1}\ |\xi-x|^{\hatt\beta_1}\ 
|\xi+x|^{\hatt\eps_1}\cr 
&\times |1-\eta|^{\hatt\lambda_2}\ |1+\eta|^{\hatt\gamma_2}\ |\eta-x|^{\hatt\beta_2}\ 
|\eta+x|^{\hatt\eps_2}\ 
|\xi+\eta|^{\hatt\kappa}\ \Pi(x,\xi,\eta)\cr
&\times (1-\xi)^{n_1}\ (1+\xi)^{n_3}\ (\xi-x)^{n_5}\ (\xi+x)^{n_7}\cr
&\times  (1-\eta)^{n_2}\ (1+\eta)^{n_4}\ (\eta-x)^{n_6}\ (\eta+x)^{n_8}\ (\xi+\eta)^{\tilde\kappa}\ .}}
The hatted quantities refer to their non--integer part, while a tilde on them 
denotes their integer part, \ie $\kappa=\hatt\kappa+\tilde\kappa$, with $\hatt\kappa=s_5,$ etc.
In \GENNTO\ the phase factor $\Pi(x,\xi,\eta)$ following from \PHASE\ 
accounts for the correct branch of the integrand. 
The coordinate $x$ enters this phase factor as a parameter.
For a given range of $x\in\IR$ the phases, which  
are analyzed in Appendix \appD, are accommodated in the integrand by choosing the 
respective contours in the complex $\eta$--plane.
More precisely, for a given pair of $x,\xi\in\IR$ we may consider the
$\eta$--integral as an integration in the complex $\eta$--plane and the phases
$\Pi(x,\xi,\eta)$ give rise to the integration contours in the complex
$\eta$--plane as shown in Appendix \appD. 
The result for \boilto\ is presented in Subsection~3.6.

\subsubsec{Three open strings and two closed strings}

For an amplitude of four open and one closed string in \Startwith\ in total we 
consider the  partial amplitudes $\Ac(1,2,3;4,5)$.
Due to $PSL(2,\IR)$ invariance on the disk we may fix the three open string 
vertex positions as \fixi:
\eqn\fixHUU{
x_1=-\infty\ \ \ ,\ \ \ x_2=0\ \ \ ,\ \ \ x_3=1\ ,}
 for the two closed strings. 
For this choice in the amplitude \Startwith\ we are left with two complex
integrations of the two closed string positions $z_1,z_2:=w$, with  
$z_1,z_2\in{\bf H}_+$.
In the double cover the closed string positions $z,w$ are integrated over the full complex plane $\IC$.
The choice \fixHUU\ implies the $c$--ghost contribution \cghost:
\eqn\cghostHUU{
\vev{c(x_1)c(x_2)c(x_3)}=(x_1-x_2)\ (x_1-x_3)\ (x_2-x_3)=z_\infty^2\ .}
Eventually, the partial amplitude \Startwith\ becomes:
\eqn\startwithHUU{\eqalign{
&\Ac(1,2,3;4,5)=\vev{c(-\infty)c(0)c(1)}\cr 
&\hskip1cm\times\int_{\IC}\ d^2z_1\ \int_{\IC}\ d^2z_2\ 
\vev{\ :V_o(-\infty):\ :V_o(0):\  :V_o(1):\  :V_c(\ov z_1,z_1):\  :V_c(\ov z_2, z_2):\ }\ .}}
For the choice \fixHU\ the correlator \Exp\ becomes:
\eqn\expHUU{\eqalign{
\vev{e^{2ip_{1\mu}X^\mu(-\infty)}&e^{2ip_{2\mu}X^\mu(0)}e^{2ip_{3\mu}X^\mu(1)}e^{iq_{1\mu}X^\mu(\ov z_1,z_1)}e^{iq_{2\mu}X^\mu(\ov z_2,z_2)}}\cr
&= z_1^{2p_2q_1}\ \ov z_1^{2p_2q_1}\ (1-z_1)^{2p_3q_1}\ (1-\ov z_1)^{2p_3q_1}\cr 
&\times z_2^{2p_2q_2}\ \ov z_2^{2p_2q_2}\ (1-z_2)^{2p_3q_2}\ (1-\ov z_2)^{2p_3q_2}\cr
&\times (z_1- z_2)^{q_1q_2}\ (\ov z_1-\ov z_2)^{q_1q_2}\ (z_1-\ov z_2)^{q_1q_2}\ 
(\ov z_1-z_2)^{q_1q_2}\ .}}
We have the five kinematic invariants:
\eqn\invviii{
s_1=4\ p_1p_2\ ,\ s_2=4\ p_2p_3\ ,\ s_3=2\ p_3q_1\ ,\ s_4=q_1q_2\ ,\ s_5=2\ p_1q_2\ .}
Furthermore, we have $p_1p_3=-\fc{s_1}{4}-\fc{s_2}{4}+s_4,\ 
p_1q_1=\fc{s_2}{4}-s_4-\fc{s_5}{2},\
p_2q_1=-\fc{s_2}{4}-\fc{s_3}{2}+\fc{s_5}{2},$ and 
$p_2q_2=-\fc{s_1}{4}+\fc{s_3}{2}-\fc{s_5}{2},\ p_3q_2=\fc{s_1}{4}-\fc{s_3}{2}-s_4$.
After performing all Wick contractions for any kinematics $\Kc_I$ 
the amplitude \startwithHUU\ boils down to 
\eqn\boilviii{
\Ac^I(1,2,3;4,5)=J^{(\kappa^I_1,\kappa^I_2)}\lf[{\lambda^I_1,\gamma^I_1,\tilde\lambda^I_1,
\tilde\gamma^I_1,\delta^I_1,\tilde\delta^I_1\atop \lambda^I_2,\gamma^I_2,\tilde\lambda^I_2,
\tilde\gamma^I_2,\delta^I_2,\tilde\delta^I_2}\ri]\ .}
In \boilviii\ the class of two--dimensional complex integrals is given by
\eqn\ANGELHUU{\eqalign{
J^{(\kappa_1,\kappa_2)}\lf[{\lambda_1,\gamma_1,\tilde\lambda_1,\tilde\gamma_1,
\delta_1,\tilde\delta_1\atop \lambda_2,\gamma_2,\tilde\lambda_2,\tilde\gamma_2,
\delta_2,\tilde\delta_2}\ri]=
\int_\IC d^2z_1 &\int_\IC d^2z_2\ 
z_1^{\lambda_1}\ \ov z_1^{\lambda_2}\ (1-z_1)^{\gamma_1}\ (1-\ov z_1)^{\gamma_2}\ 
(z_1-\ov z_1)^{\kappa_1}\cr 
&\times z_2^{\tilde\lambda_1}\ \ov z_2^{\tilde\lambda_2}\ (1-z_2)^{\tilde\gamma_1}\ (1-\ov z_2)^{\tilde\gamma_2}\ (z_2-\ov z_2)^{\kappa_2}\cr
&\times (z_1- z_2)^{\delta_1}\ (\ov z_1-\ov z_2)^{\delta_2}\ 
(z_1-\ov z_2)^{\tilde\delta_1}\ (\ov z_1-z_2)^{\tilde\delta_2}\ ,}}
with the assignments
\eqn\parameterHUU{
\matrix{
&&\lambda_1=-\fc{s_2}{2}-s_3+s_5+n_1\ ,\  
&\lambda_2=-\fc{s_2}{2}-s_3+s_5+n_2\ ,\cr\cr
&&\gamma_1=s_3+n_3\ ,\ &\gamma_2=s_3+n_4\ ,\cr\cr
&&\tilde\lambda_1=-\fc{s_1}{2}+s_3-s_5+n_5\ ,\  
&\tilde\lambda_2=-\fc{s_1}{2}+s_3-s_5+n_6\ ,\cr\cr
&&\tilde\gamma_1=\fc{s_1}{2}-s_3-2s_4+n_7\ ,\ &\tilde\gamma_2=\fc{s_1}{2}-s_3-2s_4+n_8\ ,\cr\cr
&&\delta_1=s_4+n_9\ ,\ &\delta_2=s_4+n_{10}\ ,\cr\cr 
&&\tilde\delta_1=s_4+n_{11}\ ,\ &\tilde\delta_2=s_4+n_{12}\ ,}}
and the integers integers $n_i,\kappa_i\in \IZ$.
Again, we define the non--integer part of these quantities by putting a hat on them.

Let us now discuss the evaluation of the integral \ANGELHUU.
We split it up into holomorphic and anti--holomorphic contour integrals
and apply the methods described in  the previous cases.
After introducing the parameterization $z_i=z_{1i}+iz_{2i},\ i=1,2$ 
the integrand may be considered as an analytic function in $z_{21}$ and $z_{22}$.
We then deform the $z_{2i}$--integrals along the real axis $\im(z_{2i})=0$ to the
pure imaginary axis $\re(z_{2i})=0$, \ie $iz_{2i}\in\IR$.
This way, the variables $\xi_i=z_{1i}+i z_{2i}\equiv z_i,\ \eta_i=z_{1i}-i z_{2i}\equiv \ov z_i$
become real quantities, \ie $\xi_i,\eta_i\in\IR$, $i=1,2$. 
With the Jacobian $\det\fc{\p(z_{1i},z_{2j})}{\p(\xi_k,\eta_l)}=\lf(\fc{i}{2}\ri)^2$ we arrive at:
\eqn\GENNHUL{\eqalign{
J^{(\kappa_1,\kappa_2)}\lf[{\lambda_1,\gamma_1,\tilde\lambda_1,\tilde\gamma_1,
\delta_1,\tilde\delta_1\atop \lambda_2,\gamma_2,\tilde\lambda_2,\tilde\gamma_2,
\delta_2,\tilde\delta_2}\ri]
&=-\fc{1}{4}\ \int_{-\infty}^\infty \hskip-0.15cm
d\xi_1\int_{-\infty}^\infty \hskip-0.15cm d\xi_2
\int_{-\infty}^\infty \hskip-0.15cm d\eta_1 \int_{-\infty}^\infty \hskip-0.15cm d\eta_2\ \Pi(\xi_1,\xi_2,\eta_1,\eta_2)\cr
&\times |\xi_1|^{\hatt\lambda_1}\ |\eta_1|^{\hatt\lambda_2}\ 
|1-\xi_1|^{\hatt\gamma_1}\ |1-\eta_1|^{\hatt\gamma_2}\cr
&\times |\xi_2|^{\hatt{\tilde\lambda}_1}\ 
|\eta_2|^{\hatt{\tilde\lambda}_2}\ |1-\xi_2|^{\hatt{\tilde\gamma}_1}\ 
|1-\eta_2|^{\hatt{\tilde\gamma}_2}\cr
&\times |\xi_1- \xi_2|^{\hatt\delta_1}\ |\eta_1-\eta_2|^{\hatt\delta_2}\ 
|\xi_1-\eta_2|^{\hatt{\tilde\delta}_1}\
|\eta_1-\xi_2|^{\hatt{\tilde\delta}_2}\cr
&\times \xi_1^{n_1}\ \eta_1^{n_2}\ (1-\xi_1)^{n_3}\
(1-\eta_1)^{n_4}\ (\xi_1-\eta_1)^{\kappa_1}\cr 
&\times\xi_2^{n_5}\ \eta_2^{n_6}\ (1-\xi_2)^{n_7}\ (1-\eta_2)^{n_8}
\ (\xi_2-\eta_2)^{\kappa_2}\cr
&\times (\xi_1- \xi_2)^{n_9}\ (\eta_1-\eta_2)^{n_{10}}\ 
(\xi_1-\eta_2)^{n_{11}}\ (\eta_1-\xi_2)^{n_{12}}\ .}}
In \GENNHUL\ the phase factor $\Pi(\xi_1,\xi_2,\eta_1,\eta_2)$ following from \PHASE\ 
accounts for the correct branch of the integrand. 
In the integrand these phases are accommodated by choosing the 
respective contours in the complex $\eta_1,\eta_2$--planes for given $\xi_1,\xi_2$.
The result for \boilviii\ is presented in Subsection~3.7.

\newsec{Disk amplitudes of open $\&$ closed strings vs. pure open string
disk amplitudes}

In this Section we establish the relation between 
a disk amplitude with $N_o$ open $\&$~$N_c$~closed strings to 
a disk amplitude of $N_o+2N_c$ open strings.
By reducing disk amplitudes of open $\&$ closed strings
to disk amplitudes involving only open strings the task  of computing 
those amplitudes reduces to the problem of computing
pure open string amplitudes. This map reveals important relations between open $\&$ closed string disk amplitudes and pure open string disk amplitudes.
For the latter a great deal of detailed results exists 
\refs{\Brazil,\Dan,\STi,\STii,\August,\Potsdam}, in contrast to disk
amplitudes of open and closed strings due to the more involved integrations over the full complex plane.

\subsec{Basis of complex contour integrals and open string subamplitude relations}

For each open string ordering $\si$ in the amplitude \AMPLITUDE\ 
the phase factor $\Pi(x_l,\xi_i,\eta_j)$, which is given in  \PHASE,
accounts for the correct branch of the integrand. 
For any given ordering $\Si$ of the insertions $x_l,\xi_i,\eta_j$ 
this phase is fixed  and independent on the points $x_l,\xi_i,\eta_j$.
Hence the amplitude \AMPLITUDE\ decomposes into a sum over
partial ordered $N_o+2N_c$ open string amplitudes times a specific phase factor 
$\Pi(\Sigma)$ determined by the ordering of the points $x_l,\xi_i,\eta_j$:
\eqn\SPLIT{
\Ac_\si(N_o,N_c)=\lf(\fc{i}{2}\ri)^{N_c}\ \sum_{\Si\in\Pc}
\Pi(\Sigma)\ \ A(1,\Si(2),\ldots,\Si(N_o+2N_c))\ ,}
with:
$$\Pc=\{\Si\in S_{N_o+2N_c}\bigl/\IZ_{N_o+2N_c}\ |\ \Si(i_l)=\si(l)\ ,\ \l=2,\ldots,N_o\ ;\ 
1<i_2<\ldots<i_{N_o}\leq N_{o}+2N_c\}.$$
The factor $\Pi(\Sigma)$ is the phase factor from \PHASE\ for the insertions
$x_l,\xi_i,\eta_j$ with the ordering $\Sigma$ of the $N_o+2N_c$ open string positions. Note, that the  kinematical factors $\Kc_I$ and the integers $n_{ij}^I$ have no effect on this phase.
Hence, the following discussion holds for any chosen kinematics. This is way we have dropped the index $I$ in \SPLIT.

For a given (initial) ordering $\si$ of the $N_o$ open string positions $x_l$ the sum runs over 
\eqn\Bronst{
\nu(N_o,N_c):=\fc{(N_o+2N_c-1)!}{(N_o-1)!}}  
terms accounting for all possible orderings of the points $\xi_i,\eta_j$ related
to the closed string positions. 
For a given ordering of the positions $x_k,\xi_i\in\IR$ we  consider the $\eta_j$--integrals in \AMPLITUDE\ as an integration in the complex $\eta_j$--plane and the phase
$\Pi(x_l,\xi_i,\eta_j)\equiv \Pi(\Si)$ gives rise 
to the corresponding integration contours in the complex $\eta_j$--plane.
Then the amount of terms in the sum \SPLIT\ may be drastically reduced by deforming the contour 
integrals in the complex $\eta_j$--planes.
This procedure is equivalent to finding relations among the partial amplitudes
$A(1,\Si(1),\ldots,\Si(N_o+2N_c))$ and express the sum \SPLIT\ in terms of a minimal basis.
Stated differently, identities for higher open string amplitudes may
be generated by deforming and relating the complex
integrals appearing in \AMPLITUDE\ or \SPLIT. This concept is pursued in Section 4.

The positions of the open string points $x_l$ relative to the point $\xi_j$
determine the contour in the $\eta_j$--plane around these points.
Similarly, the positions of the closed string points $\xi_i$ 
relative to the point $\xi_j$
determine the contour in the $\eta_j$--plane around the points $\eta_j=\eta_i$, 
\cf next Figure.
\ifig\below{Contours at $\eta_j=x_l$ and $\eta_j=\eta_i$ in the complex 
$\eta_j$--plane.}
{\epsfxsize=0.75\hsize\epsfbox{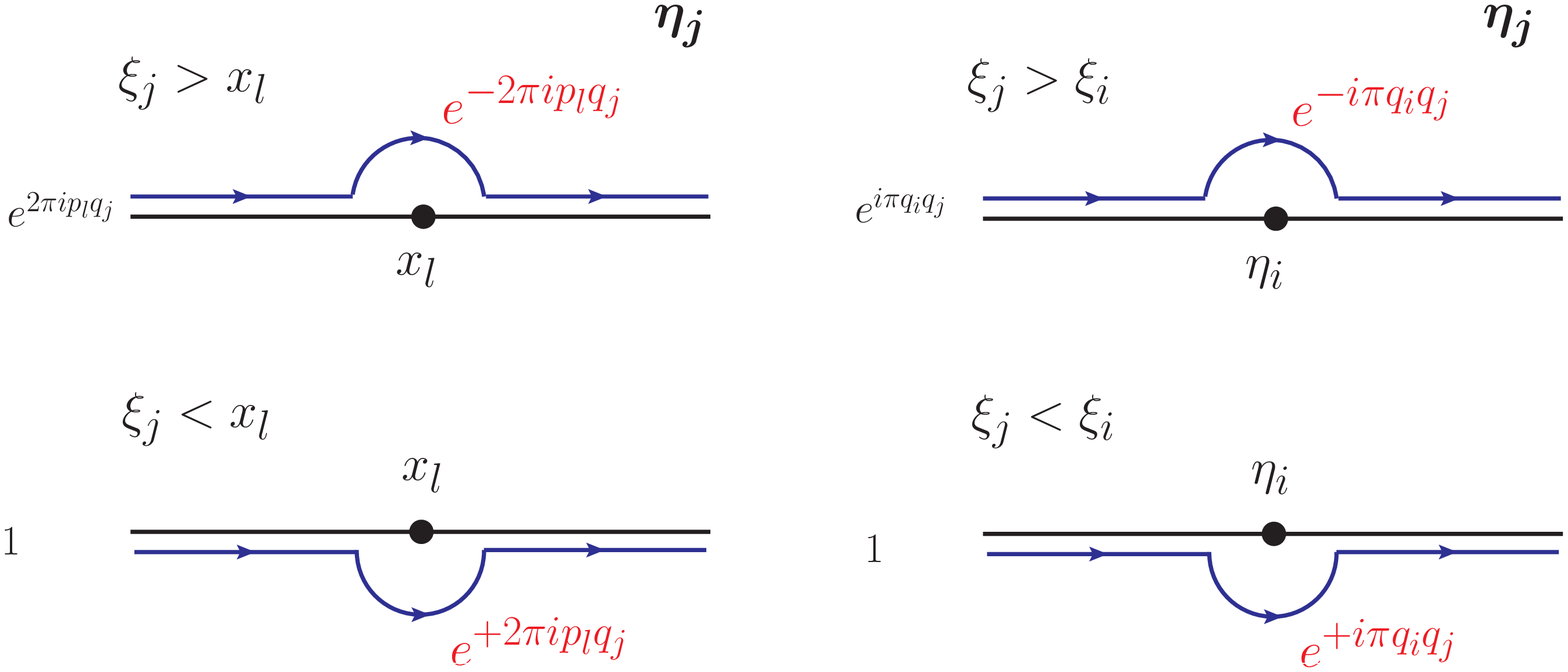}} 
\noindent
Analyzing the phases \PHASE\ related to the polynomials $(x_l-\xi_j)^{2p_lq_j}$ and 
$(x_l-\eta_j)^{2p_lq_j}$  it follows, 
that in the complex $\eta_j$--plane the points $\eta_j=x_l$  are encircled above
for $\xi_j>x_l$ and below for $\xi_j<x_l$. 
Similarly, analyzing the phases \PHASE\ related the polynomials $(\xi_i-\xi_j)^{q_iq_j}$ and $(\eta_i-\eta_j)^{q_iq_j}$  it follows, that the points $\eta_j=\eta_i$ are encircled above for $\xi_j>\xi_i$ and below for $\xi_j<\xi_i$.
Hence, the point $\eta_i=\eta_j$ is always avoided by the contours 
in the complex $\eta_i,\eta_j$--planes. 
In contrast to the closed string case on the sphere \Kawai\ in \eqq \AMPLITUDE\ the 
$\xi_i$ and $\eta_j$ integrations do not decouple due to the mixed polynomials
$(\eta_i-\xi_j)^{q_iDq_j},\ (\xi_i-\eta_j)^{q_iDq_j}$ and $(\xi_i-\eta_i)^{2q_i^2}$.
The polynomials $(\eta_i-\xi_j)^{q_iDq_j},\ (\xi_i-\eta_j)^{q_iDq_j}$ give rise to 
additional branchings in the complex $\eta_j$--plane.
For $\xi_j>\eta_i$ in the $\eta_j$--plane the point $\eta_j=\xi_i$ is avoided
above and for $\xi_j<\eta_i$ the latter is avoided below, \cf next Figure.
\ifig\above{Contours at $\eta_j=\xi_i$ in the complex $\eta_j$--plane.}
{\epsfxsize=0.35\hsize\epsfbox{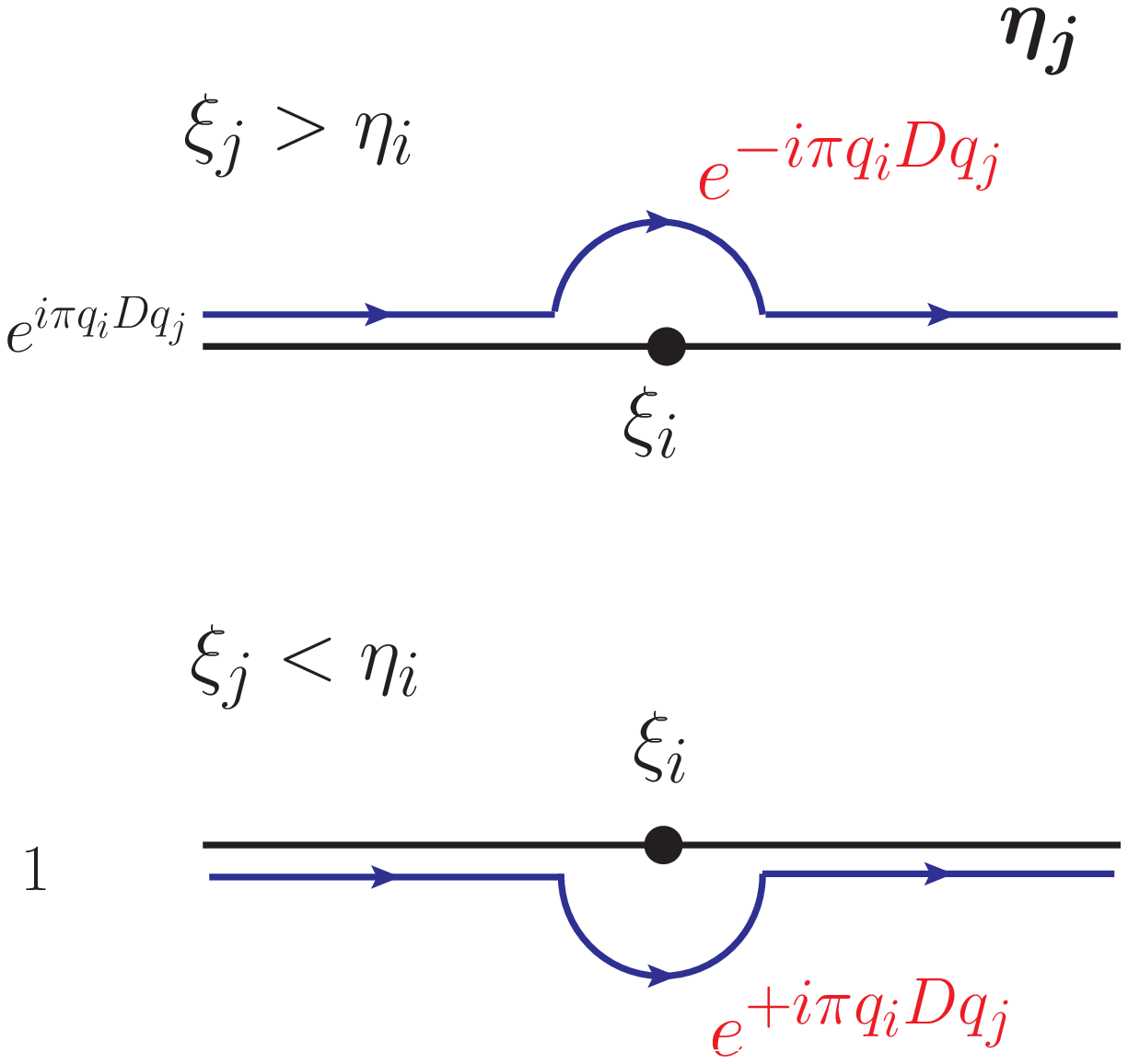}} 
\noindent
Furthermore, the polynomials $(\xi_j-\eta_j)^{2q_{j\parallel}^2}$ implying 
the difference of the two closed string points $\xi_j$ and $\eta_j$
give rise to additional branchings in the complex $\eta_j$--plane. Hence
an additional contour  appears  at the points $\eta_j=\xi_j$ 
in the complex $\eta_j$--plane, \cf next Figure.
\ifig\middl{Contour at $\eta_j=\xi_j$ in the complex $\eta_j$--plane.}
{\epsfxsize=0.375\hsize\epsfbox{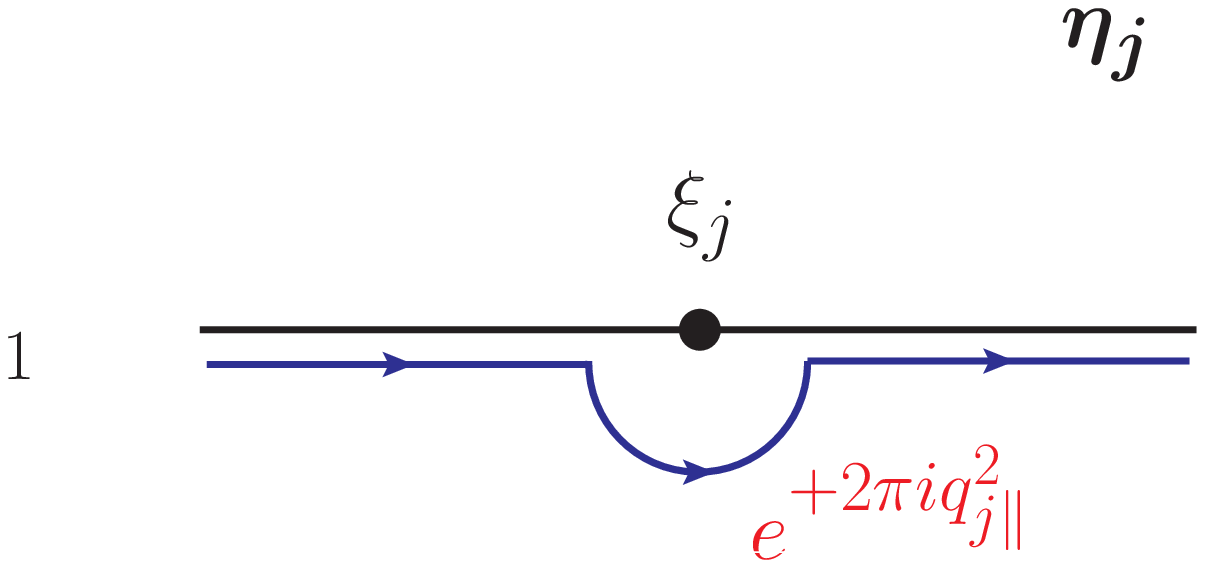}} 
\noindent

Let us consider the contour in the complex $\eta_j$--plane, with 
$j\in\{1,\ldots,N_c\}$.
If for this $j$ the following inequalities holds
\eqn\condi{
\xi_j<x_{l}\ \ \ ,\ \ \ \xi_j<\eta_i,\ \xi_i\ \ \ ,\ \ \ i\in\{1,\ldots,\hatt j,\ldots,N_c\}\ ,}
in the complex $\eta_j$--plane all the contours avoid the points $\eta_j=x_l,\xi_i,\xi_j,\eta_i$ below and hence they can be deformed away to infinity. 
Therefore in that case, there  is no contribution to the sum \SPLIT, see the next Figure.
\ifig\nulle{Contour in the complex $\eta_j$--plane.}
{\epsfxsize=0.65\hsize\epsfbox{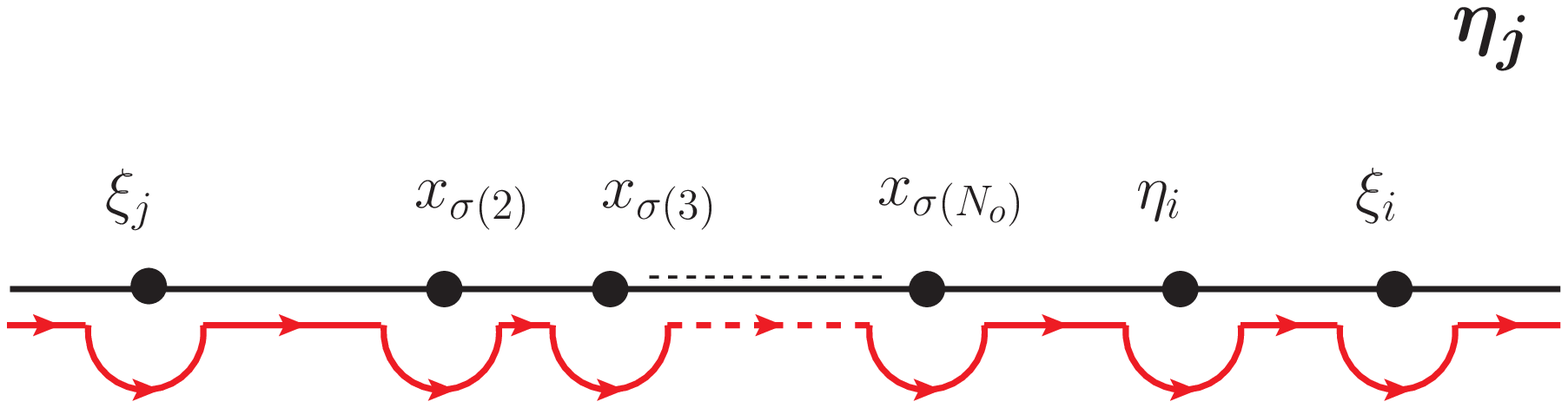}} 
\noindent
In the sum \SPLIT\ the contour in \nulle\ sums up $N_o+2N_c-1$ subamplitude contributions, which give a vanishing contribution.
By changing the positions $\eta_i,\xi_i$ such that \condi\ stays fulfilled, 
we find more vanishing contributions to \SPLIT. Furthermore, by repeating 
this analysis in different complex $\eta$--planes further vanishing contributions
are discovered.

On the other hand, if \condi\ is not met, there a contributions
to \SPLIT\ from non--vanishing contours in all complex $\eta$--planes. 
Their analysis in the complex $\eta$--planes 
is quite tedious due to the  additional branchings described in \above\ and \middl.
An illustrative example is shown in the next Figure.
We consider the integration region $x_{\si(2)}<\xi_j<\xi_k<x_{\si(3)}<\ldots<x_{\si(N_o)}<\xi_i$, $\xi_i<\eta_i$ and $\eta_j<\eta_k<\eta_i$.
The contour in the $\eta_j$--plane can be deformed to give a contribution from
$-\infty<\eta_j<x_{\si(2)}$, the contour in the $\eta_i$--plane can be deformed
to give a contribution from $\xi_i<\eta_i<\infty$, while the contour in the 
$\eta_k$--plane can be deformed to give a contribution for the range
$\eta_j<\eta_k<\xi_j$:
\eqn\totalfinal{
\sin(2\pi p_{\si(2)}q_j)\ \sin(2\pi q_{i\parallel}^2)\ 
\lf\{\ \sin(\pi q_kq_j)\ A(\Si_1)+\sin[\pi(q_kq_j+2q_kp_{\si(2)})]\ A(\Si_2)\  \ri\}\ .}
\ifig\nulle{Contours in the complex $\eta_j,\eta_i$ and $\eta_k$--planes.}
{\epsfxsize=0.76\hsize\epsfbox{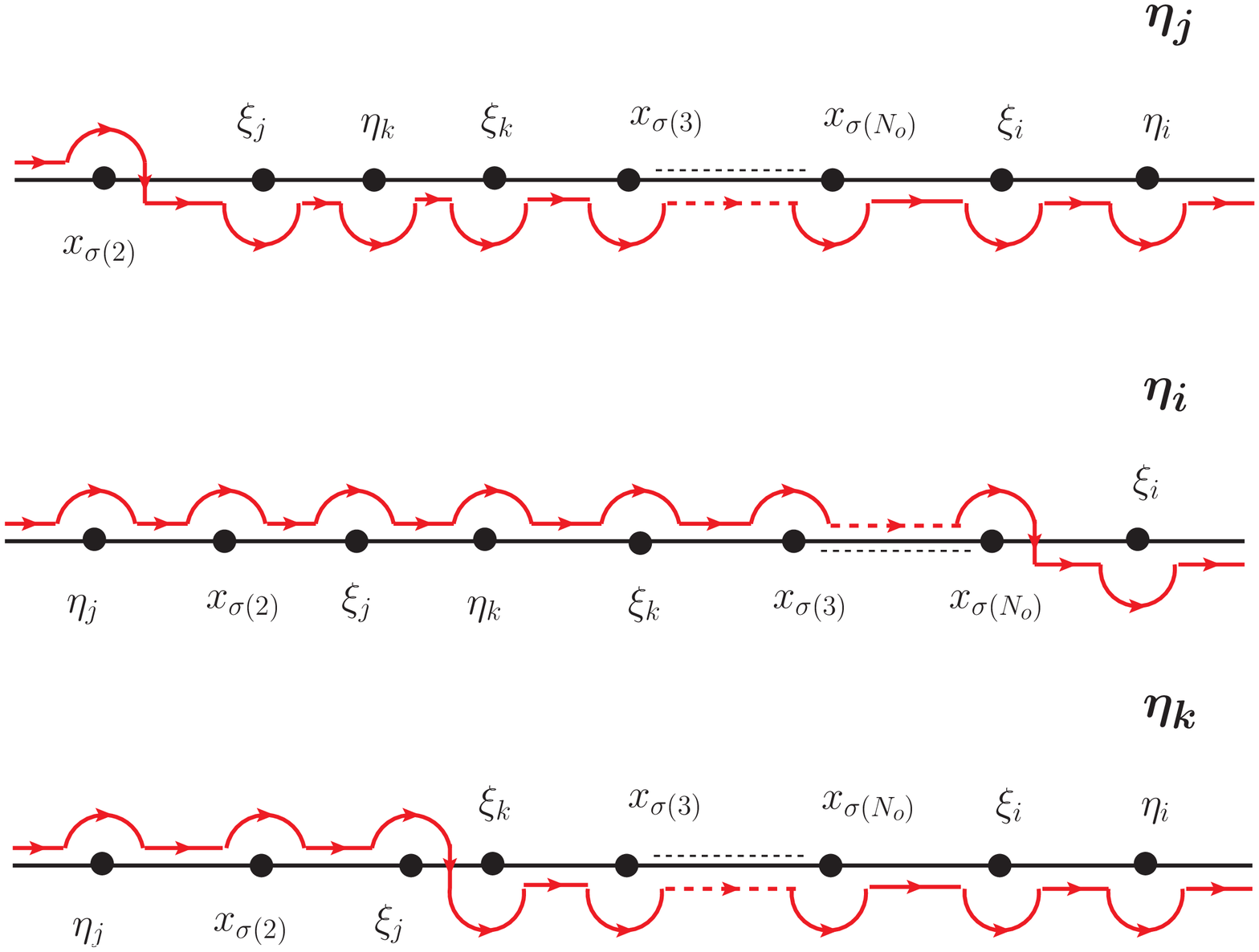}}
\noindent
Here, the integrals $A(\Si)$ represent open string subamplitudes refering
to the following ordering $\Si$ of the $N_o+2N_c$ open string positions:
\eqn\withordering{\eqalign{
\Si_1&:-\infty<\eta_j<x_{\si(2)}<\eta_k<\xi_j<\xi_k<x_{\si(3)}<\ldots<x_{\si(N_o)}<
\xi_i<\eta_i<\infty\ ,\cr
\Si_2&:-\infty<\eta_j<\eta_k<x_{\si(2)}<\xi_j<\xi_k<x_{\si(3)}<\ldots<x_{\si(N_o)}<
\xi_i<\eta_i<\infty\ .}}
The piece \totalfinal\ comprises into the sum \SPLIT.

From the previous discussion it is evident, how to treat the general case $(N_o,N_c)$
and how to reduce the sum \SPLIT.
For the four cases $(N_o,N_c)=(3,1), (2,2), (4,1)$ and 
$(0,3)$ the complex splitting \AMPLITUDE\ and the phase \PHASE\ give rise to the 
phase structure displayed in Table~1 of Appendix \appA, Tables 2,3 of 
Appendix \appB, Tables 4,5 of Appendix \appC, and Table 6 of Appendix \appD, respectively.
The corresponding contours in the complex $\eta$--plane are displayed in the
figures attached there.
To outline the methods described above in the following we shall explicitly
work out the six cases $(N_o,N_c)=(2,1), (3,1), (2,2), (4,1), (0,3)$, and~$(3,2)$.

\subsec{Two open $\&$ one closed string versus four open strings on the disk}

In this Subsection we establish the relation between 
a disk amplitude with two open $\&$ one closed string, given in \firenze, to 
a disk amplitude of four open strings.
On the other hand, the general expression of a (partial ordered) four open string amplitude is 
given by the Gaussian hypergeometric function
\eqn\fouropen{\eqalign{
A(1,2,3,4)&:=
\int_0^1 dx\ (2x)^{\hatt s_1+\hatt n_1}\ (1+x)^{\hatt s_2+\hatt n_2}\ (1-x)^{\hatt s_3+\hatt n_3}\cr
&=2^{\hatt s_1+\hatt n_1}\ B(\hatt s_1+\hatt n_1+1,\hatt s_3+\hatt n_3+1)\ \FF{2}{1}\lf[{\hatt s_1+\hatt n_1+1,-\hatt s_2-\hatt n_2\atop 2+\hatt s_1+\hatt s_3+\hatt n_1+\hatt n_3+2};-1\ri]\ ,}}
with the three integers $\hatt n_i\in\IZ$, the three kinematic invariants $\hatt s_i=\ap(k_i+k_{i+1})^2$, 
subject to the cyclic identification
$i+4\equiv i$ and the momenta $k_i$ of the four external open strings. 
The three invariants $\hatt s_i$, which fulfill $\hatt s_1+\hatt s_2+\hatt s_3=0$ 
are related to the invariants \INV\ as follows:
$\hatt s_1=\hatt s_{12}=4k_1k_2,\ \hatt s_2=\hatt s_{13}=4k_1k_3,\ \hatt s_3=\hatt s_{14}=4k_1k_4$.
The representation \fouropen\ refers to the partial ordering $(1234)$ of the
four open string vertex operators  along the boundary of the disk. 
Furthermore, we have the constraint $\hatt n_1+\hatt n_2+\hatt n_3+2=0$.
This allows to write \fouropen\ as follows:
\eqn\Fouropen{
A(1,2,3,4)=\h\ 
\int_0^1 dx\ x^{\hatt s_1+\hatt n_1}\ (1-x)^{\hatt s_3+\hatt n_3}=\h\ 
B(\hatt s_1+\hatt n_1+1\ ,\ \hatt s_3+\hatt n_3+1)\ .}
On the other hand, after some manipulations involving relations between Gamma--functions, the amplitude \firenze\ involving two open and one closed string may be cast into:
\eqn\look{
\Ac(1,2;3)=-\h\ \sin(\pi t)\ B\lf(s-2n_1-1\ ,\ t+n_1+1\ri)\ .}
In \look\ the Euler beta function may be considered as world--sheet integral \Fouropen\ 
describing a (partial ordered) four open string disk amplitude. More precisely
the latter, which originates from \fouropen\ corresponds to the choice of vertex operator positions 
\eqn\Posigluoni{
z_1=-\infty ,\ z_2=1\ ,\ z_3=-x\ ,\ z_4=x\ ,}
with $-1<x<0$.
With the identifications
\eqn\consiv{\eqalign{
\hatt s_1&=s\ \ \ ,\ \ \ \hatt s_2=\hatt s_3=t\ ,\cr
\hatt n_1&=-2-2n_1\ \ \ ,\ \ \ \hatt n_2=\hatt n_3=n_1\ ,}}
\ie $\hatt s_1=-2\hatt s_2$, the two Euler beta functions in \Fouropen\ and \look\ match and we have the following relation
\eqn\Obtainiv{
\mathboxit{
\Ac(1,2;3)=-\sin(\pi t)\ A(1,2,3,4)}}
between the amplitude $\Ac(1,2,3)$ of two open and one closed string and an amplitude involving four open strings $A(1,2,3,4)$.
According to \momenta\ and \inviv\ the relations \consiv\ are fulfilled by the 
following assignment of the four open string momenta:
\eqn\assigniv{
k_1=p_1\ ,\ k_2=p_2\ ,\ k_3=\h\ Dq\ ,\ k_4=\h\ q\ .}

Finally, the alternative expression \Firenze\ may be cast into a sum of four open string amplitudes:
\eqn\obtain{
\Ac(1,2;3)=\fc{i}{2}\ \lf\{\ e^{i\pi t}\ A(1,4,3,2)+A(1,3,2,4)+e^{i\pi t}\ A(1,3,4,2)\ \ri\}\ .}
The equivalence of the two expressions \Obtainiv\ and  \obtain\ can be shown by using the
relations
\eqn\UCSB{
A(1,3,2,4)=-2\ \cos(\pi t)\ A(1,3,4,2)\ \ \ ,\ \ \ A(1,3,4,2)=A(1,4,3,2)\ ,}
which are a consequence of \KITP. See also Subsection 4.3.1.

\subsec{Three open $\&$ one closed string versus five open strings on the disk}

In this Subsection we establish the relation between 
a disk amplitude with three open $\&$ one closed string to 
a disk amplitude of five open strings.
The generic expression of a disk amplitude $\Ac(1,2,3;4)$ 
involving three open and closed strings is given in \boilv\ and \Provee.
On the other hand, the general expression of a (partial ordered) five 
open string amplitude is 
given by the Gaussian hypergeometric function
\doubref\Dan\STii 
\eqn\fiveopen{
A(1,2,3,4,5)=
\int_0^1 dx \int_0^1 dy\ x^{\hatt s_2+n_1}\ y^{\hatt s_5+n_2}\ 
(1-x)^{\hatt s_3+n_3}\ (1-y)^{\hatt s_4+n_4}\ (1-xy)^{\hatt s_1-\hatt s_3-\hatt s_4+n_5}\ ,}
with the momenta $k_i$ of the five external open strings and the five
kinematic invariants $\hatt s_i=\ap(k_i+k_{i+1})^2$, subject to the cyclic 
identification $i+5\equiv i$.
For $\ap=2$ the five invariants $\hatt s_i$ can be related to \INV\ as follows:
\eqn\Invv{
\hatt s_1=\hatt s_{12}\ ,\ \hatt s_2=\hatt s_{23}\ ,\ \hatt s_3=\hatt s_{34}\ ,\ 
\hatt s_4=\hatt s_{45}\ ,\ \hatt s_5=\hatt s_{51}\ .}
The above representation refers to the partial ordering $(12345)$ of the
five open string vertex operators  along the boundary of the disk.

Our task is to express \Provee\ as a sum over integrals of the type
\fiveopen. By writing $z=\xi+i\eta$ and deforming the 
$\eta$--integration contour the integral \Provee\ may be written as an
integral with $z$ and $\ov z$ treated as independent real variables, \cf
Appendix \appA. With this information the expression \Provee\ may be cast 
into the form
\eqn\Obtainv{
G^{(\al)}\lf[{\lambda_1,\gamma_1\atop\lambda_2,\gamma_2}\ri]=
\sin(\pi\hatt\lambda_2)\ A(1,5,2,4,3)+\sin(\pi\hatt\alpha)\ A(1,2,3,4,5)\ ,}
with the two integrals: 
\eqn\Ints{\eqalign{
A(1,5,2,4,3):&=(-1)^{n_2}
\int_0^1 d\xi\ \int_{-\infty}^0 d\eta\ \xi^{\lambda_1}\ (1-\xi)^{\gamma_1}\ 
(-\eta)^{\lambda_2}\ (1-\eta)^{\gamma_2}\ (\xi-\eta)^\alpha\ ,\cr
A(1,2,3,4,5):&=(-1)^{m_1+m_2+\tilde\al}\ 
\int_1^\infty d\xi\ \int_\xi^\infty d\eta\ \xi^{\lambda_1}\ (\xi-1)^{\gamma_1}\ 
\eta^{\lambda_2}\ (\eta-1)^{\gamma_2}\ (\eta-\xi)^\alpha\ .}}
Quantities with a hat refer to the non--integer
parts of the parameter introduced in \withv, \ie
$\al=\hatt\al+\tilde\al,\ \lambda_i=\hatt\lambda_i+n_i,\ \gamma_i=\hatt\gamma_i+m_i$,
with $\hatt\lambda_i=t,\ \hatt\gamma_i=s,\ \hatt\al=2q_\parallel^2$ and 
$n_i,m_i,\tilde\al\in\IZ$. With \Obtainv\ any disk amplitude involving three 
open and one closed string \startwithv\ can be written as: 
\eqn\Obtainv{
\mathboxit{
\Ac(1,2,3;4)=
\sin(\pi t)\ A(1,5,2,4,3)+\sin(2\pi q_\parallel^2)\ A(1,2,3,4,5)\ .}}

The two real integrals \Ints\ appear as world--sheet integrals 
describing (partial ordered) five open string disk amplitudes.
Indeed, the integrals \Ints\ correspond to partial ordered
amplitudes of a five open string disk amplitude with the 
the choice of vertex operator positions 
\eqn\Posigluoni{
z_1=-\infty\ ,\ z_2=0\ ,\ z_3=1\ ,\ z_4=\xi\ ,\ z_5=\eta\ ,}
with $\xi,\eta\in\IR$.
More precisely, the expressions
\Ints\ describe the following ordering of vertex positions:
\eqn\corresp{\eqalign{
A(1,5,2,4,3):\ \ \ &0<\xi<1\ ,\ -\infty<\eta<0\cr 
&\Longleftrightarrow\ \ \ z_1<z_5<z_2<z_4<z_3\ ,\cr
A(1,2,3,4,5):\ \ \ &1<\xi<\infty\ ,\ \xi<\eta<\infty\cr
&\Longleftrightarrow\ \ \ z_1<z_2<z_3<z_4<z_5\ .}}
According to \momenta\ we have the assignment:
\eqn\assignv{
k_1=p_1\ ,\ k_2=p_2\ ,\ k_3=p_3\ \ \ ,\ \ \ k_4=\h\ q\ \ \ \ ,\ \ \ k_5=\h\ Dq\ .}
As a consequence from \invv\ and \momenta\ we have
\eqn\consv{
\hatt s_1=2\ s+2\ q_\parallel^2\ \ ,\ \ \hatt s_2=2\ u+2\ q_\parallel^2\ \ ,\ \ \hatt s_3=s\ \ ,\ \ \hatt s_4=2\ q_\parallel^2\ \
,\ \ \hatt s_5=u\ ,}
with $\hatt s_1+\hatt s_2=2\hatt s_3+2\hatt s_4+2\hatt s_5$.

Finally after some coordinate transformations 
the two partial amplitudes \Ints\ may be brought into the canonical
form \fiveopen, subject to the choice \consv. Indeed, with 
\eqn\transi{
\xi\ra\fc{(1-x)\ y}{1-xy}\ \ \ ,\ \ \ \eta\ra-\fc{x\ y}{1-xy}}
the integral $A(15243)$ of \Ints\ becomes:
\eqn\Assumeintei{\eqalign{
A(1,5,2,4,3)&=(-1)^{n_2}\ \int_0^1 dx \int_0^1  dy\ x^{\lambda_2}\ 
y^{\lambda_1+\lambda_2+\al+1}\ (1-x)^{\lambda_1}\ (1-y)^{\gamma_1}\cr
&\times (1-xy)^{-\lambda_1-\lambda_2-\gamma_1-\gamma_2-\alpha-3}\cr
&=(-1)^{n_2}\ \fc{\Gamma(1+\lambda_1)\ \Gamma(1+\lambda_2)\ \Gamma(1+\gamma_1)\ 
\Gamma(\lambda_1+\lambda_2+\alpha+2)}{\Gamma(2+\lambda_1+\lambda_2)\ 
\Gamma(\lambda_1+\lambda_2+\gamma_1+\alpha+3)}\cr
&\times\FF{3}{2}\lf[{1+\lambda_2\ ,\ 2+\alpha+\lambda_1+\lambda_2\ ,\ 
3+\alpha+\lambda_1+\lambda_2+\gamma_1+\gamma_2\atop
2+\lambda_1+\lambda_2\ ,\ 3+\alpha+\lambda_1+\lambda_2+
\gamma_1}\ri]\ .}}
On the other hand, with the transformations
\eqn\transii{
\xi\ra\fc{1}{x}\ \ \ ,\ \ \ \eta\ra\fc{1}{xy}}
the integral $A(12345)$ of \Ints\ assumes the form:
\eqn\Assumeinteii{\eqalign{
A(1,2,3,4,5)
&=(-1)^{m_1+m_2+\tilde\al}\ 
\int_0^1 dx \int_0^1 dy\ x^{-\lambda_1-\lambda_2-\gamma_1-\gamma_2-\alpha-3}\ 
y^{-\lambda_2-\gamma_2-\alpha-2}\ (1-x)^{\gamma_1}\cr  
&\times (1-y)^{\alpha}\ (1-xy)^{\gamma_2}\cr
&\hskip-2cm=(-1)^{m_1+m_2+\tilde\al}\ \fc{\Gamma(1+\alpha)\ \Gamma(1+\gamma_1)\
\Gamma(-1-\alpha-\lambda_2-\gamma_2)\ 
\Gamma(-2-\alpha-\lambda_1-\lambda_2-\gamma_1-\gamma_2)}{\Gamma(-\lambda_2-\gamma_2)
\ \Gamma(-1-\alpha-\lambda_1-\lambda_2-\gamma_2)}\cr
&\times\FF{3}{2}\lf[{-\gamma_2,-1-\alpha-\lambda_2-\gamma_2\ ,\
-2-\alpha-\lambda_1-\lambda_2-\gamma_1-\gamma_2\atop
-\lambda_2-\gamma_2\ ,\ -1-\alpha-\lambda_1-\lambda_2-\gamma_2}\ri]\ .}}
Eventually, with \withv\ the subamplitudes \Assumeintei\ and \Assumeinteii\
reduce to:
\eqn\assummetheform{\eqalign{
A(1,5,2,4,3)&=(-1)^{n_2}\ \int_0^1 dx \int_0^1  dy\ x^{t+n_2}\ 
y^{2t+\alpha+1+n_1+n_2}\ (1-x)^{t+n_1}\ (1-y)^{s+m_1}\cr 
&\times (1-xy)^{2u-\alpha-3-n_1-n_2-m_1-m_2}\ ,\cr
A(1,2,3,4,5)&=(-1)^{m_1+m_2+\tilde\al}\ 
\int_0^1 dx \int_0^1 dy\ x^{2u-\alpha-3-n_1-n_2-m_1-m_2}\ 
y^{u-\alpha-2-n_2-m_2}\cr 
&\times (1-x)^{s+m_1}\ (1-y)^{\alpha}\ (1-xy)^{s+m_2}\ .}}

\subsec{Two open $\&$ two closed strings versus six open strings on the disk}

In this Subsection we establish the relation between 
a disk amplitude with two open $\&$ two closed strings to 
a disk amplitude of six open strings.
The generic expression of a disk amplitude $\Ac(1,2;3,4)$
involving two open and two closed strings is given in \boilvi\ and \GEN.
On the other hand, the general expression of a (partial ordered) 
six open string amplitude is 
given by the generalized Euler integral \doubref\Dan\STii 
\eqn\sixopen{\eqalign{
A(1,2,3,4,5,6)&=
\int_0^1 dx \int_0^1 dy \int_0^1 dz\
x^{\hatt s_2}\ y^{\hatt t_2}\ z^{\hatt s_6}\
(1-x)^{\hatt s_3}\ (1-y)^{\hatt s_4}\ (1-z)^{\hatt s_5}\cr
&\times(1-xy)^{\hatt t_3-\hatt s_3-\hatt s_4}\ (1-yz)^{\hatt t_1-\hatt s_4-\hatt
s_5}\ (1-xyz)^{\hatt s_1+\hatt s_4-\hatt t_1-\hatt t_3}\ ,}}
with the six momenta $k_i$ of the six external open strings and the  nine
kinematic invariants $\hatt s_i=\ap(k_i+k_{i+1})^2$ and $\hatt
t_j=\ap(k_j+k_{j+1}+k_{j+2})^2$, subject to the cyclic identification $i+6\equiv i$.
For $\ap=2$ the nine invariants $\hatt s_i$ can be related to \INV\ as follows:
\eqn\Invvi{\eqalign{
\hatt s_1&=\hatt s_{12}\ ,\ \hatt s_2=\hatt s_{23}\ ,\ \hatt s_3=\hatt s_{34}\ ,\ 
\hatt s_4=\hatt s_{45}\ ,\ \hatt s_5=\hatt s_{56}\ ,\ \hatt s_6=\hatt s_{61}\ ,\cr 
\hatt t_1&=\hatt s_{12}+\hatt s_{23}+\hatt s_{13}\ ,\ 
\hatt t_2=\hatt s_{23}+\hatt s_{24}+\hatt s_{34}\ ,\ 
\hatt t_3=\hatt s_{12}+\hatt s_{26}+\hatt s_{16}\ .}}
The above representation refers to the partial ordering $(123456)$ of the
six open string vertex operators  along the boundary of the disk. 
The integrals \sixopen\ integrate to triple hypergeometric functions. 
The latter belong to the class of multiple Gaussian hypergeometric functions \Dan.

Our task is to express \GEN\ as a sum over integrals of the type
\sixopen. This is achieved by converting the complex integration in $z$
into two real integrals by splitting the complex $z$--integral up into holomorphic and
anti--holomorphic contour integrals. In addition, the $x$--integration along the real axis is converted
to an integration along the imaginary axis by analytic continuation of the integrand, given
in \eqq \GENN.
This procedure is performed explicitly in the Appendix \appB\ and allows to cast
the expression \GENN\ into the form
\eqn\Obtain{\eqalign{
W^{(\kappa,\al_0)}&\lf[{\al_1,\lambda_1,\gamma_1,\beta_1\atop
\al_2,\lambda_2,\gamma_2,\beta_2}\ri]=2^{-\al_0}\ i\cr
&\hskip0.4cm\times \Big\{\ e^{i\pi(\hatt\gamma_2+\hatt\beta_2)}\ 
\lf[\ \sin(\pi\hatt\gamma_2)\ A(1,6,3,5,4,2)+\sin(\pi\hatt\beta_2)\ A(1,6,4,5,3,2)\ \ri]\cr
&\hskip0.4cm-\sin(\pi\hatt\lambda_2)\lf[\ e^{i\pi\hatt\gamma_2}\ A(1,3,5,4,2,6)+
A(1,3,4,5,2,6)\ri.\cr
&\hskip0.4cm\lf.+ e^{i\pi\hatt\beta_2}\ A(1,4,5,3,2,6)+A(1,4,3,5,2,6)\ \ri]\cr
&\hskip0.4cm +\sin(\pi\hatt\beta_2)\ A(1,3,5,2,4,6)+
\sin(\pi\hatt\gamma_2)\ A(1,6,3,2,5,4)\ \Big\}+R\ ,}}
with the set of five integrals (\cf Appendix B.1 and B.2)
\eqn\Is{\eqalign{
A(1,6,3,5,4,2)&=2^{\al_0}\ (-1)^{n_3+n_4+n_6}\ 
\int_{0}^1 d\rho\ \int_{-\rho}^\rho d\xi\int_{-\infty}^{-\rho} 
d\eta\ \rho^{\al_0}\ (1+\rho)^{\alpha_1}\ (1-\rho)^{\alpha_2}\cr 
&\times(1-\xi)^{\lambda_1}\ (1-\eta)^{\lambda_2}\ (\rho-\xi)^{\gamma_1}\ 
(-\rho-\eta)^{\gamma_2}\ (\rho+\xi)^{\beta_1}\ (\rho-\eta)^{\beta_2}\ 
(\xi-\eta)^\kappa\ ,\cr
A(1,3,5,4,2,6)&=2^{\al_0}\ (-1)^{n_2+n_3+\kappa}\ 
\int_{0}^1 d\rho\ \int_{-\rho}^\rho d\xi\int_{1}^{\infty} 
d\eta\ \rho^{\al_0}\ (1+\rho)^{\alpha_1}\ (1-\rho)^{\alpha_2}\cr 
&\times(1-\xi)^{\lambda_1}\ (\eta-1)^{\lambda_2}\ (\rho-\xi)^{\gamma_1}\ 
(\rho+\eta)^{\gamma_2}\ (\rho+\xi)^{\beta_1}\ (\eta-\rho)^{\beta_2}\ 
(\eta-\xi)^\kappa\ ,\cr
A(1,3,4,5,2,6)&=2^{\al_0}\ (-1)^{n_2+\kappa}\ \int_0^1 d\rho\ \int_{\rho}^1 d\xi\int_{1}^{\infty} 
d\eta\ \rho^{\al_0}\ (1+\rho)^{\alpha_1}\ (1-\rho)^{\alpha_2}\cr 
&\times(1-\xi)^{\lambda_1}\ (\eta-1)^{\lambda_2}\ (\xi-\rho)^{\gamma_1}\ 
(\rho+\eta)^{\gamma_2}\ (\rho+\xi)^{\beta_1}\ (\eta-\rho)^{\beta_2}\
(\eta-\xi)^\kappa\ ,\cr
A(1,3,5,2,4,6)&=2^{\al_0}\ (-1)^{m_0+n_2+n_3+\kappa}\ 
\int_{1}^\infty d\rho\ \int_{-\rho}^1 d\xi\int_\rho^\infty 
d\eta\ \rho^{\al_0}\ (1+\rho)^{\alpha_1}\ (\rho-1)^{\alpha_2}\cr 
&\times(1-\xi)^{\lambda_1}\ (\eta-1)^{\lambda_2}\ (\rho-\xi)^{\gamma_1}\ 
(\rho+\eta)^{\gamma_2}\ (\rho+\xi)^{\beta_1}\ (\eta-\rho)^{\beta_2}\ (\eta-\xi)^\kappa\ ,\cr
A(1,6,3,2,5,4)&=2^{\al_0}\ (-1)^{m_0+n_1+n_3+n_4+n_6}\ 
\int_{1}^\infty d\rho\ \int_1^\rho d\xi\int_{-\infty}^{-\rho} 
d\eta\ \rho^{\al_0}\ (1+\rho)^{\alpha_1}\ (\rho-1)^{\alpha_2}\cr 
&\times(\xi-1)^{\lambda_1}\ (1-\eta)^{\lambda_2}\ (\rho-\xi)^{\gamma_1}\ 
(-\rho-\eta)^{\gamma_2}\ (\rho+\xi)^{\beta_1}\ (\rho-\eta)^{\beta_2}\ (\xi-\eta)^\kappa\ ,}}
and the three integrals (\cf Appendix B.3):
\eqn\Iss{\eqalign{
A(1,6,4,5,3,2)&=2^{\al_0}\ (-1)^{n_4+n_5+\al_0}\ 
\int_{-1}^0 d\rho\ \int^{-\rho}_\rho d\xi\int_{-\infty}^{\rho} 
d\eta\ (-\rho)^{\al_0}\ (1+\rho)^{\alpha_1}\ (1-\rho)^{\alpha_2}\cr 
&\times(1-\xi)^{\lambda_1}\ (1-\eta)^{\lambda_2}\ (\xi-\rho)^{\gamma_1}\ 
(-\rho-\eta)^{\gamma_2}\ (-\rho-\xi)^{\beta_1}\ (\rho-\eta)^{\beta_2}\ 
(\xi-\eta)^\kappa\ ,\cr
A(1,4,5,3,2,6)&=2^{\al_0}\ (-1)^{n_2+n_5+n_6+\kappa+\al_0}\ 
\int_{-1}^0 d\rho\ \int_\rho^{-\rho} d\xi\int_1^\infty 
d\eta\ (-\rho)^{\al_0}\ (1+\rho)^{\alpha_1}\ (1-\rho)^{\alpha_2}\cr 
&\times(1-\xi)^{\lambda_1}\ (\eta-1)^{\lambda_2}\ (\xi-\rho)^{\gamma_1}\ 
(\rho+\eta)^{\gamma_2}\ (-\rho-\xi)^{\beta_1}\ (\eta-\rho)^{\beta_2}\ 
(\eta-\xi)^\kappa\ ,\cr
A(1,4,3,5,2,6)&=2^{\al_0}\ (-1)^{n_2+\kappa+\al_0}\ 
\int_{-1}^0 d\rho\ \int_{-\rho}^1 d\xi\int_{1}^{\infty} 
d\eta\ (-\rho)^{\al_0}\ (1+\rho)^{\alpha_1}\ (1-\rho)^{\alpha_2}\cr 
&\times(1-\xi)^{\lambda_1}\ (\eta-1)^{\lambda_2}\ (\xi-\rho)^{\gamma_1}\ 
(\rho+\eta)^{\gamma_2}\ (\rho+\xi)^{\beta_1}\ (\eta-\rho)^{\beta_2}\
(\eta-\xi)^\kappa\ .}}
Furthermore, in \Obtain\ there may be a contribution from a  residuum $R$ contributing 
in the case $\al\leq -1$, \cf Appendix \appB.

The eight real integrals \Is\ and \Iss\ appear as world--sheet integrals 
describing (partial ordered) six open string disk amplitudes.
Indeed, the expressions \Is\ correspond to partial ordered
amplitudes of a six open string disk amplitude with the following 
choice of vertex operator positions 
\eqn\posigluoni{
z_1=-\infty\ \ ,\ \ z_2=1\ \ ,\ \ z_3=-\rho\ \ ,\ \ z_4=\rho\ \ ,\ \
z_5=\xi\ \ ,\ \ z_6=\eta\ ,}
with $\xi,\eta,\rho\in\IR$. More precisely, the expressions \Is\ describe the
following ordering of vertex positions
\eqn\gregioni{\eqalign{
A(1,6,3,5,4,2):\ \ \ &0<\rho<1\ ,\ -\rho<\xi<\rho\ ,\ -\infty<\eta<-\rho\cr
&\Longleftrightarrow\ \ \ z_1<z_6<z_3<z_5<z_4<z_2\ ,\cr
A(1,3,5,4,2,6):\ \ \ &0<\rho<1\ ,\ -\rho<\xi<\rho\ ,\ 1<\eta<\infty\cr
&\Longleftrightarrow\ \ \ z_1<z_3<z_5<z_4<z_2<z_6\ ,\cr
A(1,3,4,5,2,6):\ \ \ &0<\rho<1\ ,\ \rho<\xi<1\ ,\ 1<\eta<\infty\cr
&\Longleftrightarrow\ \ \ z_1<z_3<z_4<z_5<z_2<z_6\ ,\cr
A(1,3,5,2,4,6):\ \ \ &1<\rho<\infty\ ,\ -\rho<\xi<1\ ,\ \rho<\eta<\infty\cr 
&\Longleftrightarrow\ \ \ z_1<z_3<z_5<z_2<z_4<z_6\ ,\cr
A(1,6,3,2,5,4):\ \ \ &1<\rho<\infty\ ,\ 1<\xi<\rho\ ,\ -\infty<\eta<-\rho\cr 
&\Longleftrightarrow\ \ \ z_1<z_6<z_3<z_2<z_5<z_4\ ,}}
and:
\eqn\gregionii{\eqalign{
A(1,6,4,5,3,2):\ \ \ &-1<\rho<0\ ,\ \rho<\xi<-\rho\ ,\ -\infty<\eta<\rho\cr
&\Longleftrightarrow\ \ \ z_1<z_6<z_4<z_5<z_3<z_2\ ,\cr
A(1,4,5,3,2,6):\ \ \ &-1<\rho<0\ ,\ \rho<\xi<-\rho\ ,\ 1<\eta<\infty\cr
&\Longleftrightarrow\ \ \ z_1<z_4<z_5<z_3<z_2<z_6\ ,\cr
A(1,4,3,5,2,6):\ \ \ &-1<\rho<0\ ,\ -\rho<\xi<1\ ,\ 1<\eta<\infty\cr
&\Longleftrightarrow\ \ \ z_1<z_4<z_3<z_5<z_2<z_6\ .}}
According to \momenta\ we have the assignment:
\eqn\assignvi{
k_1=p_1\ ,\ k_2=p_2\ \ \ ,\ \ \ 
k_3=\h\ q_1\ \ \ ,\ \ \ k_4=\h\ q_1\ \ \ \ ,\ \ \ k_5=\h\ q_2\ \ \ ,\ \ \
k_6=\h\ q_2\ .}
As a consequence from \invvi\ and \momenta\ we have:
\eqn\consvi{
\hatt s_1=2\ s\ \ ,\ \ \hatt s_2=\hatt s_6=u\ \ ,\ \ \hatt s_3=0\ \ ,\ \ 
\hatt s_5=0\ \ ,\ \ 
\hatt s_4=\fc{s}{2}\ \ ,\ \ \hatt t_1=\hatt t_3=s\ \ ,\ \ \hatt t_2=2\ u\ .}
The specific choice \assignvi\ of the six lightlike external momenta $k_i$ is
shown in the next Figure.
The polygon in Fig. 9 specifies the scattering configuration of six open
strings. We will add some comments on this configuration in Section 5.
\iifig\Polygon{Polygon associated to the lightlike momenta}{of a six open 
string scattering.}
{\epsfxsize=0.4\hsize\epsfbox{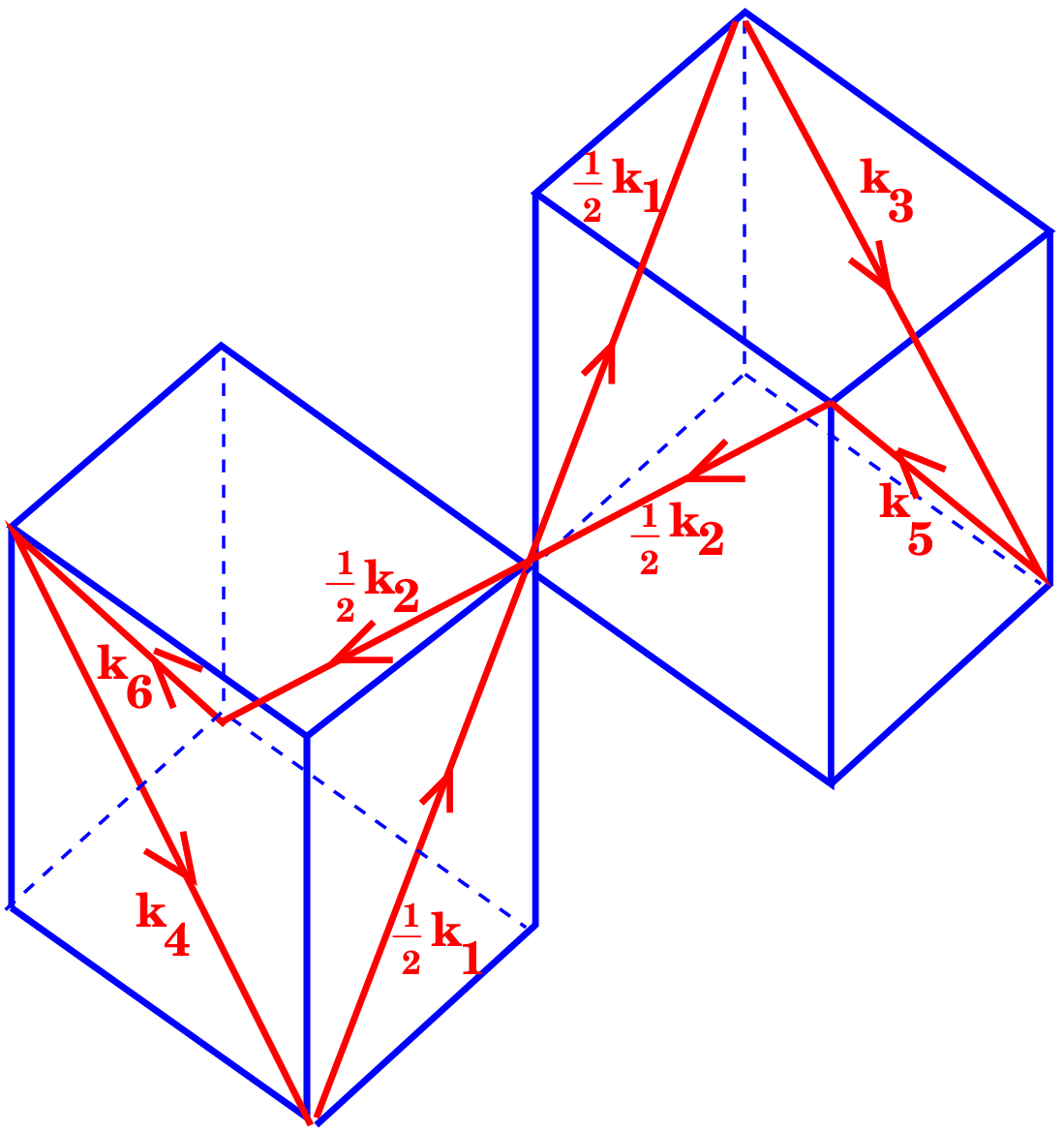}}

Not all of the eight integrals \Is\ and \Iss\ are independent. Indeed, after partial integration some relations may be found, \eg:
\eqn\partialrel{\eqalign{
&\hskip-1cm\sin(\pi s)\ [\ A(1,6,3,5,4,2)-  A(1,6,4,5,3,2)\ ]\cr
&=\sin(\pi t)\ [\ A(1,3,5,4,2,6)-A(1,4,5,3,2,6)\ ]\ ,\cr
&\hskip-1cm\sin\lf(\fc{\pi s}{2}\ri)\ [\ A(1,3,5,2,4,6)+A(1,6,3,2,5,4)\ ]\cr
&-\sin(\pi t)\ [\ A(1,3,4,5,2,6)+A(1,4,3,5,2,6)\ ]\cr
&=\cos\lf(\fc{\pi s}{2}\ri)\ \sin(\pi t)\ [\ A(1,3,5,4,2,6)+A(1,4,5,3,2,6)\ ]\cr
&-\sin\lf(\fc{\pi s}{2}\ri)\ \cos(\pi s)\ [\ A(1,6,3,5,4,2)+A(1,6,4,5,3,2)\ ]\ .}}
These two relations allow to simplify the result \Obtain\ for the relevant case \consvi.
This allows to cast the final result for the two open and two closed string disk amplitude into the following form
\eqn\Obtaini{\hskip-0.25cm
\mathboxit{
\Ac(1,2;3,4)=2\ \sin\lf(\fc{\pi s}{2}\ri)\ \sin(\pi s)\ A(1,6,3,5,4,2)-
2\ \sin\lf(\fc{\pi s}{2}\ri)\ \sin(\pi t)\ A(1,3,5,4,2,6)\ ,}} 
with the two partial six open string amplitudes $A(1,2,4,5,3,6)$ and $A(1,3,5,4,2,6)$
given in \Is. An alternative expression for \Obtaini\ is:
\eqn\ALTEROB{
\Ac(1,2;3,4)=2\ \sin\lf(\fc{\pi s}{2}\ri)\ \sin(\pi s)\ A(1,4,5,3,6,2)-
2\ \sin\lf(\fc{\pi s}{2}\ri)\ \sin(\pi u)\ A(1,5,3,6,2,4)\ .}
Comparing the two results \Obtaini\ and \ALTEROB\ 
makes manifest the symmetry of the amplitude $\Ac(1,2;3,4)$ under exchanging the labels $3$ and $4$, \ie permuting the two closed strings.

Finally, after some coordinate transformations the partial amplitudes amplitudes 
$A(1,2,4,5,3,6)$ and $A(1,3,5,4,2,6)$ may be brought into the 
canonical form \sixopen, subject to the choice \consvi. Indeed with 
\eqn\transi{
\rho\ra\fc{1-yz}{1+yz}\ \ \ ,\ \ \ \xi\ra1-\fc{2y}{1+yz}\ \ \ ,\ \ \ 
         \eta\ra1-\fc{2}{x\ (1+yz)}}
for the condition \conditionvi\ the integral $A(163542)$ of \Is\ becomes:         
\eqn\assumeintei{\eqalign{
A(1,6,3,5,4,2)&=\h\ (-1)^{n_3+n_4+n_6}\ \int_0^1 dx \int_0^1 dy \int_0^1 dz\  
x^{-2-\kappa-\lambda_2-\gamma_2-\beta_2}\  (1-x)^{\gamma_2}\cr  
&\times y^{1+\al_2+\lambda_1+\gamma_1}\ (1-y)^{\bet_1}\ 
z^{\al_2}\ (1-z)^{\gamma_1}\ (1-xy)^{\kappa}\ (1-yz)^{\al_0}\ (1-xyz)^{\beta_2}}}
On the other hand, with the transformations
\eqn\transii{
\rho\ra\fc{y}{2-y}\ \ \ ,\ \ \ \xi\ra-\fc{y\ (1-2x)}{2-y}\ \ \ ,\ \ \ 
\eta\ra\fc{2-yz}{z\ (2-y)}}
for the condition \conditionvi\ the integral $A(135426)$ of \Is\ assumes the form:
\eqn\assumeinteii{\eqalign{
A(1,3,5,4,2,6)&=(-1)^{n_2+n_3+\kappa}\ \h\int_0^1 dx \int_0^1 dy \int_0^1 dz\  
x^{\bet_1}\  (1-x)^{\gamma_1}\ y^{1+\al_0+\gamma_1+\bet_1}\cr 
&\times (1-y)^{\al_2}\ z^{-2-\lambda_2-\gamma_2-\bet_2-\kappa} \
(1-z)^{\lambda_2}\ (1-xy)^{\lambda_1}\ (1-yz)^{\beta_2}\ .}}         
With the choice \parameter\ these two integrals \assumeintei\ and \assumeinteii\ 
down to the expressions:
\eqn\assumeintei{\eqalign{
A(1,6,3,5,4,2)&=-\h\ (-1)^{n_3+n_4+n_6}\ \int_0^1 dx \int_0^1 dy \int_0^1 dz\ 
x^{u-\kappa-2-n_2-n_4-n_6}\ (1-x)^{\fc{s}{2}+n_4}\cr 
&\times y^{-\fc{s}{2}+1+m_0+n_1+n_3}\ (1-y)^{\fc{s}{2}+n_5}\ 
z^{u+m_0}\ (1-z)^{\fc{s}{2}+n_3}\ (1-xy)^\kappa\cr
&\times (1-yz)^{\al_0}\ (1-xyz)^{\fc{s}{2}+n_6}\ ,\cr
A(1,3,5,4,2,6)&=\h\ (-1)^{n_2+n_3+\kappa}\ \int_0^1 dx \int_0^1 dy \int_0^1 dz\ x^{\fc{s}{2}+n_5}\ 
(1-x)^{\fc{s}{2}+n_3}\ y^{s+1+\al_0+n_3+n_5}\cr 
&\times(1-y)^{u+m_0}\
z^{u-2-\kappa-n_2-n_4-n_6}\ (1-z)^{t+n_2}\ (1-xy)^{t+n_1}\ (1-yz)^{\fc{s}{2}+n_6}\ .}}

In superstring amplitudes the complex integral \GEN, whose final result is
given in \Obtain\ and \Obtaini, typically exhibits double
or single poles in the kinematic invariants $s,t$ and $u$.
{\it E.g.} the integral
\eqn\typically{
W^{(0,0)}\lf[{u-1,t-1,\fc{s}{2},\fc{s}{2}\atop u-1,t-1,\fc{s}{2},\fc{s}{2}}\ri]=
\int_{-\infty}^\infty dx\ |1-ix|^{2u-2}\ 
\int_\IC d^2z \ |1-z|^{2t-2}\ |z-ix|^{s}\ |z+ix|^{s}}
has poles for $z,\ov z\ra 1$.  With \Obtaini\ we find:
\eqn\Typically{
W^{(0,0)}\lf[{u-1,t-1,\fc{s}{2},\fc{s}{2}\atop u-1,t-1,\fc{s}{2},\fc{s}{2}}\ri]=
2\pi^2\ \lf(\ \fc{1}{t}+\fc{1}{u}+ \zeta(3)\ s^2+\fc{1}{4}\ \zeta(4)\ s^3\ 
\ri)+\ldots\ .}
The first term stems\foot{This may be deduced from: $\int_{|z|<\eps}d^2z\ 
|z|^{-2+k_ik_j}=\fc{2\pi}{k_ik_j}$ for $|k_ik_j|\ll-(\ln\eps)^{-1}$  \Minahan.} from the limit  $z\ra 1$ in the
integrand. 
According to \SYMMETRY\ the integral \typically\ is invariant under the
exchange of $t$ and $u$. This behaviour is manifest in the expansions \Typically.
On the other hand, there also appear world--sheet integrals without poles 
in the kinematic invariants $s,t$ and $u$.
{\it E.g.} the integral
\eqn\typicallyi{
W^{(0,0)}\lf[{u,t,\fc{s}{2}-1,\fc{s}{2}-1\atop u,t,\fc{s}{2}-1,\fc{s}{2}-1}\ri]=
\int_{-\infty}^\infty dx\ |1-ix|^{2u}\ 
\int_\IC d^2z \ |1-z|^{2t}\ |z-ix|^{s-2}\ |z+ix|^{s-2}}
has possible poles for $z\ra \pm ix$, which are cancelled after performing
the remaining $x$--integration. With \Obtaini\ we find:
\eqn\Typicallyi{\eqalign{
W^{(0,0)}\lf[{u,t,\fc{s}{2}-1,\fc{s}{2}-1\atop u,t,\fc{s}{2}-1,\fc{s}{2}-1}\ri]&=-4\pi^2\ \lf[\ 1+s+s^2+\fc{1}{4}\ \zeta(2)\ (t^2+3\ tu+u^2)\ \ri]+\ldots\ .}}
According to \SYMMETRY\ the integral \typicallyi\ is invariant under the
exchange of $t$ and $u$. Again, this behaviour is manifest in the expansions \Typicallyi.

\subsec{Four open $\&$ one closed string versus six open strings on the disk}

In this Subsection we establish the relation between a disk amplitude with four open \& one closed 
string to a disk amplitude of six open strings. 
The generic expression of a (color ordered) 
disk amplitude $\Ac(1,\si(2),\si(3),\si(4);5)$ involving four open and one closed string is given in \boilvii\ 
and \GENHU. On the other hand, the 
general expression of a (partial ordered) six open string amplitude is given in \sixopen.

Our task is to express \GENHU\ as a sum over integrals of the type
\sixopen. This is achieved by converting the complex integration
into two real integrals by splitting the complex $z$--integral up into holomorphic and
anti--holomorphic contour integrals. 
This procedure is performed in the Appendix \appC. Depending on the range of the real parameter $x$
we obtain three integrals of the type \sixopen. Eventually, we may cast
the expressions for the three partial amplitudes \GENHU\ into the following form
\eqn\ObtainVII{
\mathboxit{
\eqalign{
W_{\si_1}^{(\kappa,\al_0)}\lf[{\al_1,\lambda_1,\gamma_1,\beta_1\atop
\al_2,\lambda_2,\gamma_2,\beta_2}\ri]&=
\sin(\pi\hatt\bet_2)\ A(1,6,3,5,4,2)+\sin(\pi\hatt\kappa)\ A(1,3,4,2,5,6)\cr
&+\sin(\pi\hatt\kappa)\ A(1,3,4,5,6,2)+\sin[\pi(\hatt\kappa+\hatt\lambda_2)]\ 
A(1,3,4,5,2,6)\ ,\cr
W_{\si_2}^{(\kappa,\al_0)}\lf[{\al_1,\lambda_1,\gamma_1,\beta_1\atop
\al_2,\lambda_2,\gamma_2,\beta_2}\ri]&=\sin(\pi\hatt\gamma_2)\ 
A(1,6,4,5,3,2)+\sin(\pi\hatt\kappa)\ A(1,4,3,2,5,6)\cr
&+\sin(\pi\hatt\kappa)\ A(1,4,3,5,6,2)+\sin[\pi(\hatt\kappa+\hatt\lambda_2)\ 
A(1,4,3,5,2,6)\ ,\cr
W_{\si_3}^{(\kappa,\al_0)}\lf[{\al_1,\lambda_1,\gamma_1,\beta_1\atop
\al_2,\lambda_2,\gamma_2,\beta_2}\ri]&=\sin(\pi\hatt\bet_2)\ 
A(1,6,3,5,2,4)+\sin(\pi\hatt\kappa)\ A(1,3,2,4,5,6)\cr
&+\sin(\pi\hatt\kappa)\ A(1,3,2,5,6,4)+\sin[\pi(\hatt\kappa+\hatt\gamma_2)]\ 
A(1,3,2,5,4,6)\ ,}}}
with the four integrals for $\si_1$
\eqn\IsH{\eqalign{
A(1,6,3,5,4,2)&=2^{\al_0}\ (-1)^{n_3+n_4+n_6}\ \int_0^1 dx\int_{-x}^x d\xi\int_{-\infty}^{-x} 
d\eta\ x^{\al_0}\ (1+x)^{\alpha_1}\ (1-x)^{\alpha_2}\cr 
&\times(1-\xi)^{\lambda_1}\ (1-\eta)^{\lambda_2}\ (x-\xi)^{\gamma_1}\ 
(x-\eta)^{\gamma_2}\ (\xi+x)^{\beta_1}\ (-x-\eta)^{\beta_2}\ 
(\xi-\eta)^\kappa\ ,\cr
A(1,3,4,5,2,6)&=2^{\al_0}\ (-1)^{n_2+\tilde\kappa}\ \int_0^1 dx\int_{x}^1 d\xi\int_1^\infty 
d\eta\ x^{\al_0}\ (1+x)^{\alpha_1}\ (1-x)^{\alpha_2}\cr 
&\times(1-\xi)^{\lambda_1}\ (\eta-1)^{\lambda_2}\ (\xi-x)^{\gamma_1}\ 
(\eta-x)^{\gamma_2}\ (\xi+x)^{\beta_1}\ (\eta+x)^{\beta_2}\ (\eta-\xi)^\kappa\ ,\cr
A(1,3,4,5,6,2)&=2^{\al_0}\ (-1)^{\tilde\kappa}\ \int_0^1 dx\int_{x}^1 d\xi\int_\xi^1 
d\eta\ x^{\al_0}\ (1+x)^{\alpha_1}\ (1-x)^{\alpha_2}\cr 
&\times(1-\xi)^{\lambda_1}\ (1-\eta)^{\lambda_2}\ (\xi-x)^{\gamma_1}\ 
(\eta-x)^{\gamma_2}\ (\xi+x)^{\beta_1}\ (\eta+x)^{\beta_2}\ (\eta-\xi)^\kappa\ ,\cr
A(1,3,4,2,5,6)&=2^{\al_0}\ (-1)^{n_1+n_2+\tilde\kappa}\ \int_0^1 dx\ \int\limits_1^\infty d\xi\ \int\limits_\xi^\infty d\eta \ x^{\al_0}\ (1+x)^{\alpha_1}\ (1-x)^{\alpha_2}\cr 
&\times(\xi-1)^{\lambda_1}\ (\xi-x)^{\gamma_1}\ (\xi+x)^{\beta_1}\ (\eta-1)^{\lambda_2}\ (\eta-x)^{\gamma_2}\ (\eta+x)^{\beta_2}\ 
(\eta-\xi)^\kappa\ ,}}
the four integrals for $\si_2$
$$\eqalign{
A(1,6,4,5,3,2)&=2^{\al_0}\ (-1)^{n_4+n_5+n_6+m_0}\ 
\int_{-1}^0 dx\int^{-x}_x d\xi\int_{-\infty}^x 
d\eta\ (-x)^{\al_0}\ (1+x)^{\alpha_1}\ (1-x)^{\alpha_2}\cr 
&\times(1-\xi)^{\lambda_1}\ (1-\eta)^{\lambda_2}\ (\xi-x)^{\gamma_1}\ (x-\eta)^{\gamma_2}\ 
(-x-\xi)^{\beta_1}\ (-x-\eta)^{\beta_2}\ (\xi-\eta)^\kappa\ ,\cr
A(1,4,3,5,2,6)&=2^{\al_0}\ (-1)^{n_2+m_0+\tilde\kappa}\ 
\int_{-1}^0 dx\int_{-x}^1 d\xi\int_1^\infty 
d\eta\ (-x)^{\al_0}\ (1+x)^{\alpha_1}\ (1-x)^{\alpha_2}\cr 
&\times(1-\xi)^{\lambda_1}\ (\eta-1)^{\lambda_2}\ (\xi-x)^{\gamma_1}\ (\eta-x)^{\gamma_2}\ 
(\xi+x)^{\beta_1}\ (\eta+x)^{\beta_2}\ (\eta-\xi)^\kappa\ ,}$$
\eqn\IsHi{\eqalign{
A(1,4,3,5,6,2)&=2^{\al_0}\ (-1)^{m_0+\tilde\kappa}\ 
\int_{-1}^0 dx\int_{-x}^1 d\xi\int_\xi^1 
d\eta\ (-x)^{\al_0}\ (1+x)^{\alpha_1}\ (1-x)^{\alpha_2}\cr 
&\times(1-\xi)^{\lambda_1}\ (1-\eta)^{\lambda_2}\ (\xi-x)^{\gamma_1}\ (\eta-x)^{\gamma_2}\ 
(\xi+x)^{\beta_1}\ (\eta+x)^{\beta_2}\ (\eta-\xi)^\kappa\ ,\cr
A(1,4,3,2,5,6)&=2^{\al_0}\ (-1)^{n_1+n_2+m_0+\tilde\kappa}\ \int_{-1}^0 dx
\int\limits_1^\infty d\xi\ \int\limits_\xi^\infty d\eta\ (-x)^{\al_0}\ (1+x)^{\alpha_1}\ (1-x)^{\alpha_2}\cr 
&\times (\xi-1)^{\lambda_1}\ (\xi-x)^{\gamma_1}\ (\xi+x)^{\beta_1}\ 
(\eta-1)^{\lambda_2}\ (\eta-x)^{\gamma_2}\ (\eta+x)^{\beta_2}\ 
(\eta-\xi)^\kappa\ ,}}
the four integrals for $\si_3$:
\eqn\IsHii{\eqalign{
A(1,6,3,5,2,4)&=2^{\al_0}\ (-1)^{n_3+n_4+n_6+m_2}\ 
\int_1^\infty dx\int_{-x}^1 d\xi\int_{-\infty}^{-x} 
d\eta\ x^{\al_0}\ (1+x)^{\alpha_1}\ (x-1)^{\alpha_2}\cr 
&\times(1-\xi)^{\lambda_1}\ (1-\eta)^{\lambda_2}\ (x-\xi)^{\gamma_1}\ 
(x-\eta)^{\gamma_2}\ (\xi+x)^{\beta_1}\ (-x-\eta)^{\beta_2}\
(\xi-\eta)^\kappa\ ,\cr
A(1,3,2,5,4,6)&=2^{\al_0}\ (-1)^{n_1+n_2+n_3+m_2+\tilde\kappa}\ 
\int_{1}^\infty dx\int_1^x d\xi\int_x^\infty 
d\eta\ x^{\al_0}\ (1+x)^{\alpha_1}\ (x-1)^{\alpha_2}\cr 
&\times(\xi-1)^{\lambda_1}\ (\eta-1)^{\lambda_2}\ (x-\xi)^{\gamma_1}\ 
(\eta-x)^{\gamma_2}\ (\xi+x)^{\beta_1}\ (\eta+x)^{\beta_2}\ (\eta-\xi)^\kappa\ ,\cr
A(1,3,2,5,6,4)&=2^{\al_0}\ (-1)^{n_1+n_2+n_3+n_4+m_2+\tilde\kappa}\ 
\int_{1}^\infty dx\int_1^x d\xi\int_x^\infty 
d\eta\ x^{\al_0}\ (1+x)^{\alpha_1}\ (x-1)^{\alpha_2}\cr 
&\times(\xi-1)^{\lambda_1}\ (\eta-1)^{\lambda_2}\ (x-\xi)^{\gamma_1}\ 
(x-\eta)^{\gamma_2}\ (\xi+x)^{\beta_1}\ (\eta+x)^{\beta_2}\ (\eta-\xi)^\kappa\ ,\cr
A(1,3,2,4,5,6)&=2^{\al_0}\ (-1)^{n_1+n_2+m_2+\tilde\kappa}\ 
\int_{1}^\infty dx\int\limits_x^\infty d\xi\ \int\limits_\xi^\infty d\eta\
x^{\al_0}\ (1+x)^{\alpha_1}\ (x-1)^{\alpha_2}\cr 
&\times(\xi-1)^{\lambda_1}\ (\xi-x)^{\gamma_1}\ (\xi+x)^{\beta_1}\  (\eta-1)^{\lambda_2}\ (\eta-x)^{\gamma_2}\ (\eta+x)^{\beta_2}\ (\eta-\xi)^\kappa\ .}}
Note, that the quantities with a hat are the non--integer parts of the
variables \parameterHU. In addition, we have introduced 
$\kappa=\hatt\kappa+\tilde\kappa$, with $\hatt\kappa=2\ q_\parallel^2$, \cf also Appendix \appC.

The twelve real integrals \IsH, \IsHi\ and \IsHii\ appear as world--sheet integrals 
describing (partial ordered) six open string disk amplitudes.
Indeed, the expressions \IsH\ correspond to partial ordered
amplitudes of a six open string disk amplitude with the following 
choice of vertex operator positions 
\eqn\posigluoni{
z_1=-\infty\ \ ,\ \ z_2=1\ \ ,\ \ z_3=-x\ \ ,\ \ z_4=x\ \ ,\ \
z_5=\xi\ \ ,\ \ z_6=\eta\ ,}
with $\xi,\eta,x\in\IR$.
In the following we shall explicitly work out the case $\kappa\in\IZ$, \ie
$\hatt\kappa=2q_\parallel^2=0$ and $\kappa=\tilde\kappa$. In this case
only the first lines of \ObtainVII\ contribute.
Then expressions \IsH\ describe the following ordering of vertex positions:
\eqn\gregioni{\eqalign{
&\si_1\ \ 
\lf\{\ \ \ \eqalign{A(1,6,3,5,4,2):\ \ \ \ \ &    0<x<1\ ,\ -x<\xi<x\ ,\ -\infty<\eta<-x \cr
                               \ \ \ \ \ & \Longleftrightarrow  z_1<z_6<z_3<z_5<z_4<z_2\ ,\cr
                     A(1,3,4,5,2,6):\ \ \ \ \ &      0<x<1\ ,\ x<\xi<1\ ,\ 1<\eta<\infty  \cr
                               \ \ \ \ \ & \Longleftrightarrow  z_1<z_3<z_4<z_5<z_2<z_6\ ,}\ri.\cr\crr\crr
&\si_2\ \ 
\lf\{\ \ \ \eqalign{A(1,6,4,5,3,2):\ \ \ \ \ &   -1<x<0\ ,\ x<\xi<-x\ ,\ -\infty<\eta<x\cr
                               \ \ \ \ \ & \Longleftrightarrow z_1<z_6<z_4<z_5<z_3<z_2\ ,\cr
                     A(1,4,3,5,2,6):\ \ \ \ \ &   -1<x<0\ ,\ -x<\xi<1\ ,\ 1<\eta<\infty\ ,\cr
                               \ \ \ \ \ & \Longleftrightarrow z_1<z_4<z_3<z_5<z_2<z_6\ ,}\ri.\cr\crr\crr
&\si_3\ \ 
\lf\{\ \ \ \eqalign{A(1,6,3,5,2,4):\ \ \ \ \ &1<x<\infty\ ,\ -x<\xi<1\ ,\ -\infty<\eta<-x\cr
                                         &-\infty<x<-1\ ,\ 1<\xi<-x\ ,\ -x<\eta<\infty\cr
                                         &\Longleftrightarrow z_1<z_6<z_3<z_5<z_2<z_4\ ,\cr
                     A(1,3,2,5,4,6):\ \ \ \ \ &1<x<\infty\ ,\ 1<\xi<x\ ,\ x<\eta<\infty\cr 
                                         &-\infty<x<-1\ ,\ x<\xi<1\ ,\ -\infty<\eta<x\cr
                                         &\Longleftrightarrow z_1<z_3<z_2<z_5<z_4<z_6\ .}\ri.}}
According to \momenta\ we have the assignment:
\eqn\assignvii{
k_1=p_1\ ,\ k_2=p_2\ ,\ k_3=p_3\ ,\ k_4=p_4\ \ \ ,\ \ \ 
k_5=\h\ q\ \ \ ,\ \ \ k_6=\h\ q\ .}
As a consequence from \invvii\ and \momenta\ we have:
\eqn\consvii{\eqalign{
\hatt s_1=s_1\ \ ,\ \ \hatt s_2&=s_2\ \ ,\ \ \hatt s_3=s_3\ \ ,\ \ \hatt
s_4=s_4\ \ ,\ \ \hatt s_5=0\ \ ,\ \ \hatt s_6=s_5\ ,\cr 
\hatt t_1&=2\ s_4\ \ ,\ \ \hatt t_2=2\ s_5\ \ ,\ \ 
\hatt t_3=\h\ s_1+\h\ s_3\ .}}

Finally after some coordinate transformations the set of partial amplitudes \IsH\ 
may be brought into the canonical form \sixopen, subject to the choice \consvii.
Indeed, the following transformations
\eqn\transi{\matrix{
&A(1,6,3,5,4,2):&\ds{x\ra-1+\fc{2}{1+yz},}&\ds{\xi\ra1-\fc{2\ y}{1+yz},}
         &\ds{\eta\ra1-\fc{2}{x\ (1+yz)}}\ ,\cr\cr
&A(1,3,4,5,2,6):&\ds{x\ra\fc{xy}{2-xy},}&\ds{\xi\ra\fc{(2-x)\ y}{2-xy},}
         &\ds{\eta\ra\fc{2-xyz}{z\ (2-xy)}}\ ,\cr\cr
&A(1,6,4,5,3,2):&\ds{x\ra1-\fc{2}{1+yz},}&\ds{\xi\ra1-\fc{2\ y}{1+yz},}
         &\ds{\eta\ra1-\fc{2}{x\ (1+yz)}}\ ,}}
$$\hskip-1.25cm\matrix{
&A(1,4,3,5,2,6):&\ds{x\ra-\fc{xy}{2-xy},}&\ds{\xi\ra\fc{(2-x)\ y}{2-xy},}
         &\ds{\eta\ra\fc{2-xyz}{z\ (2-xy)}}\ ,\cr\cr
&A(1,6,3,5,2,4)=A(1,4,2,5,3,6):&\ds{x\ra\fc{1}{1-2\ yz},} &\ds{\xi\ra\fc{1-2\ y}{1-2\ yz},}
         &\ds{\eta\ra-\fc{2-x}{x\ (1-2\ yz)}}\ ,\cr\cr
&A(1,3,2,5,4,6)=A(1,6,4,5,2,3):&\ds{x\ra-\fc{1}{1-2\ xy},}&\ds{\xi\ra\fc{1-2\ y}{1-2\ xy},}
          &\ds{\eta\ra-\fc{2-z}{z\ (1-2\ xy)}}}$$
acting on their integrands bring for the case \conditionvii\ 
the corresponding integrals \IsH\ into the form:
\eqn\assumeintei{\eqalign{
A(1,6,3,5,4,2)&=\h\ (-1)^{n_3+n_4+n_6}\ \int_0^1 dx \int_0^1 dy \int_0^1 dz\  
x^{-2-\kappa-\lambda_2-\gamma_2-\beta_2}\  (1-x)^{\beta_2}\cr  
&\times y^{1+\al_2+\lambda_1+\gamma_1}\ (1-y)^{\bet_1}\ z^{\al_2}\ 
(1-z)^{\gamma_1}\ (1-xy)^{\kappa}\ (1-yz)^{\al_0}\ (1-xyz)^{\gamma_2}\ ,\cr
A(1,3,4,5,2,6)&=\h\ (-1)^{n_2+\tilde\kappa}\ \int_0^1 dx \int_0^1 dy \int_0^1 dz\  
x^{\al_0}\  (1-x)^{\gamma_1}\  y^{1+\al_0+\gamma_1+\bet_1}\ (1-y)^{\lambda_1}\cr 
&\times z^{-2-\kappa-\lambda_2-\gamma_2-\bet_2}\ 
(1-z)^{\lambda_2}\ (1-xy)^{\al_2}\ (1-yz)^\kappa\ (1-xyz)^{\gamma_2}\ ,\cr
A(1,6,4,5,3,2)&=\h\ (-1)^{n_4+n_5+n_6+m_0}\ \int_0^1 dx \int_0^1 dy \int_0^1 dz\  
x^{-2-\kappa-\lambda_2-\gamma_2-\beta_2}\  (1-x)^{\gamma_2}\cr  
&\times y^{1+\al_1+\lambda_1+\beta_1}\ (1-y)^{\gamma_1}\ 
z^{\al_1}\ (1-z)^{\bet_1}\ (1-xy)^{\kappa}\ (1-yz)^{\al_0}\ (1-xyz)^{\bet_2}\ ,\cr
A(1,4,3,5,2,6)&=\h\ (-1)^{n_2+m_0+\tilde\kappa}\ \int_0^1 dx \int_0^1 dy \int_0^1 dz\  
x^{\al_0}\  (1-x)^{\bet_1}\  y^{1+\al_0+\gamma_1+\bet_1}\ (1-y)^{\lambda_1}\cr 
&\times z^{-2-\kappa-\lambda_2-\gamma_2-\bet_2}\ 
(1-z)^{\lambda_2}\ (1-xy)^{\al_1}\ (1-yz)^\kappa\ (1-xyz)^{\beta_2},\cr
A(1,6,3,5,2,4)&=\h\ (-1)^{n_3+n_4+n_6+m_2}\ \int_0^1 dx \int_0^1 dy \int_0^1 dz\  
x^{-2-\kappa-\lambda_2-\gamma_2-\beta_2}\  (1-x)^{\beta_2}\cr
&\times y^{1+\al_2+\lambda_1+\gamma_1}\ (1-y)^{\bet_1}\  z^{\al_2}\ 
(1-z)^{\lambda_1}\ (1-xy)^{\kappa}\ (1-yz)^{\al_1}\ (1-xyz)^{\lambda_2}\ ,\cr
A(1,3,2,5,4,6)&=\h\ (-1)^{n_1+n_2+n_3+m_2+\tilde\kappa}\ 
\int_0^1 dx \int_0^1 dy \int_0^1 dz\  
x^{\al_1}\  (1-x)^{\lambda_1}\  y^{1+\al_1+\lambda_1+\bet_1}\cr 
&\times (1-y)^{\gamma_1}\ z^{-2-\kappa-\lambda_2-\gamma_2-\bet_2}\ 
(1-z)^{\gamma_2}\ (1-xy)^{\al_2}\ (1-yz)^\kappa\ (1-xyz)^{\lambda_2}\ .}}
Recall, that the parameter in \assumeintei\ are listed in  \parameterHU.
Eventually, the final expressions for the amplitude \boilvii\ become
for integer $\kappa$, \ie $ \hatt\kappa=0,\kappa=\tilde\kappa$
\eqn\Obtainvii{\mathboxit{\eqalign{
\Ac(1,3,4,2;5)&=\sin\lf[\pi\lf(\fc{s_1}{2}-\fc{s_3}{2}-s_4\ri)\ri]\ A(1,6,3,5,4,2)\cr
&-\sin\lf[\pi\lf(\fc{s_1}{2}-\fc{s_3}{2}+s_5\ri)\ri]\ A(1,3,4,5,2,6)\ ,\cr
\Ac(1,4,3,2;5)&=\sin(\pi s_4)\ 
A(1,6,4,5,3,2)-\sin\lf[\pi\lf(\fc{s_1}{2}-\fc{s_3}{2}+s_5\ri)\ri]
\ A(1,4,3,5,2,6)\ ,\cr
\Ac(1,4,2,3;5)&=\sin\lf[\pi\lf(\fc{s_1}{2}-\fc{s_3}{2}-s_4\ri)\ri]\ 
A(1,6,3,5,2,4)+\sin(\pi s_4)\ A(1,3,2,5,4,6)\ ,}}}
with the six partial amplitudes given in \assumeintei.
The second amplitude $\Ac(1,4,3,2;5)$ follows from the first amplitude 
$\Ac(1,3,4,2;5)$ by permuting the open string labels $3$ and $4$ and replacing 
$s_2\ra-s_2-s_3+2s_5,\ s_4\ra\fc{s_1}{2}-\fc{s_3}{2}-s_4$. Furthermore, the 
third amplitude $\Ac(1,4,2,3;5)$ is obtained from the second by relabeling the 
open strings $2$ and $3$ and performing $s_1\ra-s_1-s_2+2s_4,\ s_3\ra-s_2-s_3+2s_5$.

\subsec{Three closed strings versus six open strings on the disk}

In this Subsection we establish the relation between a disk amplitude with 
three closed strings to a disk amplitude of six open strings. 
The generic expression of a disk amplitude $\Ac(1,2,3)$ of three closed strings 
is given in \boilto\ and \GENto. On the other hand, the 
general expression of a (partial ordered) six open string amplitude  can be found 
in \sixopen.

Our task is to express \GENto\ as a sum over integrals of the type
\sixopen. This is achieved by converting the complex integration
into two real integrals \GENNTO\ 
by splitting the complex $z$--integral up into holomorphic and
anti--holomorphic contour integrals. 
This procedure is performed in the Appendix \appD. 
The contributions of the four contours $I_a, I_d, I_{c_1}$ and $I_{c_2}$, given in
\eqqs (D.3) and (D.4), give
\eqn\zwischen{\eqalign{
0<x<1:&\ \ \sin(\pi s_4)\ A(1,2,5,4,6,3)-\sin\lf[\pi\lf(\fc{s_1}{2}-\fc{s_3}{2}-\fc{s_5}{2}+s_6\ri)\ri]\ A(1,3,4,6,2,5)\ ,\cr
-1<x<0:&\ \ \sin\lf[\pi\lf(\fc{s_1}{2}-\fc{s_3}{2}-\fc{s_5}{2}-s_4\ri)\ri]\ A(1,2,5,3,6,4)\cr
&-\sin\lf[\pi\lf(\fc{s_1}{2}-\fc{s_3}{2}-\fc{s_5}{2}+s_6\ri)\ri]\ A(1,4,3,6,2,5)}}
as a result of applying the amplitude relations:
\eqn\birkar{\eqalign{
\sin(\pi s_5)\ A(1,3,4,2,6,5)-\sin[\pi(s_4+s_6)]\ A(1,5,2,4,6,3)-\sin(\pi s_6)&\ A(1,5,2,4,3,6)\cr
-\sin\lf[\pi\lf(\fc{s_1}{2}-\fc{s_3}{2}-\fc{s_5}{2}+s_6\ri)\ri]&\ A(1,3,4,6,2,5)=0\ ,\cr\crr
\sin(\pi s_5)\ A(1,4,3,2,6,5)-\sin\lf[\pi\lf(\fc{s_1}{2}-\fc{s_3}{2}-\fc{s_5}{2}-s_4+s_6\ri)\ri]&\ A(1,5,2,3,6,4)\cr
-\sin(\pi s_6)\ A(1,5,2,3,4,6)-\sin\lf[\pi\lf(\fc{s_1}{2}-\fc{s_3}{2}-\fc{s_5}{2}+s_6\ri)\ri]&\ A(1,4,3,6,2,5)=0\ .}}
The terms \zwischen\ combine with the contributions (D.5) 
from the contours $I_{b_1}$ and $I_{b_2}$ into:
\eqn\Johannesburg{\eqalign{
W^{(\kappa,\al_0,\al_3)}&\lf[{\alpha_1,\lambda_1,\gamma_1,\beta_1,\eps_1\atop
\alpha_2,\lambda_2,\gamma_2,\beta_2,\eps_2}\ri]\cr
&=\sin(\pi s_4)\ A(1,2,5,4,6,3)
+\sin\lf[\pi\lf(\fc{s_1}{2}-\fc{s_3}{2}-\fc{s_5}{2}-s_4\ri)\ri]\ A(1,2,5,3,6,4)\cr
&-e^{i\pi s_5}\ \sin\lf[\pi\lf(\fc{s_1}{2}-\fc{s_3}{2}+\fc{s_5}{2}+s_6\ri)\ri]\ 
\lf[\ A(1,3,4,6,2,5)+A(1,4,3,6,2,5)\ \ri] ,\cr
&-e^{\h i\pi(s_1-s_3+s_5)}\ \sin(\pi s_5)\ 
\lf[\ A(1,3,4,6,5,2)+A(1,4,3,6,5,2)\ \ri]\ .}}
Eventually, the amplitude \boilto\
can be expressed in terms of a six--dimensional basis of 
partial ordered six open string amplitudes 
\sixopen\ as:
\eqn\ObtainTO{
\mathboxit{
\eqalign{\Ac(1,2,3)&=\Big\{\ \sin(\pi s_4)\ A(1,3,6,4,5,2)
+\sin\lf[\pi\lf(\fc{s_1}{2}-\fc{s_3}{2}-\fc{s_5}{2}-s_4\ri)\ri]\ A(1,4,6,3,5,2)\cr
&+\sin\lf[\pi\lf(-\fc{s_1}{2}+\fc{s_3}{2}-\fc{s_5}{2}-s_6\ri)\ri]
\ \lf[\ A(1,3,4,6,2,5)+A(1,4,3,6,2,5)\ \ri]\cr
&+\lf(\ \sin\lf[\pi\lf(\fc{s_1}{2}-\fc{s_3}{2}-\fc{s_5}{2}\ri)\ri]-
\sin\lf[\pi\lf(\fc{s_1}{2}-\fc{s_3}{2}+\fc{s_5}{2}\ri)\ri]\ \ri)\cr 
&\hskip2.5cm\times\lf[\ A(1,3,4,6,5,2)+A(1,4,3,6,5,2)\ \ri]\ \Big\}\ 
(-1)^{m_3+n_6+n_8}\ .}}}
In \ObtainTO\ the basis of six integrals (\cf Appendix \appDi\ and \appDii) 
is given by:
\def\inte#1#2#3#4#5#6{\int_{#1}^{#2} dx\int_{#3}^{#4} d\xi\int_{#5}^{#6} d\eta\ }
$$\eqalign{
A(1,3,6,4,5,2)&=2^{1+\al_0+\al_3}\ (-1)^{n_5}
\inte{0}{1}{-x}{x}{x}{1} 
x^{\al_3}\ (1+x)^{1+\al_1}\ (1-x)^{1+\al_2}\cr  
&\times (1-\xi)^{\lambda_1}\ (1+\xi)^{\gamma_1}\ (x-\xi)^{\bet_1}\  (\xi+x)^{\eps_1}\cr
&\times(1-\eta)^{\lambda_2}\ (1+\eta)^{\gamma_2}\ (\eta-x)^{\bet_2}\ 
(\eta+x)^{\eps_2}\ (\xi+\eta)^\kappa\  \cr
A(1,3,4,6,2,5)&=2^{1+\al_0+\al_3}\ (-1)^{n_2+n_5+n_7}
\inte{0}{1}{-1}{-x}{1}{\infty} 
x^{\al_3}\ (1+x)^{1+\al_1}\cr 
&\times (1-x)^{1+\al_2}\ (1-\xi)^{\lambda_1}\ (1+\xi)^{\gamma_1}\ 
(x-\xi)^{\bet_1}\ (-x-\xi)^{\eps_1}\cr 
&\times (\eta-1)^{\lambda_2}\ (1+\eta)^{\gamma_2}\ (\eta-x)^{\bet_2}\ 
(\eta+x)^{\eps_2}\ (\xi+\eta)^\kappa\ ,\cr
A(1,3,4,6,5,2)&=2^{1+\al_0+\al_3}\, (-1)^{n_5+n_7}
\inte{0}{1}{-1}{-x}{-\xi}{1}x^{\al_3}\, (1+x)^{1+\al_1}\, (1-x)^{1+\al_2}\cr 
&\times (1-\xi)^{\lambda_1}\ (1+\xi)^{\gamma_1}\ (x-\xi)^{\bet_1}\ 
(-\xi-x)^{\eps_1}\cr 
&\times (1-\eta)^{\lambda_2}\ (1+\eta)^{\gamma_2}\ (\eta-x)^{\bet_2}\  
(\eta+x)^{\eps_2}\ (\xi+\eta)^\kappa\ ,\cr
A(1,4,6,3,5,2)&=2^{1+\al_0+\al_3}\ (-1)^{m_0+n_7}
\inte{-1}{0}{x}{-x}{-x}{1} 
(-x)^{\al_3}\, (1+x)^{1+\al_1}\cr 
&\times (1-x)^{1+\al_2}\ 
(1-\xi)^{\lambda_1}\ (1+\xi)^{\gamma_1}\ (\xi-x)^{\bet_1}\ 
(-\xi-x)^{\eps_1}\cr 
&\times (1-\eta)^{\lambda_2}\ (1+\eta)^{\gamma_2}\ (\eta-x)^{\bet_2}\ 
(\eta+x)^{\eps_2}\ (\xi+\eta)^\kappa\ ,\cr
A(1,4,3,6,2,5)&=2^{1+\al_0+\al_3}\ (-1)^{m_0+n_2+n_5+n_7}
\inte{-1}{0}{-1}{x}{1}{\infty} 
(-x)^{\al_3}\ (1+x)^{1+\al_1}\cr
&\times  (1-x)^{1+\al_2}\ (1-\xi)^{\lambda_1}\ (1+\xi)^{\gamma_1}\ (x-\xi)^{\bet_1}\ 
(-\xi-x)^{\eps_1}\cr
&\times (\eta-1)^{\lambda_2}\ (1+\eta)^{\gamma_2}\ (\eta-x)^{\bet_2}\  
(\eta+x)^{\eps_2}\ (\xi+\eta)^\kappa\ ,\cr
A(1,4,3,6,5,2)&=2^{1+\al_0+\al_3}\ (-1)^{m_0+n_5+n_7}
\inte{-1}{0}{-1}{x}{-\xi}{1}(-x)^{\al_3}\ (1+x)^{1+\al_1}\cr
&\times (1-x)^{1+\al_2}\ (1-\xi)^{\lambda_1}\ 
(1+\xi)^{\gamma_1}\ (x-\xi)^{\bet_1}\ (-\xi-x)^{\eps_1}}$$
\eqn\ISTO{\eqalign{
&\times (1-\eta)^{\lambda_2}\ (1+\eta)^{\gamma_2}\ (\eta-x)^{\bet_2}\  
(\eta+x)^{\eps_2}\ (\xi+\eta)^\kappa\ .}}
The six real integrals \ISTO\ appear as world--sheet integrals 
describing (partial ordered) six open string disk amplitudes.
Indeed, the expressions \ISTO\  correspond to partial ordered
amplitudes of a six open string disk amplitude with the following 
choice of vertex operator positions 
\eqn\posigluonto{
z_1=-1\ \ ,\ \ z_2=1\ \ ,\ \ z_3=-x\ \ ,\ \ z_4=x\ \ ,\ \
z_5=\eta\ \ ,\ \ z_6=-\xi\ ,}
with $\xi,\eta\in\IR$ and $-1<x<1$.
More precisely, the expressions \ISTO\ describe the
following ordering of vertex positions
\eqn\gregionto{\eqalign{
A(1,3,6,4,5,2):\ \ \ &0<x<1\ ,\ -x<\xi<x\ ,\ x<\eta<1\cr
&\Longleftrightarrow\ \ \ z_1<z_3<z_6<z_4<z_5<z_2\ ,\cr
A(1,3,4,6,2,5):\ \ \ &0<x<1\ ,\ -1<\xi<-x\ ,\ 1<\eta<\infty\cr 
&\Longleftrightarrow\ \ \ z_1<z_3<z_4<z_6<z_2<z_5\ ,\cr
A(1,3,4,6,5,2):\ \ \ &0<x<1\ ,\ -1<\xi<-x\ ,\ -\xi<\eta<1\cr 
&\Longleftrightarrow\ \ \ z_1<z_3<z_4<z_6<z_5<z_2\ ,\cr
A(1,4,6,3,5,2):\ \ \ &-1<x<0\ ,\ x<\xi<-x\ ,\ -x<\eta<1\cr
&\Longleftrightarrow\ \ \ z_1<z_4<z_6<z_3<z_5<z_2\ ,\cr
A(1,4,3,6,2,5):\ \ \ &-1<x<0\ ,\ -1<\xi<x\ ,\ 1<\eta<\infty\cr 
&\Longleftrightarrow\ \ \ z_1<z_4<z_3<z_6<z_2<z_5\ ,\cr
A(1,4,3,6,5,2):\ \ \ &-1<x<0\ ,\ -1<\xi<x\ ,\ -\xi<\eta<1\cr 
&\Longleftrightarrow\ \ \ z_1<z_4<z_3<z_6<z_5<z_2\ .}}
According to \momenta\ we have the assignment:
\eqn\assignvii{
k_1=\h\ D q_1\ ,\ k_2=\h\ q_1\ ,\ k_3=\h\ Dq_2\ ,\ k_4=\h\ q_2\ \ \ ,\ \ \ 
k_5=\h\ Dq_3\ \ \ ,\ \ \ k_6=\h\ q_3\ .}
As a consequence from \invto\ and \INV\ we have:
\eqn\consto{\eqalign{
\hatt s_i&=s_i\ \ ,\ \ i=1,\ldots,6\ ,\cr 
\hatt t_1&=\h\ (s_1- s_3+ s_5)\ \ ,\ \ 
\hatt t_2=\h\ (- s_1+ s_3+ s_5)\ \ ,\ \ 
\hatt t_3=\h\ ( s_1+ s_3-s_5)\ .}}

Eventually, after some coordinate transformations the set of partial amplitudes 
\ISTO\ may be brought into the canonical form \sixopen, subject to the choice \consto. 
\def\intredu{\int_0^1\hskip-0.25cm dx\hskip-0.1cm\int_0^1\hskip-0.25cm dy\hskip-0.1cm\int_0^1\hskip-0.25cm dz}
$$\eqalign{
A(1,3,6,4,5,2)&=\h(-1)^{n_5}\intredu\;  
x^{\bet_1}\; (1-x)^{\eps_1}\;  y^{1+\al_0+\lambda_2+\gamma_2}\; (1-y)^{\beta_2}\cr 
&\times z^{\al_0}\ 
(1-z)^{\lambda_2}\ (1-xy)^{\kappa}\ (1-yz)^{\h(\al_2+\lambda_1-\lambda_2+\beta_1-\beta_2)}\ (1-xyz)^{\gamma_1},\ \cr}$$
\eqn\Durban{\eqalign{
A(1,3,4,6,2,5)&=\h(-1)^{n_2+n_5+n_7}\intredu\; 
x^{-2+\kappa-\al_0-\al_1-\al_2}\;  y^{1+\al_0+\lambda_2+\gamma_2}\;  z^{\gamma_2}\; (1-x)^{\eps_1}\cr 
&\times(1-y)^{\gamma_1}\; 
(1-z)^{\lambda_2}\; (1-xy)^{\h(\al_2+\lambda_1-\lambda_2+\beta_1-\beta_2)}\ (1-yz)^{\kappa}\; (1-xyz)^{\beta_2},\ \cr
A(1,3,4,6,5,2)&=\h(-1)^{n_5+n_7}\intredu\; 
x^{-2+\kappa-\al_0-\al_1-\al_2}\;  (1-x)^{\eps_1}\;  y^{1+\al_0+\lambda_2+\gamma_2}\cr 
&\times (1-y)^{\kappa}\; z^{\al_0}\; 
(1-z)^{\lambda_2}\; (1-xy)^{\beta_2}\; (1-yz)^{\gamma_1}\; (1-xyz)^{\h(\al_2+\lambda_1-\lambda_2+\beta_1-\beta_2)},\cr
A(1,4,6,3,5,2)&=\h(-1)^{m_0+n_7}\intredu\; 
x^{\eps_1}\;  (1-x)^{\bet_1}\;  y^{1+\al_0+\lambda_2+\gamma_2}\; 
(1-y)^{\eps_2}\cr 
&\times z^{\al_0}\; 
(1-z)^{\lambda_2}\; (1-xy)^{\kappa}\; 
(1-yz)^{\h(\al_1-\gamma_1+\gamma_2-\bet_1+\bet_2)}\; (1-xyz)^{\gamma_1},\ \cr
A(1,4,3,6,2,5)&=\h(-1)^{m_0+n_2+n_5+n_7}\intredu\; 
x^{-2+\kappa-\al_0-\al_1-\al_2}\;  (1-x)^{\bet_1}\;  
y^{1+\al_0+\lambda_2+\gamma_2}\cr 
&\times (1-y)^{\gamma_1}\; z^{\gamma_2}\; 
(1-z)^{\lambda_2}\; (1-xy)^{\h(\al_1-\gamma_1+\gamma_2-\beta_1+\beta_2)}\; (1-yz)^{\kappa}\; (1-xyz)^{\eps_2},\cr
A(1,4,3,6,5,2)&=\h(-1)^{m_0+n_5+n_7}\intredu\; 
x^{-2+\kappa-\al_0-\al_1-\al_2}\;  (1-x)^{\bet_1}\;  
y^{1+\al_0+\lambda_2+\gamma_2}\cr 
&\times (1-y)^{\kappa}\;  z^{\al_0}\; 
(1-z)^{\lambda_2}\; (1-xy)^{\eps_2}\; (1-yz)^{\gamma_1}\; 
(1-xyz)^{\h(\al_1-\gamma_1+\gamma_2-\beta_1+\beta_2)},}}
subject to the conditions \conditionto.

Instead of \ObtainTO\ we may also put the amplitude \boilto\ into the form:
\eqn\ObtainTOO{
\eqalign{\Ac(1,2,3)&=\Big\{\ \sin(\pi s_4)\ A(1,2,3,6,4,5)
+\sin\lf[\pi\lf(\fc{s_1}{2}-\fc{s_3}{2}-\fc{s_5}{2}-s_4\ri)\ri]\ A(1,2,4,6,3,5)\cr
&+\sin(\pi s_6)\ \lf[\ A(1,5,2,3,4,6)+A(1,5,2,4,3,6)\ \ri]\cr
&+\lf(\ \sin\lf[\pi\lf(\fc{s_1}{2}-\fc{s_3}{2}-\fc{s_5}{2}\ri)\ri]-
\sin\lf[\pi\lf(\fc{s_1}{2}-\fc{s_3}{2}+\fc{s_5}{2}\ri)\ri]\ \ri)\cr 
&\hskip2.5cm\times\lf[\ A(1,2,3,4,6,5)+A(1,2,4,3,6,5)\ \ri]\ \Big\}\ 
(-1)^{m_3+n_6+n_8}\ .}}
The two expressions \ObtainTO\ and \ObtainTOO\ are manifest symmetric under the 
permutation $3\leftrightarrow4$, with ($\bet_i\leftrightarrow\eps_i,\ \al_1\leftrightarrow\al_2$)
\eqn\permutii{
s_2\ra\h\ (-s_1-s_3+s_5)-s_2\ \ \ ,\ \ \ s_4\ra\h\ (s_1-s_3-s_5)-s_4\ .}
Furthermore, by comparing \ObtainTO\ and \ObtainTOO\ the symmetry under the permutation
$1\leftrightarrow2$, with 
($\lambda_i\leftrightarrow\gamma_i,\ \al_1\leftrightarrow\al_2$)
\eqn\permuti{
s_2\ra\h\ (-s_1-s_3+s_5)-s_2\ \ \ ,\ \ \ s_6\ra\h\ (-s_1+s_3-s_5)-s_6}
can be exhibited.
Finally, the symmetry $5\leftrightarrow6$, with ($\lambda_i\leftrightarrow\gamma_i,\ \bet_i\leftrightarrow\eps_i$)
\eqn\permutiii{
s_4\ra\h\ (s_1-s_3-s_5)-s_4\ \ \ ,\ \ \ s_6\ra\h\ (-s_1+s_3-s_5)-s_6}
can be proven by applying some relations between partial amplitudes.
Moreover, it  can be shown, that the result \ObtainTO\ is 
invariant under the exchange $(12)\leftrightarrow(34)$, with
\eqn\permutiv{
s_1\longleftrightarrow s_3\ \ \ ,\ \ \ s_4\longleftrightarrow s_6\ ,}
and the exchange $(34)\leftrightarrow(56)$, with:
\eqn\permutiv{
s_2\longleftrightarrow s_6\ \ \ ,\ \ \ s_3\longleftrightarrow s_5\ .}

Finally, let us discuss the special case $q_{3\perp}=0$. 
For this case the invariants \invto\ simplify to $s_4=\fc{1}{4}(s_1-s_3),\ s_5=0$ and 
$s_6=\fc{1}{4}(-s_1+s_3)$, \ie the assignments \parameterto\ become:
$\hatt\kappa=0,\ \hatt\lambda_i=\hatt\gamma_i=\fc{1}{4}(-s_1+s_3),\ 
\hatt\bet_i=\hatt\eps_i=\fc{1}{4}(s_1-s_3)$ and $\hatt\lambda_i+\hatt\beta_i=0$.
The four contours giving rise to the integrals 
$I_a,I_{c_2},I_{b_1}$ and $I_{b_2}$ of Appendix \appD\ 
do not contribute, while the two contours corresponding to the integrals 
$I_d$ and $I_{c_1}$ amount to:
\eqn\ObtainTOOOOOOO{\eqalign{
\Ac(1,2,3)=\sin\lf[\fc{\pi}{4}(s_1-s_3)\ri]& 
\lf\{\ A(1,3,6,4,5,2)+A(1,4,6,3,5,2)\ri.\cr
&\lf.-A(1,3,4,6,2,5)-A(1,4,3,6,2,5)\ \ri\}\ ,}}
with the four partial ordered six open string amplitudes $A(1,3,6,4,5,2), 
A(1,4,6,3,5,2),$\ $A(1,3,4,6,2,5)$ and $A(1,4,3,6,2,5)$ given in \ISTO\ or in 
canonical form in \Durban.
Note, that for this case we have  the identities
\eqn\wehaveIdi{
A(1,6,3,4,2,5)=A(1,3,4,6,2,5)\ \ \ ,\ \ \ A(1,6,4,3,2,5)=A(1,4,3,6,2,5)}
relating the two contour integrals $I_d$ and $I_{b_1}$.

\subsec{Three open $\&$ two closed strings versus seven open strings on the disk}

In this Subsection we establish the relation between 
a disk amplitude with three open $\&$ two closed strings to 
a disk amplitude of seven open strings.
The generic expression of a disk amplitude $\Ac(1,2,3;4,5)$
involving three open and two closed strings is given in \boilviii\ and  \ANGELHUU.
On the other hand, the general expression of a (partial ordered) 
seven open string amplitude is 
given by the generalized Euler integral \doubref\Dan\August 
\eqn\sevenopen{\eqalign{
A(1,2,3,4,5,6,7)&=
\int_0^1 dx\int_0^1 dy \int_0^1 dz\int_0^1 dw\
x^{\hatt s_2}\ y^{\hatt t_2}\ z^{\hatt t_6}\ w^{\hatt s_7}
(1-x)^{\hatt s_3}\ (1-y)^{\hatt s_4}\ (1-z)^{\hatt s_5}\cr
&\times  (1-w)^{\hatt s_6}\ 
(1-xy)^{\hatt t_3-\hatt s_3-\hatt s_4}\ (1-yz)^{\hatt t_4-\hatt s_4-\hatt s_5}\
(1-xyz)^{\hatt s_4-\hatt t_3-\hatt t_4+\hatt t_7}\cr
&\times (1-zw)^{\hatt t_5-\hatt s_5-\hatt s_6}\ 
(1-yzw)^{\hatt s_5+\hatt t_1-\hatt t_4-\hatt t_5}\
(1-xyzw)^{\hatt s_1-\hatt t_1+\hatt t_4-\hatt t_7}\ ,}}
with the seven momenta $k_i$ of the seven external open strings and the $14$
kinematic invariants $\hatt s_i=\ap(k_i+k_{i+1})^2$ and $\hatt
t_j=\ap(k_j+k_{j+1}+k_{j+2})^2$, subject to the cyclic identification $i+7\equiv i$.
For $\ap=2$ the nine invariants $\hatt s_i$ can be related to \INV\ as follows:
\eqn\Invvii{\eqalign{
\hatt s_1&=\hatt s_{12}\ ,\ \hatt s_2=\hatt s_{23}\ ,\ \hatt s_3=\hatt s_{34}\ ,\ 
\hatt s_4=\hatt s_{45}\ ,\ \hatt s_5=\hatt s_{56}\ ,\ \hatt s_6=\hatt s_{67}\ ,\ 
\hatt s_7=\hatt s_{71}\cr 
\hatt t_1&=\hatt s_{12}+\hatt s_{23}+\hatt s_{13}\ ,\ 
\hatt t_2=\hatt s_{23}+\hatt s_{24}+\hatt s_{34}\ ,\ 
\hatt t_3=\hatt s_{34}+\hatt s_{35}+\hatt s_{45}\ ,\cr
\hatt t_4&=\hatt s_{45}+\hatt s_{46}+\hatt s_{56}\ ,\ 
\hatt t_5=\hatt s_{56}+\hatt s_{57}+\hatt s_{67}\ ,\ 
\hatt t_6=\hatt s_{16}+\hatt s_{17}+\hatt s_{67}\ ,\ 
\hatt t_7=\hatt s_{12}+\hatt s_{17}+\hatt s_{27}\ .}}
The above representation refers to the partial ordering $(1234567)$ of the
seven open string vertex operators  along the boundary of the disk. 
The integrals \sevenopen\ integrate to  multiple Gaussian hypergeometric functions \Dan.

Our task is to express \ANGELHUU\ as a sum over integrals of the type
\sevenopen. This is achieved by converting the two complex integrations in $z_1$
and $z_2$ into four real integrals by splitting the two complex integrals up into holomorphic and
anti--holomorphic contour integrals, \cf \eqq \GENNHUL.
The expression \GENNHUL\ decomposes into a sum of partial ordered
amplitudes describing a seven open string disk amplitude with the following 
choice of vertex operator positions 
\eqn\posigluoni{
z_1=-\infty\ \ ,\ \ z_2=0\ \ ,\ \ z_3=1\ \ ,\ \ z_4=\xi_1\ \ ,\ \
z_5=\eta_1\ \ ,\ \ z_6=\xi_2\ \ ,\ \ z_7=\eta_2,}
with $\xi_1,\xi_2,\eta_1,\eta_2\in\IR$. 

According to \momenta\ we have the assignment:
\eqn\assignviii{
k_1=p_1\ ,\ k_2=p_2\ \ \ ,\ \ \ 
k_3=p_3\ \ \ ,\ \ \ k_4=\h\ q_1\ \ \ \ ,\ \ \ k_5=\h\ q_1\ \ \ ,\ \ \
k_6=\h\ q_2\ \ \ ,\ \ \ k_7=\h\ q_2\ .}
As a consequence from \invviii\ and \momenta\ we have:
\eqn\consviii{\eqalign{
&\hatt s_1=s_1\ \ ,\ \ \hatt s_2=s_2\ \ ,\ \ \hatt s_3=s_3\ \ ,\ \ \hatt s_4=0\ \ ,\ \ 
\hatt s_5=s_4\ \ ,\ \ \hatt s_6=0\ \ ,\ \ 
\hatt s_7=s_5\ ,\cr
&\hatt t_1=4\ s_4\ \ ,\ \ \hatt t_2=\h\ s_2+s_5\ \ ,\ \ \hatt t_3=2\ s_3\ \ ,\ \ 
\hatt t_4=\hatt t_5=2\ s_4\ \ ,\ \ \hatt t_6=2\ s_5\ \ ,\ \ \hatt t_7=\h\ s_1+s_3\ .}}

Eventually, the amplitude \boilviii\ can be expressed as sum over partial ordered
disk amplitudes involving seven open strings with integrals of the type \sevenopen: 
\eqn\Obtainviii{\hskip-0.5cm\mathboxit{\eqalign{
\Ac(1,2,3;4,5)&=-2\ \sin\lf[\pi\lf(\fc{s_1}{2}-s_3+s_5\ri)\ri]\ \sin(\pi s_3)\  [\ 
A(1,5,3,4,6,2,7)+A(1,4,3,5,6,2,7)\ ]\cr
&+2\ \sin\lf(\fc{\pi s_2}{2}\ri)\ \sin(\pi s_4)\  
[\ A(1,5,6,3,2,4,7)-A(1,5,6,2,3,4,7)\ ]\cr
&+2\ \sin\lf[\pi\lf(\fc{s_2}{2}+s_3-s_5\ri)\ri]\ 
\sin\lf[\pi\lf(\fc{s_1}{2}-s_3-2s_4\ri)\ri]\cr
&\times [\ A(1,4,2,5,6,3,7)+A(1,5,2,4,6,3,7)\ ]+\ldots\ .}}}

\newsec{Open string subamplitude relations}
\def\h{\fc{1}{2}}

In the previous Section we have seen, that any disk amplitude of $N_o$ open and 
$N_c$ closed strings decomposes into a sum over many color ordered $N_o+2N_c$--point open string partial subamplitudes, \cf \SPLIT. 
In the previous Section we have derived relation between those subamplitudes
by explicitly deforming the contour integrals in the complex $\eta_i$--planes.
In this Section we develop a different method of deriving relations between subamplitudes of open string amplitudes.

We derive (string) relations between partial subamplitudes involving 
$N$ open strings. For $N=N_o+2N_c$ and the condition \momenta\
these relations can be used to simplify \SPLIT.

\subsec{String theory generalization of Kleiss--Kuijf and Bern--Carrasco--Johansson
relations}

At tree--level with states all in the adjoint representation the full $N$--gluon 
amplitude ${\cal A}$ can be decomposed as:
\eqn\QCD{
{\cal A}(1,2,\ldots,N)=g_{YM}^{N-2}\ \sum_{\sigma\in S_{N-1}}
\Tr(T^{a_1}T^{a_{\sigma(2)}}\ldots T^{a_{\sigma(N)}})\ A(1,\sigma(2),\ldots,\sigma(N))\ ,}
with $S_{N-1}=S_N/{\bf Z}_N$ and $A(1,2,\ldots,N)$ the tree--level color--ordered 
$N$--leg partial amplitude (helicity subamplitude).
The $T^a$ are color--group generators encoding the color of each external states.
The sum is over all $(N-1)!$ cyclic inequivalent permutations of external states, 
which is equivalent to all permutations with the first state kept fixed.
The whole kinematics, \ie helicities and polarizations are encoded in the partial 
amplitudes $A(1,2,\ldots,N)$. By construction the latter are 
independent on color indices.

In \QCD\ the $(N-1)!$ subamplitudes are not all independent. In fact, 
in addition to cyclic symmetries by applying reflection and parity symmetries
\eqn\BASIC{\eqalign{
\ds{A(1,2,\ldots,N)}&=\ds{(-1)^N\ A(N,N-1,\ldots,2,1)}}}
we may reduce the number of independent partial amplitudes from $(N-1)!$ 
to~$\h~(N~-~1)!$.
So far these results  are just inherited from the properties of the group traces.
In addition, in field--theory these results are a consequence of studying the sum of Feynman diagrams which 
contribute to each subamplitude, while in string theory they follow from the properties of the string world--sheet. 
Hence these relations \BASIC\ hold both in field and string theory.

Moreover, in field theory there holds the dual Ward identity \Mangano
\eqn\dualWard{\eqalign{
A_{FT}(1,2,3,\ldots,N-1,N)&+A_{FT}(2,1,3,\ldots,N-1,N)+A_{FT}(2,3,1,\ldots,N-1,N)\cr
&+\ldots+A_{FT}(2,3,\ldots,N-1,1,N)=0\ ,}}
which is {\it not} satisfied by the string subamplitudes. 
Furthermore, Kleiss and Kuijf have found a new set of relations, 
which allows to express all subamplitudes in terms of a minimal basis of $(N-2)!$ 
elements \refs{\Kleiss,\DelDuca}.
Recently, in \Bern\ further relations have been derived which allow
to boost this number from $(N-2)!$ elements to $(N-3)!$ independent basis elements.
The proof of these relations is completely based  on a field theory derivation 
\refs{\DelDuca,\Bern}.

In the following, we demonstrate, that also in string theory relations between
various subamplitudes may be derived. These relations, which are {\it different} 
than the field--theory relations, may be considered 
as the string theory upgrade of the dual Ward identity \dualWard, 
Kleiss--Kuijf and Bern--Carrasco--Johansson
relations, which then hold to all orders in $\ap$. 
These identities also allow to reduce the number of independent subamplitudes to $(N-3)!$.
Clearly, in the field--theory limit our relations boil down to the 
Kleiss--Kuijf and Bern--Carrasco--Johansson
relations.
However, it should be stressed that the full string theory amplitude generically {\it do not}
fulfill neither dual Ward nor Kleiss--Kuijf nor Bern--Carrasco--Johansson
relations, however they {\it do} fulfill modified relations, 
which boil down to the former in the field--theory limit.
A prominent example is the photon--decoupling identity or subcyclic property:
\eqn\subcyclic{
\sum_{\sigma\in S_{N-1}} A_{FT}(1,\sigma(2),\sigma(3),\ldots,\sigma(N))=0\ ,}
which is {\it not} fulfilled by the full string amplitude.
Hence, for $\ap\rightarrow 0$ our string theory relations provide  
a prove of the Kleiss--Kuijf and Bern--Carrasco--Johansson
relations for any color group and for arbitrary space--time dimension.

\subsec{World--sheet derivation of open string subamplitude relations}

In this section by applying world--sheet string techniques 
we derive new algebraic identities between subamplitudes relevant for string theory.
Recall, that the expression for partial amplitude $A(1,\ldots,N)$ has the general form  (with $\ap=2$)
\eqn\SUBAMP{
A(1,\ldots,N)=V_{\rm CKG}^{-1}\ \int\limits_{z_1<\ldots<z_N}\
\lf(\ \prod_{j=1}^{N}\ dz_j\ \ri)\ \sum_{\Kc_I}\Kc_I\ \prod_{i<j}^{N}   
|z_{i}-z_{j}|^{\hatt s_{ij}}\ (z_{i}-z_{j})^{n^I_{ij}}\ ,}
\cf \STartwith\ for the case $N_c=0, N_o=N$. Here, the integers $n^I_{ij}$ refer to the kinematical factor $\Kc_I$ and the invariants $\hatt s_{ij}$ are defined in \INV.
The field theory identity \dualWard\ has the following generalization in string
theory \Plahte:
\eqn\DUAL{\eqalign{
A(1,2,\ldots,N)&+e^{i\pi \hatt s_{12}}\ A(2,1,3,\ldots,N-1,N)+
e^{i\pi(\hatt s_{12}+\hatt s_{13})}\ A(2,3,1,\ldots,N-1,N)\cr
&+\ldots+e^{i\pi(\hatt s_{12}+\hatt s_{13}+\ldots+\hatt s_{1N-1})}\ A(2,3,\ldots,N-1,1,N)=0\ .}}
This relation can be derived by analytically continuing the $z_1$--integration
in \SUBAMP\ to the whole complex plane and integrating $z_1$ along the 
contour integral depicted in the next Figure.
\ifig\plahte{Contour integral in the complex $z_1$--plane.}{\epsfxsize=0.65\hsize\epsfbox{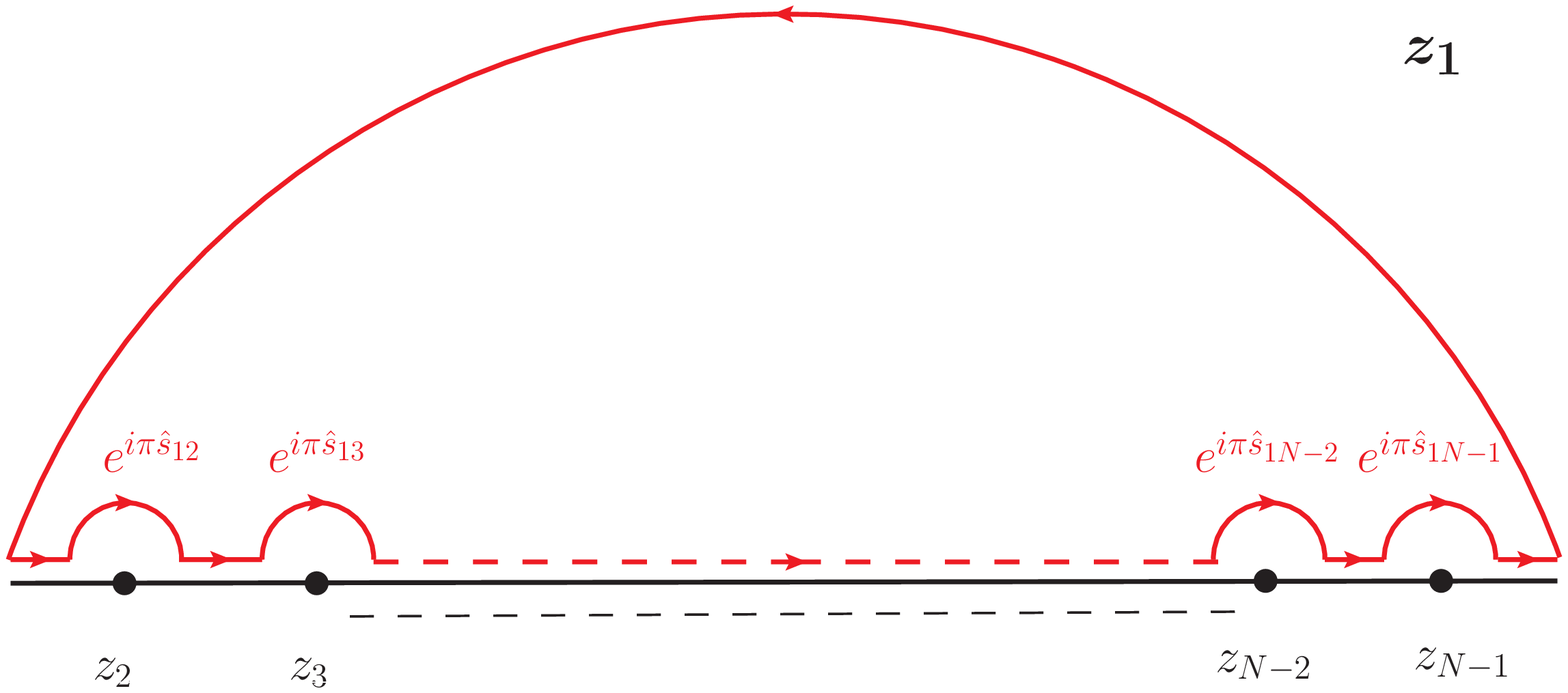}}
\noindent
The $z_1$--integration along the real axis gives a phase factor each 
time when encircling one open string vertex position $z_j,\ j=2,\ldots,N$.
On the other hand, the semicircle can be deformed to infinity. There the integrand
behaves as $z_1^{-2h_1}$ with $h_1$ the conformal weight of the vertex
operator $V_o(z_1)$. Since we consider massless open strings, we have $h_1=1$
and thus no contribution from the semicircle.
Of course, since the partial amplitudes $A(1,\ldots,N)$ are real we obtain two 
relations from \DUAL.

Note, that the integers $n_{ij}^I$ in \SUBAMP\ do not enter in the derivation of the
result \DUAL\ as the branching is caused by the factors 
$|z_i-z_j|^{\hatt s_{ij}}$ only. However, the factors~$(z_i-z_j)^{n_{ij}^I}$~do not affect the phases.
The same is true for the kinematical factors $\Kc_I$. More precisely, 
to each given kinematics $\Kc_I$ a specific set of integers $n_{ij}^I$ is assigned, but the latter are independent on the ordering $\si$, since the contractions of all 
fields yielding the factors $(z_i-z_j)^{n_{ij}^I}$ are taken w.r.t. to one specific ordering for any partial amplitude.
It is only the  integration region $\Ic_\si$, which changes for each subamplitude \SUBAMP. 
Hence, the kinematical factors $\Kc_I$ and the integers $n_{ij}^I$ have no 
effect on the result \DUAL.
Essentially the derivation of the subamplitude relations
could be restricted to Neveu--Schwarz states only as 
the amplitudes involving Ramond states may be simply obtained from the latter by
using supersymmetric Ward identities~\August.

In the field--theory limit $\ap\ra0$, \ie $e^{i\pi\hatt s_{ij}}\ra 1$, the 
expression \DUAL\ boils down to the dual Ward identity \dualWard.
Of course, by analytically continuing any other position $z_j,\ j=2,\ldots,N$
we obtain further relations. This way including permutations we can generate $N!$ complex relations of the type \DUAL, which allow for expressing all subamplitudes in terms of a minimal basis.
In fact, these relations allow for a complete reduction\foot{On the day when submitting this project for publication the letter \VH\ has appeared, with results
similar to what we have found in this Subsection 4.2.
However, prior to this publication our results have been made already 
public in various seminars, \cf \eg \FISI.} of the $(N-1)!$ string
subamplitudes of an open string $N$--point amplitude to a minimal basis of 
\eqn\BRONST{
\nu_o(N)=(N-3)!}
subamplitudes just like in field--theory.
This number is identical to the dimension of a minimal basis of generalized Gaussian 
hypergeometric functions describing the full $N$--point open string amplitude 
\refs{\Dan,\STii,\August}.

Note, that the proof of the relation \DUAL\ and its permutations 
do not rely on any kinematic properties of the subamplitudes, on the amount of supersymmetry or the space--time dimension. Moreover, they hold for any type
of massless string states both from the NS and R sector.
Hence, these relations hold in any space--time dimension $D$, for 
any amount of supersymmetry and any gauge group.

\subsec{Examples}

In this Subsection we discuss some examples and show how the solutions 
of the string relations of the type \DUAL\ 
reproduce the known field--theory relations in their $\ap\ra0$--limit.

\subsubsec{$N=4$}

The subamplitudes are of the form \Fouropen.
Cyclicity reduces the number of subamplitudes to $(4-1)!=6$.
Starting at the relation \DUAL
\eqn\DUALiv{
A(1,2,3,4)+e^{i\pi s}\ A(2,1,3,4)+e^{i\pi (s+t)} A(2,3,1,4)=0}
and its permutations give rise to $24$ relations, which allow to express all
six subamplitudes in terms of one ($\nu_o(4)=1$). 
As basis we choose the subamplitude $A(1,2,3,4)$ and obtain the result: 
\eqn\RELi{
\fc{A(1,2,4,3)}{A(1,2,3,4)}=\fc{\sin(\pi u)}{\sin(\pi t)}\ \ \ ,\ \ \ 
\fc{A(1,3,2,4)}{A(1,2,3,4)}=\fc{\sin(\pi s)}{\sin(\pi t)}\ .}
As a result these relations allow to express all six partial amplitudes in terms of
one, say $A(1,2,3,4)$:
\eqn\allow{\eqalign{
\ds{A(1,4,3,2)}&=\ds{A(1,2,3,4)\ ,}\cr
\ds{A(1,2,4,3)}&=\ds{A(1,3,4,2)=\fc{\sin(\pi u)}{\sin(\pi t)}\ A(1,2,3,4)\ ,}\cr
\ds{A(1,3,2,4)}&=\ds{A(1,4,2,3)=\fc{\sin(\pi s)}{\sin(\pi t)}\ A(1,2,3,4)\ .}}}

Clearly, in the field--theory limit the relations \RELi\ simply reduce to the
well--known identities \Bern:
\eqn\RELii{
\fc{A_{FT}(1,2,4,3)}{A_{FT}(1,2,3,4)}=\fc{u}{t}\ \ \ ,\ \ \ 
\fc{A_{FT}(1,3,2,4)}{A_{FT}(1,2,3,4)}=\fc{s}{t}\ .}
and the subcyclic property 
$$A_{FT}(1,2,3,4)+A_{FT}(1,3,4,2)+A_{FT}(1,4,2,3)=0$$ 
can be proven.

\subsubsec{$N=5$}

The subamplitudes are of the form \fiveopen.
Cyclicity, reflection and parity transformations \BASIC\ reduce the number of subamplitudes
to $\h(5-1)!=12$. However, our new relations of the type \DUAL\ 
allow for a reduction of this number to a minimal  basis of two ($\nu_o(5)=2$).
Starting at the relation \DUAL
\eqn\DUALv{
A(1,2,3,4,5)+e^{i\pi\hatt s_1}\ A(2,1,3,4,5)+e^{i\pi(\hatt s_4-\hatt s_2)}\  A(2,3,1,4,5)+e^{-i\pi \hatt s_5}\ A(2,3,4,1,5)=0}
and its permutations give $120$ relations, which allow to express all
twelve or $24$ subamplitudes in terms of two. The five kinematic invariants are defined in \Invv. As basis we choose the two subamplitudes  $A(1,2,3,4,5)=-A(1,5,4,3,2)$ 
and $A(1,3,2,4,5)=-A(1,5,4,2,3)$. For completeness we write down all
$24$ subamplitudes:
\eqn\extremelynontrival{\eqalign{
\ds{A(1,2,5,4,3)}&=\ds{-A(1,3,4,5,2)=\sin[\pi(\hatt s_3-\hatt s_1-\hatt s_5)]^{-1}}\cr
&\times\ds{\left\{\ \sin[\pi(\hatt s_3-\hatt s_5)]\ A(1,2,3,4,5)+
\sin[\pi(\hatt s_2+\hatt s_3-\hatt s_5)]\ A(1,3,2,4,5)\ \right\}\ ,}\cr
\ds{A(1,3,4,2,5)}&=\ds{-A(1,5,2,4,3)=\sin[\pi(\hatt s_3-\hatt s_1-\hatt s_5)]^{-1}}\cr
&\times\ds{\left\{\ \sin(\pi \hatt s_1)\ A(1,2,3,4,5)-
\sin[\pi(\hatt s_1+\hatt s_2)]\ A(1,3,2,4,5)\ \right\}\ ,}\cr
\ds{A(1,2,4,5,3)}&=\ds{-A(1,3,5,4,2)=\sin[\pi(\hatt s_1-\hatt s_3-\hatt s_4)]^{-1}}\cr
&\times\ds{\left\{\ \sin(\pi \hatt s_3)\ A(1,2,3,4,5)+\sin[\pi(\hatt s_2+\hatt s_3)]\ 
A(1,3,2,4,5)\ \right\}\ ,}}}
\eqn\extremelynontrivali{\eqalign{
\ds{A(1,2,4,3,5)}&=\ds{-A(1,5,3,4,2)= \sin[\pi(\hatt s_3+\hatt s_4-\hatt s_1)]^{-1}}\cr
&\times\ds{\left\{\ \sin[\pi(\hatt s_1-\hatt s_4)]\ A(1,2,3,4,5)+\sin[\pi(\hatt s_1+\hatt s_2-\hatt s_4)]\ 
A(1,3,2,4,5)\ \right\}\ ,}\cr
\ds{A(1,3,2,5,4)}&=\ds{-A(1,4,5,2,3)=\sin[\pi(\hatt s_3-\hatt s_1-\hatt s_5)]^{-1}\ 
\sin[\pi(\hatt s_2-\hatt s_4-\hatt s_5)]^{-1}}\cr
&\times\ds{\left\{\ \sin(\pi \hatt s_1)\ \sin(\pi \hatt s_3)\ A(1,2,3,4,5)\ri.}\cr
&+\ds{\lf.\sin[\pi(\hatt s_1+\hatt s_5)]\ \sin[\pi(\hatt s_2+\hatt s_3-\hatt s_5)]\ A(1,3,2,4,5)\ \right\}\ ,}\cr
\ds{A(1,4,2,3,5)}&=\ds{-A(1,5,3,2,4)=-\sin[\pi(\hatt s_1-\hatt s_3-\hatt s_4)]^{-1}\ 
\sin[\pi(\hatt s_2-\hatt s_4-\hatt s_5)]^{-1}}\cr
&\times\ds{\left\{\ \sin(\pi \hatt s_1)\ \sin(\pi \hatt s_3)\ A(1,2,3,4,5)\ri.}\cr
&+\ds{\lf.\sin[\pi(\hatt s_3+\hatt s_4)]\ \sin[\pi(\hatt s_1+\hatt s_2-\hatt s_4)]\ A(1,3,2,4,5)\ \right\}\ ,}}}
\eqn\extremelynontrivalii{\eqalign{
\ds{A(1,4,3,2,5)}&=\ds{-A(1,5,2,3,4)=-\sin[\pi(\hatt s_3-\hatt s_1-\hatt s_5)]^{-1}\ 
\sin[\pi(\hatt s_2-\hatt s_4-\hatt s_5)]^{-1}}\cr
&\times\ds{\lf\{\ \sin[\pi(\hatt s_1+\hatt s_2-\hatt s_4)]\ \sin[\pi(\hatt s_2+\hatt s_3-\hatt s_5)]\ 
A(1,2,3,4,5)\ri.}\cr
&+\ds{\lf.\sin(\pi \hatt s_1)\ \sin[\pi(\hatt s_2+\hatt s_3-\hatt s_4-\hatt s_5)]\ A(1,3,2,4,5)\ \ri\}\ ,}\cr
\ds{A(1,4,5,3,2)}&=\ds{-A(1,2,3,5,4)=-\sin[\pi(\hatt s_1-\hatt s_3-\hatt s_4)]^{-1}\ 
\sin[\pi(\hatt s_2-\hatt s_4-\hatt s_5)]^{-1}}\cr
&\times\ds{\lf\{\ \sin[\pi(\hatt s_1+\hatt s_2-\hatt s_4)]\ \sin[\pi(\hatt s_2+\hatt s_3-\hatt s_5)]\ 
A(1,2,3,4,5)\ri.}\cr
&+\ds{\lf.\sin(\pi \hatt s_3)\ \sin[\pi(\hatt s_1+\hatt s_2-\hatt s_4-\hatt s_5)]\ A(1,3,2,4,5)\ \ri\}\ ,}}}
\eqn\extremelynontrivalixx{\eqalign{
A(1,2,5,3,4)&=-A(1,4,3,5,2)\cr
&\ds{=\sin[\pi(\hatt s_1-\hatt s_3-\hatt s_4)]^{-1}\ 
\sin[\pi(\hatt s_3-\hatt s_1-\hatt s_5)]^{-1} \ \sin[\pi (\hatt s_2-\hatt s_4-\hatt s_5)]^{-1}}\cr
&\times\ds{\Big\{\ -\fc{1}{4}\  
\left(\ \sin[\pi( \hatt s_1-\hatt s_2-\hatt s_3)]-
\sin[\pi( \hatt s_1+\hatt s_2-\hatt s_3)]+
\sin[\pi (\hatt s_1+\hatt s_2+\hatt s_3)]\right.}\cr
&\ds{+\sin[\pi (\hatt s_1+\hatt s_2-\hatt s_3-2 \hatt s_4)]+\sin[\pi (-\hatt s_1+\hatt s_2+\hatt s_3-2 \hatt s_5)]}\cr
&\ds{\lf.-\sin[\pi (\hatt s_1+\hatt s_2+\hatt s_3-2 \hatt s_4-2 \hatt s_5)]\ \right)\ \ A(1,2,3,4,5)} \cr
&+\ds{\sin[\pi (\hatt s_1+\hatt s_2-\hatt s_4)] \ \sin[\pi (\hatt s_2+\hatt s_3-\hatt s_5)] \ \sin[\pi(\hatt s_4+\hatt s_5)]\ 
A(1,3,2,4,5)\ \Big\}\ ,}}}
\eqn\extremelynontrivaliii{\eqalign{
\ds{A(1,3,5,2,4)}&=\ds{-A(1,4,2,5,3)}\cr
&\ds{=\sin[\pi(\hatt s_1-\hatt s_3-\hatt s_4)]^{-1}\ 
\sin[\pi(\hatt s_3-\hatt s_1-\hatt s_5)]^{-1} \ \sin[\pi (\hatt s_2-\hatt s_4-\hatt s_5)]^{-1}}\cr
&\times\ds{\Big\{\ \sin(\pi \hatt s_1)\ \sin(\pi \hatt s_3)\ \sin[\pi(\hatt s_4+\hatt s_5)]\ 
A(1,2,3,4,5)}\cr
&-\ds{\fc{1}{4}\ \left(\ \sin[\pi (\hatt s_1+\hatt s_2+\hatt s_3+\hatt s_4+\hatt s_5)]-
\sin[\pi( \hatt s_1+\hatt s_2+\hatt s_3-\hatt s_4-\hatt s_5)]\ri.}\cr
&\ds{+\sin[\pi(\hatt s_1+\hatt s_2-\hatt s_3-\hatt s_4-\hatt s_5)]-\sin[\pi (\hatt s_1+\hatt s_2-\hatt s_3-\hatt s_4+\hatt s_5)]}\cr
&\hskip-1cm\ds{\lf.-\sin[\pi (\hatt s_1-\hatt s_2-\hatt s_3+\hatt s_4+\hatt s_5)]-\sin[\pi (-\hatt s_1+\hatt s_2+\hatt s_3+\hatt s_4-\hatt s_5)]
\ \right)\ A(1,3,2,4,5)\ \Big\}.}}}

Again, the field--theory limit of these  relations, which is simply obtained by 
replacing the $\sin$--function by its argument boils down to a system of 
identities, which
solves the Kleiss--Kuijf and Bern--Carrasco--Johansson identities.
E.g. the full string amplitudes fulfill the following relations:
\eqn\subcycupgrade{\eqalign{
&\ds{\sin[\pi (\hatt s_2-\hatt s_4)]\ A(1,2,3,4,5)+\lf\{\ \sin[\pi (\hatt s_1+\hatt s_2-\hatt s_4)]-
\sin(\pi \hatt s_1)\ \ri\}\ A(1,3,4,5,2)}\cr
&+\ds{\sin[\pi(\hatt s_2-\hatt s_4)]\ A(1,4,5,2,3) +\lf\{\ \sin(\pi \hatt s_5)+
\sin[\pi (\hatt s_2-\hatt s_4-\hatt s_5)]\ \ri\}\ A(1,5,2,3,4)=0}\ ,\cr
&\ds{\lf[\sin(\pi \hatt s_1)+\sin(\pi \hatt s_5)\ri]\ A(1,2,3,4,5)+\sin[\pi (\hatt s_1+\hatt s_5)]\ 
A(1,3,4,5,2)}\cr
&+\lf\{\ \sin[\pi (\hatt s_1+\hatt s_2-\hatt s_4)]- \sin[\pi (\hatt s_2-\hatt s_4-\hatt s_5)]\ \ri\}\ 
A(1,4,5,2,3)\cr
&+ \sin[\pi (\hatt s_1+\hatt s_5)]\ A(1,5,2,3,4)=0\ .}}
Clearly, in the field theory limit, these two relations boil down to the 
the subcyclic identity~\Bern:
\eqn\BERi{
A_{FT}(1,2,3,4,5)+ A_{FT}(1,3,4,5,2)+  A_{FT}(1,4,5,2,3)+  A_{FT}(1,5,2,3,4)=0\ .}
In these lines also  the Kleiss--Kuijf relation
\eqn\BERii{
A_{FT}(1,2,3,5,4)+ A_{FT}(1,2,3,4,5)+ A_{FT}(1,2,4,3,5)+  A_{FT}(1,4,2,3,5)=0\ ,}
the Bern--Carrasco--Johansson relations
\eqn\BERiii{\eqalign{
&\ds{A_{FT}(1,3,4,2,5)\ \hatt s_{13}\ \hatt s_{24}+\hatt s_{12}\ \hatt s_{45}\ A_{FT}(1,2,3,4,5)-\hatt s_{14}\ 
(\hatt s_{24}+\hatt s_{25})\ 
A_{FT}(1,4,3,2,5)=0\ ,}\cr
&\ds{A_{FT}(1,2,4,3,5)\ \hatt s_{35}\ \hatt s_{24}+\hatt s_{14}\ \hatt s_{25}\ A_{FT}(1,4,3,2,5)-\hatt s_{45}\ 
(\hatt s_{12}+\hatt s_{24})\ 
A_{FT}(1,2,3,4,5)=0\ ,}\cr  
&\ds{A_{FT}(1,4,2,3,5]\ \hatt s_{35}\ \hatt s_{24}+\hatt s_{12}\ \hatt s_{45}\ A_{FT}(1,2,3,4,5)-\hatt s_{25}\ 
(\hatt s_{14}+\hatt s_{24})\ 
A_{FT}(1,4,3,2,5)=0\ ,}\cr 
&\ds{A_{FT}(1,3,2,4,5)\ \hatt s_{13}\ \hatt s_{24}+\hatt s_{14}\ \hatt s_{25}\ A_{FT}(1,4,3,2,5)-\hatt s_{12}\ 
(\hatt s_{24}+\hatt s_{45})\ 
A_{FT}(1,2,3,4,5)=0\ ,} }}
and
\eqn\subz{
A_{FT}(1,2,4,3,5)\ \hatt s_{24}-(\hatt s_{14}+\hatt s_{45})\ A_{FT}(1,2,3,4,5)-\hatt s_{14}\ A_{FT}(1,2,3,5,4)=0\ .}
can be trivially deduced.

\subsubsec{$N=6$}

The subamplitudes are of the form \sixopen.
Cyclicity, reflection and parity transformations \BASIC\ reduce the number of subamplitudes
to $\h(6-1)!=60$. However, our new relations of the type \DUAL\ 
allow for a reduction of this number to a minimal  basis of six ($\nu_o(6)=6$).
Starting at the relation \DUAL
\eqn\DUALvi{\eqalign{
&A(1,2,3,4,5,6)+e^{i\pi \hatt s_1}\ A(2,1,3,4,5,6)+e^{i\pi (\hatt t_1-\hatt s_2)}\  A(2,3,1,4,5,6)\cr
&+e^{i \pi (\hatt s_5-\hatt t_2)}\ A(2,3,4,1,5,6)+e^{-i\pi \hatt  s_6}\  A(2,3,4,5,1,6)=0}}
and its permutations give $720$ relations, which allow to express all
$60$ subamplitudes in terms of six. 
The nine kinematic invariants are defined in \Invvi.
We choose the six subamplitudes  $A(1,2,3,4,5,6),\ A(1,2,3,4,6,5),$
$A(1,2,3,5,4,6),\ A(1,2,4,3,5,6)$,\ $A(1,2,4,3,6,5)$ and $A(1,2,4,5,3,6)$
as basis and derive the relations
$$\eqalign{
A(1,2,3,5,6,4)&=-\sin[\pi(\hatt s_2+\hatt s_5-\hatt t_1-\hatt t_2)]^{-1}\  
\lf\{\ \sin[\pi(\hatt s_2-\hatt t_2)]\ 
A(1,2,3,4,5,6)\ri.\cr
&+\sin[\pi(\hatt s_2+\hatt s_3-\hatt t_2)]\ A(1,2,4,3,5,6)\cr 
&\lf.+\sin[\pi (\hatt s_2-\hatt s_4-\hatt t_2)]\ 
A(1,2,3,5,4,6)\ \ri\}\ ,\cr
A(1,2,3,6,4,5)&=-\sin[\pi(\hatt s_2+\hatt s_4-\hatt s_6-\hatt t_1)]^{-1}\ 
\lf\{\ \sin[\pi (\hatt s_2-\hatt s_6)]\ A[1,2,3,4,5,6]\ri.\cr
&+\sin[\pi (\hatt s_2+\hatt s_3-\hatt s_6)]\ A(1,2,4,3,5,6)\cr 
&+\sin[\pi (\hatt s_2-\hatt s_5-\hatt s_6)]\ A(1,2,3,4,6,5)\cr
&+\sin[\pi (\hatt s_2+\hatt s_3-\hatt s_5-\hatt s_6)]\ A(1,2,4,3,6,5) \cr
&\lf.+\sin[\pi(\hatt s_2-\hatt s_4-\hatt s_6+\hatt t_3)]\ A(1,2,4,5,3,6)\ \ri\}\ ,\cr
A(1,2,5,3,4,6)&=\sin[\pi (\hatt s_3+\hatt s_6-\hatt t_2-\hatt t_3)]^{-1}\ \lf\{\ 
-\sin[\pi(\hatt s_6-\hatt t_2)]\ A(1,2,3,4,5,6)\ri.\cr
&-\sin[\pi (\hatt s_5+\hatt s_6-\hatt t_2)]\ A(1,2,3,4,6,5)\cr  
&\lf.+\sin[\pi (\hatt s_4-\hatt s_6+\hatt t_2)]\ A(1,2,3,5,4,6)\ \ri\}\ ,}$$
\eqn\RELvi{\eqalign{
A(1,2,5,3,6,4)&=\sin[\pi(\hatt s_2+\hatt s_5-\hatt t_1-\hatt t_2)]^{-1}\ 
\sin[\pi(\hatt s_3+\hatt s_6-\hatt t_2-\hatt t_3)]^{-1}\cr 
&\times \lf\{\ \sin[\pi (\hatt s_2-\hatt s_4-\hatt t_2)]\ 
\sin[\pi (\hatt s_6-\hatt t_2)]\ A(1,2,3,4,5,6)\ri.\cr
&+\sin[\pi (\hatt s_2+\hatt s_3-\hatt s_4-\hatt t_2)]\ 
\sin[\pi (\hatt s_6-\hatt t_2)]\ A(1,2,4,3,5,6) \cr
&+\sin[\pi (\hatt s_2-\hatt s_4-\hatt t_2)]\ 
\sin[\pi (\hatt s_5+\hatt s_6-\hatt t_2)]\ A(1,2,3,4,6,5) \cr
&+\sin[\pi (\hatt s_2+\hatt s_3-\hatt s_4-\hatt t_2)]\ 
\sin[\pi(\hatt s_5+\hatt s_6-\hatt t_2)]\ A(1,2,4,3,6,5) \cr
&-\sin[\pi (\hatt s_2-\hatt s_4-\hatt t_2)]\ 
\sin[\pi (\hatt s_4-\hatt s_6+\hatt t_2)]\ A(1,2,3,5,4,6)\cr
&\lf.+\sin(\pi \hatt s_4) 
\sin[\pi (\hatt s_2-\hatt s_4-\hatt s_6+\hatt t_3)]\ A(1,2,4,5,3,6)\ \ri\}\ ,\cr
A(1,3,2,4,5,6)&=-\sin[\pi(\hatt s_1+\hatt s_2-\hatt t_1)]\ \lf\{\ 
\sin[\pi (\hatt s_1-\hatt t_1)]\ A(1,2,3,4,5,6)\ri.\cr
&+\sin[\pi (\hatt s_1-\hatt s_3-\hatt t_1)]\ A(1,2,4,3,5,6)\cr
&\lf.+ \sin[\pi\ (\hatt s_1+\hatt s_4-\hatt t_1-\hatt t_3)]\ A(1,2,4,5,3,6)\ \ri\}\ ,\cr
A(1,4,2,3,5,6)&=\sin[\pi(\hatt s_2+\hatt s_5-\hatt t_1-\hatt t_2)]^{-1}\ \lf\{\ 
-\sin[\pi (\hatt s_5-\hatt t_1)]\ A(1,2,3,4,5,6) \ri.\cr
&-\sin[\pi(\hatt s_4+\hatt s_5-\hatt t_1)]\ A(1,2,3,5,4,6)\cr  
&\lf. +\sin[\pi(\hatt s_3-\hatt s_5+\hatt t_1)]\ A(1,2,4,3,5,6)\ \ri\},\cr
A(1,5,4,6,3,2)&=\sin[ (\hatt s_2+\hatt s_4-\hatt s_6-\hatt t_1)]^{-1}\ \lf\{\ 
-\sin[\pi(\hatt s_2-\hatt s_6)]\ A(1,2,3,4,5,6)\ri.\cr
& -\sin[\pi (\hatt s_2+\hatt s_3-\hatt s_6)]\ A(1,2,4,3,5,6) \cr
&-\sin[\pi (\hatt s_2-\hatt s_5-\hatt s_6)]\ A(1,2,3,4,6,5)\cr
&-\sin[\pi (\hatt s_2+\hatt s_3-\hatt s_5-\hatt s_6)]\ A(1,2,4,3,6,5) \cr 
&\lf.-\sin[\pi (\hatt s_2-\hatt s_4-\hatt s_6+\hatt t_3)]\ A(1,2,4,5,3,6)\ \ri\}\ ,\cr
&\vdots}}

\subsec{Open string subamplitude relations and amplitudes of open and closed strings}
\def\g2{e^{i\pi\hatt\gamma_2}}
\def\b2{e^{i\pi\hatt\beta_2}}
\def\l2{e^{i\pi\hatt\lambda_2}}
\def\d2{e^{i\pi\hatt\al}}

In this Subsection we explicitly work out the sum \SPLIT\ by taking into account the 
the phase $\Pi(\Si)$ given in \eqq \PHASE.
As already described in \eqq \Bronst\ for a given ordering $\si$ of open strings
this sum gives $\nu(N_o,N_c)$ partial ordered
amplitudes of $N_o+2N_c$ open strings.
However, by using the open string subamplitude relations derived above 
we are able to reduce the number of terms drastically.
As discussed in Subsection 3.1 this manipulation is equivalent to deforming
the complex contours and reduce to the independent contours as performed
in Section 3.
The open string subamplitude relations for $N$ open strings from above 
can be applied by imposing on them the momentum constraint \momenta. 
Since for a disk amplitude of $N$ open strings the number of independent 
subamplitudes is given by $\nu_o(N)=(N-3)!$ this means that for a disk amplitude 
of $N_o$ open strings
and $N_c$ closed strings the final result can be written in terms of 
a basis of $\nu_o(N_o+2N_c)=(N_o+2N_c-3)!$ subamplitudes.
Hence in \SPLIT\ the number $\nu(N_o,N_c)$ of subamplitudes can be reduced to
$(N_o+2N_c-3)!$ subamplitudes. In other words the final result
for the a disk amplitude of $N_o$ open strings
and $N_c$ closed strings can be written in terms of 
\eqn\Bronste{
\tilde\nu(N_o,N_c)=(N_o+2N_c-3)!}
partial amplitudes of a pure open string disk amplitude involving $N_o+2N_c$
open strings.
For special configurations of the closed string momenta as $q_{i\parallel}^2=0$
this number is further
reduced due to additional relations within the basis of subamplitudes. 
In fact, in Section~3 we have seen, that:
\eqn\specialconf{
\lf.\tilde\nu_0(3,1)\ri|_{special \atop
config.}=1\ \ \ ,\ \ \  \lf.\tilde\nu_0(2,2)\ri|_{special \atop
config.}=2\ \ \ ,\ \ \ \lf.\tilde\nu_0(4,1)\ri|_{special \atop config.}=4\ ,\ldots
\ .}

\subsubsec{Three open and one closed string}

According to \Bronst\ for this case in the sum \SPLIT\ we expect $\nu(3,1)=12$ partial ordered
amplitudes of five open strings.
The phases \PHASE\ give
\eqn\phasev{
\Pi(\xi,\eta)=\cases{e^{i\pi\hatt\gamma_2}\ ,&  $(1-\xi)\ (1-\eta)<0\ ,$\cr
                     e^{i\pi\hatt\lambda_2}\ ,& $\xi\ \eta<0\ ,$\cr
                     e^{i\pi\hatt\al}\ ,&$\xi<\eta$\ ,}}
with $\hatt\lambda_2,\hatt\gamma_2,\hatt\al$ given in \withv\ and referring to the
non--integer parts, \ie $\hatt\lambda_2=t,\ \hatt\gamma_2=s,\ 
\hatt\al=2q_\parallel^2$.
With \phasev\ the sum \SPLIT\ becomes:
\eqn\ukv{\eqalign{
\Ac(1,2,3;4)&=A(1,5,4,2,3)+\d2\ A(1,4,5,2,3)+\d2\l2\ A(1,4,2,5,3)\cr
&+\d2\l2\g2\ A(1,4,2,3,5)+\l2\ A(1,5,2,4,3)+A(1,2,5,4,3)\cr
&+\d2\ A(1,2,4,5,3)+\d2\g2\ A(1,2,4,3,5)+\l2\g2\ A(1,5,2,3,4)\cr
&+\g2\ A(1,2,5,3,4)+A(1,2,3,5,4)+\d2\ A(1,2,3,4,5)\ .}}
The phase factor \phasev\ is in correspondence to the phases displayed
in Table 1.
With \consv\ the solutions (4.13)--(4.17) boil down to:
$$\eqalign{
A(1,3,4,2,5)&=-A_1\ \ \ ,\ \ \ A(1,5,4,3,2)=-A_2\ ,\cr
A(1,5,3,4,2)&=-A(1,2,4,3,5)\cr
&=-\h\ 
\fc{1}{\sin(\pi s)}\ \lf\{\fc{\sin[\pi(2t+\hatt\al)]}{\cos(\pi t)}\ A_1
-2\ \fc{\sin(\pi\hatt\al)\ \sin[\pi(2u+\hatt\al)]}{\sin(2\pi t)}\ A_2\ri\},\cr
A(1,4,3,5,2)&=-A(1,2,5,3,4)=-\h\ 
\fc{1}{\sin(\pi s)\ \sin(\pi u)}\ 
\lf\{\fc{\sin[\pi(u+\hatt\al)\ \sin[\pi(2t+\hatt\al)]}{\cos(\pi t)}\ A_1\ri.\cr
&\lf.-2\ \fc{\sin(\pi\hatt\al)\ \sin[\pi(2u+\hatt\al)]\ \sin[\pi(s-t)]}{\sin(2\pi t)}\ A_2\ri\}\ ,\cr}$$
\eqn\RELLv{\eqalign{
A(1,5,2,3,4)&=-A(1,4,3,2,5)\cr
&=\h\ 
\fc{1}{\sin(\pi u)}\ \lf\{\fc{\sin[\pi(2t+\hatt\al)]}{\cos(\pi t)}\ A_1
-2\ \fc{\sin(\pi\hatt\al)\ \sin[\pi(2s+\hatt\al)]}{\sin(2\pi t)}\ A_2\ri\},\cr
A(1,4,2,3,5)&=-A(1,5,3,2,4)=\h\ 
\fc{1}{\sin(\pi s)\ \sin(\pi u)}\ 
\lf\{\fc{\sin[\pi(s+\hatt\al)\ \sin[\pi(2t+\hatt\al)]}{\cos(\pi t)}\ A_1\ri.\cr
&\lf.+2\ \fc{\sin(\pi\hatt\al)\ \sin[\pi(2s+\hatt\al)\ \sin[\pi(t-u)]}{\sin(2\pi t)}\ A_2\ri\}\ ,\cr
A(1,2,4,5,3)&=-A(1,3,5,4,2)\cr
&=-\h\ 
\fc{1}{\sin(\pi s)}\ \lf\{\fc{\sin[\pi(t-u)]}{\cos(\pi t)}\ A_1
-2\ \fc{\sin[\pi(2u+\hatt\al)]\ \sin[\pi(s+\hatt\al)]}{\sin(2\pi t)}\ A_2\ri\},
\cr
A(1,2,5,4,3)&=-A(1,3,4,5,2)=-\h\ \fc{1}{\cos(\pi t)}\  A_1-
\fc{\sin[\pi(2 u+\hatt\al)]}{\sin(2 \pi t)}\ A_2\  ,\cr
A(1,4,5,2,3)&=-A(1,3,2,5,4)\cr
&=\h\ \fc{1}{\sin(\pi u)}\ \lf\{\fc{\sin[\pi(s-t)]}{\cos(\pi t)}\ A_1
+2\ \fc{\sin[\pi(u+\hatt\al)]\ \sin[\pi(2s+\hatt\al)]}{\sin(2\pi t)}\ A_2\ri\}\ ,\cr
A(1,4,5,3,2)&=-A(1,2,3,5,4)=-\fc{1}{4}\ \fc{1}{\sin(\pi s)\ \sin(\pi u)\ \cos(\pi t)}\cr
&\times\lf\{\ 2\ \sin(\pi t)\ \sin[\pi(2t+\hatt\al)]\ A_1+[\cos(2\pi t)-\cos(2\pi s)
+2\ \sin(\pi u)^2]\ A_2\ \ri\},\cr
A(1,5,4,2,3)&=-A(1,3,2,4,5)=-\h\  \fc{1}{\cos(\pi t)}\ A_1-\fc{\sin[\pi(2s+\hatt\al)]}{\sin(2\pi t)}\ A_2\ ,\cr
A(1,3,5,2,4)&=-A(1,4,2,5,3)=-\fc{1}{4}\ \fc{1}{\sin(\pi s)\ \sin(\pi u)\ \cos(\pi t)}\cr
&\times\Big\{\ [\cos(2\pi s)+\cos(2\pi u)-2\ \cos(\pi t)^2]\ A_1\cr
&+2\ \fc{\sin(\pi\hatt\al)\ 
\sin[\pi(2s+\hatt\al)]\ \sin[\pi(2u+\hatt\al)]}{\sin(\pi t)}\ \ A_2\ \Big\}\ ,}}
with $A_1=A(1,5,2,4,3)$ and $A_2=A(1,2,3,4,5)$.
After inserting these solutions into \ukv\ we find (with $s+t+u=-\hatt\al$):
\eqn\ukv{
\Ac(1,2,3;4)=
2i\ \sin(\pi\hatt\lambda_2)\ A(1,5,2,4,3)+2i\ \sin(\pi\hatt\alpha)\ A(1,2,3,4,5)\ .}For $\hatt\al=0$ this result has also been derived by the same methods
in \LM. However, here  we have considered the most general case 
involving $\alpha\notin\IZ$, \ie $\hat\al\neq0$. For this case the phases 
\phasev\ have to be considered, \cf Appendix \appA. As a result, in this case the full 
amplitude is expressed by two independent partial amplitudes
$A(1,5,2,4,3)$ and $A(1,2,3,4,5)$ in agreement with the counting formula \Bronste.

\subsubsec{Two open and two closed strings}

According to \Bronst\ for this case in the sum \SPLIT\ we expect $\nu(2,2)=120$ partial ordered
amplitudes of six open strings.
The phases \PHASE\ give
\eqn\phasevi{
\Pi(\rho,\xi,\eta)=\cases{e^{i\pi\hatt\gamma_2}\ ,&  $(\xi-\rho)\ (\eta+\rho)<0\ ,$\cr
                     e^{i\pi\hatt\beta_2}\ ,&  $(\xi+\rho)\ (\eta-\rho)<0\ ,$\cr
                     e^{i\pi\hatt\lambda_2}\ ,&  $(1-\xi)\ (1-\eta)<0\ ,$\cr
                     e^{i\pi\hatt\alpha_2}\ ,&  $|x|>1\ ,$}}
with $\hatt\alpha_2,\hatt\lambda_2,\hatt\gamma_2,\hatt\beta_2$ given 
in \parameter\ and referring to the non--integer parts, \ie
$\hatt\alpha_2=u,\ \hatt\lambda_2=t,\ \hatt\gamma_2=\hatt\beta_2=\fc{s}{2}$.
With \phasevi\ the sum \SPLIT\ becomes:
$$\eqalign{
&\Ac(1,2;3,4)=[\ A(1,6,5,3,4,2)+A(1,5,6,3,4,2)\ ]+\g2\ A(1,5,3,6,4,2)\cr
&+\g2\ \b2\ A(1,5,3,4,6,2)+\l2\ \g2\b2\ A(1,5,3,4,2,6)+\b2\ A(1,6,3,5,4,2)\cr
&+\g2\b2\ [\ A(1,3,6,5,4,2)+A(1,3,5,6,4,2)\ ]+\g2\ A(1,3,5,4,6,2)\cr
&+\l2\g2\ A(1,3,5,4,2,6)+\g2\b2\ A(1,6,3,4,5,2)+\b2\ A(1,3,6,4,5,2)\cr
&+[\ A(1,3,4,6,5,2)+A(1,3,4,5,6,2)\ ]+\l2\ A(1,3,4,5,2,6)\cr
&+\l2\b2\g2\ A(1,6,3,4,2,5)+\l2\b2\ A(1,3,6,4,2,5)+\l2\ A(1,3,4,6,2,5)\cr
&+[\ A(1,3,4,2,6,5)+A(1,3,4,2,5,6)\ ]+[\ A(1,6,5,4,3,2)+A(1,5,6,4,3,2)\ ]\cr
&+\b2\ A(1,5,4,6,3,2)+\g2\b2\ A(1,5,4,3,6,2) +\l2\g2\b2\ A(1,5,4,3,2,6)\cr
&+\g2\ A(1,6,4,5,3,2)+\g2\b2\ [\ A(1,4,6,5,3,2)+A(1,4,5,6,3,2)\ ]\cr
&+\b2\ A(1,4,5,3,6,2)+\l2\b2\ A(1,4,5,3,2,6)+\g2\b2\ A(1,6,4,3,5,2)\cr
&+\g2\ A(1,4,6,3,5,2)+[A(1,4,3,6,5,2)+A(1,4,3,5,6,2)]\cr
&+\l2\ A(1,4,3,5,2,6)+\l2\b2\g2\ A(1,6,4,3,2,5)+\l2\g2\ A(1,4,6,3,2,5)\cr
&+\l2\ A(1,4,3,6,2,5)+[\ A(1,4,3,2,6,5)+A(1,4,3,2,5,6)\ ]\cr
&+e^{i\pi\hatt\alpha_2}\ \Big\{\ [\ A(1,6,5,3,2,4)+A(1,5,6,3,2,4)\ ]+\g2\ 
A(1,5,3,6,2,4)\cr
&+\l2\g2\ A(1,5,3,2,6,4)+\l2\g2\b2\ A(1,5,3,2,4,6)+\b2\ A(1,6,3,5,2,4)\cr
&+\b2\g2\ [\ A(1,3,6,5,2,4)+A(1,3,5,6,2,4)\ ]+\l2\g2\b2\ A(1,3,5,2,6,4)\cr
&+\l2\g2\ A(1,3,5,2,4,6)+\l2\b2\ A(1,6,3,2,5,4)+\l2\g2\b2\ A(1,3,6,2,5,4)\cr
&+\b2\g2\ [A(1,3,2,6,5,4)+A(1,3,2,5,6,4)]+\g2\ A(1,3,2,5,4,6)}$$
\eqn\ukvi{\eqalign{
&+\l2\g2\b2\ A(1,6,3,2,4,5)+\l2\b2\ A(1,3,6,2,4,5)+\b2\ A(1,3,2,6,4,5)\cr
&+[\ A(1,3,2,4,6,5)+A(1,3,2,4,5,6)\ ]\ \Big\}\ .}}
The phase factor \phasevi\ is in correspondence to the phases displayed
in Tables 2 and 3.
With \consvi\ the solution \RELvi\ boils down to:
\eqn\RELLvi{\eqalign{
A(1,2,3,5,6,4)&=\sin(\pi t)^{-1}\ \{\ \sin(\pi u)\ A(1,2,3,4,5,6)+\sin(\pi u)\ A(1,2,4,3,5,6)\cr
&-\sin\lf[\pi\lf(\fc{s}{2}+t\ri)\ri]\ A(1,2,3,5,4,6)\ \}\ ,\cr
A(1,2,3,6,4,5]&=A(1,2,4,5,3,6)\ ,\cr
A(1,2,5,3,4,6)&=\sin(\pi t)^{-1}\ \{\ \sin(\pi u)\ [\ A(1,2,3,4,5,6)+A(1,2,3,4,6,5)\ ]\cr
&+\sin\lf[\pi\lf(\fc{s}{2}+u\ri)\ri]\ A(1,2,3,5,4,6)\ \}\ ,\cr
A(1,2,5,3,6,4)&=\sin(\pi t)^{-2}\ \{ \sin\lf(\fc{\pi s}{2}\ri)^2\ A(1,2,4,5,3,6)\cr
& -\sin(\pi u) \sin\lf[\pi\lf(\fc{s}{2}+t\ri)\ri]\ [\ A(1,2,3,4,5,6) +A(1,2,3,4,6,5)\ ]\cr
&-\sin(\pi u)\ \sin\lf[\pi\lf(\fc{s}{2}+t\ri)\ri] 
\ [\ A(1,2,4,3,5,6) +A(1,2,4,3,6,5)\ ]\cr
&+  \sin\lf[\pi\lf(\fc{s}{2}+t\ri)\ri]^2 \ A(1,2,3,5,4,6)\ \}\ ,\cr
A(1,3,2,4,5,6)&=\sin(\pi t)^{-1}\ \{\ \sin\lf(\fc{\pi s}{2}\ri)\ A(1,2,4,5,3,6)\cr
&+\sin(\pi s)\ [\ A(1,2,3,4,5,6) + A(1,2,4,3,5,6)\ ]\ \}\ ,\cr
A(1,4,2,3,5,6)&=\sin(\pi t)^{-1}\ \{\ \sin\lf(\fc{\pi s}{2}\ri)\ A(1,2,3,5,4,6)\cr 
&+\sin(\pi s)\ [\ A(1,2,3,4,5,6)+ A(1,2,4,3,5,6)\ ]\ \}\ ,\cr
A(1,5,4,6,3,2)&=A(1,2,4,5,3,6)\ ,\cr
A(1,2,4,6,3,5)&=A(1,2,3,5,4,6)\ ,\cr
A(1,5,3,6,4,2)&=A(1,2,3,5,4,6)\ ,\cr
&\vdots}}
and the whole result \ukvi\ becomes \Obtain\ or \Obtaini:
\eqn\UKvi{
\Ac(1,2;3,4)=4\ \sin\lf(\fc{\pi s}{2}\ri)\ \sin(\pi s)\ A(1,6,3,5,4,2)-
4\ \sin\lf(\fc{\pi s}{2}\ri)\ \sin(\pi t)\ A(1,3,5,4,2,6)\ .}
Note, that this expression is different than the one given in \LM.

\subsubsec{Four open and one closed string}

According to \Bronst\ for a given open string ordering $\si$, described in \Orderings, 
in the sum \SPLIT\ we expect $\nu(4,1)=20$ partial ordered amplitudes of six open strings.
The open string coordinate $x$ represents an ordering coordinate for the 
phases \PHASE
\eqn\phasevii{
\Pi(x,\xi,\eta)=\cases{e^{i\pi\hatt\lambda_2}\ ,&  $(1-\xi)\ (1-\eta)<0\ ,$\cr
                     e^{i\pi\hatt\gamma_2}\ ,&  $(\xi-x)\ (\eta-x)<0\ ,$\cr
                     e^{i\pi\hatt\beta_2}\ ,&  $(\xi+x)\ (\eta+x)<0\ ,$}}
with $\hatt\lambda_2,\hatt\gamma_2,\hatt\beta_2$ given in \parameterHU\ and
referring to the non--integer parts, \ie
$\hatt\lambda_2=-\h s_1+\h s_3-s_5,\ \hatt\gamma_2=s_4,\ \hatt\beta_2=\h s_1-\h s_3-s_4$.
We have worked out the general case $\kappa\notin\IZ$ in Subsection 3.5 
and Appendix \appC.
Here we shall restrict to the case $\kappa\in\IZ$, \ie no branching from 
the term $(z-\ov z)^\kappa$ is considered.
For the ordering $\si_1$ the sum \SPLIT\ decomposes as:
\eqn\ukviia{\eqalign{
\Ac(1,3,4,2;5)&=[\ A(1,6,5,3,4,2)+A(1,5,6,3,4,2)\ ]+\b2\ A(1,5,3,6,4,2)\cr
&+\g2\b2\ A(1,5,3,4,6,2)+\l2\g2\b2\ A(1,5,3,4,2,6)\cr
&+\b2\ A(1,6,3,5,4,2)+[\ A(1,3,6,5,4,2)+A(1,3,5,6,4,2)\ ]\cr
&+\g2\ A(1,3,5,4,6,2)+\l2\g2\ A(1,3,5,4,2,6)\cr
&+\g2\b2\ A(1,6,3,4,5,2)+\g2\ A(1,3,6,4,5,2)\cr
&+[\ A(1,3,4,6,5,2)+A(1,3,4,5,6,2)\ ]+\l2\ A(1,3,4,5,2,6)\cr
&+\l2\b2\g2\ A(1,6,3,4,2,5)+\l2\g2\ A(1,3,6,4,2,5)\cr
&+\l2\ A(1,3,4,6,2,5)+[\ A(1,3,4,2,6,5)+A(1,3,4,2,5,6)\ ]\ .}}
With the subamplitude relations \RELvi\ subject to \consvii
$$\eqalign{
A(1,3,6,4,2,5)&=-cs_1\ sn_1\ A(1,3,4,5,2,6)-cs_1\ sin(\pi s_5)\ A(1,5,3,4,2,6) \cr
&-\fc{sn_4}{\sin(\pi s_4)}\ [\ 
A(1,5,2,4,3,6) +A(1,5,3,4,2,6)\ ]\ ,\cr
A(1,2,4,3,5,6)&=-A(1,2,4,3,6,5)-cs_2\ sn_1\ A(1,5,2,4,3,6)\cr
&-cs_2\ \sin(\pi s_4)\ A(1,2,4,5,3,6)\ ,\cr
A(1,2,4,5,6,3)&=-A(1,2,4,6,5,3)-\fc{sn_4}{\sin(\pi s_4)}\ A(1,2,4,5,3,6) \cr
&+\fc{cs_1 sn_1^2}{
sin(\pi s_4)}\ A(1,3,4,5,2,6) +\fc{cs_1\ sn_1\ \sin(\pi s_5)}{\sin(\pi s_4)}\ A(1,5,3,4,2,6) \ ,\cr
A(1,2,4,6,3,5)&=A(1,2,4,5,3,6)+\fc{sn_1}{\sin(\pi s_4)}\ [\
A(1,5,2,4,3,6)-A(1,5,3,4,2,6)\ ]\ ,\cr
A(1,2,5,4,3,6)&=-cs_2\ [\ sn_2\ A(1,2,4,5,3,6)-\sin(\pi s_5)\ A(1,5,2,4,3,6)\ ]\ ,\cr
A(1,2,5,4,6,3)&=\sin(\pi s_4)^{-1}\ [\ sn_2\ A(1,2,4,5,3,6)-sn_1\ A(1,3,4,5,2,6)\cr
&-\sin(\pi s_5)\ A(1,5,2,4,3,6)\ ]\ ,}$$
\eqn\RELviia{\eqalign{
A(1,2,5,6,4,3)&=-A(1,3,4,5,6,2)-\fc{cs_2\ sn_2^2}{\sin(\pi s_4)}\
A(1,2,4,5,3,6) \cr
&+\fc{sn_3}{\sin(\pi s_4)}\ A(1,3,4,5,2,6) +
\fc{cs_2\ sn_2\ \sin(\pi s_5)}{\sin(\pi s_4)}\ A(1,5,2,4,3,6)\ ,\cr
A(1,2,6,4,3,5)&=-cs_2\ sn_2\ A(1,2,4,5,3,6)-\fc{cs_2\ sn_1 sn_2}{\sin(\pi
     s_4)}\ A(1,5,2,4,3,6)\cr
&+\fc{sn_3}{\sin(\pi s_4)}\ A(1,5,3,4,2,6)\ ,\cr
A(1,3,4,6,2,5)&=A(1,3,4,5,2,6)+\fc{sn_2}{\sin(\pi s_4)}\ 
[\ A(1,5,2,4,3,6) -A(1,5,3,4,2,6)\ ]\ ,\cr
A(1,3,5,4,6,2)&=\sin(\pi s_4)^{-1} [\ sn_2\ A(1,2,4,5,3,6)-sn_1
A(1,3,4,5,2,6)\cr
&-\sin(\pi s_5)\ A(1,5,3,4,2,6) \ ]\ ,\cr
A(1,3,4,2,5,6)&=-A(1,3,4,2,6,5)-cs_1\ sn_2\ A(1,5,3,4,2,6)\cr
&+cs_1\ \sin(\pi s_4)\ A(1,3,4,5,2,6)\ ,\cr
A(1,3,5,4,2,6)&=-cs_1 [\ sn_1\ A(1,3,4,5,2,6)+\sin(\pi s_5)\ A(1,5,3,4,2,6)\ ]\ ,}}
with
\eqn\shortlast{\eqalign{
sn_1&=\sin\lf[\fc{\pi}{2}\lf(s_1-s_3+2 s_5\ri)\ri]\ \ \ ,\ \ \ 
sn_2=\sin\lf[\fc{\pi}{2}\lf(s_1-s_3-2 s_4\ri)\ri]\ ,\cr
sn_3&=cs_1^{-1}=\sin\lf[\fc{\pi}{2}\lf(s_1-s_3-2 s_4+2s_5\ri)\ri]\ \ \ ,\ \ \ 
sn_4=cs_2^{-1}=\sin\lf[\fc{\pi}{2}\lf(s_1-s_3\ri)\ri]\ ,}}
and \BASIC\ we arrive at:
\eqn\UKviia{
\Ac(1,3,4,2;5)=2i\ \sin(\pi\hatt\beta_2)\ A(1,2,4,5,3,6)+2i\ \sin(\pi\hatt\lambda_2)\ 
A(1,3,4,5,2,6)\ .} 
For the ordering $\si_2$ the sum \SPLIT\ gives:
\eqn\ukviib{\eqalign{
\Ac(1,4,3,2;5)&=[\ A(1,6,5,4,3,2)+A(1,5,6,4,3,2)\ ]+\g2\ A(1,5,4,6,3,2)\cr
&+\g2\b2\ A(1,5,4,3,6,2)+\l2\g2\b2\ A(1,5,4,3,2,6)\cr
&+\g2\ A(1,6,4,5,3,2)+[\ A(1,4,6,5,3,2)+A(1,4,5,6,3,2)\ ]\cr
&+\b2\ A(1,4,5,3,6,2)+\l2\b2\ A(1,4,5,3,2,6)\cr
&+\g2\b2\ A(1,6,4,3,5,2)+\b2\ A(1,4,6,3,5,2)\cr
&[\ A(1,4,3,6,5,2)+A(1,4,3,5,6,2)\ ]+\l2\ A(1,4,3,5,2,6)\cr
&+\l2\b2\g2\ A(1,6,4,3,2,5)+\l2\b2\ A(1,4,6,3,2,5)\cr
&+\l2\ A(1,4,3,6,2,5)+[\ A(1,4,3,2,6,5)+A(1,4,3,2,5,6)\ ]\ .}}
With the subamplitude relations \RELvi\ subject to \consvii\ 
and \BASIC\ we arrive at:
\eqn\UKviiaa{
\Ac(1,4,3,2;5)=2i\ \sin(\pi\hatt\gamma_2)\ A(1,2,3,5,4,6)+2i\ \sin(\pi\hatt\lambda_2)\ 
A(1,4,3,5,2,6)\ .} 
Finally for the ordering $\si_3$ the sum \SPLIT\ becomes:
\eqn\ukviic{\eqalign{
\Ac(1,4,2,3;5)&=[\ A(1,6,5,3,2,4)+A(1,5,6,3,2,4)\ ]+\b2\ A(1,5,3,6,2,4)\cr
&+\l2\b2\ A(1,5,3,2,6,4)+\l2\g2\b2\ A(1,5,3,2,4,6)\cr
&+\b2\ A(1,6,3,5,2,4)+[\ A(1,3,6,5,2,4)+A(1,3,5,6,2,4)\ ]\cr
&+\l2\ A(1,3,5,2,6,4)+\l2\g2\ A(1,3,5,2,4,6)\cr
&+\l2\b2\ A(1,6,3,2,5,4)+\l2\ A(1,3,6,2,5,4)\cr
&+[\ A(1,3,2,6,5,4)+A(1,3,2,5,6,4)\ ]+\g2\ A(1,3,2,5,4,6)\cr
&+\l2\g2\b2\ A(1,6,3,2,4,5)+\l2\g2\ A(1,3,6,2,4,5)\cr
&+\g2\ A(1,3,2,6,4,5)+[\ A(1,3,2,4,6,5)+A(1,3,2,4,5,6)\ ]\ .}}
With the subamplitude relations \RELvi\ subject to \consvii\ 
and \BASIC\ we arrive at:
\eqn\UKviiaaa{
\Ac(1,4,2,3;5)=2i\ \sin(\pi\hatt\beta_2)\ A(1,4,2,5,3,6)+2i\ \sin(\pi\hatt\gamma_2)\ 
A(1,3,2,5,4,6)\ .} 
The case $x<-1$ gives exactly the same expression.

According to the results \UKviia, \UKviiaa\ and \UKviiaaa\ we might conclude
that the basis is six--dimensional for this case, in agreement with \Bronste.
However, due to \consvii\  some  momenta are restricted. Hence, we should 
expect a smaller basis. In fact, for \consvii\ two of the six partial amplitudes
may be expressed by a basis of four: 
\eqn\CPUSA{\eqalign{
\sin[\pi (s_2-2s_4-2s_5)]\ &\sin\lf[\fc{\pi}{2} (s_1-s_3-2 s_4)\ri]\ A(1,4,2,5,3,6)\cr  
&=-\sin[\pi (s_1-s_3-2 s_4)]\ \sin(\pi s_4)\ A(1,2,3,5,4,6)\cr
&-\sin[\pi (s_1+s_2-s_3-2 s_4)]\ \sin(\pi s_4)\ A(1,3,2,5,4,6)\cr
&+\sin\lf[\fc{\pi}{2} (s_1-s_3-2 s_4)\ri]\ \sin[\pi (s_3+2 s_4)]\ A(1,2,4,5,3,6)\cr 
&-\sin(\pi s_3)\ \sin\lf[\fc{\pi}{2} (s_1-s_3+2 s_5)\ri]\ A(1,3,4,5,2,6)\ ,\cr
\sin[\pi (s_2-2 s_4-2s_5)]\ &\sin\lf[\fc{\pi}{2} (s_1-s_3+2 s_5)\ri]\ A(1,4,3,5,2,6)\cr 
&=\sin\lf[\fc{\pi}{2} (s_1-s_3-2 s_4)\ri]\ \sin[\pi (s_2+s_3-2 s_5)]\ A(1,2,4,5,3,6)\cr 
&-\sin(\pi s_4)\ \sin[\pi (s_1-s_3+2 s_5)]\ A(1,3,2,5,4,6) \cr
&- \sin(\pi s_4)\ \sin[\pi (s_1-s_2-s_3+2 s_5)]\ A(1,2,3,5,4,6)\cr
&- \sin\lf[\fc{\pi}{2} (s_1-s_3+2 s_5)\ri]\sin[\pi (s_2+s_3-2s_4-2s_5)]A(1,3,4,5,2,6).}}
This confirms \specialconf.
The phase factor \phasevii\ is in correspondence to the phases displayed
in Tables~4~and~5.

\subsubsec{Three closed strings}

\def\e2{e^{i\pi\hatt\epsilon_2}}
\def\k2{e^{i\pi\hatt\kappa}}
\def\Wc{{\cal W}}

Although we have worked out this case in full generality in Subsection 3.6 
and Appendix \appD, we find it instructive to also work out the sum \SPLIT.
According to \Bronst\ in the sum \SPLIT\ we expect $\nu(0,3)=40$ 
partial ordered amplitudes of six open strings.
The open string coordinate $x$ represents an ordering coordinate for the 
phases \PHASE
\eqn\phaseto{
\Pi(x,\xi,\eta)=\cases{e^{i\pi\hatt\lambda_2}\ ,&  $(1-\xi)\ (1-\eta)<0\ ,$\cr
                     e^{i\pi\hatt\gamma_2}\ ,&  $(1+\xi)\ (1+\eta)<0\ ,$\cr
                     e^{i\pi\hatt\bet_2}\ ,&  $(\xi-x)\ (\eta-x)<0\ ,$\cr
                     e^{i\pi\hatt\eps_2}\ ,&  $(\xi+x)\ (\eta+x)<0\ ,$\cr
                     e^{i\pi\hatt\kappa}\ ,&  $\xi+\eta<0\ ,$}}
with $\hatt\lambda_2,\hatt\gamma_2,\hatt\beta_2,\hatt\eps_2,\hatt\kappa$ defined in \parameterto\ and
referring to their non--integer parts, \ie
$\hatt\lambda_2=s_6,\ \hatt\gamma_2=\h(-s_1+s_3-s_5)-s_6,\ \hatt\beta_2=s_4$,\ 
$\hatt\eps_2=\h(s_1-s_3-s_5)-s_4$ and $\hat\kappa=s_5$.
With \phaseto\ for $0<x<1$ the sum decomposes \SPLIT\ as:
\eqn\uktoi{\eqalign{
\Wc(1,2,3)_+&=\l2\g2\b2\e2\ A(6,5,1,3,4,2)+\l2\b2\e2\ A(6,1,5,3,4,2)\cr
&+\l2\b2\ A(6,1,3,5,4,2)+\l2\ A(6,1,3,4,5,2)+A(6,1,3,4,2,5)\cr 
&+\g2\k2\ A(1,5,3,4,2,6)+\g2\e2\k2\ A(1,3,5,4,2,6)\cr
&+\b2\g2\e2\k2\ A(1,3,4,5,2,6)+\l2\b2\g2\e2\k2\ A(1,3,4,2,5,6)\cr
&+\l2\b2\g2\e2\ A(1,3,4,2,6,5)+\b2\e2\k2\ A(1,5,6,3,4,2)\cr
&+\b2\e2\ A(1,6,5,3,4,2)+\b2\ A(1,6,3,5,4,2)+A(1,6,3,4,5,2)\cr
&+\l2\ A(1,6,3,4,2,5)+\e2\k2\ A(1,5,3,6,4,2)+\k2\ A(1,3,5,6,4,2)\cr
&+A(1,3,6,5,4,2)+\b2\ A(1,3,6,4,5,2)+\l2\b2\ A(1,3,6,4,2,5)\cr
&+\k2\ A(1,5,3,4,6,2)+\e2\k2\ A(1,3,5,4,6,2)\cr
&+\b2\e2\k2\ A(1,3,4,5,6,2)+\b2\e2\ A(1,3,4,6,5,2)\cr
&+\l2\b2\e2\ A(1,3,4,6,2,5)\ .}}
With the subamplitude relations \RELvi\ subject to \consto, \eg
\eqn\subthree{\eqalign{
A(5,6,1,3,4,2)&+\k2\ A(5,1,3,4,2,6)+\b2\g2\e2\k2\ A(5,1,6,3,4,2)\cr
&+\g2\e2\k2\ A(5,1,3,6,4,2)+\g2\k2\ A(5,1,3,4,6,2)=0\ ,}}
\eqq \uktoi\ can be brought into the form:
\eqn\arrivetoi{\eqalign{
\Wc(1,2,3)_+&=2i\ \Big\{\ -\sin(\pi s_5)\ A(1,3,4,2,6,5)+\sin(\pi s_6)\ A(1,5,2,4,3,6)\cr
&+\sin(\pi s_4)\ A(1,2,5,4,6,3)+\sin[\pi(s_4+s_6)]\ A(1,5,2,4,6,3)\cr
&-e^{-i\pi (\hatt\gamma_2+\hatt\lambda_2)}\ \sin(\pi s_5)\ 
\lf[\ A(1,3,4,6,5,2)+e^{i\pi \hatt\lambda_2}\ A(1,3,4,6,2,5)\ \ri]\ \Big\}\ .}}
On the other hand, for $-1<x<0$ we obtain:
\eqn\uktoii{\eqalign{
\Wc(1,2,3)_-&=\l2\g2\b2\e2\ A(6,5,1,4,3,2)+\l2\b2\e2\ A(6,1,5,4,3,2)\cr
&+\l2\e2\ A(6,1,4,5,3,2)+\l2\ A(6,1,4,3,5,2)+A(6,1,4,3,2,5)\cr
&+\g2\k2\ A(1,5,4,3,2,6)+\g2\b2\k2\ A(1,4,5,3,2,6)\cr
&+\b2\g2\e2\k2\ A(1,4,3,5,2,6)+\l2\b2\g2\e2\k2\ A(1,4,3,2,5,6)\cr
&+\l2\b2\g2\e2\ A(1,4,3,2,6,5)+\b2\e2\k2\ A(1,5,6,4,3,2)\cr
&+\b2\e2\ A(1,6,5,4,3,2)+\e2\ A(1,6,4,5,3,2)+A(1,6,4,3,5,2)\cr
&+\l2\ A(1,6,4,3,2,5)+\b2\k2\ A(1,5,4,6,3,2)+\k2\ A(1,4,5,6,3,2)\cr
&+A(1,4,6,5,3,2)+\e2\ A(1,4,6,3,5,2)+\l2\e2\ A(1,4,6,3,2,5)\cr
&+\k2\ A(1,5,4,3,6,2)+\b2\k2\ A(1,4,5,3,6,2)\cr
&+\b2\e2\k2\ A(1,4,3,5,6,2)+\b2\e2\ A(1,4,3,6,5,2)\cr
&+\l2\b2\e2\ A(1,4,3,6,2,5)\ .}}
With the subamplitude relations \RELvi\ subject to \consto, \eg
\eqn\subthree{\eqalign{
A(5,6,1,4,3,2)&+\k2\ A(5,1,4,3,2,6)+\b2\g2\e2\k2\ A(5,1,6,4,3,2)\cr
&+\g2\b2\k2\ A(5,1,4,6,3,2)+\g2\k2\ A(5,1,4,3,6,2)=0\ ,}}
\eqq \uktoii\ reduces to:
\eqn\arrivetoii{\eqalign{
\Wc(1,2,3)_-&=2i\ \Big\{\ -\sin(\pi s_5)\ A(1,4,3,2,6,5)+\sin(\pi s_6)\ A(1,5,2,3,4,6)\cr
&+\sin\lf[\pi\lf(\fc{s_1}{2}-\fc{s_3}{2}-\fc{s_5}{2}-s_4\ri)\ri]\ A(1,2,5,3,6,4)\cr
&+\sin\lf[\pi\lf(\fc{s_1}{2}-\fc{s_3}{2}-\fc{s_5}{2}-s_4+s_6\ri)\ri]\ A(1,5,2,3,6,4)\cr
&-e^{-i\pi (\hatt\gamma_2+\hatt\lambda_2)}\ \sin(\pi s_5)\ 
\lf[\ A(1,4,3,6,5,2)+e^{i\pi \hatt\lambda_2}\ A(1,4,3,6,2,5)\ \ri]\ \Big\}\ .}}
Eventually, the two expressions \arrivetoi\ and \arrivetoii\ sum up to the result \Johannesburg:
\eqn\sumupto{
W^{(\kappa,\al_0,\al_3)}\lf[{\alpha_1,\lambda_1,\gamma_1,\beta_1,\eps_1\atop
\alpha_2,\lambda_2,\gamma_2,\beta_2,\eps_2}\ri]=\Wc(1,2,3)_++\Wc(1,2,3)_+\ .}
The phase factor \phaseto\ is in correspondence to the phases displayed
in Table~6.
Finally as a remark 
for $x>1$ one obtains exactly the same expression \arrivetoi\ as for $0<x<1$, 
while the case $x<-1$ yields \arrivetoii.

\newsec{Couplings of brane $\&$ bulk string states  vs. pure brane couplings}

In this Section we explicitly compute disk amplitudes involving open and 
closed strings and express these amplitudes as pure open string disk amplitudes.
This provides the map between couplings of brane $\&$ bulk string states onto 
pure brane couplings. We would like to point out, that disk amplitudes involving only 
closed strings are somewhat special. For these amplitudes
the pole structure is completely inherited from the pure open string amplitudes.
As a consequence these amplitudes furnish a divergence due to a dilaton tadpole.

Essentially the relation between couplings of brane $\&$ bulk string states to pure brane couplings could be restricted to Neveu--Schwarz fields only as 
the couplings involving Ramond fields may be simply obtained from the latter by
using supersymmetric Ward identities \August.

\subsec{Two brane $\&$ one bulk field versus four brane fields}

The three--point amplitude \startwithiv\ involving two open strings and one closed string
is identical to the four--point amplitude of involving four open strings after suitable identification of polarizations and momenta.
This fact has been anticipated at the level of the kinematics in \HKi.
However, in this Subsection for completeness and as illustrative example  we want to 
derive the full relation: We present the three--point amplitude
involving two vectors \gaugevertexzero\ and one
massless closed string field \dilaton\ and equate it with the appropriate four--point open superstring amplitude with four vectors \fieldsiii\ and \gaugevertexzero.

The amplitude under consideration follows from the expression \startwithiv
\eqn\Startwithiv{\eqalign{
\Ac(1,2;3)&=x_\infty^2\ \int_{-\infty}^\infty dx\ (1+ix)\cr
&\times \vev{:V_{A^a}^{(0)}(x_\infty,\zeta_1,2p_1):\ :V_{A^a}^{(0)}(1,\zeta_2,2p_2):\ :V_G^{(-1,-1)}(-ix,ix,\eps,q):}\ ,}}
with the two open string vertex operators $V_o$ given in \gaugevertexzero\ and the closed string vertex in \dilaton. With \expi\ and \inviv\ the full amplitude may be cast into the form \firenze
\eqn\finaliv{
\Ac(1,2;3)=2^{s-1}\ B\lf(\h,\h-t\ri)\ K(1,2,3)\ ,}
with the kinematical factor:
\eqn\kiniv{\eqalign{
K(1,2,3)&=\{\ t\ (\zeta_1\zeta_2)\ \tr(\eps D)-4\ t\ (\zeta_1\eps D\zeta_2)+\h\ 
\tr(\eps D)\ (\zeta_1k_3)\ (\zeta_2 k_4)+2\ (\zeta_1\zeta_2)\ (k_1\eps Dk_2)\cr
&-\ \lf[\ (\zeta_1\eps D p_1)\ (\zeta_2q)+(\zeta_2\eps D p_2)\ (\zeta_1q)
+(\zeta_1\eps D p_1)\ (\zeta_2Dq)+(\zeta_2\eps D p_2)\ (\zeta_1Dq)\ \ri]\cr
&+\lf[\ (\zeta_1\eps Dk_3)\ (\zeta_2k_1)+(\zeta_2\eps Dk_3)\ (\zeta_1k_2) \ri]\ \}\ .}}
Note, that we have worked with a symmetric tensor $\eps$ taking into account the degrees
of freedom of the graviton and dilaton field. This is way the kinematical factor appears to be symmetric in the indices $3$ and $4$.

On the other hand, the (partial ordered) four open superstring amplitude involving four 
vectors \fieldsiii\ and \gaugevertexzero\ 
\eqn\FFinaliv{\eqalign{
A(1,2,3,4)&=x_\infty^2\ \int_{1}^\infty dx\ (1+x)\cr 
&\hskip-0.5cm\times\vev{\ :V_{A^a}^{(0)}(x_\infty,\xi_1,2k_1):\ :V_{A^a}^{(0)}(1,\xi_2,2k_2):\ 
:V_{A^a}^{(-1)}(-x,\xi_3,2k_3):\ :V_{A^a}^{(-1)}(x,\xi_4,2k_4):}}}
assumes the form
\eqn\Finaliv{
A(1234)=\fc{1}{\hatt s_2}\ B(\hatt s_1,\hatt s_3)\ K(\xi_1,k_1;\xi_2,k_2;\xi_3,k_3;\xi_4,k_4)\ ,}
with $\hatt s_1=4k_1k_2,\ \hatt s_2=4k_1k_3,\ \hatt s_3=4k_1k_4$ and the kinematical factor \JSCH:
\eqn\Kiniv{\eqalign{
&K(\xi_1,k_1;\xi_2,k_2;\xi_3,k_3;\xi_4,k_4)=\hatt s_2\hatt s_3\ (\xi_1\xi_2)\ (\xi_3\xi_4)
+\hatt s_1\hatt s_2\ (\xi_1\xi_4)\ (\xi_2\xi_3)+\hatt s_1\hatt s_3\ (\xi_1\xi_3)\ (\xi_2\xi_4)\cr
&\hskip0.25cm+\hatt s_1\ \lf[(\xi_1\xi_3)(\xi_2k_3)(\xi_4k_1)+(\xi_1\xi_4)(\xi_2k_4)(\xi_3k_1) +
(\xi_2\xi_3)(\xi_1k_3)(\xi_4k_2)+(\xi_2\xi_4)(\xi_1k_4)(\xi_3k_2)\ri]\cr
&\hskip0.25cm+\hatt s_2\ \lf[(\xi_1\xi_2)(\xi_3k_2)(\xi_4k_1)+(\xi_1\xi_4)(\xi_2k_1)(\xi_3k_4) +
(\xi_2\xi_3)(\xi_1k_2)(\xi_4k_3)+(\xi_3\xi_4)(\xi_1k_4)(\xi_2k_3)\ri]\cr
&\hskip0.25cm+\hatt s_3\ \lf[(\xi_1\xi_2)(\xi_3k_1)(\xi_4k_2)+(\xi_1\xi_3)(\xi_2k_1)(\xi_4k_3) +
(\xi_2\xi_4)(\xi_1k_2)(\xi_3k_4)+(\xi_3\xi_4)(\xi_1k_3)(\xi_2k_4)\ri].}} 

With the assignment of momenta \assigniv\ and 
\eqn\Assigniv{
\xi_1=\zeta_1\ \ \ ,\ \ \ \xi_2=\zeta_2\ \ \ ,\ \ \ \xi_3\otimes\xi_4=\eps D\ ,}
we have the following relation between the amplitude \finaliv\ of two open and one closed string and the four open string partial amplitude \Finaliv
\eqn\Agreeiv{
\mathboxit{\ \Ac(1,2;3)=\sin(\pi t)\ A(1,2,3,4)\ ,\ }}
in agreement with the proposition \Obtainiv.
Note, that the relation \Agreeiv\ holds for both NS and R states as external states.

\subsec{Two closed strings versus four brane fields}

A correspondence between a disk amplitude of two closed strings and a disk amplitude of four open strings has been observed in {\it Refs.} \doubref\GM\HK.
We briefly exhibit this result to show the difference to the relation \Agreeiv.

We consider the disk amplitude
\eqn\StartW{
\Ac(1,2)=\int_0^1 dy\ (1-y^2)\ 
\vev{:\ V_G^{(-1,-1)}(-iy,iy,\eps_1,q_1):\ :V_G^{(0,0)}(-i,i,\eps_2,q_2):}\ ,}
with the two closed string vertex operators given in \dilaton.
From momentum conservation \conservation\ we have $q_{1\parallel}+q_{2\parallel}=0$, with
$q_\parallel=\h(q+Dq)$. The kinematic invariants are 
$s=2q_{1\parallel}^2=2q_{2\parallel}^2,\ t=q_1q_2$ and $u=q_1Dq_2$ \HK.
The amplitude \StartW\ can be related to the four open string amplitude $A(1234)$, given in \Finaliv, as
\eqn\AgreeIV{
\mathboxit{\ \Ac(1,2)=\ A(1,2,3,4)\ ,\ }}
subject to the identifications \doubref\GM\HK
\eqn\subject{\eqalign{
&k_1=\h Dq_1\ \ ,\ \ k_2=q_1\ \ ,\ \ k_3=\h Dq_2\ \ ,\ \ k_4=\h q_2\ ,\cr
&\hatt s_1=s\ \ \ ,\ \ \ \hatt s_2=t\ \ \ ,\ \ \ \hatt s_3=u\ ,}}
and:
\eqn\subjecti{
\xi_1\otimes\xi_2=\eps_1D\ \ \ ,\ \ \ \xi_3\otimes\xi_4=\eps_2D\ .}
In the double cover with the relations \allow\ the result \AgreeIV\ becomes:
\eqn\AgreeIVa{
\Ac(1,2)=\lf(1+\fc{\sin(\pi u)}{\sin(\pi t)}\ri)\ A(1,2,3,4)\ ,\ }

Note, that in contrast to the equation \Agreeiv, in which a $\sin$--factor enters, 
in the relation \AgreeIV\  an amplitude of only closed strings $\Ac(1,2)$ is 
directly equated with an amplitude of four open strings. As a consequence, the singularity
structure of the disk amplitude of two closed strings  is described by the four open string amplitude. 
A mathematical explanation for this behaviour follows from the fact, that for the choice
$z_1=iy$ and $z_2=i$ of the closed string 
vertex positions all factors of complex $i$  drop out in the correlator
\StartW.
Hence, the integral \StartW\ becomes a real integral  of the type \Fouropen\ and no analytic continuation producing $\sin$--factors is necessary. Furthermore, in the case
of three closed strings on the disk only one complex coordinate has to be 
analytically continued. As a consequence only one $\sin$--factor appears 
in the relation \ObtainTO\ to six open string amplitudes.

\subsec{Three open $\&$ one closed strings versus five brane fields}

The four--point amplitude \startwithvi\ involving three open strings and one closed string
is identical to the five--point amplitude of five open strings after suitable 
identification of polarizations and momenta.
In this Subsection we want to present the four--point amplitude
involving three vectors \gaugevertexzero\ and one
massless closed string Ramond field \TEN\ and equate it with the 
appropriate five--point 
open superstring amplitude with five vectors \fieldsiii\ and \gaugevertexzero.

The amplitude under consideration\foot{We perform the computation for some
$D=4$ superstring compactification with $f^\mu$ being the vector field strength
of a scalar with some internal  index $j$ \oocc. Furthermore, the polarizations 
$\zeta_1,\zeta_2$ of the first two vectors are aligned w.r.t. to some internal directions $i_1$ and $i_2$.}
follows from the expression \startwithv
\eqn\Startwithv{\eqalign{
\Ac(1,2,3;4)&=x_\infty^2\cr 
&\hskip-0.65cm\times\int_{\IC} d^2z\  
\vev{V_{A^{a_1}}^{(0)}(x_\infty,\zeta_1,2p_1)\ V_{A^{a_2}}^{(0)}(0,\zeta_2,2p_2)\ 
V_{A^{a_3}}^{(-1)}(1,\zeta_3,2p_3)\ V^{(-\h,-\h)}_{F}(\ov z,z,f,q)}\ ,}}
with the three open string vertex operators $V_o$ given in \fieldsiii\ and 
\gaugevertexzero\  and the Ramond closed string vertex \TEN.
After performing the Wick contractions, applying the Greens
functions from Section 2, with {\it Eqs.} 
\expi, \invv\ and with the internal correlator \oocc 
\eqn\internalcorr{
\vev{\Lambda^{i_1}(z_1)\ov\Lambda^{i_2}(z_2)\Sigma^j(z,\ov z)}=(M^j+\ov M^j)\ 
\fc{\delta^{i_1i_2}}{z_{12}(z-\ov z)^{3/4}}\ 
\lf[\fc{(z_1-z)(z_2-\ov z)}{(z_1-\ov z)(z_2-z)}\ri]^{1/2}}
the whole amplitude \Startwithv\ may be written in terms of a single function $F$
\eqn\finalv{
\Ac(1,2,3;4)=\fc{i}{\sqrt 2}\ 
\Tr(T^{a_1}T^{a_2}T^{a_3})\ (M^j+\ov M^j)\ \eps_{\lambda_1\lambda_2\mu_3\mu}\ 
p_1^{\lambda_1}p_2^{\lambda_2}\zeta_3^{\mu_3}\ f^\mu\ F\ ,}
with the function: 
\eqn\Integralss{
F=\fc{1}{2}\ \lf(\ G^{(-1)}\lf[{t-1,s-1\atop
t,s}\ri]-G^{(-1)}\lf[{t-1,s\atop t,
s-1}\ri]\ \ri)=\h\ V\lf[{t-1,s-1\atop t,s-1}\ri]\ .}
To cast the amplitude \Startwithv\ into the short form \finalv\ the following identities
\eqn\Relationsident{\eqalign{
\ \ \ &u\ \lf(\ G^{(-1)}\lf[{t,s-1\atop t,s}\ri]+
G^{(-1)}\lf[{t-1,s\atop t+1,s-1}\ri]\
\ri)\cr
&=-t\ \lf(\ G^{(-1)}\lf[{t-1,s\atop t,s-1}\ri]+
G^{(-1)}\lf[{t-1,s-1\atop t,s}\ri]\ \ri)\cr
&=-s\ G^{(-1)}\lf[{t-1,s-1\atop t+1,s-1}\ri]
+(1-s)\ G^{(-1)}\lf[{t,s-1\atop t,s-1}\ri]}}
have been used. These relations  may be proven by inserting the explicit result
(A.17) and using various  hypergeometric functions identities.

On the other hand, from the  five open superstring amplitude
involving three vectors \fieldsiii\ and \gaugevertexzero\ and 
two gauginos \fieldsii\ we may extract the partial ordered amplitude
\eqn\FFinalv{\eqalign{
A(1,5,2,4,3)&=x_\infty^2\ \int_0^1 d\xi\ \int_{-\infty}^0 d\eta\  
\vev{\ :V_{A^{a_1}}^{(0)}(x_\infty,\xi_1,2k_1):
:V_{A^{a_2}}^{(0)}(1,\xi_2,2k_2):\cr
&\times :V_{A^{a_3}}^{(-1)}(1,\xi_3,2k_3):\ :V_{\chi}^{(-1/2)}(\xi,u_4,2k_4):\ :V_{\ov\chi}^{(-1/2)}(\eta,\ov v_5,2k_5):}\ .}}
After performing all contractions  and using \Assumeintei\  the subamplitude
\FFinalv\ becomes
\eqn\FFinalv{\eqalign{
A(1,5,2,4,3)&=\eps_{\lambda_1\lambda_2\mu_3\mu}\ 
k_1^{\lambda_1}k_2^{\lambda_2}\xi_3^{\mu_3}\ 
(u^\al_4\si_{\al\dot\beta}^\mu \ov v^{\dot\beta}_5)\ 
\int_0^1 d\xi\ \int_{-\infty}^0 d\eta\  
\xi^{-\hatt s_2-\hatt s_3+\hatt s_5-1}\ (1-\xi)^{\hatt s_3-1}\cr
&\times (-\eta)^{-\hatt s_1+\hatt s_3-\hatt s_5}\ (1-\eta)^{\hatt s_1-\hatt s_3-\hatt s_4-1}\ (\xi-\eta)^{\hatt s_4}\cr
&=\fc{\Gamma(-\hatt s_2-\hatt s_3+\hatt s_5)\ \Gamma(1-\hatt s_1+\hatt s_3-\hatt s_5)\ \Gamma(\hatt s_3)\ 
\Gamma(1-\hatt s_1-\hatt s_2+\hatt s_4)}{\Gamma(1-\hatt s_1-\hatt s_2)\ 
\Gamma(1-\hatt s_1-\hatt s_2+\hatt s_3+\hatt s_4)}\cr
&\times\FF{3}{2}\lf[{1-\hatt s_1+\hatt s_3-\hatt s_5\ ,\ 1-\hatt s_1-\hatt s_2
+\hatt s_4\ ,\ -\hatt s_2\atop
1-\hatt s_1-\hatt s_2\ ,\ 1-\hatt s_1-\hatt s_2+\hatt s_3+\hatt s_4}\ri]\cr &\times\eps_{\lambda_1\lambda_2\mu_3\mu}\ 
k_1^{\lambda_1}k_2^{\lambda_2}\xi_3^{\mu_3}\ 
(u^\al_4\si_{\al\dot\beta}^\mu \ov v^{\dot\beta}_5)\ ,}}
with the five invariants $\hatt s_i$ given in \Invv.

With the assignment of momenta \assignv, \consv, and (note $\ov\si^{\mu\dot\beta\al}\si^\nu_{\al\dot\bet}=-2\delta^{\mu\nu}$) 
\eqn\Assignvi{
\xi_1=\zeta_1\ \ \ ,\ \ \ \xi_2=\zeta_2\ \ \ ,\ \ \ \xi_3=\zeta_3\ \ \ ,\ \ \ 
u_4^\al\otimes\ov v_5^{\dot\bet}=f_\mu\ \ov\sigma^{\mu\al\dot\bet}\ ,}
we have the following relation between the amplitude \Startwithv\ of three open and one closed string and the  
five open string partial amplitudes \FFinalv
\eqn\Agreev{
\mathboxit{\ \Ac(1,2,3;4)=\sin(\pi t)\ A(1,5,2,4,3)\ ,\ }}
in agreement with the proposition \Obtainv.

The function $F$ encodes the $\ap$--dependence of the string $S$--matrix. 
It is given by
\eqn\Lowexpp{
F=\fc{1}{s}-\fc{\zeta(3)}{4}\ tu-\fc{\zeta(5)}{32}\ tu\ (u^2-st)+\ldots\ .}

\subsec{Two open $\&$ two closed strings versus six brane fields}

The four--point amplitude \startwithvi\ involving two open strings and two closed strings
is identical to the six--point amplitude of six open strings after suitable 
identification of polarizations and momenta.
In this Subsection we want to present the four--point amplitude
involving two vectors \gaugevertexzero\ and two
massless closed string fields \dilaton\ and equate it with the appropriate six--point 
open superstring amplitude with six vectors \fieldsiii\ and \gaugevertexzero.

The amplitude under consideration follows from the expression \startwithvi
\eqn\Startwithvi{\eqalign{
\Ac(1,2;3,4)&=x_\infty^2\ \int_{-\infty}^\infty dx\ (1+ix)\ \int_{\IC} d^2z\cr 
&\hskip-1.5cm 
\vev{\ :V_{A^{a_1}}^{(0)}(x_\infty,\zeta_1,2p_1):\ :V_{A^{a_2}}^{(0)}(1,\zeta_2,2p_2):\ :V_{G_1}^{(0,0)}(-ix,ix,\eps_1,q_1):\ :V_{G_2}^{(-1,-1)}(\ov z,z,\eps_2,q_2):\ },}}
with the two open string vertex operators $V_o$ given in \gaugevertexzero\ and the 
two closed string vertices \dilaton. With \exp\ and \invvi\ the full amplitude may be cast\foot{We perform the computation for a setup with the polarizations 
$\zeta_1,\zeta_2$ of the  two vectors aligned orthogonal 
w.r.t. the D--brane world--volume. Hence these states describe D--brane positions.}
into the form \boilvi
\eqn\finalvi{
\Ac(1,2;3,4)=F_1\ K_1(1,2,3,4)+F_2\ K_2(1,2,3,4)\ ,}
with the two kinematical factors
\eqn\kinvi{\eqalign{
K_1(1,2,3,4)&=tu\ (\zeta_1\zeta_2)\ \lf\{\ \tr(\eps_1\eps_2)+ \fc{2}{t}\ (p_1\eps_1\eps_2p_2)+\fc{2}{u}\  (p_2\eps_1\eps_2p_1)\ri.\cr
&+\lf.\fc{1}{u^2}\ (p_2\eps_1p_2)\ (p_1\eps_2p_1)+\fc{1}{t^2}\ (p_1\eps_1p_1)\ (p_2\eps_2p_2)+\fc{2}{tu}\ (p_1\eps_1p_2)\ (p_1\eps_2p_2)\ \ri\}\ ,\cr\cr
K_2(1,2,3,4)&=(\zeta_1\zeta_2)\ \lf\{\ (p_1\eps_1\eps_2p_1) +(p_2\eps_1\eps_2p_2) -\fc{u}{t}\ (p_1\eps_1\eps_2p_2) -\fc{t}{u}\ (p_2\eps_1\eps_2p_1)\ri.\cr
&+\fc{1}{u}\ \lf[(p_1\eps_1p_2)\ (p_1\eps_2p_1)+(p_2\eps_1p_2)\ (p_1\eps_2p_2)\ri]\cr
&+\fc{1}{t}\ \lf[(p_1\eps_1p_1)\ (p_1\eps_2p_2)+(p_1\eps_1p_2)\ (p_2\eps_2p_2)\ri]\cr
&\lf.-\fc{t}{u^2}\ (p_2\eps_1p_2)\ (p_1\eps_2p_1)-\fc{u}{t^2}\ (p_1\eps_1p_1)\ (p_2\eps_2p_2)
-\lf(\fc{1}{t}+\fc{1}{u}\ri)\ (p_1\eps_1p_2)\ (p_1\eps_2p_2)\ \ri\},}}
and the complex integrals, \cf \eqq \GEN:
\eqn\FUNCTIONS{\eqalign{
F_1&=\int_{-\infty}^\infty dx\ x^{2}\ (1+ix)^{u-1}\ 
(1-ix)^{u-1}\cr
&\times \int_\IC d^2z \ (1-z)^{t}\ (1-\ov z)^{t}\ 
(z-ix)^{\fc{s}{2}-1}\ (\ov z-ix)^{\fc{s}{2}-1}\ (z+ix)^{\fc{s}{2}-1}\ (\ov z+ix)^{\fc{s}{2}-1}\ ,\cr
F_2&=\fc{u}{2}\ \int_{-\infty}^\infty dx\ (1+ix)^{u-1}\ (1-ix)^{u-1}\cr
&\times \int_\IC d^2z \ (z+\ov z)\ (1-z)^{t}\ (1-\ov z)^{t}\ 
(z-ix)^{\fc{s}{2}-1}\ (\ov z-ix)^{\fc{s}{2}-1}\ (z+ix)^{\fc{s}{2}-1}\ (\ov z+ix)^{\fc{s}{2}-1}\ .}}
The amplitude \Startwithvi\ is completely specified by the basis of two generalized hypergeometric functions $F_1,F_2$ and the two kinematical factors $K_1,K_2$. 
Since the two functions $F_1,F_2$ and the kinematics $K_1,K_2$
are invariant under the exchange $t\leftrightarrow u$, $\eps_1\leftrightarrow \eps_2$ \cf also \SYMMETRY, this symmetry is manifest in the amplitude \finalvi.

According to \Obtain\ and \Obtaini\ 
the two integrals can be evaluated and written in terms of two
six open string partial amplitudes \sixopen
\eqn\FUNCTIONSi{\eqalign{
F_1&=\sin\lf(\fc{\pi s}{2}\ri)\ \sin(\pi s)\ A_1(163542)-\sin\lf(\fc{\pi s}{2}\ri)\ \sin(\pi t)\ A_1(135426)\ ,\cr
F_2&=\sin\lf(\fc{\pi s}{2}\ri)\ \sin(\pi s)\ A_2(163542)-\sin\lf(\fc{\pi s}{2}\ri)\ 
\sin(\pi t)\ A_2(135426)\ ,}}
with the four world--sheet integrals:
\eqn\resuuua{\eqalign{
A_1(163542)
&=-\fc{1}{8}\ \int_0^1dx\int_0^1dy\int_0^1dz\ x^{\fc{u}{2}}\ (1-x)^{\fc{s}{4}}\ 
y^{-\fc{s}{4}}\ (1-y)^{\fc{s}{4}}\ z^{\fc{u}{2}}\ (1-z)^{\fc{s}{4}}\ (1-xyz)^{\fc{s}{4}}\cr 
&\times \fc{(1-yz)^2}{(1-x)\ y\ (1-y)\ z\ (1-z)\ (1-xyz)}\ ,\cr
A_1(135426)
&=-\fc{1}{8}\ \int_0^1dx\int_0^1dy\int_0^1dz\    (1-x)^{\fc{s}{4}}\ y^{\fc{s}{2}}\ (1-y)^{\fc{t}{2}}\ z^{\fc{u}{2}}\ (1-z)^{\fc{t}{2}}\cr 
&\times (1-xy)^{\fc{u}{2}}\  (1-xyz)^{\fc{s}{4}}\ \fc{y}{x\ (1-x)\ (1-y)\ (1-yz)},}}
\eqn\resuuub{\eqalign{
A_2(163542)
&=\fc{u}{8}\ \int_0^1dx\int_0^1dy\int_0^1dz\ x^{\fc{u}{2}}\ (1-x)^{\fc{s}{4}}\ y^{-\fc{s}{4}}\ (1-y)^{\fc{s}{4}}\ z^{\fc{u}{2}}\ (1-z)^{\fc{s}{4}}\ (1-xyz)^{\fc{s}{4}}\cr
&\times\lf\{\ -\fc{1}{x\ (1-x)\ y\ (1-y)\ z\ (1-z)}-\fc{y}{(1-x)\ (1-y)\ (1-xyz)}\ri.\cr
&\lf.-\fc{1}{x\ (1-y)\ (1-z)\ (1-xyz)}+\fc{1}{(1-x)\ y\ z\ (1-z)\ (1-xyz)}\ \ri\}\ ,\cr
A_2(135426)
&=\fc{u}{8}\ \int_0^1dx\int_0^1dy\int_0^1dz\ (1-x)^{\fc{s}{4}}\ y^{\fc{s}{2}}\ 
(1-y)^{\fc{t}{2}}\ z^{\fc{u}{2}}\ (1-z)^{\fc{t}{2}}\cr
&\times (1-xy)^{\fc{u}{2}}\ (1-xyz)^{\fc{s}{4}}\ 
\fc{2-y}{(1-x)\ (1-y)}\ \lf\{\ \fc{1}{x\ y\ z}+\fc{1}{1-yz}\ \ri\}\ .}}

On the other hand, from the six open superstring amplitude involving six
vectors \fieldsiii\ and\ \gaugevertexzero\ 
we may extract the two (partial ordered) amplitudes 
$$\eqalign{
A(1,6,3,5,4,2)&=x_\infty^2\ \int_0^1 d\rho\ (1+\rho)\ \int_{-\rho}^\rho d\xi\int_{-\infty}^{-\rho} d\eta\ \vev{\ :V_{A^a}^{(0)}(x_\infty,\zeta_1,2k_1):\ :V_{A^a}^{(0)}(1,\zeta_2,2k_2):\cr 
&\times :V_{A^a}^{(-1)}(-\rho,\zeta_3,2k_3):\ :V_{A^a}^{(-1)}(\rho,\zeta_4,2k_4):\ :
V_{A^a}^{(0)}(\xi,\zeta_5,2k_5):\ :V_{A^a}^{(0)}(\eta,\zeta_6,2k_6)}\ ,}$$
\eqn\FFinalvi{\eqalign{
A(1,3,5,4,2,6)&=x_\infty^2\ \int_{0}^1 d\rho\ (1+\rho)\ \int_{-\rho}^\rho d\xi\int_{1}^{\infty} d\eta\ \vev{\ :V_{A^a}^{(0)}(x_\infty,\zeta_1,2k_1):\ :V_{A^a}^{(0)}(1,\zeta_2,2k_2):\cr 
&\times :V_{A^a}^{(-1)}(-\rho,\zeta_3,2k_3):\ :V_{A^a}^{(-1)}(\rho,\zeta_4,2k_4):\ :
V_{A^a}^{(0)}(\xi,\zeta_5,2k_5):\ :V_{A^a}^{(0)}(\eta,\zeta_6,2k_6)}\ .}}
The final result for  the two partial amplitudes \FFinalvi\ can be written as 
\eqn\Resuuu{\eqalign{
A(1,6,3,5,4,2)&=K_1\ A_1(163542)+K_2\ A_2(163542)\ ,\cr
A(1,3,5,4,2,6)&=K_1\ A_1(135426)+K_2\ A_2(135426)\ ,}}
with the kinematical factors $K_1,K_2$ given in \kinvi\ and the four six open string disk integrals:
\eqn\Resuuua{\eqalign{
A_1(163542)&=-2^{\hatt s_3}\ 
\int_{0}^1 d\rho\ \int_{-\rho}^\rho d\xi\int_{-\infty}^{-\rho} 
d\eta\ \rho^{\hatt s_3+2}\ (1+\rho)^{\hatt s_2-1}\ 
(1-\rho)^{-\hatt s_2-\hatt s_3+\hatt t_2-1}\cr 
&\times(1-\xi)^{\hatt s_3+\hatt s_6-\hatt t_2-\hatt t_3}\ (1-\eta)^{-\hatt s_1-\hatt s_6+\hatt t_3}\ (\rho-\xi)^{\hatt s_4-1}\ 
(-\rho-\eta)^{\hatt s_1+\hatt s_4-\hatt t_1-\hatt t_3-1}\cr
&\times (\rho+\xi)^{-\hatt s_3-\hatt s_4+\hatt t_3-1}\ (\rho-\eta)^{-\hatt s_4-\hatt s_5+\hatt t_1-1}\ (\xi-\eta)^{\hatt s_5}\ ,\cr
A_1(135426)&=-2^{\hatt s_3}\ \int_{0}^1 d\rho\ \int_{-\rho}^\rho d\xi\int_{1}^{\infty} 
d\eta\ \rho^{\hatt s_3+2}\ (1+\rho)^{\hatt s_2-1}\ (1-\rho)^{-\hatt s_2-\hatt s_3+\hatt t_2-1}\cr 
&\times(1-\xi)^{\hatt s_3+\hatt s_6-\hatt t_2-\hatt t_3}\ (\eta-1)^{-\hatt s_1-\hatt s_6+\hatt t_3}\ (\rho-\xi)^{\hatt s_4-1}\ 
(\rho+\eta)^{\hatt s_1+\hatt s_4-\hatt t_1-\hatt t_3-1}\cr
&\times (\rho+\xi)^{-\hatt s_3-\hatt s_4+\hatt t_3-1}\ (\eta-\rho)^{-\hatt s_4-\hatt s_5+\hatt t_1-1}\ (\eta-\xi)^{\hatt s_5}\ ,}}
\eqn\Resuuub{\eqalign{
A_2(163542)
&=2^{\hatt s_3-1}\ \hatt s_6\ \int_{0}^1 d\rho\ \int_{-\rho}^\rho d\xi\int_{-\infty}^{-\rho} 
d\eta\ \rho^{\hatt s_3}\ (1+\rho)^{\hatt s_2-1}\ (1-\rho)^{-\hatt s_2-\hatt s_3+\hatt t_2-1}\cr 
&\times(1-\xi)^{\hatt s_3+\hatt s_6-\hatt t_2-\hatt t_3}\ (1-\eta)^{-\hatt s_1-\hatt s_6+\hatt t_3}\ (\rho-\xi)^{\hatt s_4-1}\ 
(-\rho-\eta)^{\hatt s_1+\hatt s_4-\hatt t_1-\hatt t_3-1}\cr
&\times (\rho+\xi)^{-\hatt s_3-\hatt s_4+\hatt t_3-1}\ (\rho-\eta)^{-\hatt s_4-\hatt s_5+\hatt t_1-1}\ 
\ (\eta-\xi)^{\hatt s_5}\ (\xi+\eta)\ ,\cr
A_2(135426)
&=2^{\hatt s_3-1}\ \hatt s_6\ \int_{0}^1 d\rho\ \int_{-\rho}^\rho d\xi\int_{1}^{\infty} 
d\eta\ \rho^{\hatt s_3}\ (1+\rho)^{\hatt s_2-1}\ (1-\rho)^{-\hatt s_2-\hatt s_3+\hatt t_2-1}\cr 
&\times(1-\xi)^{\hatt s_3+\hatt s_6-\hatt t_2-\hatt t_3}\ (\eta-1)^{-\hatt s_1-\hatt s_6+\hatt t_3}\ (\rho-\xi)^{\hatt s_4-1}\ 
(\rho+\eta)^{\hatt s_1+\hatt s_4-\hatt t_1-\hatt t_3-1}\cr
&\times (\rho+\xi)^{-\hatt s_3-\hatt s_4+\hatt t_3-1}\ (\eta-\rho)^{-\hatt s_4-\hatt s_5+\hatt t_1-1}\ 
\ (\eta-\xi)^{\hatt s_5}\ (\eta+\xi)\ .}}
Clearly, with \consvi\ and the transformations \transi\ and \transii\ the two sets of expressions \resuuua, \resuuub\ and \Resuuua, \Resuuub\ may be identified.

With the assignment of momenta \assignvi\ and
\eqn\Assignvi{
\xi_1=\zeta_1\ \ \ ,\ \ \ \xi_2=\zeta_2\ \ \ ,\ \ \ 
\xi_3\otimes\xi_4=\eps_1\ \ \ ,\ \ \ \xi_5\otimes\xi_6=\eps_2\ ,}
we have the following relation between the amplitude \Startwithvi\ of two open and two closed strings and the pair of 
six open string partial amplitudes \FFinalvi
\eqn\Agreevi{
\mathboxit{\ \Ac(1,2;3,4)=\sin\lf(\fc{\pi s}{2}\ri)\ \sin(\pi s)\
A(1,6,3,5,4,2)-\sin\lf(\fc{\pi s}{2}\ri)\ \sin(\pi t)\ A(1,3,5,4,2,6)\ ,\ }}
in agreement with the proposition \Obtaini.

The $\ap$--expansion of the amplitude \Startwithiv\ is completely 
determined by the $\ap$--expansions of the two functions \FUNCTIONS
\eqn\ExpansionsFI{\eqalign{
F_1&=-\fc{1}{s}-\fc{1}{4}\ \zeta(2)\ s+\fc{1}{16}\ \lf(3\ s^2+2\ tu\ri)\ \zeta(3)
-\fc{1}{64}\ \lf(19\ s^3-5\ stu\ri)\ \zeta(4)\cr
&+\fc{1}{64}\ s^2\ \lf(3\ s^2+5\ tu\ri)\ 
\zeta(2)\ \zeta(3)+\fc{3}{768}\ \lf(45\ s^4+22\ s^2\ tu+8\ t^2u^2\ri)\ \zeta(5)+\ldots\ ,\cr\cr
F_2&=\fc{1}{4}\ \zeta(2)\ tu-\fc{3}{16}\ stu\ \zeta(3)
+\fc{1}{64}\ \lf(19\ s^2tu-4\ t^2u^2\ri)\ \zeta(4)\cr
&-\fc{1}{128}\ stu\ \lf(6\ s^2+8\ tu\ri)\ \zeta(2)\ \zeta(3)-\fc{5}{256}\ stu\ 
\lf(9\ s^2-4\ tu\ri)\ \zeta(5)+\ldots\ .}}
Note, that the integrals \resuuua\ and \resuuub\ allow for at most a triple pole, which is converted to
a single pole by the two $sin$--factors in \FUNCTIONS.
The lowest (field--theory) order of the amplitude \Startwith\ is given by the first term
of \finalvi, \ie $\lf.\Ac(1,2;3,4)\ri|_{\ap^0}=\fc{K_1}{s}$. This order
has been already determined in \fotoii. On the other hand,
the NLO string correction at $\ap^2$ stems from both $K_1$ and $K_2$.
Note, that this order gives non--trivial gravitational corrections to the D--brane
world volume couplings at the order $\ap^2$.

\newsec{Concluding remarks}

We have made explicit\foot{The general expressions on the disk world--sheet are given in \eqqs \AMPLITUDE, \DISENT\ and \momenta, while the integrated amplitudes are presented in \eqqs \Obtainiv,\ \Obtainv,\ \Obtaini,\ \Obtainvii,\ \ObtainTO\ and \Obtainviii, respectively.} 
the link between disk amplitudes of open $\&$ closed
strings and amplitudes of pure open strings.
 This map represents a sort of generalized KLT relation on the disk.
It relates a disk amplitude of $N_o$ open and $N_c$ closed strings to a sum of
disk amplitudes involving $N_o+2N_c$ open strings only.
After analytic continuation the complex world--sheet integrations  over the closed string positions $z_i$ become real integrals along the boundary of the disk.
This way each closed string state is divided up into two (interacting)
open string fields at real positions $\eta_i,\xi_i$: 
\eqn\LAST{
\matrix{
&&&\cr
&V_{\rm closed}(\ov z_i,z_i)&\simeq V_{\rm open}(\ov z_i)\ V_{\rm open}(z_i)
\simeq &V_{\rm open}(\eta_i)\ V_{\rm open}(\xi_i)\ .\cr
&{\scriptstyle z_i\ \in\ \IC}&&{\scriptstyle\eta_i,\ \xi_i\ \in\ \IR}\cr
&&&}}
The correspondence between open $\&$ closed
strings and amplitudes of pure open strings is made on the string world--sheet.
Therefore this map holds for amplitudes in any space--time dimensions
and gives relations between couplings of brane $\&$ bulk 
string states  vs. pure brane couplings. Some illustrative examples
for these relations, which have been derived in this work, are:
\eqn\ILLU{\eqalign{
\Ac(1,2)&=A(1,2,3,4)\ ,\cr
\Ac(1,2;3)&=\sin(\pi t)\ A(1,2,3,4)\ ,\cr
\Ac(1,2,3;4)&=\sin(\pi t)\ A(1,5,2,4,3)\ ,\cr
\Ac(1,2;3,4)&=\sin\lf(\fc{\pi s}{2}\ri)\ \sin(\pi s)\
A(1,6,3,5,4,2)-\sin\lf(\fc{\pi s}{2}\ri)\ \sin(\pi t)\ A(1,3,5,4,2,6)\ ,\cr
\Ac(1,2,3,4;5)&=\sin(\pi s_4)\ A(1,6,4,5,3,2)-\sin\pi\lf(\fc{s_1}{2}-\fc{s_3}{2}+s_5\ri)\ A(1,4,3,5,2,6)\ ,\cr
\Ac(1,2,3)&=\sin\pi\lf(\fc{s_1}{4}-\fc{s_3}{4}\ri)\ 
\lf\{\ A(1,3,6,4,5,2)+A(1,4,6,3,5,2)\ri.\cr
&\hskip3cm\lf.-A(1,3,4,6,2,5)-A(1,4,3,6,2,5)\ \ri\}\ ,\cr
&\vdots}}
Here, the open/closed disk amplitudes are denoted by 
$\Ac(1,\ldots,N_o;N_{o}+1,\ldots,N_o+N_c)$, while the partial ordered open string amplitudes are given by $A(1,\ldots,N_o+2N_c)$ and permutations thereof. 
 
This sort of generalized KLT relation on the disk gives rise to direct relations
between gauge and gravitational amplitudes and pure gauge amplitudes.
At the level of the effective action the map \LAST, \ie the correspondence between
disk amplitudes of open \& closed strings and pure open string amplitudes
reveals important relations between brane and bulk couplings and pure brane couplings, \cf Section 5.
Note, that this map is different than in the KLT case, where tree--level
couplings of closed strings are mapped to squares of open string couplings.
Hence this map does not directly translate into a map at the level of 
couplings in a Lagrangian.

Formally for a disk amplitude of two vectors and two massless Neveu--Schwarz
closed string states we obtain
\eqn\nice{\eqalign{
&\vev{A_{\mu_1}(x_1)\ A_{\mu_2}(x_2)\ G_{\mu_3\mu_4}(\ov z_1,z_1)\ 
G_{\mu_5\mu_6}(\ov z_2,z_2)}\  \zeta^{\mu_1}\ \zeta^{\mu_2}\ \eps_1^{\mu_3\mu_4}\ \eps_2^{\mu_5\mu_6}\ \cr
&\hskip1.75cm\simeq \vev{A_{\mu_1}(x_1)\ A_{\mu_2}(x_2)\ A_{\mu_3}(x_3)\ A_{\mu_4}(x_4)\ 
A_{\mu_5}(x_5)\ A_{\mu_6}(x_6)}\ \xi^{\mu_1}\xi^{\mu_2}\xi^{\mu_3}\xi^{\mu_4}\xi^{\mu_5}\xi^{\mu_6}\ ,}}
with the identifications \Assignvi\ and assignment of momenta \assignvi.
For a special choice of polarizations the first amplitude of \nice\ has been computed in Subsection 5.3.
For a disk amplitude of two vectors and two massless Ramond $p$-- and $q$--forms
the correspondence
\eqn\nicei{\eqalign{
&\vev{A_{\mu_1}(x_1)\ A_{\mu_2}(x_2)\ F_{\al\bet}(\ov z_1,z_1)\ 
F_{\gamma\delta}(\ov z_2,z_2)}\cr  
&\hskip1.75cm\times \zeta^{\mu_1}\ \zeta^{\mu_2}\ 
f^1_{\mu_0\ldots\mu_p}\ \lf(P^+\Gamma^{[\mu_0}\ldots\Gamma^{\mu_p]}\ri)^{\al\bet}\ 
f^2_{\nu_0\ldots\nu_q}\ \lf(P^+\Gamma^{[\nu_0}\ldots\Gamma^{\nu_q]}\ri)^{\gamma\delta}\cr
&\hskip1.75cm\simeq\vev{A_{\mu_1}(x_1)\ A_{\mu_2}(x_2)\ \chi_{\al}(x_3)\ \chi_{\bet}(x_4)\ 
\chi_{\gamma}(x_5)\ \chi_{\delta}(x_6)}\ \xi^{\mu_1}\xi^{\mu_2}\ 
u^\alpha\ v^\beta\ u^\gamma\ v^\delta}}
can be established.

After compactification the relations \nice\ and \nicei\ give rise to various
relations between open and closed string moduli
and pure open string moduli. These identities establish 
non--trivial relations between disk correlators involving internal open $\&$ 
closed conformal fields of open and closed string vertex operators and
correlators involving only internal open conformal fields of open string vertex
operators \oocc. This way disk amplitudes involving open and closed string 
moduli are mapped to amplitudes of only open string moduli \oocc. 
See also {\it Refs.} \Herbst\ for a possible relation.

With the techniques developed in this work any disk amplitude with an
arbitrary number of open and closed strings can be computed.
It should be possible to derive similar relations as the one we have found at the disk tree--level for other Riemann surfaces with boundaries. 
{\it E.g.} at the one--loop--level a cylinder amplitude of open $\&$ closed
strings should be expressible as sum of cylinder amplitudes involving only open strings.

Finally, we find it interesting to mention, 
that the polygon \Polygon\ relevant to the scattering of two open and two closed strings appears in a different context \AM.
The polygon made of the sequence of lightlike segments specifies the
scattering configuration of six gluons with self--crossing in the $T$--dual coordinates. 
The polygon defines the boundary conditions for the string world--sheet 
relevant for the leading exponential behaviour of the six--point scattering
amplitude in field--theory.
The polygon contains the kinematic information about the momenta of the six gluons.
Furthermore, the same polygon as \Polygon\  specifies the
scattering configuration of two gluons and two quarks in the $T$--dual
coordinates \McGreevy\ and the configuration space is a subspace of
the one of the six--gluon worldsheet with crossings.


\centerline{\noindent{\bf Acknowledgments} }

I am indebted to Lance Dixon, and Peter Mayr for many valuable discussions. 
In addition, I thank Angelos Fotopoulos, Hans Jockers,
and Johanna Knapp for sharing their insights with me.
Furthermore, I thank the Kavli Institute for Theoretical Physics in Santa Barbara, 
the Stanford Institute For Theoretical Physics at Stanford University and 
the Galileo Galilei Institute for Theoretical Physics in Firenze 
for hospitality and financial support during completion  of this work.
This research was supported in part by the National Science Foundation 
under Grant No. PHY05-51164 and INFN.
The diagrams have been created by the program JaxoDraw~\Jaxo.

\break
\appendix\appA{Complex world--sheet integrals 
$\ss{G^{(\al)}\lf[{\lambda_1,\gamma_1\atop\lambda_2,\gamma_2}\ri]}$}

In this appendix we compute the complex integral \Provee
\eqn\Prove{
G^{(\al)}\lf[{\lambda_1,\gamma_1\atop\lambda_2,\gamma_2}\ri]:=
\int_\IC d^2z \ z^{\lambda_1}\ \ov z^{\lambda_2}\ 
(1-z)^{\gamma_1}\ (1-\ov z)^{\gamma_2}\ (z-\ov z)^\al\ ,}
for $\al\in\IR$. 
In analogy to the procedure applied in Subsection 2.3.3 we split the complex integral
into holomorphic and anti--holomorphic pieces by rotating the $z_2$--integration from the real axis to the pure imaginary axis, \ie
$iz_2\in\IR$. This way, the variables  $\xi=z_1+iz_2$ and $\eta=z_1-iz_2$ become two real independent 
quantities, \ie $\xi,\eta\in\IR$ and 
\Prove\ becomes
\eqn\angel{\eqalign{
G^{(\al)}\lf[{\lambda_1,\gamma_1\atop\lambda_2,\gamma_2}\ri]&=
\h\int_{-\infty}^\infty d\xi\ \int_{-\infty}^\infty d\eta \ |\xi|^{\hatt\lambda_1}\ 
|1-\xi|^{\hatt\gamma_1}\ 
|\eta|^{\hatt\lambda_2}\ |1-\eta|^{\hatt\gamma_2}\ |\xi-\eta|^{\hatt\alpha}\cr
&\times \xi^{n_1}\ (1-\xi)^{m_1}\ \eta^{n_2}\ (1-\eta)^{m_2}\
(\xi-\eta)^{\tilde\alpha}\ \Pi(\xi,\eta)\ ,}}
with the decomposition 
$\lambda_i=\hatt \lambda_i+n_i,\ \gamma_i=\hatt\gamma_i+m_i,\ \al=\hatt\al+\tilde\al$
separating integer and non--integer parts and 
a phase factor $\Pi$ given in the Table 1. \br

{\vbox{\ninepoint{$$
\vbox{\offinterlineskip\tabskip=0pt
\halign{\strut\vrule#
&~$#$~\hfil 
&\vrule$#$ 
&~$#$~\hfil 
&\vrule$#$ 
&~$#$~\hfil 
&\vrule$#$ 
&~$#$~\hfil 
&\vrule$#$
&~$#$~\hfil 
&\vrule$#$ 
&~$#$~\hfil 
&\vrule$#$
&~$#$~\hfil 
&\vrule$#$\cr
\noalign{\hrule}
&  &&(\xi,\eta)  
&& \eta<\xi && \xi<\eta<0   &&  0<\eta<1&& \eta>1 && {\rm total\ phase} &\cr
\noalign{\hrule}
&(i)&&\xi<0 &&1 && e^{i\pi\hat\alpha} &&e^{i\pi(\hat\alpha+\hat\lambda_2)}&& 
e^{i\pi(\hat\alpha+\hat\lambda_2+\hat\gamma_2)} &&\sigma_\lambda &\cr
\noalign{\hrule}}}$$}}}
{\vbox{\ninepoint{$$
\vbox{\offinterlineskip\tabskip=0pt
\halign{\strut\vrule#
&~$#$~\hfil 
&\vrule$#$ 
&~$#$~\hfil 
&\vrule$#$ 
&~$#$~\hfil 
&\vrule$#$ 
&~$#$~\hfil 
&\vrule$#$
&~$#$~\hfil 
&\vrule$#$ 
&~$#$~\hfil 
&\vrule$#$
&~$#$~\hfil 
&\vrule$#$\cr
\noalign{\hrule}
&  &&(\xi,\eta)  
&& \eta<0 && 0<\eta<\xi   &&  \xi<\eta<1&& \eta>1 && {\rm total\ phase} &\cr
\noalign{\hrule}
&(ii)&&0<\xi<1   &&e^{i\pi\hat\lambda_2} &&
1 &&e^{i\pi\hat\alpha}&&e^{i\pi(\hat\alpha+\hat\gamma_2)}&&1 &\cr
\noalign{\hrule}}}$$}}}
{\vbox{\ninepoint{$$
\vbox{\offinterlineskip\tabskip=0pt
\halign{\strut\vrule#
&~$#$~\hfil 
&\vrule$#$ 
&~$#$~\hfil 
&\vrule$#$ 
&~$#$~\hfil 
&\vrule$#$ 
&~$#$~\hfil 
&\vrule$#$
&~$#$~\hfil 
&\vrule$#$ 
&~$#$~\hfil 
&\vrule$#$
&~$#$~\hfil 
&\vrule$#$\cr
\noalign{\hrule}
&  &&(\xi,\eta)  
&& \eta<0 && 0<\eta<1   &&  1<\eta<\xi&& \eta>\xi && {\rm total\ phase} &\cr
\noalign{\hrule}
&(iii)&&\xi>1 &&e^{i\pi(\hat\lambda_2+\hat\gamma_2)}&&e^{i\pi\hat\gamma_2} &&1 &&
e^{i\pi\hat\alpha} &&\sigma_\gamma &\cr
\noalign{\hrule}}}$$
\vskip-6pt
\centerline{\noindent{\bf Table 1:}
{\sl Phases $\Pi(\xi,\eta)$ along the integration region $(\xi,\eta)$.}}
\vskip10pt}}}

\noindent
The phases $\sigma_\lambda,\ \sigma_\gamma$ are defined as: 
$\sigma_\lambda=e^{i\pi(n_1+n_2)},\ \sigma_\gamma=e^{i\pi(m_1+m_2)}$.

\subsec{Case\ $\al=-2$}

\noindent
Let us first consider the case $\al=-2$. For this case, we find
\eqn\Prove{\eqalign{
G^{(-2)}\lf[{\lambda_1,\gamma_1\atop\lambda_2,\gamma_2}\ri]&
=-\sin(\pi\gamma_2)\ \fc{\Gamma(1+\lambda_1)\ \Gamma(1-\lambda_2-\gamma_2)\ 
\Gamma(1+\gamma_1)\ \Gamma(1+\gamma_2)}{\Gamma(2+\lambda_1+\gamma_1)\ 
\Gamma(2-\lambda_2)}\cr 
&\times\FF{3}{2}\lf[{1+\lambda_1,1-\lambda_2-\gamma_2,2\atop 2+\lambda_1+
\gamma_1,2-\lambda_2};1\ri]-\pi\ (\lambda_1\gamma_2-\lambda_2\gamma_1)\cr
&\times\lf[\fc{\Gamma(\gamma_1+\gamma_2)\ \Gamma(\lambda_1+\lambda_2)}
{\Gamma(1+\lambda_1+\lambda_2+\gamma_1+\gamma_2)}+\si_\gamma\ 
\fc{\Gamma(\gamma_1+\gamma_2)\ \Gamma(-\lambda_1-\lambda_2-\gamma_1-\gamma_2)}
{\Gamma(1-\lambda_1-\lambda_2)}\ri]\ ,}}
Obviously, the result is symmetric under the exchange of $\lambda_i\leftrightarrow
\gamma_i$.

After analyzing the structure of the contour integrals, displayed in the next two figures, 
we find that only the 
case $(ii)$ contributes to \angel. Furthermore, there is a contribution
from the residuum at $\eta=\xi$ of case $(iii)$.
\ifig\contiii{\ The complex $\eta$--plane and the contour integrals for the two 
cases~$(i)$~and~$(ii)$. }{\epsfxsize=0.42\hsize\epsfbox{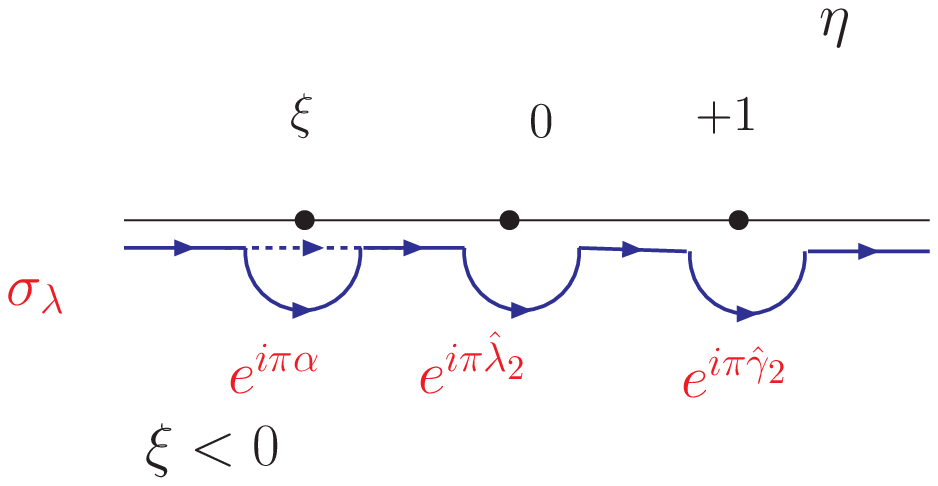}
\epsfxsize=0.42\hsize\epsfbox{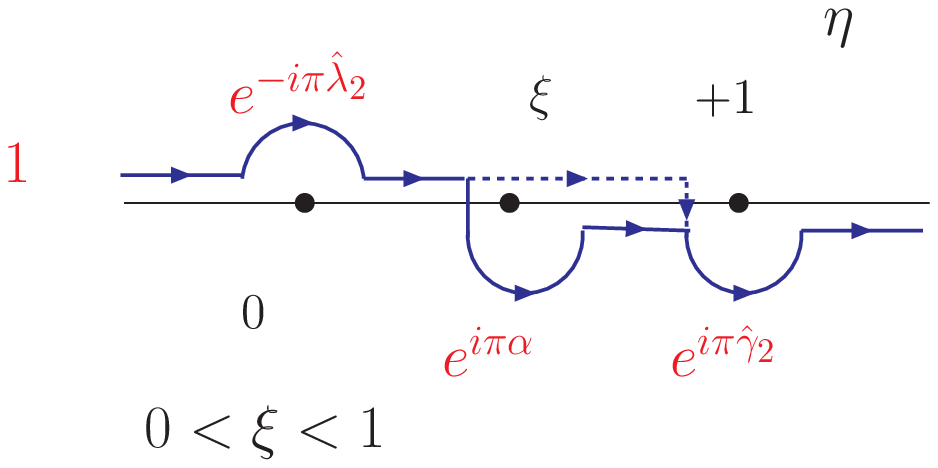}}
\ifig\contiii{\ The complex $\eta$--plane and the contour integrals for the 
cases $(iii)$. }{\epsfxsize=0.42\hsize\epsfbox{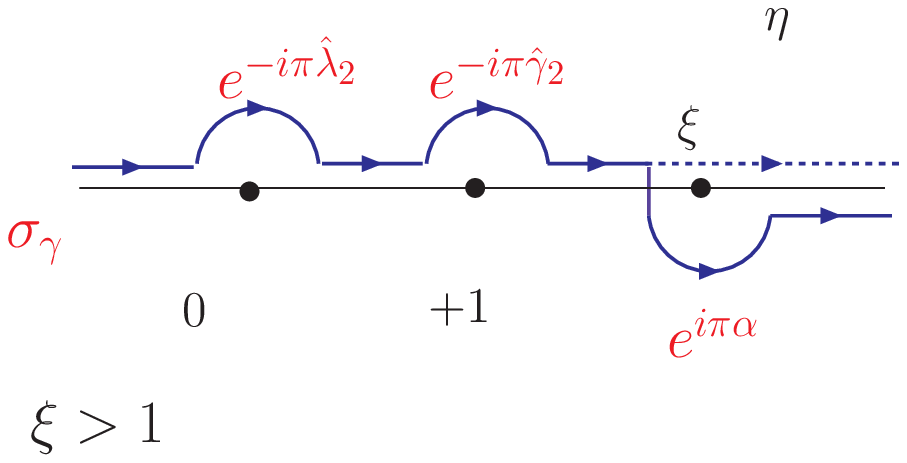}}
\noindent
We may deform the contour in the 
complex $\eta$--plane and integrate along the real $\eta$--axis from $-\infty$ until 
$0$:
\eqn\Angell{
G^{(-2)}_{ii}\lf[{\lambda_1,\gamma_1\atop\lambda_2,\gamma_2}\ri]=
\sin(\pi\lambda_2)\ \int_0^1 d\xi\ \xi^{\lambda_1}\ (1-\xi)^{\gamma_1}\ \int_0^\infty
d\eta\ \eta^{\lambda_2}\ (1+\eta)^{\gamma_2}\ (\xi+\eta)^{-2}}
After writing $(\xi+\eta)^{-2}=\xi^{-2}(1+\eta/\xi)^{-2}$ 
the $\eta$--integration is performed using the integral 3.259 (3) of \Gradst.
Then the identity 9.131 (2) of \Gradst\ allows to write 
the hypergeometric function $\FF{2}{1}[1-\xi]$ as a sum
of two hypergeometric functions  $\FF{2}{1}[\xi]$ with argument $\xi$.
Finally the $\xi$--integral is performed with 7.512 (12) of \Gradst\
and \Angell\ gives:
\eqn\Angel{\eqalign{
G^{(-2)}_{ii}\lf[{\lambda_1,\gamma_1\atop\lambda_2,\gamma_2}\ri]&=-\pi\ 
\fc{\Gamma(1+\gamma_1)\ \Gamma(1+\lambda_1)\ \Gamma(1-\gamma_2-\lambda_2)}
{\Gamma(-\gamma_2)\ \Gamma(2+\lambda_1+\gamma_1)\ \Gamma(2-\lambda_2)}\ 
\FF{3}{2}\lf[{1+\lambda_1,1-\lambda_2-\gamma_2,2\atop
2+\lambda_1+\gamma_1,2-\lambda_2};1\ri]\cr
&-\pi\ \lambda_2\ \fc{\Gamma(1+\gamma_1)\ \Gamma(\lambda_1+\lambda_2)}{
\Gamma(1+\lambda_1+\lambda_2+\gamma_1)}\ \FF{3}{2}\lf[{\lambda_1+\lambda_2,-\gamma_2,
1+\lambda_2\atop 1+\lambda_1+\lambda_2+\gamma_1,\lambda_2};1\ri]\ .}}
Alternatively, we may integrate 
from $1$ until $\infty$ including the residuum at $\eta=\xi$:
\eqn\Angela{\eqalign{
G^{(-2)}_{ii}\lf[{\lambda_1,\gamma_1\atop\lambda_2,\gamma_2}\ri]&=
\sin(\pi\gamma_2)\ \int_0^1 d\xi\ \xi^{\lambda_1}\ (1-\xi)^{\gamma_1}\ \int_1^\infty
d\eta\ \eta^{\lambda_2}\ (\eta-1)^{\gamma_2}\ (\eta-\xi)^{-2}\cr
&+\int_0^1 d\xi\ \xi^{\lambda_1}\ (1-\xi)^{\gamma_1}\ \oint_{\eta=\xi}d\eta\ 
\eta^{\lambda_2}\ (1-\eta)^{\gamma_2}\ (\xi-\eta)^{-2}\cr
&=-\pi\ 
\fc{\Gamma(1+\gamma_1)\ \Gamma(1+\lambda_1)\ \Gamma(1-\gamma_2-\lambda_2)}
{\Gamma(-\gamma_2)\ \Gamma(2+\lambda_1+\gamma_1)\ \Gamma(2-\lambda_2)}\ 
\FF{3}{2}\lf[{1+\lambda_1,1-\lambda_2-\gamma_2,2\atop
2+\lambda_1+\gamma_1,2-\lambda_2};1\ri]\cr
&+\pi\ (\lambda_1\gamma_2-\lambda_2\gamma_1)\ \fc{\Gamma(\lambda_1+\lambda_2)\ 
\Gamma(\gamma_1+\gamma_2)}{\Gamma(1+\lambda_1+\lambda_2+\gamma_1+\gamma_2)}\ .}}
The residuum of \Angela\ is evaluated by taking into account the expansions:
\eqn\expforres{\eqalign{
\eta^{\lambda_2}&=\xi^{\lambda_2}+\lambda_2\ \xi^{\lambda_2-1}\ (\eta-\xi)+\ldots\ ,\cr
(1-\eta)^{\gamma_2}&=(1-\xi)^{\gamma_2}-\gamma_2\ (1-\xi)^{\gamma_2-1}\ (\eta-\xi)+
\ldots\ .}}
It is straightforward to show, that the two expressions \Angel\ and \Angela\ 
are indeed identical and yield the first two terms of \Prove.
Finally, for the case $(iii)$ the residuum at $\eta=\xi$ gives:
\eqn\finres{\eqalign{
R^{(-2)}_{iii}\lf[{\lambda_1,\gamma_1\atop\lambda_2,\gamma_2}\ri]&=\h\ 
\si_\gamma\ \int_1^\infty d\xi\ \xi^{\lambda_1}\ (\xi-1)^{\gamma_1}\ \oint_{\eta=\xi}d\eta\ 
\eta^{\lambda_2}\ (\eta-1)^{\gamma_2}\ (\xi-\eta)^{-2}\cr
&=\pi\ \si_\gamma\ (\lambda_1\gamma_2-\lambda_2\gamma_1)\ 
\fc{\Gamma(\gamma_1+\gamma_2)\ \Gamma(-\lambda_1-\lambda_2-\gamma_1-\gamma_2)}
{\Gamma(1-\lambda_1-\lambda_2)}\ ,}}
which is the last term of \Prove. 

To conclude, the total result for \Prove\ is the sum of \Angela\ and
\finres, \ie
$$G^{(-2)}\lf[{\lambda_1,\gamma_1\atop\lambda_2,\gamma_2}\ri]=
G^{(-2)}_{ii}\lf[{\lambda_1,\gamma_1\atop\lambda_2,\gamma_2}\ri]+
R^{(-2)}_{iii}\lf[{\lambda_1,\gamma_1\atop\lambda_2,\gamma_2}\ri]$$
and combines into  \Prove.

By partial integration we may deduce some relations between the two integrals
$\ss{G^{(-2)}\lf[{\lambda_1,\gamma_1\atop\lambda_2,\gamma_2}\ri]}$ and 
$\ss{V\lf[{\lambda_1,\gamma_1\atop\lambda_2,\gamma_2}\ri]}$. The latter function
has been introduced in \Special. {\it E.g.} the following relation may be proven:
$$G^{(-2)}\lf[{\lambda,\gamma\atop\lambda,\gamma}\ri]=
-\h\ \lf(\ \lambda\ V\lf[{\lambda-1,\gamma\atop\lambda-1,\gamma}\ri]+\gamma\ 
V\lf[{\lambda,\gamma-1\atop\lambda,\gamma-1}\ri]\ri)\ .$$
For the case $\gamma_i=\gamma,\ \lambda_i=\lambda$ \eqq \Prove\ reduces to:
\eqn\Provepartial{
G^{(-2)}\lf[{\lambda,\gamma\atop\lambda,\gamma}\ri]:=\int_\IC d^2z \ 
|z|^{2\lambda}\ |1-z|^{2\gamma}\ \fc{1}{(z-\ov z)^2}
=\fc{\pi}{2}\ \fc{\Gamma(\lambda)\ \Gamma(\gamma)\ \Gamma(-\gamma-\lambda)}
{\Gamma(-\lambda)\ \Gamma(-\gamma)\ \Gamma(\gamma+\lambda)}\ ,}
which my be used to prove the previous relation explicitly.

\subsec{General case $\al\neq -2$}

\noindent
In this subsection we shall compute \Prove\ for general $\al\in\IR$.
After introducing the parameterization $\xi=z_1+iz_2\equiv z,\
\eta=z_1-iz_2\equiv \ov z$ the integral \Prove\ may be written
as \angel\ with the phase factor $\Pi$ given in the Table 3. 
After analyzing the structure of the contour integrals we find that the two
cases $(ii)$ and $(iii)$ contribute to \angel. 

\ifig\contiii{\ The complex $\eta$--plane and the contour integrals for the two 
cases~$(i)$~and~$(ii)$. }{\epsfxsize=0.42\hsize\epsfbox{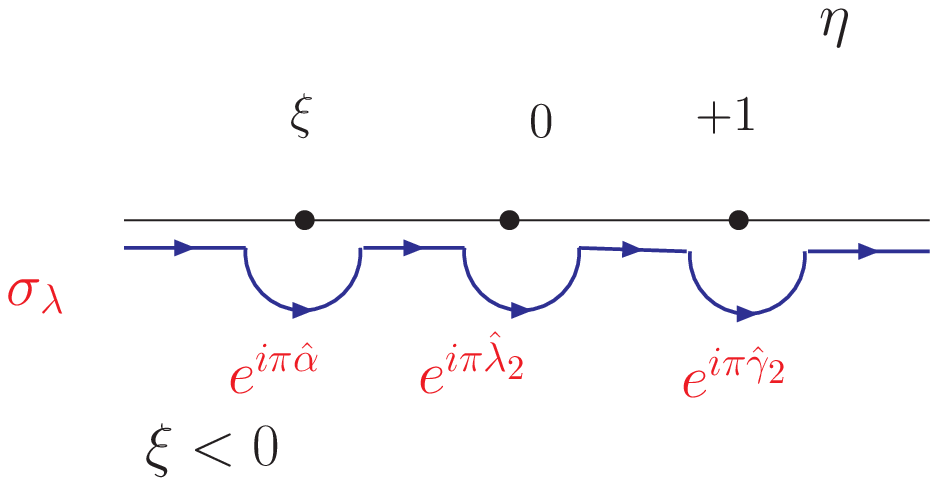}
\epsfxsize=0.42\hsize\epsfbox{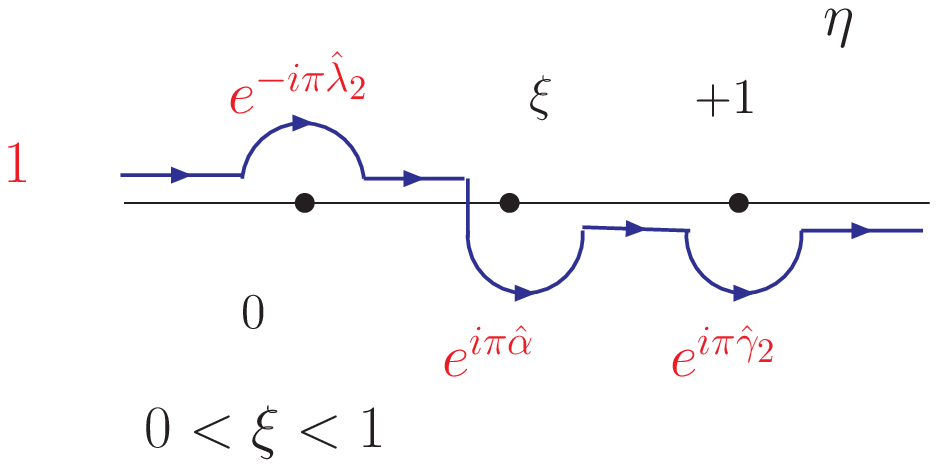}}
\ifig\contiii{\ The complex $\eta$--plane and the contour integrals for the 
case $(iii)$. }{\epsfxsize=0.42\hsize\epsfbox{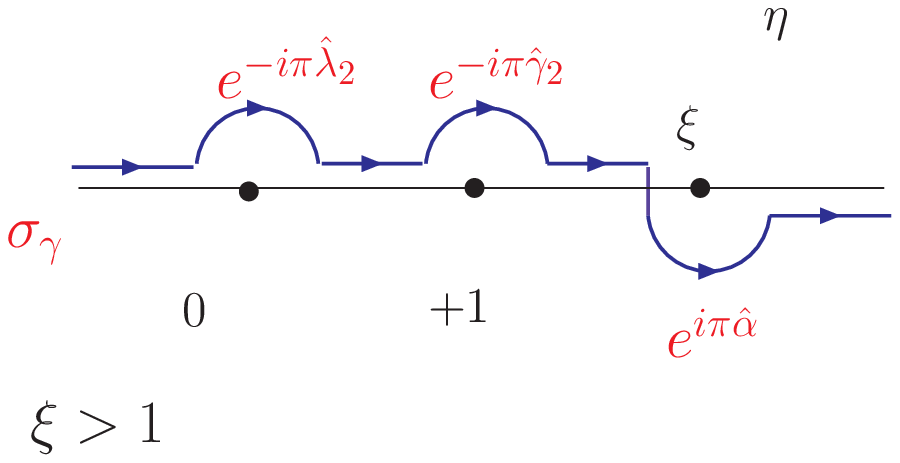}}

\noindent
For case $(ii)$ we may deform the contour in the 
complex $\eta$--plane to the left hand side and integrate along the 
real $\eta$--axis from $-\infty$ until $0$. 
Alternatively, as we shall demonstrate, we may also deform the contour to the 
right hand side and integrate along the real $\eta$--axis from $\xi$ until 
$\infty$, see the next Figure.
\ifig\Dontiia{\ Deformed contour of the case ~$(ii)$.}
{\epsfxsize=0.42\hsize\epsfbox{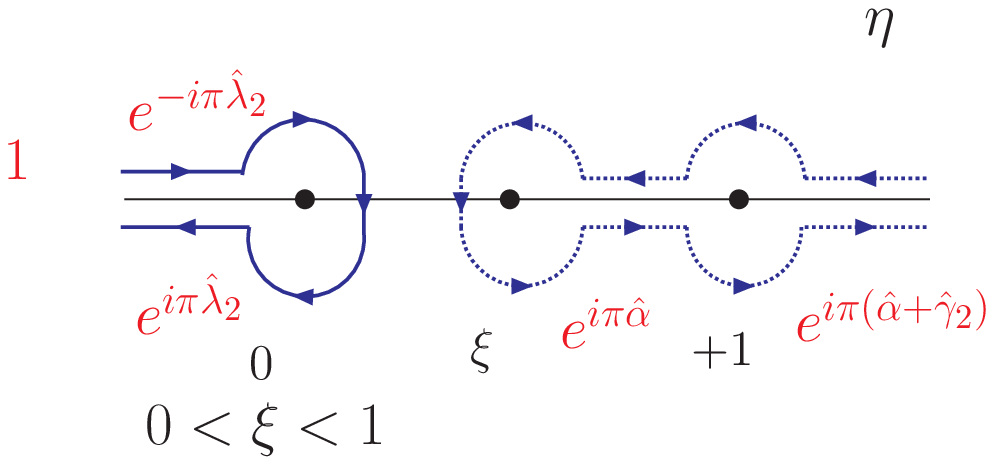}}

\noindent
Hence the total contributions from $(ii)$ is:
\eqn\AAngell{
G^{(\al)}_{ii}\lf[{\lambda_1,\gamma_1\atop\lambda_2,\gamma_2}\ri]=
\sin(\pi\lambda_2)\ \int_0^1 d\xi\ \xi^{\lambda_1}\ (1-\xi)^{\gamma_1}\ \int_0^\infty
d\eta\ \eta^{\lambda_2}\ (1+\eta)^{\gamma_2}\ (\xi+\eta)^{\al}\ .}
After writing $(\xi+\eta)^{-\al}=\xi^{-\al}(1+\eta/\xi)^{-\al}$ we proceed
as described below \Angell\ and obtain
\eqn\Angeli{\eqalign{
G^{(\al)}_{ii}\lf[{\lambda_1,\gamma_1\atop\lambda_2,\gamma_2}\ri]&=
\sin(\pi\lambda_2)\ 
\fc{\Gamma(1+\al+\lambda_2)\Gamma(1+\gamma_1)\ \Gamma(1+\lambda_1)\ 
\Gamma(-\al-1-\gamma_2-\lambda_2)}
{\Gamma(-\gamma_2)\ \Gamma(2+\lambda_1+\gamma_1)}\cr 
&\times\FF{3}{2}\lf[{1+\lambda_1,-\al-1-\lambda_2-\gamma_2,-\al\atop
2+\lambda_1+\gamma_1,-\al-\lambda_2};1\ri]\cr
&+\sin(\pi\lambda_2)\ 
\fc{\Gamma(1+\lambda_2)\ \Gamma(1+\gamma_1)\ \Gamma(-\al-1-\lambda_2)\ 
\Gamma(2+\al+\lambda_1+\lambda_2)}{\Gamma(-\al)\ 
\Gamma(3+\al+\lambda_1+\lambda_2+\gamma_1)}\cr 
&\times\FF{3}{2}\lf[{2+\al+\lambda_1+\lambda_2,-\gamma_2,
1+\lambda_2\atop 3+\al+\lambda_1+\lambda_2+\gamma_1,2-\al+\lambda_2};1\ri]\cr
&=\sin(\pi\lambda_2)\ \fc{\Gamma(1+\lambda_2)\ \Gamma(1+\gamma_1)\ 
\Gamma(-\al-1-\lambda_2-\gamma_2)\
\Gamma(2+\al+\lambda_1+\lambda_2)}{\Gamma(-\al-\gamma_2)\
\Gamma(3+\al+\lambda_1+\lambda_2+\gamma_1)}\cr
&\times\FF{3}{2}\lf[{-\gamma_2,1+\lambda_2,1+\gamma_1\atop 
-\al-\gamma_2,3+\al+\lambda_1+\lambda_2+\gamma_1};1\ri]\ ,}}
which gives the first two lines of \Provee\ and reduces to \Angel\ for $\al=-2$.
Alternatively, we may integrate from $\xi$ until $1$ with phase factor
$e^{i\pi \hatt\al}$ and from $1$ until
$\infty$ with phase factor $e^{i\pi(\hatt\al+\hatt\gamma_2)}$:
\eqn\Angelaa{\eqalign{
G^{(\al)}_{ii}\lf[{\lambda_1,\gamma_1\atop\lambda_2,\gamma_2}\ri]&=
\sin(\pi\al)\ \int_0^1 d\xi\ 
\xi^{\lambda_1}\ (1-\xi)^{\gamma_1}\ \int_\xi^1 d\eta\ 
\eta^{\lambda_2}\ (1-\eta)^{\gamma_2}\ (\eta-\xi)^{\al}\cr
&+\sin[\pi(\gamma_2+\al)]\ \int_0^1 d\xi\ \xi^{\lambda_1}\ (1-\xi)^{\gamma_1}\ 
\int_1^\infty d\eta\ \eta^{\lambda_2}\ (\eta-1)^{\gamma_2}\
(\eta-\xi)^{\al}\cr
&=\sin[\pi(\gamma_2+\al)]\ 
\fc{\Gamma(1+\lambda_1)\ \Gamma(1+\gamma_1)\ \Gamma(1+\gamma_2)\ 
\Gamma(-1-\al-\lambda_2-\gamma_2)}
{\Gamma(2+\lambda_1+\gamma_1)\ \Gamma(-\al-\lambda_2)}\cr 
&\times\FF{3}{2}\lf[{1+\lambda_1,-\al-1-\lambda_2-\gamma_2,-\al\atop
2+\lambda_1+\gamma_1,-\al-\lambda_2};1\ri]\cr
&+\sin(\pi\al)\ \fc{\Gamma(1+\al)\ \Gamma(1+\gamma_2)\ \Gamma(1+\lambda_1)\
\Gamma(2+\al+\gamma_1+\gamma_2)}{\Gamma(2+\al+\gamma_2)\ 
\Gamma(3+\al+\lambda_1+\gamma_1+\gamma_2)}\cr 
&\times\FF{3}{2}\lf[{-\lambda_2,1+\gamma_2,2+\al+\gamma_1+\gamma_2\atop
2+\al+\gamma_2,3+\al+\lambda_1+\gamma_1+\gamma_2};1\ri]\ .}}
It is straightforward to show, that both expressions \Angelaa\ and \Angeli\ 
are indeed identical. Furthermore, for $\al=-2$ both 
expressions reduce to \Angel\ and \Angela, respectively.

Again, for $\al\in\IZ$ the second term of \Angelaa\ may be interpreted as the 
residuum at $\eta=\xi$. With the expansions 
\eqn\Expforres{\eqalign{
\eta^{\lambda_2}&=\xi^{\lambda_2}\ \sum_{n=0}^\infty (-1)^n\ \xi^{-n}\ 
\fc{(-\lambda_2,n)}{(1,n)}\ (\eta-\xi)^n\ ,\cr
(1-\eta)^{\gamma_2}&=(1-\xi)^{\gamma_2}\ \sum_{n=0}^\infty (1-\xi)^{-n}\ 
\fc{(-\gamma_2,n)}{(1,n)}\ (\eta-\xi)^n\ ,}}
we compute the residuum at $\eta=\xi$
$$\eqalign{
\h\int_0^1 d\xi\ &\xi^{\lambda_1} (1-\xi)^{\gamma_1}\ \oint_{\eta=\xi}d\eta\ 
\eta^{\lambda_2}\ (1-\eta)^{\gamma_2}\ (\xi-\eta)^{\al}\cr
&=-\pi\ \fc{\Gamma(1+\lambda_1+\lambda_2)\ \Gamma(2+\al+\gamma_1+\gamma_2)\ 
\Gamma(-1-\al-\gamma_2)}{\Gamma(-\al)\ \Gamma(-\gamma_2)\ 
\Gamma(3+\al+\lambda_1+\lambda_2+\gamma_1+\gamma_2)}\cr 
&\times\FF{3}{2}\lf[{-\lambda_2,2+\al+\gamma_1+\gamma_2,1+\al\atop
2+\al+\gamma_2,-\lambda_1-\lambda_2};1\ri]\ ,}$$
which indeed agrees with the second term of \Angelaa\ and reduces to the second term
of \Angela\ for $\al=-2$.

Finally for the case $(iii)$ we may deform the contour in the 
complex $\eta$--plane and integrate along the real $\eta$--axis from $\xi$ until 
$\infty$, see the next Figure.
\ifig\Dontiiia{\ Deformed contour of the case ~$(iii)$.}
{\epsfxsize=0.42\hsize\epsfbox{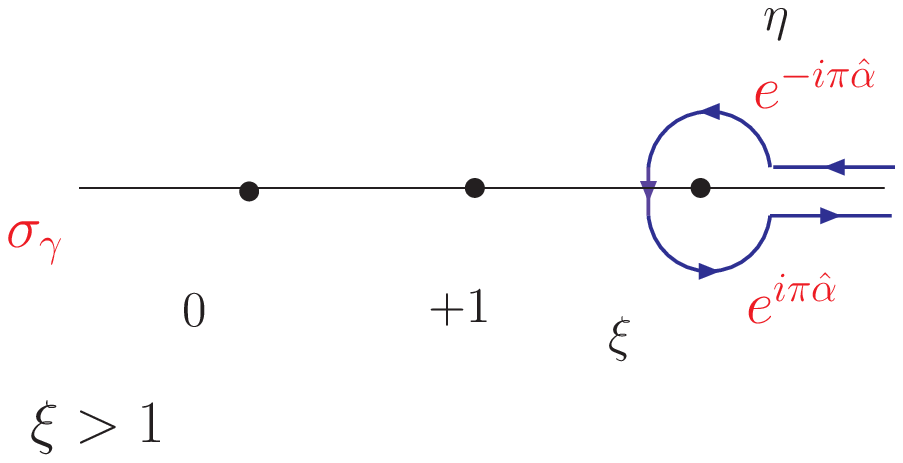}}
\noindent
Hence the total contributions from $(iii)$ is:
\eqn\AAngell{\eqalign{
G^{(\al)}_{iii}\lf[{\lambda_1,\gamma_1\atop\lambda_2,\gamma_2}\ri]&=
\sin(\pi\al)\ \si_\gamma\ \int_1^\infty d\xi\ \xi^{\lambda_1}\ (\xi-1)^{\gamma_1}\ \int_\xi^\infty
d\eta\ \eta^{\lambda_2}\ (\eta-1)^{\gamma_2}\ (\eta-\xi)^{\al}\cr
&=\sin(\pi\al)\ \si_\gamma\ 
\FF{3}{2}\lf[{-\gamma_2,-1-\al-\lambda_2-\gamma_2,-\al-2-\lambda_1-
\lambda_2-\gamma_1-\gamma_2\atop
-\lambda_2-\gamma_2,-1-\al-\lambda_1-\lambda_2-\gamma_2};1\ri]\cr
&\times\fc{\Gamma(1+\al)\ \Gamma(1+\gamma_1)\ 
\Gamma(-1-\al-\lambda_2-\gamma_2)\ \Gamma(-2-\al-\lambda_1-\lambda_2-\gamma_1-\gamma_2)}
{\Gamma(-1-\al-\lambda_1-\lambda_2-\gamma_2)\ \Gamma(-\lambda_2-\gamma_2)}\ .}}
Note, that for the case $\al=-2$, the above expression gives exactly the residuum
\finres:
$$G^{(-2)}_{iii}\lf[{\lambda_1,\gamma_1\atop\lambda_2,\gamma_2}\ri]=
R^{(-2)}_{iii}\lf[{\lambda_1,\gamma_1\atop\lambda_2,\gamma_2}\ri]\ .$$

For $\al\in\IZ$, with \Expforres\ and 
\eqn\Expforresi{
(\eta-1)^{\gamma_2}=(\xi-1)^{\gamma_2}\ \sum_{n=0}^\infty (\xi-1)^{-n}\
\fc{(-\gamma_2,n)}{(1,n)}\ (\xi-\eta)^n}
we may compute the general expression for the residuum at $\eta=\xi$ in the
case of $(iii)$:
\eqn\generalRES{\eqalign{
R^{(\al)}_{iii}\lf[{\lambda_1,\gamma_1\atop\lambda_2,\gamma_2}\ri]&=\h\ 
\si_\gamma\ \int_1^\infty d\xi\ \xi^{\lambda_1}\ (\xi-1)^{\gamma_1}\ \int_{\eta=\xi}
d\eta\ \eta^{\lambda_2}\ (\eta-1)^{\gamma_2}\ (\xi-\eta)^{\al}\cr
&\hskip-2cm=\pi\ \si_\gamma\ \fc{\Gamma(-\al-\gamma_2)\ \Gamma(1+\al+\gamma_2)\ 
\Gamma(2+\al+\gamma_1+\gamma_2)\ \Gamma(-2-\al-\lambda_1-\lambda_2-\gamma_1-\gamma_2)}
{\Gamma(-\al)\ \Gamma(-\gamma_2)\ \Gamma(-\lambda_1-\lambda_2)\ 
\Gamma(2+\al+\gamma_2)}\cr
&\times\FF{3}{2}\lf[{-\lambda_2,1+\al,2+\al+\gamma_1+\gamma_2\atop
-\lambda_1-\lambda_2,2+\al+\gamma_2};1\ri]}\ ,}
which reduces to the expression \finres\ for $\al=-2$ and 
agrees with \AAngell\ for $\al\in\IZ$, \ie
$$G^{(\al)}_{iii}\lf[{\lambda_1,\gamma_1\atop\lambda_2,\gamma_2}\ri]=
R^{(\al)}_{iii}\lf[{\lambda_1,\gamma_1\atop\lambda_2,\gamma_2}\ri]\ \ ,\ \ 
\al\in\IZ\ .$$

To conclude, the total result for \Provee\ is the sum of \Angelaa\ and
\AAngell, \ie
\eqn\lastequ{\mathboxit{
G^{(\al)}\lf[{\lambda_1,\gamma_1\atop\lambda_2,\gamma_2}\ri]=\cases{
G^{(\al)}_{ii}\lf[{\lambda_1,\gamma_1\atop\lambda_2,\gamma_2}\ri]+
R^{(\al)}_{iii}\lf[{\lambda_1,\gamma_1\atop\lambda_2,\gamma_2}\ri]\ , &
$\al\in\IZ\ ,$\cr\cr\cr
G^{(\al)}_{ii}\lf[{\lambda_1,\gamma_1\atop\lambda_2,\gamma_2}\ri]+
G^{(\al)}_{iii}\lf[{\lambda_1,\gamma_1\atop\lambda_2,\gamma_2}\ri]\ , &
$\al\notin\IZ\ ,$}}}
which combines into  \Provee.

\subsec{Relations}

It is straightforward to prove the following relations
$$\eqalign{
&G^{(\al)}\lf[{\lambda_1+1,\gamma_1\atop\lambda_2,\gamma_2}\ri]-
G^{(\al)}\lf[{\lambda_1,\gamma_1\atop\lambda_2+1,\gamma_2}\ri]=
G^{(\al+1)}\lf[{\lambda_1,\gamma_1\atop\lambda_2,\gamma_2}\ri]\ ,\cr
&G^{(\al)}\lf[{\lambda_1+2,\gamma_1\atop\lambda_2,\gamma_2}\ri]+
G^{(\al)}\lf[{\lambda_1,\gamma_1\atop\lambda_2+2,\gamma_2}\ri]-
2\ G^{(\al)}\lf[{\lambda_1+1,\gamma_1\atop\lambda_2+1,\gamma_2}\ri]=
G^{(\al+2)}\lf[{\lambda_1,\gamma_1\atop\lambda_2,\gamma_2}\ri]\ ,\cr
&G^{(\al)}\lf[{\lambda_1,\gamma_1+1\atop\lambda_2,\gamma_2}\ri]-
G^{(\al)}\lf[{\lambda_1,\gamma_1\atop\lambda_2,\gamma_2+1}\ri]=-
G^{(\al+1)}\lf[{\lambda_1,\gamma_1\atop\lambda_2,\gamma_2}\ri]\ ,\cr
&G^{(\al)}\lf[{\lambda_1,\gamma_1+1\atop\lambda_2,\gamma_2}\ri]+
G^{(\al)}\lf[{\lambda_1+1,\gamma_1\atop\lambda_2,\gamma_2}\ri]=
G^{(\al)}\lf[{\lambda_1,\gamma_1\atop\lambda_2,\gamma_2}\ri]\ ,\cr
&G^{(\al)}\lf[{\lambda_1,\gamma_1\atop\lambda_2,\gamma_2+1}\ri]+
G^{(\al)}\lf[{\lambda_1,\gamma_1\atop\lambda_2+1,\gamma_2}\ri]=
G^{(\al)}\lf[{\lambda_1,\gamma_1\atop\lambda_2,\gamma_2}\ri]\ ,\cr
&G^{(\al)}\lf[{\lambda_1-1,\gamma_1\atop\lambda_2,\gamma_2}\ri]-
G^{(\al)}\lf[{\lambda_1,\gamma_1\atop\lambda_2-1,\gamma_2}\ri]=-
G^{(\al+1)}\lf[{\lambda_1-1,\gamma_1\atop\lambda_2-1,\gamma_2}\ri]\ ,}$$
for generic $\al\in\IR$. Furthermore, we may prove the identity
$$G^{(\al)}\lf[{\lambda-1,\gamma\atop\lambda,\gamma}\ri]=-\fc{\al}{\lambda}\ 
G^{(\al-1)}\lf[{\lambda,\gamma\atop\lambda,\gamma}\ri]+\fc{\gamma}{\lambda}\ 
G^{(\al)}\lf[{\lambda,\gamma-1\atop\lambda,\gamma}\ri]\ ,$$
which follows from partial integration.

\subsec{Alternative derivation for $\lambda_i=\lambda,\ \gamma_i=\gamma$}

For $\lambda_i=\lambda,\ \gamma_i=\gamma$ there is yet an other way to prove \Provee:
With 
\eqn\trivial{
\Gamma(-\lambda)^{-1}\ 
\int_0^\infty d\xi\ \xi^{-\lambda-1}\ e^{-\xi|z|^2}=|z|^{2\lambda}\ ,}
we may write the integral $\ss{G^{(\al)}\lf[{\lambda,\gamma\atop\lambda,\gamma}\ri]}$
$$\eqalign{
G^{(\al)}\lf[{\lambda,\gamma\atop\lambda,\gamma}\ri]&=(2i)^\al\ 
\lf[1+(-1)^\al\ri]\ \Gamma(-\lambda)^{-1}\ \Gamma(-\gamma)^{-1}\ 
\int_{-\infty}^\infty dz_1 \int_{0}^\infty
dz_2\ z_2^\al\cr
&\times \int_0^\infty d\xi\ \int_0^\infty d\eta\ \xi^{-\lambda-1}\ \eta^{-\gamma-1}\ 
e^{-\xi(z_1^2+z_2^2)-\eta[ (1-z_1)^2+z_2^2 ]}\ ,}$$
which gives:
$$\eqalign{
G^{(\al)}\lf[{\lambda,\gamma\atop\lambda,\gamma}\ri]&=\fc{\sqrt\pi}{2}\ (2i)^\al\ 
\lf[1+(-1)^\al\ri]\ \fc{\Gamma\lf(\h+\fc{\al}{2}\ri)}{\Gamma(-\lambda)\ 
\Gamma(-\gamma)}\cr 
&\times\int_0^\infty d\xi\ \int_0^\infty d\eta\ \xi^{-\lambda-1}\ \eta^{-\gamma-1}\ 
(\xi+\eta)^{-1-\fc{\al}{2}}\ e^{-\fc{\xi\eta}{\xi+\eta}}\ .}$$
With the change of coordinates $\eta=\fc{x}{1-x}\xi$ and afterwards 
$\xi=\fc{y}{x}$ (\cf \GHMR), and 
performing the $x$-- and $y$--integrations we arrive at:
$$\eqalign{
G^{(\al)}\lf[{\lambda,\gamma\atop\lambda,\gamma}\ri]&=
\fc{\sqrt\pi}{2}\ (2i)^\al\ \lf[1+(-1)^\al\ri]\ \fc{\Gamma\lf(\h+\fc{\al}{2}\ri)}
{\Gamma(-\lambda)\ \Gamma(-\gamma)}\cr 
&\times \fc{\Gamma\lf(1+\fc{\al}{2}+\gamma\ri)
\ \Gamma\lf(-1-\fc{\al}{2}-\lambda-\gamma\ri)\ \Gamma\lf(1+\fc{\al}{2}+\lambda\ri)}{
\Gamma\lf(2+\al+\lambda+\gamma\ri)}\ .}$$
It is straightforward to show, that for $\lambda_i=\lambda,\ \gamma_i=\gamma$ \eqq
\Provee\ boils down to the above~result.

\goodbreak
\appendix\appB{Complex world--sheet integral  
$W^{(\kappa,\al_0)}\lf[{\alpha_1,\lambda_1,\gamma_1,\beta_1\atop
\alpha_2,\lambda_2,\gamma_2,\beta_2}\ri]$}

In this Appendix we shall accomplish the computation of the complex integral \GEN.
It has been argued in Subsection 2.4.3, that this amounts to consider the integral \GENN
\eqn\GENN{\eqalign{
W^{(\kappa,\al_0)}\lf[{\alpha_1,\lambda_1,\gamma_1,\beta_1\atop
\alpha_2,\lambda_2,\gamma_2,\beta_2}\ri]&=\h\ \int_{-\infty}^{\infty} d\rho\
|\rho|^{\al_0}\ |1+\rho|^{\hatt\alpha_1}\ |1-\rho|^{\hatt\alpha_2}\cr
&\times (1+\rho)^{n_0}\ (1-\rho)^{m_0}\ 
I^{(\kappa)}\lf[{\lambda_1,\gamma_1,\beta_1\atop\lambda_2,\gamma_2,\beta_2}\ri](\rho)\ ,}}
with:
\eqn\ANGELL{\eqalign{
I^{(\kappa)}\lf[{\lambda_1,\gamma_1,\beta_1\atop\lambda_2,\gamma_2,\beta_2}\ri](\rho)&=
\int_{-\infty}^\infty d\xi\ \int_{-\infty}^\infty d\eta \ 
|1-\xi|^{\hatt\lambda_1}\ |\xi-\rho|^{\hatt\gamma_1}\ |\xi+\rho|^{\hatt\beta_1}\cr 
&\times |1-\eta|^{\hatt\lambda_2}\ |\eta+\rho|^{\hatt\gamma_2}\ |\eta-\rho|^{\hatt\beta_2}\ 
|\xi-\eta|^\kappa\ \Pi(\rho,\xi,\eta)\cr 
&\times (1-\xi)^{n_1}\ (\xi-\rho)^{n_3}\ (\xi+\rho)^{n_5}\ 
(1-\eta)^{n_2}\ (\eta+\rho)^{n_4}\ (\eta-\rho)^{n_6}\ .}}
Here we have defined the non--integer part of the parameter \parameter\ by
putting 
hat. In the following we shall first analyze the phase factor
$\Pi(\rho,\xi,\eta)$, which has been introduced in \ANGELL.
In what follows we have to discuss the three cases $(1)\ 0<\rho<1$, 
$(2)\ -1<\rho<0$, and $(3)\ \rho>1$, separately. We restrict to the case
$\kappa\in\IZ$, \ie no branching arises from the term $(\xi-\eta)^\kappa$

\subsec{Case\  $0<\rho < 1$}

For this case  the analysis of the  phase factor $\Pi(\rho,\xi,\eta)$
is summarized in Table 2.
\vskip0.1cm
{\vbox{\ninepoint{$$
\vbox{\offinterlineskip\tabskip=0pt
\halign{\strut\vrule#
&~$#$~\hfil 
&\vrule$#$ 
&~$#$~\hfil 
&\vrule$#$ 
&~$#$~\hfil 
&\vrule$#$ 
&~$#$~\hfil 
&\vrule$#$
&~$#$~\hfil 
&\vrule$#$
&~$#$~\hfil 
&\vrule$#$
&~$#$~\hfil 
&\vrule$#$\cr
\noalign{\hrule}
&  &&(\xi,\eta)  
&& \eta<-\rho  &&  -\rho<\eta<\rho&& \rho<\eta<1 && \eta>1&& {\rm total\ phase} &\cr
\noalign{\hrule}
&(i)&&\xi<-\rho &&1  &&e^{i\pi\hat\gamma_2} && e^{i\pi(\hat\beta_2+\hat\gamma_2)}
&& e^{i\pi(\hat\beta_2+\hat\gamma_2+\hat\lambda_2)}  &&\sigma_\gamma \sigma_\beta&\cr
\noalign{\hrule}}}$$
$$\vbox{\offinterlineskip\tabskip=0pt
\halign{\strut\vrule#
&~$#$~\hfil 
&\vrule$#$ 
&~$#$~\hfil 
&\vrule$#$ 
&~$#$~\hfil 
&\vrule$#$ 
&~$#$~\hfil 
&\vrule$#$ 
&~$#$~\hfil 
&\vrule$#$
&~$#$~\hfil 
&\vrule$#$
&~$#$~\hfil 
&\vrule$#$\cr
\noalign{\hrule}
&  &&(\xi,\eta)  
&& \eta<-\rho && -\rho<\eta<\rho&& \rho<\eta<1 && \eta>1&& {\rm total\ phase} &\cr
\noalign{\hrule}
&(ii)&&-\rho<\xi<\rho &&e^{i\pi\hat\beta_2} && e^{i\pi(\hat\gamma_2+\hat\beta_2)}  &&e^{i\pi\hat\gamma_2}&&
e^{i\pi(\hat\gamma_2+\hat\lambda_2)}&&\sigma_\gamma &\cr
\noalign{\hrule}}}$$
$$\vbox{\offinterlineskip\tabskip=0pt
\halign{\strut\vrule#
&~$#$~\hfil 
&\vrule$#$ 
&~$#$~\hfil 
&\vrule$#$ 
&~$#$~\hfil 
&\vrule$#$ 
&~$#$~\hfil 
&\vrule$#$
&~$#$~\hfil 
&\vrule$#$
&~$#$~\hfil 
&\vrule$#$
&~$#$~\hfil 
&\vrule$#$\cr
\noalign{\hrule}
&  &&(\xi,\eta)  
&& \eta<-\rho && -\rho<\eta<\rho   &&  \rho<\eta<1 && \eta>1&& 
{\rm total\ phase} &\cr
\noalign{\hrule}
&(iii)&&\rho<\xi<1 &&e^{i\pi(\hat\gamma_2+\hat\beta_2)} && e^{i\pi\hat\beta_2} &&
1&& e^{i\pi\hat\lambda_2} &&1&\cr
\noalign{\hrule}}}$$
$$\vbox{\offinterlineskip\tabskip=0pt
\halign{\strut\vrule#
&~$#$~\hfil 
&\vrule$#$ 
&~$#$~\hfil 
&\vrule$#$ 
&~$#$~\hfil 
&\vrule$#$ 
&~$#$~\hfil 
&\vrule$#$
&~$#$~\hfil 
&\vrule$#$
&~$#$~\hfil 
&\vrule$#$
&~$#$~\hfil 
&\vrule$#$\cr
\noalign{\hrule}
&  &&(\xi,\eta)  
&& \eta<-\rho && -\rho<\eta<\rho   &&  \rho<\eta<1&& 1<\eta&& 
{\rm total\ phase} &\cr
\noalign{\hrule}
&(iv)&&\xi>1 &&e^{i\pi(\hat\lambda_2+\hat\gamma_2+\hat\beta_2)} && 
e^{i\pi(\hat\lambda_2+\hat\beta_2)} &&
e^{i\pi\hat\lambda_2}&& 1 &&\sigma_\lambda&\cr
\noalign{\hrule}}}$$
\vskip0pt
\centerline{\noindent{\bf Table 2:}
{\sl Phases $\Pi(\rho,\xi,\eta)$ along the integration region $(\xi,\eta)$ for $0<\rho<1$.}}
\vskip10pt}}}
\vskip-0.5cm
\br
We have introduced the total phases
$\sigma_\lambda:=e^{i\pi(n_1+n_2)},\sigma_\gamma:=e^{i\pi(n_3+n_4)}$
and $\sigma_\beta:~=~e^{i\pi(n_5+n_6)}$. 
The different phase structures in the 
complex $\eta$--plane are shown in the next four figures. More precisely, these figures
display the way, how to integrate in the 
complex $\eta$--plane to take into account the phases of Table 2.
\ifig\contii{The complex $\eta$--plane and the contour integrals for the two 
cases $(i)$ and~$(ii)$. }{\epsfxsize=0.5\hsize\epsfbox{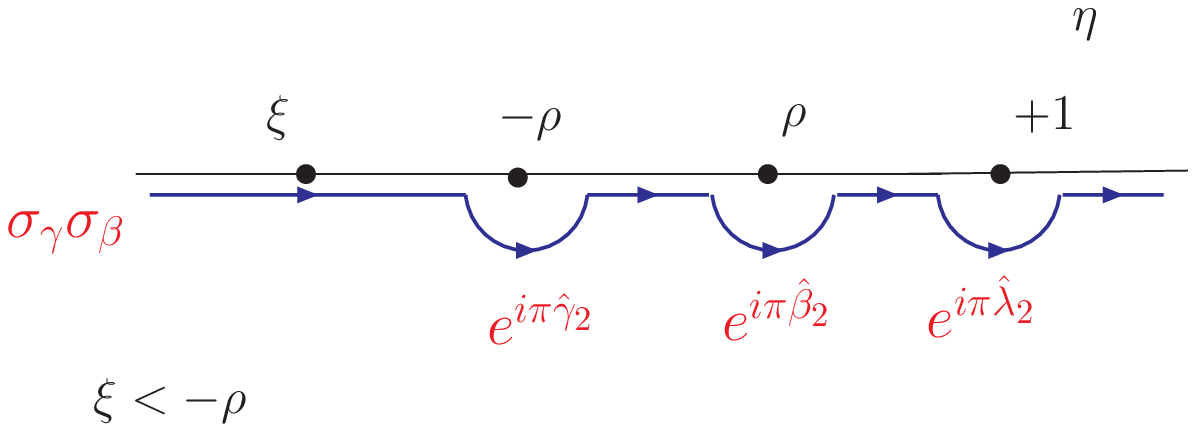}
\epsfxsize=0.5\hsize\epsfbox{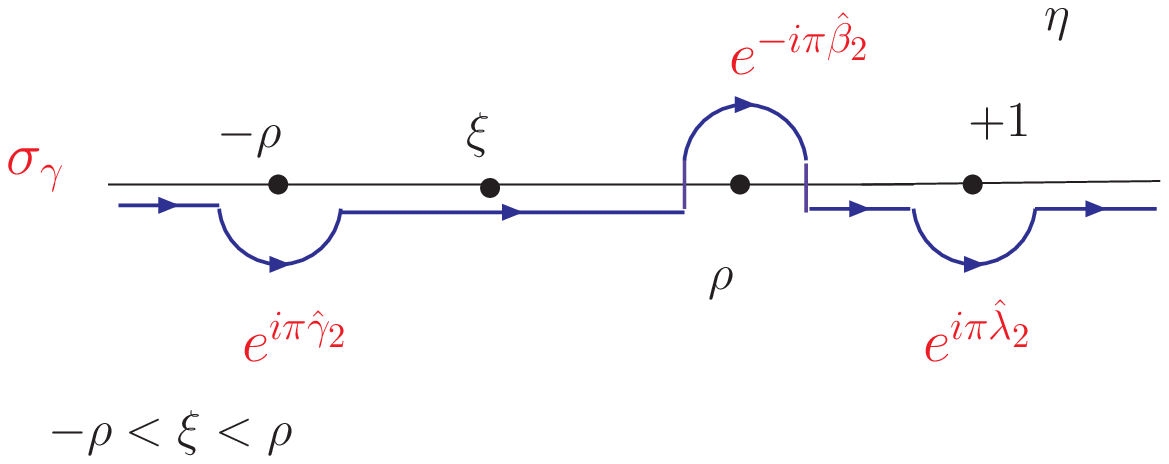}}
\ifig\contii{The complex $\eta$--plane and the contour integrals for the two 
cases~$(iii)$~and~$(iv)$.}{\epsfxsize=0.5\hsize\epsfbox{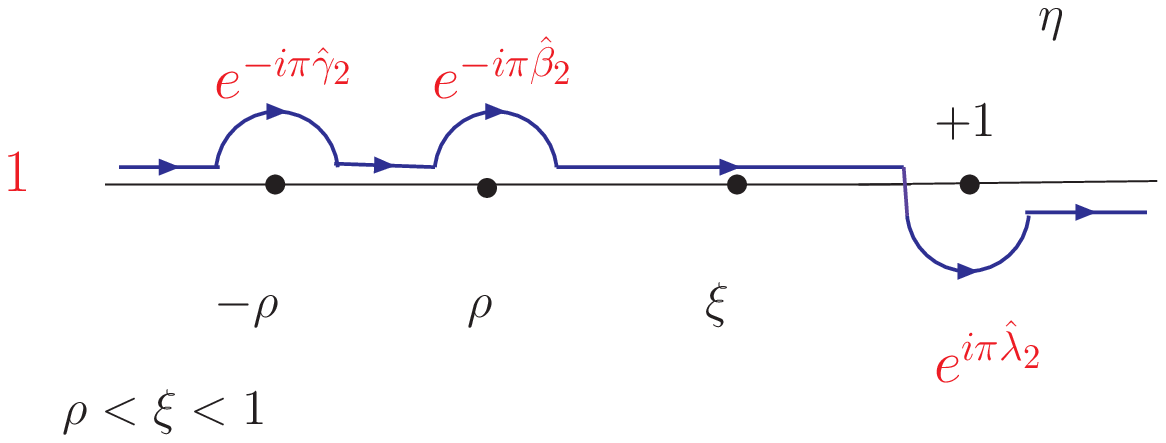}
\epsfxsize=0.5\hsize\epsfbox{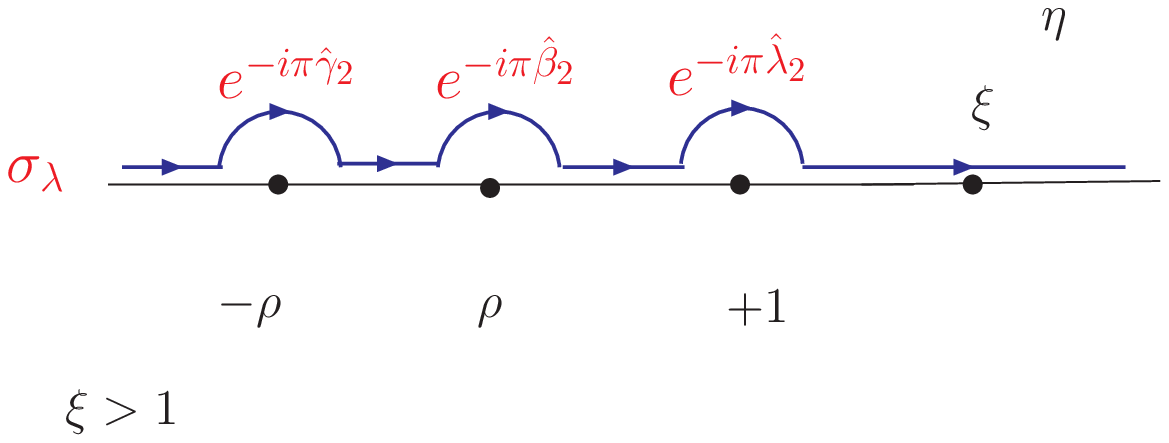}}
\noindent
After analyzing the structure of the contour integrals we find that the 
two cases $(ii)$ and $(iii)$ contribute to \ANGELL. 
For case $(ii)$ we may deform the contour in the complex $\eta$--plane into two pieces.
One piece along the real $\eta$--axis from $-\infty$ to $-\rho$ and reverse taking into 
account the phase factors $e^{-i\pi \hatt\gamma_2}$ and $e^{i\pi \hatt\gamma_2}$, respectively.
The second piece goes along the real $\eta$--axis from $+1$ to $\infty$ and 
reverse taking into account the phase factors $e^{i\pi \hatt\lambda_2}$ and 
$e^{-i\pi \hatt\lambda_2}$, respectively, see the next Figure.
\ifig\Fontii{\ Deformed contours of the case $(ii)$.}
{\epsfxsize=0.5\hsize\epsfbox{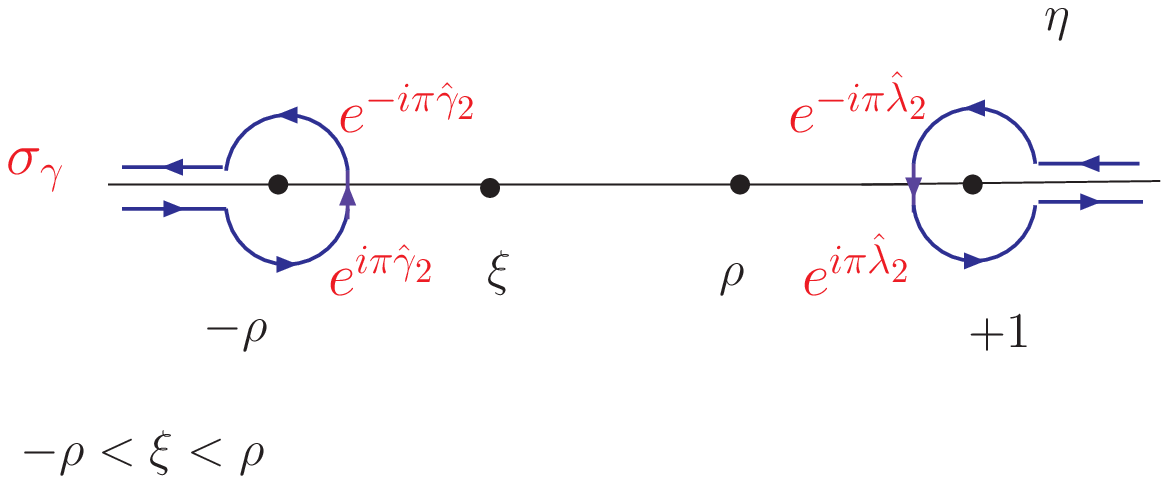}}
\noindent
Hence the two contributions $I_a$ and $I_{c_1}$ from $(ii)$ give
\eqn\Iiia{\eqalign{
I^{(\kappa)}_{a}
\lf[{\lambda_1,\gamma_1,\beta_1\atop\lambda_2,\gamma_2,\beta_2}\ri](\rho)&=
\sin(\pi\gamma_2)\ e^{i\pi(\gamma_2+\beta_2)}\ \si_\gamma\ 
\int\limits_{-\rho}^\rho d\xi\ \int\limits_{-\infty}^{-\rho} d\eta \ 
(1-\xi)^{\lambda_1}\ (\rho-\xi)^{\gamma_1}\ (\rho+\xi)^{\beta_1}\cr 
&\times (1-\eta)^{\lambda_2}\ (-\rho-\eta)^{\gamma_2}\ (\rho-\eta)^{\beta_2}\ 
(\xi-\eta)^\kappa\ ,\cr
I^{(\kappa)}_{c_1}
\lf[{\lambda_1,\gamma_1,\beta_1\atop\lambda_2,\gamma_2,\beta_2}\ri](\rho)&=
-\sin(\pi\lambda_2)\  e^{i\pi(\gamma_2+\kappa)}\ \si_\gamma\ 
\int\limits_{-\rho}^\rho d\xi\ \int\limits_1^\infty d\eta \ 
(1-\xi)^{\lambda_1}\ (\rho-\xi)^{\gamma_1}\ (\rho+\xi)^{\beta_1}\cr 
&\times (\eta-1)^{\lambda_2}\ (\rho+\eta)^{\gamma_2}\ (\eta-\rho)^{\beta_2}\ 
(\eta-\xi)^\kappa\ ,}}
respectively.
On the other hand, for case $(iii)$ we may first deform the whole contour in the 
complex $\eta$--plane and integrate along the real $\eta$--axis from $+1$ until 
$\infty$ and reverse respecting 
the phase factors $e^{-i\pi \hatt\lambda_2}$ and $e^{i\pi \hatt\lambda_2}$, respectively, 
see the next Figure. 
\ifig\defConti{\ Deformed contour of the case ~$(iii)$.}
{\epsfxsize=0.5\hsize\epsfbox{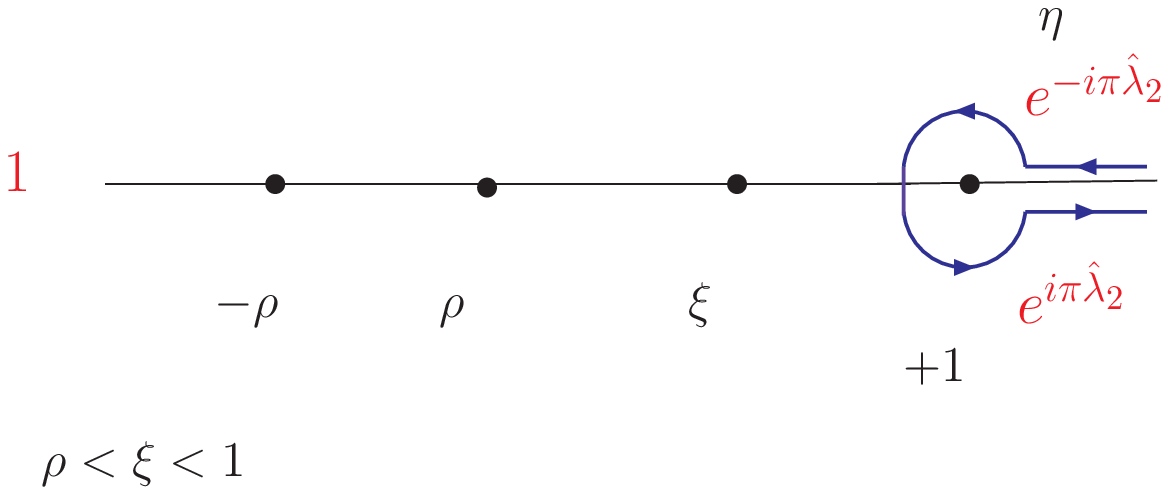}}
\noindent
Therefore, the contribution $I_{c_2}$ from $(iii)$ gives:
\eqn\Iiiia{\eqalign{
I^{(\kappa)}_{c_2}\lf[{\lambda_1,\gamma_1,\beta_1\atop\lambda_2,\gamma_2,\beta_2}\ri]
(\rho)&=-\sin(\pi\lambda_2)\ (-1)^\kappa\ 
\int\limits_{\rho}^1 d\xi\ \int\limits_1^\infty d\eta \ 
(1-\xi)^{\lambda_1}\ (\xi-\rho)^{\gamma_1}\ (\xi+\rho)^{\beta_1}\cr 
&\times (\eta-1)^{\lambda_2}\ (\eta+\rho)^{\gamma_2}\ (\eta-\rho)^{\beta_2}\ 
(\eta-\xi)^\kappa\ .}}

\subsec{Case\  $-1<\rho < 0$}

For this case  the analysis of the  phase factor $\Pi(\rho,\xi,\eta)$
boils down to the previous Subsection B.1.
By using the exchange symmetry \fulfill\ we can apply the previous results with
the following regrouping $\gamma_i\leftrightarrow\beta_i$ and $\alpha_1\leftrightarrow\alpha_2$.

Hence the two contributions $I_a'$ and $I_{c_1'}$ from $(ii)$ yield
\eqn\Iiias{\eqalign{
I^{(\kappa)}_{a'}
\lf[{\lambda_1,\gamma_1,\beta_1\atop\lambda_2,\gamma_2,\beta_2}\ri](\rho)&=
\sin(\pi\beta_2)\ e^{i\pi(\gamma_2+\beta_2)}\ \si_\beta\ 
\int\limits_{-\rho}^\rho d\xi\ \int\limits_{-\infty}^{-\rho} d\eta \ 
(1-\xi)^{\lambda_1}\ (\xi-\rho)^{\gamma_1}\ (-\rho-\xi)^{\beta_1}\cr 
&\times (1-\eta)^{\lambda_2}\ (-\rho-\eta)^{\gamma_2}\ (\rho-\eta)^{\beta_2}\  
(\xi-\eta)^\kappa\ ,\cr
I^{(\kappa)}_{c_1'}
\lf[{\lambda_1,\gamma_1,\beta_1\atop\lambda_2,\gamma_2,\beta_2}\ri](\rho)&=
-\sin(\pi\lambda_2)\ e^{i\pi(\beta_2+\kappa)}\ \si_\beta\ 
\int\limits_{-\rho}^\rho d\xi\ \int\limits_1^\infty d\eta \ 
(1-\xi)^{\lambda_1}\ (\xi-\rho)^{\gamma_1}\ (-\rho-\xi)^{\beta_1}\cr 
&\times (\eta-1)^{\lambda_2}\ (\rho+\eta)^{\gamma_2}\ (\eta-\rho)^{\beta_2}\ 
(\eta-\xi)^\kappa\ ,}}
respectively.
Furthermore, the contribution from case $(iii)$ gives:
\eqn\Iiiias{\eqalign{
I^{(\kappa)}_{c_2'}\lf[{\lambda_1,\gamma_1,\beta_1\atop\lambda_2,\gamma_2,\beta_2}\ri]
(\rho)&=-\sin(\pi\lambda_2)\ (-1)^\kappa\ 
\int\limits_{\rho}^1 d\xi\ \int\limits_1^\infty d\eta \ 
(1-\xi)^{\lambda_1}\ (\xi-\rho)^{\gamma_1}\ (\rho+\xi)^{\beta_1}\cr 
&\times (\eta-1)^{\lambda_2}\ (\rho+\eta)^{\gamma_2}\ (\eta-\rho)^{\beta_2}\ 
(\eta-\xi)^\kappa\ .}}

\subsec{Case\  $\rho>1$}

For this case  the analysis of the  phase factor $\Pi(\rho,\xi,\eta)$
is summarized in Table 3.
\vskip0.1cm
{\vbox{\ninepoint{$$
\vbox{\offinterlineskip\tabskip=0pt
\halign{\strut\vrule#
&~$#$~\hfil 
&\vrule$#$ 
&~$#$~\hfil 
&\vrule$#$ 
&~$#$~\hfil 
&\vrule$#$ 
&~$#$~\hfil 
&\vrule$#$
&~$#$~\hfil 
&\vrule$#$
&~$#$~\hfil 
&\vrule$#$
&~$#$~\hfil 
&\vrule$#$\cr
\noalign{\hrule}
&  &&(\xi,\eta)  
&& \eta<-\rho  &&  -\rho<\eta<1&& 1<\eta<\rho && \eta>\rho&& 
{\rm total\ phase} &\cr
\noalign{\hrule}
&(i)&&\xi<-\rho &&1  &&
e^{i\pi\hat\gamma_2} && e^{i\pi(\hat\gamma_2+\hat\lambda_2)}
&& e^{i\pi(\hat\gamma_2+\hat\lambda_2+\hat\gamma_2)} 
 &&\sigma_\gamma \sigma_\beta& \cr
\noalign{\hrule}}}$$
$$\vbox{\offinterlineskip\tabskip=0pt
\halign{\strut\vrule#
&~$#$~\hfil 
&\vrule$#$ 
&~$#$~\hfil 
&\vrule$#$ 
&~$#$~\hfil 
&\vrule$#$ 
&~$#$~\hfil 
&\vrule$#$
&~$#$~\hfil 
&\vrule$#$ 
&~$#$~\hfil 
&\vrule$#$
&~$#$~\hfil 
&\vrule$#$\cr
\noalign{\hrule}
&  &&(\xi,\eta)  
&& \eta<-\rho && -\rho<\eta<1&& 1<\eta<\rho && \eta>\rho&& 
{\rm total\ phase} &\cr
\noalign{\hrule}
&(ii)&&-\rho<\xi<1 &&e^{i\pi\hat\beta_2} && e^{i\pi(\hat\beta_2+\hat\gamma_2)} &&
e^{i\pi(\hat\beta_2+\hat\gamma_2+\hat\lambda_2)} && e^{i\pi(\hat\lambda_2+\hat\gamma_2)} &&\sigma_\gamma &\cr
\noalign{\hrule}}}$$
$$\vbox{\offinterlineskip\tabskip=0pt
\halign{\strut\vrule#
&~$#$~\hfil 
&\vrule$#$ 
&~$#$~\hfil 
&\vrule$#$ 
&~$#$~\hfil 
&\vrule$#$ 
&~$#$~\hfil 
&\vrule$#$
&~$#$~\hfil 
&\vrule$#$
&~$#$~\hfil 
&\vrule$#$
&~$#$~\hfil 
&\vrule$#$\cr
\noalign{\hrule}
&  &&(\xi,\eta)  
&& \eta<-\rho && -\rho<\eta<1   &&  1<\eta<\rho && \eta>\rho&& 
{\rm total\ phase} &\cr
\noalign{\hrule}
&(iii)&&1<\xi<\rho &&e^{i\pi(\hat\lambda_2+\hat\beta_2)} && e^{i\pi(\hat\lambda_2+\hat\gamma_2+\hat\beta_2)} &&
 e^{i\pi(\hat\gamma_2+\hat\beta_2)}&& e^{i\pi\hat\gamma_2} &&\sigma_\lambda\sigma_\gamma&\cr
\noalign{\hrule}}}$$
$$\vbox{\offinterlineskip\tabskip=0pt
\halign{\strut\vrule#
&~$#$~\hfil 
&\vrule$#$ 
&~$#$~\hfil 
&\vrule$#$ 
&~$#$~\hfil 
&\vrule$#$ 
&~$#$~\hfil 
&\vrule$#$
&~$#$~\hfil 
&\vrule$#$ 
&~$#$~\hfil 
&\vrule$#$
&~$#$~\hfil 
&\vrule$#$\cr
\noalign{\hrule}
&  &&(\xi,\eta)  
&& \eta<-\rho && -\rho<\eta<1   &&  1<\eta<\rho&& \rho<\eta&& 
{\rm total\ phase} &\cr
\noalign{\hrule}
&(iv)&&\xi>\rho &&e^{i\pi(\hat\lambda_2+\hat\beta_2+\hat\gamma_2)} && 
e^{i\pi(\hat\lambda_2+\hat\beta_2)} &&e^{i\pi\hat\beta_2} && 1&&\sigma_\lambda&\cr
\noalign{\hrule}}}$$
\vskip0pt
\centerline{\noindent{\bf Table 3:}
{\sl Phases $\Pi(\rho,\xi,\eta)$ along the integration region $(\xi,\eta)$ for $\rho>1$.}}
\vskip10pt}}}
\vskip-0.5cm
\br
The different phase structures in the complex $\eta$--plane are shown in the next four figures.
\ifig\contii{The complex $\eta$--plane and the contour integrals for the two 
cases $(i)$ and~$(ii)$. }{\epsfxsize=0.5\hsize\epsfbox{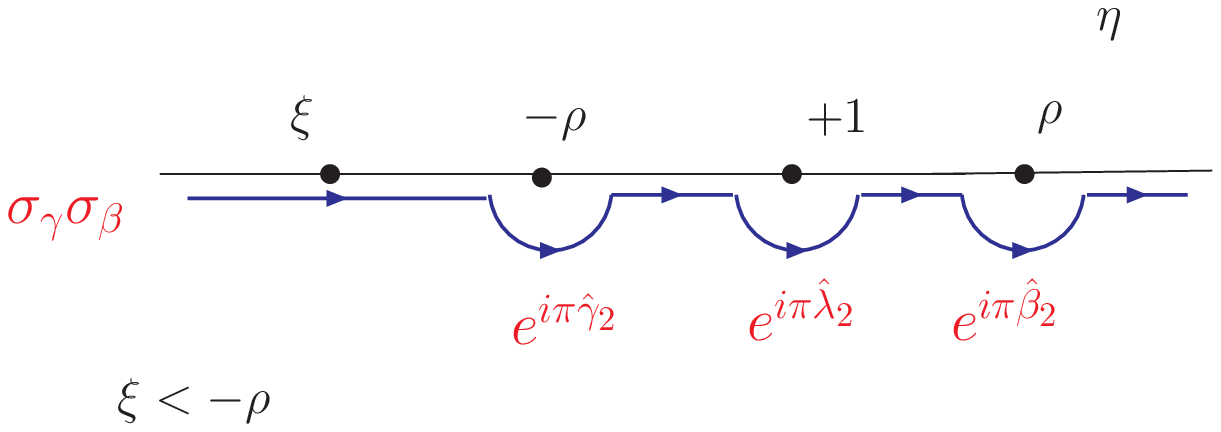}
\epsfxsize=0.5\hsize\epsfbox{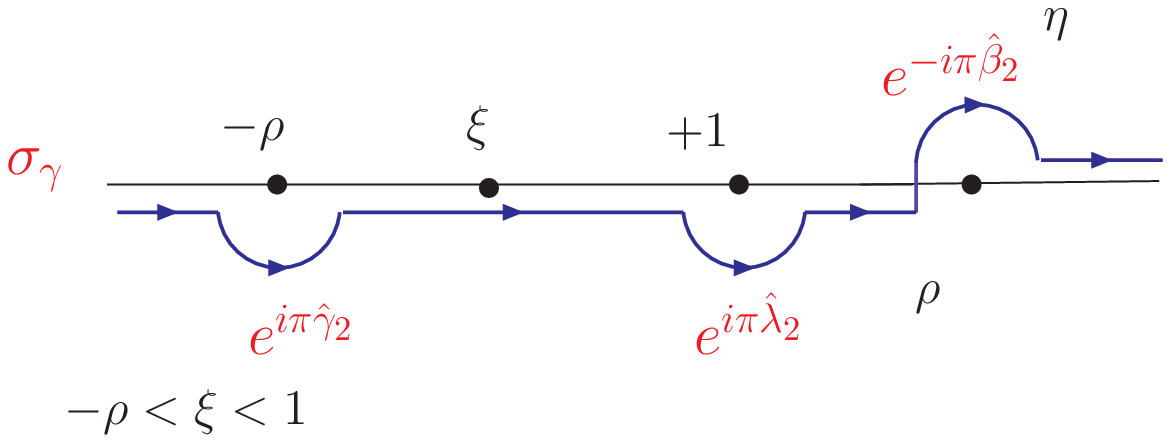}}
\ifig\contii{The complex $\eta$--plane and the contour integrals for the two 
cases~$(iii)$~and~$(iv)$.}{\epsfxsize=0.5\hsize\epsfbox{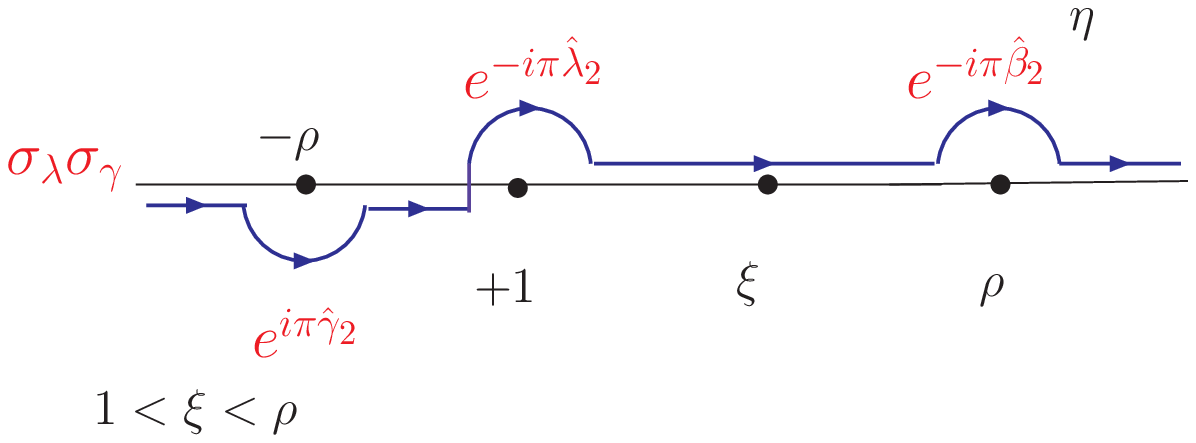}
\epsfxsize=0.5\hsize\epsfbox{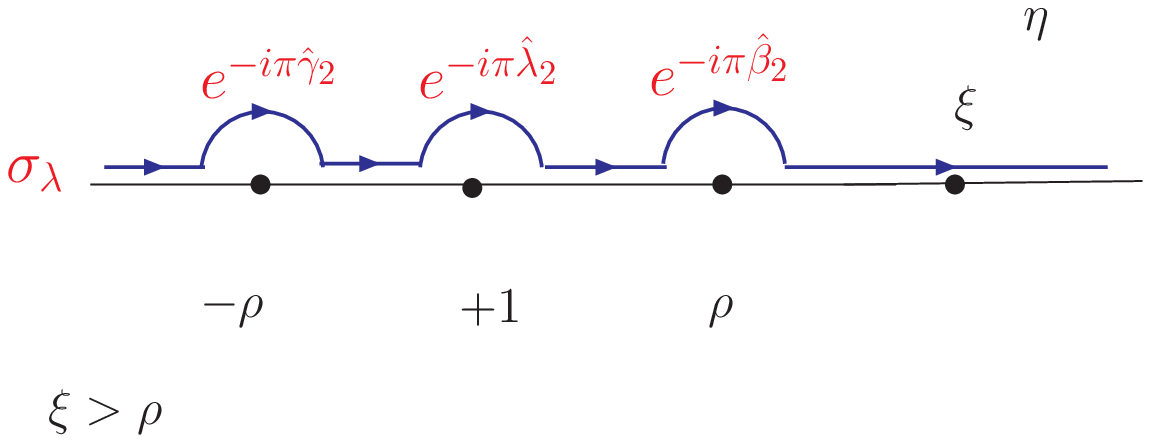}}
\noindent
After analyzing the structure of the contour integrals we find that the 
two cases $(ii)$ and $(iii)$ contribute to \ANGELL. 
For case $(ii)$ we may deform the contour in the complex $\eta$--plane and
integrate along the real $\eta$--axis from $\rho$ to $\infty$ and reverse taking into 
account the phase factors $e^{-i\pi \hatt\beta_2}$ and $e^{i\pi \hatt\beta_2}$, respectively.
On the other hand in the case of $(iii)$ we may deform the contour and
integrate along the real $\eta$--axis from $-\infty$ to the point $-\rho$, 
encircling $-\rho$ and moving backwards to $-\infty$, see the next Figure.
\ifig\Fontii{Deformed contours of the two 
cases~$(ii)$~and~$(iii)$.}{\epsfxsize=0.5\hsize\epsfbox{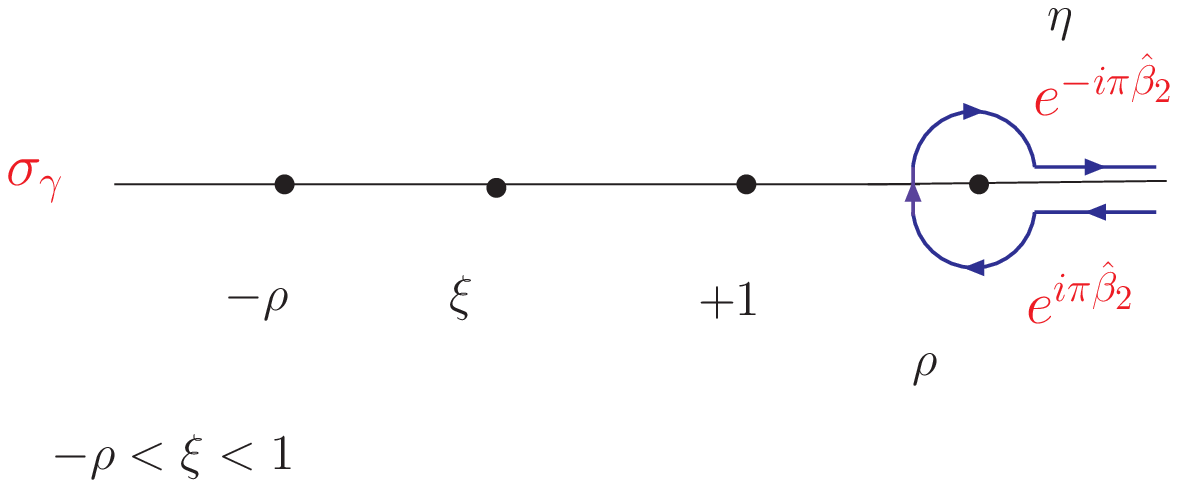}\ \ \ 
\epsfxsize=0.5\hsize\epsfbox{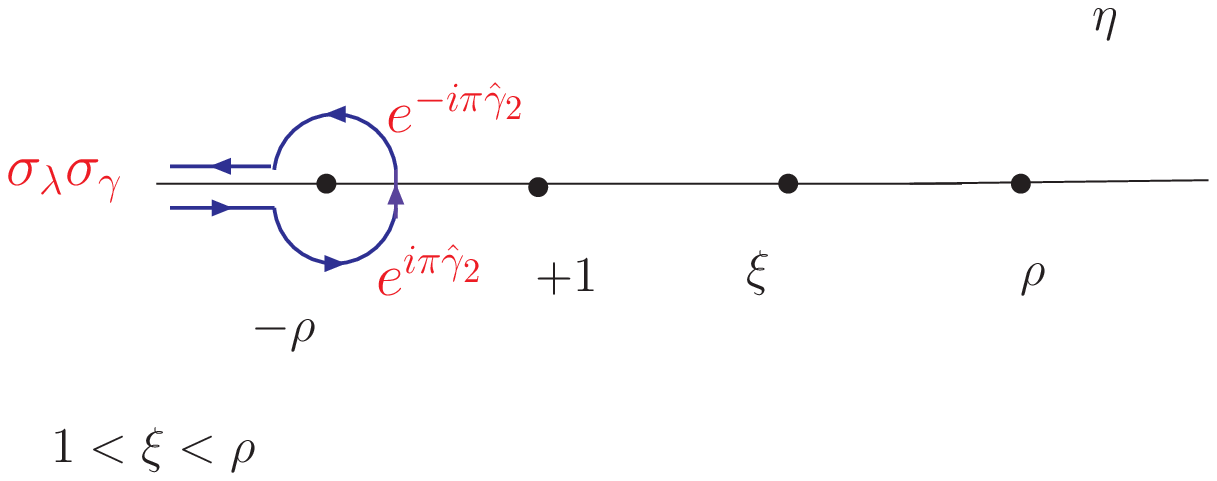}}
\noindent
Hence the contributions $I_{b_1}$ and $I_{b_2}$ from $(ii)$ and $(iii)$ give
\eqn\Iii{\eqalign{
I^{(\kappa)}_{b_1}\lf[{\lambda_1,\gamma_1,\beta_1\atop\lambda_2,\gamma_2,\beta_2}\ri](\rho)&=\si_\gamma\ \sin(\pi\beta_2)\ e^{i\pi(\lambda_2+\gamma_2+\beta_2+\kappa)}\  
\int\limits_{-\rho}^1 d\xi\ \int\limits^{\infty}_{\rho} d\eta\ 
(1-\xi)^{\lambda_1}\ (\rho-\xi)^{\gamma_1}\ (\rho+\xi)^{\beta_1}\cr 
&\times (\eta-1)^{\lambda_2}\ (\rho+\eta)^{\gamma_2}\ (\eta-\rho)^{\beta_2}\ 
(\eta-\xi)^\kappa\ ,\cr
I^{(\kappa)}_{b_2}\lf[{\lambda_1,\gamma_1,\beta_1\atop\lambda_2,\gamma_2,\beta_2}\ri](\rho)&=\si_\lambda\si_\gamma\ 
\sin(\pi\gamma_2)\ e^{i\pi(\lambda_2+\gamma_2+\beta_2)}\
\int\limits_1^\rho d\xi\ \int\limits_{-\infty}^{-\rho} d\eta \ 
(\xi-1)^{\lambda_1}\ (\rho-\xi)^{\gamma_1}\ (\rho+\xi)^{\beta_1}\cr 
&\times (1-\eta)^{\lambda_2}\ (-\rho-\eta)^{\gamma_2}\ (\rho-\eta)^{\beta_2}\ 
(\xi-\eta)^\kappa\ ,}}
respectively.

Finally, for the case $-\infty<\rho<-1$ we obtain exactly the same contributions.

\subsec{Total contribution in complex $\rho$--plane}

So far, we have reduced the $\xi$ and $\eta$ integrations from \ANGELL\ to 
contour integrals  in the complex $\eta$--plane. These integrals, which are given in \eqqs
\Iiia, \Iiiia, \Iiias, \Iiiias\ and \Iii\ depend analytically on the
complex variable $\rho$. Now we have to analyze the phase structure in the complex 
$\rho$--plane and properly take into account all phases. The analysis in the
complex $\rho$--plane is identical to the discussions in Subsections 2.4.1. and 3.2.
After some work and putting all parts together we find the final result given in \Obtain.

\subsec{Special cases}

In this Subsection we discuss some interesting special cases for the complex 
integral~$W^{(\kappa,\al_0)}\lf[{\alpha_1,\lambda_1,\gamma_1,\beta_1\atop
\alpha_2,\lambda_2,\gamma_2,\beta_2}\ri]$.

\vskip0.5cm
\noindent
$\underline{\lambda_1,\ \lambda_2=0:}$

\noindent
For this case the integral \ANGEL\ can be reduced to the integral \Special.
The whole integral \GEN\ factorizes into an integral of the type \baini\ and 
\Special:
\eqn\insti{
W^{(0,\al_0)}\lf[{\alpha_1,0,\gamma_1,\beta_1\atop
\alpha_2,0,\gamma_2,\beta_2}\ri]=2^{-\al_0}\ 
V\lf[{\beta_1, \gamma_1\atop\beta_2,\gamma_2}\ri]\int_{0}^\infty dx\ (2x)^{2+\al_0+\gamma_1+\gamma_2+\bet_1+\bet_2}
\ (1+ix)^{\al_1}(1-ix)^{\al_2}.}
On the other hand, for $\lambda_1,\lambda_2,\kappa=0$ the non--vanishing terms in 
\Obtain\ are: 
\eqn\kitpi{\eqalign{
2^{-\al_0}\ e^{i\pi(\gamma_2+\beta_2)}\  \sin(\pi\gamma_2)\ &A(1,6,3,5,4,2)\ra 2^{-\al_0}\ e^{i\pi(\gamma_2+\beta_2)}\ V\lf[{\beta_1, \gamma_1\atop\beta_2,\gamma_2}\ri]\cr 
&\times\int_{0}^1 d\rho\ (2\rho)^{2+\al_0+\gamma_1+\gamma_2+\bet_1+\bet_2}
\ (1+\rho)^{\al_1}\ (1-\rho)^{\al_2}\ ,\cr
2^{-\al_0}\ [\ \sin(\pi\beta_2)\ &A(1,3,5,2,4,6)
+\sin(\pi\gamma_2)\ A(1,6,3,2,5,4)\ ]\ra2^{-\al_0}\ 
V\lf[{\beta_1, \gamma_1\atop\beta_2,\gamma_2}\ri]\cr
&\times \int_{1}^\infty d\rho\ (2\rho)^{2+\al_0+\gamma_1+\gamma_2+\bet_1+\bet_2}
\ (1+\rho)^{\al_1}\ (\rho-1)^{\al_2}\ ,\cr
2^{-\al_0}\ e^{i\pi(\gamma_2+\beta_2)}\  \sin(\pi\beta_2)\ &A(1,6,4,5,3,2)\ra2^{-\al_0}\ 
e^{i\pi(\gamma_2+\beta_2)}\ 
V\lf[{\beta_1,\gamma_1\atop\beta_2,\gamma_2}\ri]\cr 
&\times\int_{-1}^0 d\rho\ (-2\rho)^{2+\al_0+\gamma_1+\gamma_2+\bet_1+\bet_2}
\ (1+\rho)^{\al_1}\ (1-\rho)^{\al_2}\ .}}
The three expressions in \kitpi\ combine to \insti. This can be verified by
making use of the relation \obtain\ and \firenze.

\vskip0.5cm
\noindent
$\underline{\gamma_1,\ \gamma_2=0:}$

\noindent
Again for this case the complex $z$--integral \ANGEL\ can be reduced to the 
integral \Special.
The whole integral \GEN\ factorizes into an integral of the type \baini\ and \Special:
\eqn\insti{
W^{(0,\al_0)}\lf[{\alpha_1,\lambda_1,0,\beta_1\atop
\alpha_2,\lambda_2,0,\beta_2}\ri]=
V\lf[{\lambda_1, \beta_1\atop\lambda_2,\beta_2}\ri]\ \int_{0}^\infty dx\ x^{\al_0}
\ (1+ix)^{1+\al_1+\lambda_1+\beta_1}\ (1-ix)^{1+\al_2+\lambda_2+\beta_2}\ .}
On the other hand, for $\gamma_1,\gamma_2,\kappa=0$ the non--vanishing terms in 
\Obtain\ are: 
\eqn\kitpii{\eqalign{
-2^{-\al_0}\ \sin(\pi\lambda_2)\ & [\ e^{i\pi \gamma_2}\ A(1,3,5,4,2,6)+A(1,3,4,5,2,6)\ ]\ra
-2^{-\al_0}\ V\lf[{\lambda_1, \beta_1\atop\lambda_2,\beta_2}\ri]\cr
&\times \int_0^1 d\rho\ (2\rho)^{\al_0}\ 
(1+\rho)^{1+\al_1+\lambda_1+\beta_1}\ (1-\rho)^{1+\al_2+\lambda_2+\beta_2}\ ,}}
$$\eqalign{
2^{-\al_0}\ \sin(\pi\beta_2)\ &A(1,3,5,2,4,6)\ra
2^{-\al_0}\  \ V\lf[{\lambda_1, \beta_1\atop\lambda_2,\beta_2}\ri]\cr 
&\times \int_1^\infty d\rho\ (2\rho)^{\al_0}\ 
(1+\rho)^{1+\al_1+\lambda_1+\beta_1}\ (\rho-1)^{1+\al_2+\lambda_2+\beta_2}\ ,\cr
-2^{-\al_0}\ \sin(\pi\lambda_2)\ & A(1,4,3,5,2,6)\ra-2^{-\al_0}\ V\lf[{\lambda_1, \beta_1\atop\lambda_2,\beta_2}\ri]\cr 
&\times\int_{-1}^0 d\rho\ (-2\rho)^{\al_0}\ (1+\rho)^{1+\al_1+\lambda_1+\beta_1}\ 
(1-\rho)^{1+\al_2+\lambda_2+\beta_2}\ ,\cr
2^{-\al_0}\ \ [\ e^{i\pi(\gamma_2+\beta_2)}\ \sin(\pi\beta_2)\ &A(1,6,4,5,3,2)
-\sin(\pi\lambda_2)\ e^{i\pi\beta_2}\ A(1,4,5,3,2,6)\ ]\ra 0\ .}$$
Again, the three expressions in \kitpi\ combine to \insti. This can be verified by
making use of the relation \obtain\ and \firenze.

\vskip0.5cm
\noindent
$\underline{\beta_1,\ \beta_2=0:}$

\noindent
Again for this case the integral \ANGEL\ can be reduced to the integral \Special.
The whole integral \GEN\ factorizes into an integral of the type \baini\ and 
\Special:
\eqn\insti{
W^{(0,\al_0)}\lf[{\alpha_1,\lambda_2,\gamma_1,0\atop
\alpha_2,\lambda_2,\gamma_2,0}\ri]=
V\lf[{\lambda_1,\gamma_1\atop\lambda_2,\gamma_2}\ri]\ \int_{0}^\infty dx\ x^{\al_0}
\ (1+ix)^{1+\al_1+\lambda_2+\gamma_2}\ (1-ix)^{1+\al_2+\lambda_1+\gamma_1}\ .}
On the other hand, for $\beta_1,\beta_2,\kappa=0$ the non--vanishing terms in 
\Obtain\ are: 
\eqn\kitpiii{\eqalign{
2^{-\al_0}\ e^{i\pi(\gamma_2+\beta_2)}\ \sin(\pi\gamma_2)\ & A(1,6,3,5,4,2)
-2^{-\al_0}\ \sin(\pi\lambda_2)\ e^{i\pi\gamma_2}\ A(1,3,5,4,2,6)\ra 0\ ,\cr
-2^{-\al_0}\ \sin(\pi\lambda_2)\ & A(1,3,4,5,2,6)\ra
-2^{-\al_0}\ V\lf[{\lambda_1, \gamma_1\atop\lambda_2,\gamma_2}\ri]\cr 
&\times \int_0^1 d\rho\ (2\rho)^{\al_0}\ 
(1+\rho)^{1+\al_1+\lambda_2+\gamma_2}\ (1-\rho)^{1+\al_2+\lambda_1+\gamma_1}\ ,\cr
2^{-\al_0}\ \sin(\pi\gamma_2)\  &A(1,6,3,2,5,4)\ra
2^{-\al_0}\ V\lf[{\lambda_1, \gamma_1\atop\lambda_2,\gamma_2}\ri]}}
$$\eqalign{&\times \int_1^\infty d\rho\ (2\rho)^{\al_0}\ 
(1+\rho)^{1+\al_1+\lambda_2+\gamma_2}\ (\rho-1)^{1+\al_2+\lambda_1+\gamma_1}\ ,\cr
-2^{-\al_0}\ \sin(\pi\lambda_2)\ [\ e^{i\pi \beta_2}&\ A(1,4,5,3,2,6)+A(1,4,3,5,2,6)\ ]\ra
-2^{-\al_0}\ V\lf[{\lambda_1, \gamma_1\atop\lambda_2,\gamma_2}\ri]\cr
&\times \int_{-1}^0 d\rho\ (-2\rho)^{\al_0}\ (1+\rho)^{1+\al_1+\lambda_2+\gamma_2}\ 
(1-\rho)^{1+\al_2+\lambda_1+\gamma_1}\ .}$$
The three expressions in \kitpi\ combine to \insti. This can be verified by
making use of the relation \obtain\ and \firenze.

\appendix\appC{Complex world--sheet integral 
$W_\si^{(\kappa,\al_0)}\lf[{\al_1,\lambda_1,\gamma_1,\beta_1\atop\al_2,\lambda_2,\gamma_2,\beta_2}\ri]$}

In this Appendix we shall accomplish the computation of the complex integral \GENHU.
It has been argued in Subsection 2.4.4, that this amounts to consider the integral \GENNHU
\eqn\GENNHU{\eqalign{
W_\si^{(\kappa,\al_0)}\lf[{\alpha_1,\lambda_1,\gamma_1,\beta_1\atop
\alpha_2,\lambda_2,\gamma_2,\beta_2}\ri]&=\fc{i}{2}\ \int_{-\infty}^{\infty} dx\
|2x|^{\hatt\al_0}\ |1+x|^{\hatt\alpha_1}\ |1-x|^{\hatt\alpha_2}\cr 
&\times (2x)^{m_0}\ (1+x)^{m_1}\ (1-x)^{m_2}\ 
I^{(\kappa)}\lf[{\lambda_1,\gamma_1,\beta_1\atop\lambda_2,\gamma_2,\beta_2}\ri](x)\ ,}}
with:
\eqn\ANGELLHU{\eqalign{
I^{(\kappa)}\lf[{\lambda_1,\gamma_1,\beta_1\atop\lambda_2,\gamma_2,\beta_2}\ri](x)&=
\int_{-\infty}^\infty d\xi\ \int_{-\infty}^\infty d\eta \ 
|1-\xi|^{\hatt\lambda_1}\ |\xi-x|^{\hatt\gamma_1}\ |\xi+x|^{\hatt\beta_1}\cr 
&\times |1-\eta|^{\hatt\lambda_2}\ |\eta-x|^{\hatt\gamma_2}\ |\eta+x|^{\hatt\beta_2}\ 
|\xi-\eta|^{\hatt\kappa}\ \Pi(x,\xi,\eta)\cr
&\times (1-\xi)^{n_1}\ (\xi-x)^{n_3}\ (\xi+x)^{n_5}\ 
(1-\eta)^{n_2}\ (\eta-x)^{n_4}\ (\eta+x)^{n_6}\ (\xi-\eta)^{\tilde\kappa}.}}
Again, quantities with a hat refer to the non--integer part of the
parameter \parameterHU. Furthermore we write $\kappa=\hatt\kappa+\tilde\kappa$
separating the non--integer part $\hatt\kappa$.
In \GENNHU\ the integration w.r.t. the real variable $x$ is determined by the range $\Ic_\si$.
For a given range $\Ic_\si$ the phase factor $\Pi(x,\xi,\eta)$, which has been introduced in \ANGELLHU, is determined by the variables $\xi$ and $\eta$.
Hence in the following for a given $x\in\Ic_\si$ we shall  analyze the phase factor
$\Pi(x,\xi,\eta)$
In what follows we have to discuss the three cases $(1)\ 0<x<1$,\ 
$(2)\ -1<x<0$, and $(3)\ -\infty<x<-1 \cup 1<x<\infty$, separately.

\subsec{Case\  $0<x< 1$}

According to \Orderings\ this case describes the ordering $\si_1$ of the four 
open strings along the real $x$--axis.
The values of the phase $\Pi(x,\xi,\eta)$ are displayed in the next Table.

\vskip0.1cm
{\vbox{\ninepoint{$$
\vbox{\offinterlineskip\tabskip=0pt
\halign{\strut\vrule#
&~$#$~\hfil 
&\vrule$#$ 
&~$#$~\hfil 
&\vrule$#$ 
&~$#$~\hfil 
&\vrule$#$ 
&~$#$~\hfil 
&\vrule$#$
&~$#$~\hfil 
&\vrule$#$ 
&~$#$~\hfil 
&\vrule$#$
&~$#$~\hfil 
&\vrule$#$
&~$#$~\hfil 
&\vrule$#$\cr
\noalign{\hrule}
&  &&(\xi,\eta)  
&& \eta<\xi && \xi<\eta<-x  &&  -x<\eta<x&& x<\eta<1 && \eta>1&& 
{\rm total\ phase} &\cr
\noalign{\hrule}
&(i)&&\xi<-x &&1 && e^{i\pi\hat\kappa}  &&
e^{i\pi(\hat\kappa+\hat\beta_2)} && e^{i\pi(\hat\kappa+\hat\beta_2+\hat\gamma_2)}
&& e^{i\pi(\hat\kappa+\hat\beta_2+\hat\gamma_2+\hat\lambda_2)} 
 &&\sigma_\gamma \sigma_\beta&\cr
\noalign{\hrule}}}$$
$$\vbox{\offinterlineskip\tabskip=0pt
\halign{\strut\vrule#
&~$#$~\hfil 
&\vrule$#$ 
&~$#$~\hfil 
&\vrule$#$ 
&~$#$~\hfil 
&\vrule$#$ 
&~$#$~\hfil 
&\vrule$#$
&~$#$~\hfil 
&\vrule$#$ 
&~$#$~\hfil 
&\vrule$#$
&~$#$~\hfil 
&\vrule$#$
&~$#$~\hfil 
&\vrule$#$\cr
\noalign{\hrule}
&  &&(\xi,\eta)  
&& \eta<-x && -x<\eta<\xi   &&  \xi<\eta<x&& x<\eta<1 && \eta>1&& 
{\rm total\ phase} &\cr
\noalign{\hrule}
&(ii)&&-x<\xi<x &&e^{i\pi\hat\beta_2} && 1 &&
e^{i\pi\hat\kappa}&& e^{i\pi(\hat\kappa+\hat\gamma_2)} 
&& e^{i\pi(\hat\kappa+\hat\gamma_2+\hat\lambda_2)} &&\sigma_\gamma &\cr
\noalign{\hrule}}}$$
$$\vbox{\offinterlineskip\tabskip=0pt
\halign{\strut\vrule#
&~$#$~\hfil 
&\vrule$#$ 
&~$#$~\hfil 
&\vrule$#$ 
&~$#$~\hfil 
&\vrule$#$ 
&~$#$~\hfil 
&\vrule$#$
&~$#$~\hfil 
&\vrule$#$ 
&~$#$~\hfil 
&\vrule$#$
&~$#$~\hfil 
&\vrule$#$
&~$#$~\hfil 
&\vrule$#$\cr
\noalign{\hrule}
&  &&(\xi,\eta)  
&& \eta<-x && -x<\eta<x   &&  x<\eta<\xi&& \xi<\eta<1 && \eta>1&& 
{\rm total\ phase} &\cr
\noalign{\hrule}
&(iii)&&x<\xi<1 &&e^{i\pi(\hat\gamma_2+\hat\beta_2)} && e^{i\pi\hat\gamma_2} &&
1&& e^{i\pi\hat\kappa}  && e^{i\pi(\hat\kappa+\hat\lambda_2)} &&1&\cr
\noalign{\hrule}}}$$
$$\vbox{\offinterlineskip\tabskip=0pt
\halign{\strut\vrule#
&~$#$~\hfil 
&\vrule$#$ 
&~$#$~\hfil 
&\vrule$#$ 
&~$#$~\hfil 
&\vrule$#$ 
&~$#$~\hfil 
&\vrule$#$
&~$#$~\hfil 
&\vrule$#$ 
&~$#$~\hfil 
&\vrule$#$
&~$#$~\hfil 
&\vrule$#$
&~$#$~\hfil 
&\vrule$#$\cr
\noalign{\hrule}
&  &&(\xi,\eta)  
&& \eta<-x && -x<\eta<x   &&  x<\eta<1&& 1<\eta<\xi && \eta>\xi&& 
{\rm total\ phase} &\cr
\noalign{\hrule}
&(iv)&&\xi>1 &&e^{i\pi(\hat\lambda_2+\hat\gamma_2+\hat\beta_2)} && 
e^{i\pi(\hat\lambda_2+\hat\gamma_2)} &&
e^{i\pi\hat\lambda_2}&& 1
&& e^{i\pi\hat\kappa} &&\sigma_\lambda&\cr
\noalign{\hrule}}}$$
\vskip0pt
\centerline{\noindent{\bf Table 4:}
{\sl Phases $\Pi(x,\xi,\eta)$ along the integration region $(\xi,\eta)$ for $0<x<1$.}}
\vskip10pt}}}
\vskip-0.5cm
\noindent We have introduced the total phases
$\sigma_\lambda:=e^{i\pi(n_1+n_2)},\sigma_\gamma:=e^{i\pi(n_3+n_4)}$
and $\sigma_\beta:~=~e^{i\pi(n_5+n_6)}$. 
The different phase structures in the 
complex $\eta$--plane are shown in the next four figures. More precisely, these figures
display the way, how to integrate in the 
complex $\eta$--plane to take into account the phases of Table 4.
\ifig\contii{The complex $\eta$--plane and the contour integrals for the two 
cases $(i)$ and~$(ii)$. }{\epsfxsize=0.5\hsize\epsfbox{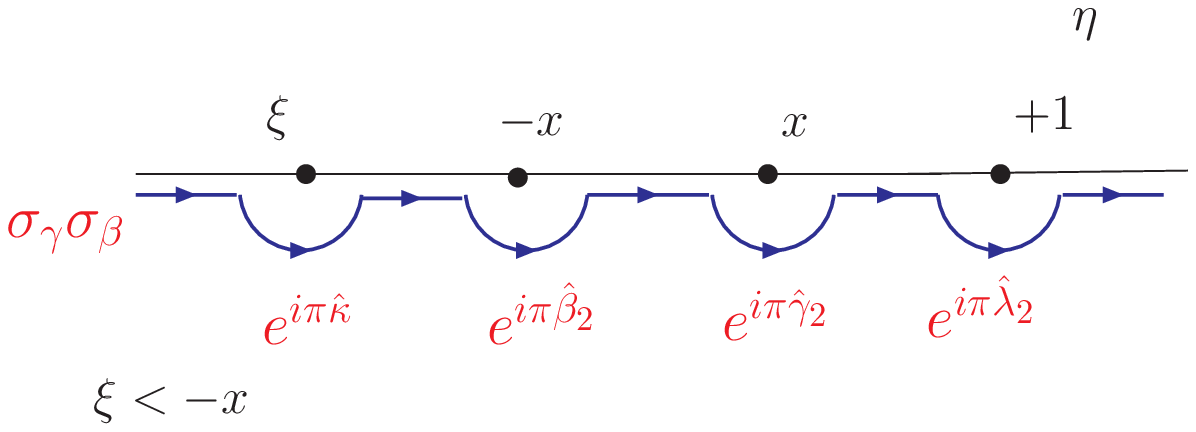}
\epsfxsize=0.5\hsize\epsfbox{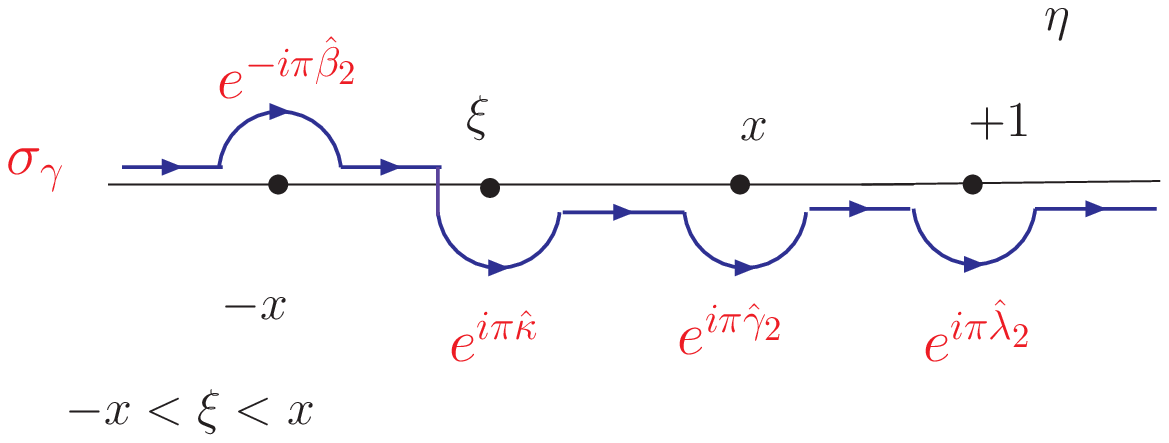}}
\ifig\contii{The complex $\eta$--plane and the contour integrals for the two 
cases~$(iii)$~and~$(iv)$.}{\epsfxsize=0.5\hsize\epsfbox{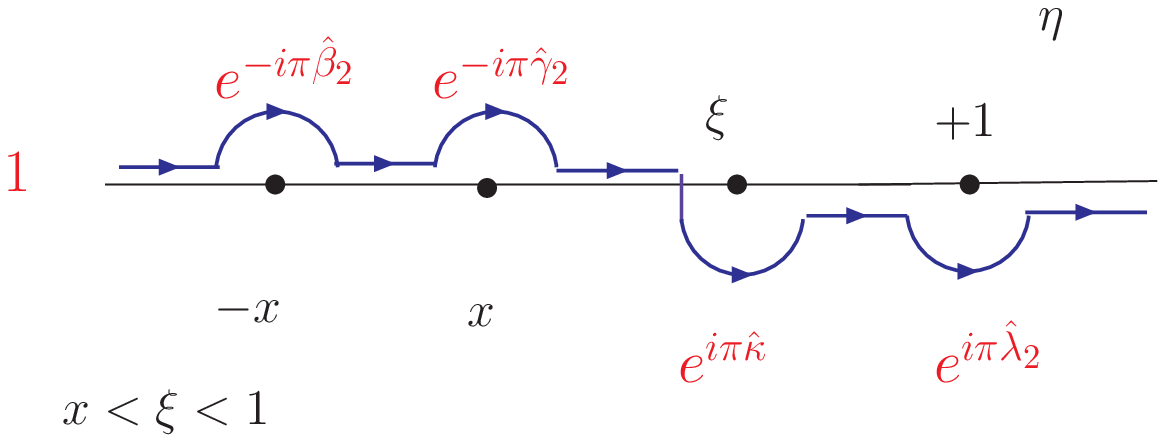}
\epsfxsize=0.5\hsize\epsfbox{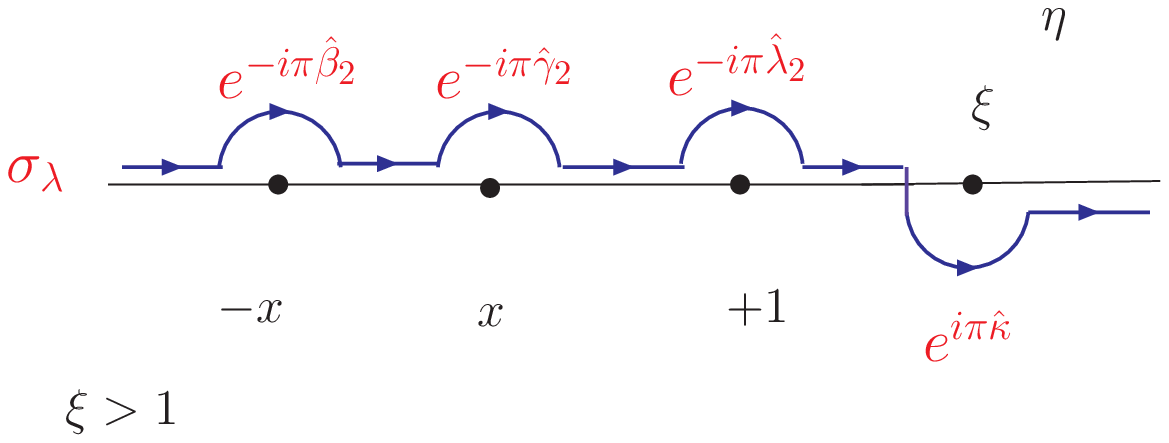}}
\noindent
After analyzing the structure of the contour integrals we find that the 
three cases $(ii),\ (iii)$ and $(iv)$ contribute to \ANGELLHU. 
Furthermore, in the two cases $(iii)$ and $(iv)$ there are contributions from
a residuum at $\eta=\xi$. 
For case $(ii)$ we may deform the contour in the complex $\eta$--plane and
integrate along the real $\eta$--axis from $-\infty$ to $-x$ and reverse taking into 
account the phase factors $e^{-i\pi \hatt\beta_2}$ and $e^{i\pi \hatt\beta_2}$, respectively.
On the other hand, in the case of $(iv)$ we may deform the contour and
integrate along the real $\eta$--axis from $\xi$ until $\infty$ and reverse
taking into account the corresponding phases
$e^{-i\pi \hatt\kappa}$ and $e^{i\pi \hatt\kappa}$, respectively, see the next Figure.
\ifig\Fontii{Deformed contours of the two 
cases~$(ii)$~and~$(iv)$.}{\epsfxsize=0.5\hsize\epsfbox{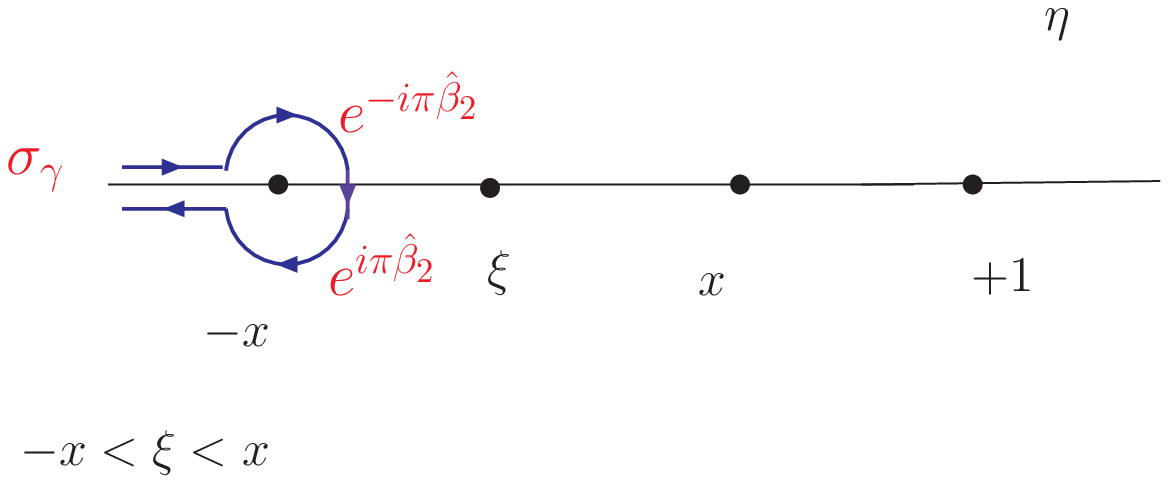}
\epsfxsize=0.5\hsize\epsfbox{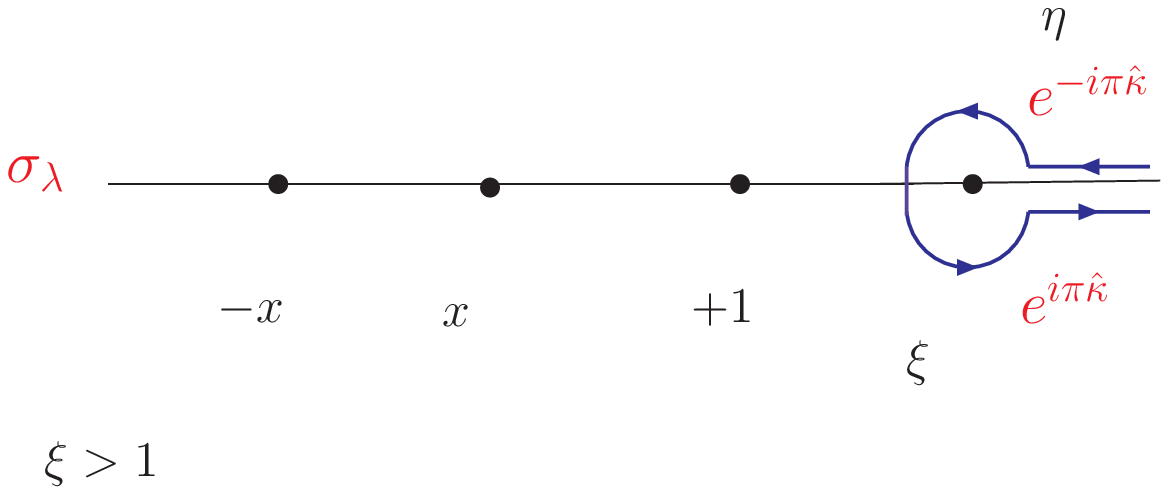}}
\noindent
Hence the contributions from $(ii)$ and $(iv)$ give:
\eqn\Iiia{\eqalign{
I^{(\kappa)}_{\si_1,ii}\lf[{\lambda_1,\gamma_1,\beta_1\atop\lambda_2,\gamma_2,\beta_2}\ri](x)&=
\sin(\pi\beta_2)\ \sigma_\gamma\ 
\int\limits_{-x}^x d\xi\ \int\limits_{-\infty}^{-x} d\eta \ 
(1-\xi)^{\lambda_1}\ (x-\xi)^{\gamma_1}\ (\xi+x)^{\beta_1}\cr 
&\times (1-\eta)^{\lambda_2}\ (-\eta+x)^{\gamma_2}\ (-\eta-x)^{\beta_2}\ 
(\xi-\eta)^\kappa\ ,\cr
I^{(\kappa)}_{\si_1,iv}\lf[{\lambda_1,\gamma_1,\beta_1\atop\lambda_2,\gamma_2,\beta_2}\ri](x)&=
\sin(\pi\kappa)\ \sigma_\lambda\ 
\int\limits_1^\infty d\xi\ \int\limits_\xi^\infty d\eta \ 
(\xi-1)^{\lambda_1}\ (\xi-x)^{\gamma_1}\ (\xi+x)^{\beta_1}\cr 
&\times (\eta-1)^{\lambda_2}\ (\eta-x)^{\gamma_2}\ (\eta+x)^{\beta_2}\ 
(\eta-\xi)^\kappa\ .}}
On the other hand, for case $(iii)$ we may first deform the whole contour in the 
complex $\eta$--plane and integrate along the real $\eta$--axis from $\xi$ until 
$\infty$ and reverse respecting 
the phase factors at $\eta=\xi$ and $\eta=1$, respectively, see the next Figure. 
\ifig\defConti{\ Deformed contour of the case ~$(iii)$.}
{\epsfxsize=0.5\hsize\epsfbox{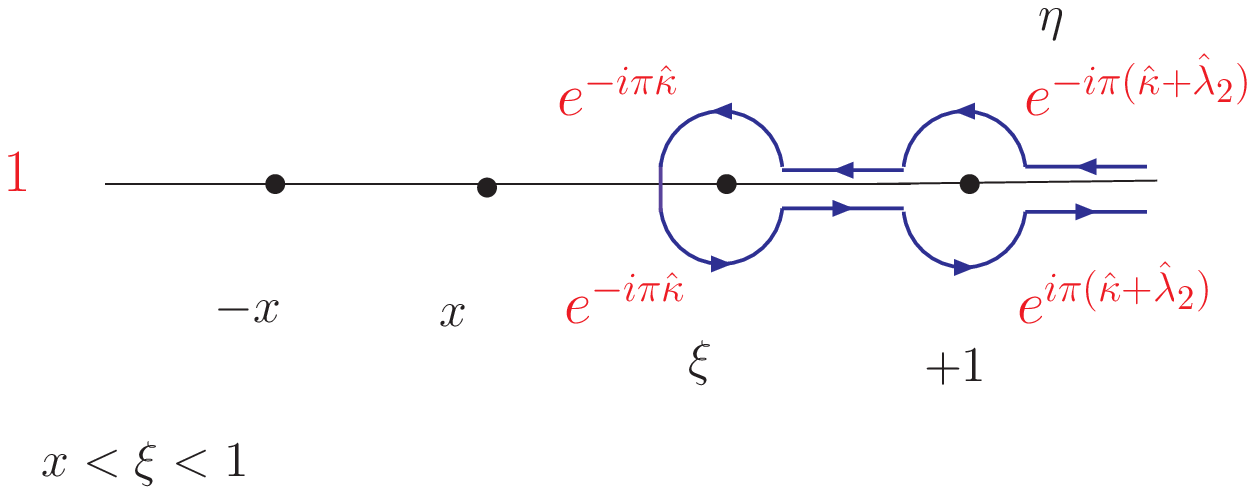}}
\noindent
Therefore, the contribution from $(iii)$ gives:
\eqn\Iiiia{\eqalign{
I^{(\kappa)}_{\si_1,iii}\lf[{\lambda_1,\gamma_1,\beta_1\atop\lambda_2,\gamma_2,\beta_2}\ri](x)&=
\sin[\pi(\kappa+\lambda_2)]\ 
\int\limits_{x}^1 d\xi\ \int\limits_1^\infty d\eta \ 
(1-\xi)^{\lambda_1}\ (\xi-x)^{\gamma_1}\ (\xi+x)^{\beta_1}\cr 
&\times (\eta-1)^{\lambda_2}\ (\eta-x)^{\gamma_2}\ (\eta+x)^{\beta_2}\ 
(\eta-\xi)^\kappa}}
$$\eqalign{&+\sin(\pi\kappa)\ \int\limits_{x}^1 d\xi\ \int_\xi^1 d\eta \ 
(1-\xi)^{\lambda_1}\ (\xi-x)^{\gamma_1}\ (\xi+x)^{\beta_1}\cr 
&\times 
(1-\eta)^{\lambda_2}\ (\eta-x)^{\gamma_2}\ (\eta+x)^{\beta_2}\ 
(\eta-\xi)^\kappa\ .}$$

After taking into account the $x$--integration of \GENNHU\ the integrals \Iiia\ and \Iiiia\ give rise to \Obtain.
\subsec{Case\ $-1<x < 0$}

According to \Orderings\ this case describes the ordering $\si_2$ of the four 
open strings along the real $x$--axis.
For this case  the analysis of the  phase factor $\Pi(x,\xi,\eta)$
boils down to the previous Subsection C.1.
By using the exchange symmetry \fulfillHU\ we can apply the previous results with
the following regrouping 
$\gamma_i\leftrightarrow\beta_i$ and $\alpha_1\leftrightarrow\alpha_2$.

Hence the two contributions $I_{ii}$ and $I_{iv}$ from the cases $(ii)$ and $(iv)$
yield
\eqn\Iiiaa{\eqalign{
I^{(\kappa)}_{\si_2,ii}\lf[{\lambda_1,\gamma_1,\beta_1\atop\lambda_2,\gamma_2,\beta_2}\ri](x)&=\sin(\pi\gamma_2)\ \sigma_\beta\ 
\int\limits_{x}^{-x} d\xi\ \int\limits_{-\infty}^{x} d\eta \ 
(1-\xi)^{\lambda_1}\ (\xi-x)^{\gamma_1}\ (-x-\xi)^{\beta_1}\cr 
&\times (1-\eta)^{\lambda_2}\ (x-\eta)^{\gamma_2}\ (-x-\eta)^{\beta_2}\ 
(\xi-\eta)^\kappa\ ,\cr
I^{(\kappa)}_{\si_2,iv}\lf[{\lambda_1,\gamma_1,\beta_1\atop\lambda_2,\gamma_2,\beta_2}\ri](x)&=\sin(\pi\kappa)\ \sigma_\lambda\ 
\int\limits_1^\infty d\xi\ \int\limits_\xi^\infty d\eta \ 
(\xi-1)^{\lambda_1}\ (\xi-x)^{\gamma_1}\ (\xi+x)^{\beta_1}\ \cr 
&\times (\eta-1)^{\lambda_2}\ (\eta-x)^{\gamma_2}\ (\eta+x)^{\beta_2}\ 
(\eta-\xi)^\kappa\ .}}
Furthermore the contribution from case $(iii)$ gives:
\eqn\Iiiiaa{\eqalign{
I^{(\kappa)}_{\si_2,iii}\lf[{\lambda_1,\gamma_1,\beta_1\atop\lambda_2,\gamma_2,\beta_2}\ri](x)&=\sin[\pi(\kappa+\lambda_2)]\ 
\int\limits_{-x}^1 d\xi\ \int\limits_1^\infty d\eta \ 
(1-\xi)^{\lambda_1}\ (\xi-x)^{\gamma_1}\ (\xi+x)^{\beta_1}\cr 
&\times (\eta-1)^{\lambda_2}\ (\eta-x)^{\gamma_2}\ (\eta+x)^{\beta_2}\ 
(\eta-\xi)^\kappa\cr
&+\sin(\pi\kappa)\ \int\limits_{-x}^1 d\xi\ \int_\xi^1 d\eta \ 
(1-\xi)^{\lambda_1}\ (\xi-x)^{\gamma_1}\ (\xi+x)^{\beta_1}\ \cr 
&\times 
(1-\eta)^{\lambda_2}\ (\eta-x)^{\gamma_2}\ (\eta+x)^{\beta_2}\ 
(\eta-\xi)^\kappa\ .}}

After taking into account the $x$--integration of \GENNHU\ the integrals \Iiiaa\ and \Iiiiaa\ give rise to \Obtain.
\subsec{Case\  $x>1$ and $x<-1$}

According to \Orderings\ this case describes the ordering $\si_3$ of the four 
open strings along the real $x$--axis.
For $x>1$ the values of the phase $\Pi(x,\xi,\eta)$ are displayed in the next Table.

\vskip0.1cm
{\vbox{\ninepoint{$$
\vbox{\offinterlineskip\tabskip=0pt
\halign{\strut\vrule#
&~$#$~\hfil 
&\vrule$#$ 
&~$#$~\hfil 
&\vrule$#$ 
&~$#$~\hfil 
&\vrule$#$ 
&~$#$~\hfil 
&\vrule$#$
&~$#$~\hfil 
&\vrule$#$ 
&~$#$~\hfil 
&\vrule$#$
&~$#$~\hfil 
&\vrule$#$
&~$#$~\hfil 
&\vrule$#$\cr
\noalign{\hrule}
&  &&(\xi,\eta)  
&& \eta<\xi && \xi<\eta<-x  &&  -x<\eta<1&& 1<\eta<x && \eta>x&& 
{\rm total\ phase} &\cr
\noalign{\hrule}
&(i)&&\xi<-x &&1 && e^{i\pi\hat\kappa}  &&
e^{i\pi(\hat\kappa+\hat\beta_2)} && e^{i\pi(\hat\kappa+\hat\beta_2+\hat\lambda_2)}
&& e^{i\pi(\hat\kappa+\hat\beta_2+\hat\lambda_2+\hat\gamma_2)} 
 &&\sigma_\gamma \sigma_\beta&\cr
\noalign{\hrule}}}$$
$$\vbox{\offinterlineskip\tabskip=0pt
\halign{\strut\vrule#
&~$#$~\hfil 
&\vrule$#$ 
&~$#$~\hfil 
&\vrule$#$ 
&~$#$~\hfil 
&\vrule$#$ 
&~$#$~\hfil 
&\vrule$#$
&~$#$~\hfil 
&\vrule$#$ 
&~$#$~\hfil 
&\vrule$#$
&~$#$~\hfil 
&\vrule$#$
&~$#$~\hfil 
&\vrule$#$\cr
\noalign{\hrule}
&  &&(\xi,\eta)  
&& \eta<-x && -x<\eta<\xi   &&  \xi<\eta<1&& 1<\eta<x && \eta>x&& 
{\rm total\ phase} &\cr
\noalign{\hrule}
&(ii)&&-x<\xi<1 &&e^{i\pi\hat\beta_2} && 1 &&
e^{i\pi\hat\kappa}&& e^{i\pi(\hat\kappa+\hat\lambda_2)} 
&& e^{i\pi(\hat\kappa+\hat\lambda_2+\hat\gamma_2)} &&\sigma_\gamma &\cr
\noalign{\hrule}}}$$
$$\vbox{\offinterlineskip\tabskip=0pt
\halign{\strut\vrule#
&~$#$~\hfil 
&\vrule$#$ 
&~$#$~\hfil 
&\vrule$#$ 
&~$#$~\hfil 
&\vrule$#$ 
&~$#$~\hfil 
&\vrule$#$
&~$#$~\hfil 
&\vrule$#$ 
&~$#$~\hfil 
&\vrule$#$
&~$#$~\hfil 
&\vrule$#$
&~$#$~\hfil 
&\vrule$#$\cr
\noalign{\hrule}
&  &&(\xi,\eta)  
&& \eta<-x && -x<\eta<1   &&  1<\eta<\xi&& \xi<\eta<x && \eta>x&& 
{\rm total\ phase} &\cr
\noalign{\hrule}
&(iii)&&1<\xi<x &&e^{i\pi(\hat\lambda_2+\hat\beta_2)} && e^{i\pi\hat\lambda_2} &&
1&& e^{i\pi\hat\kappa} && e^{i\pi(\hat\kappa+\hat\gamma_2)} &&\sigma_\lambda\sigma_\gamma&\cr
\noalign{\hrule}}}$$
$$\vbox{\offinterlineskip\tabskip=0pt
\halign{\strut\vrule#
&~$#$~\hfil 
&\vrule$#$ 
&~$#$~\hfil 
&\vrule$#$ 
&~$#$~\hfil 
&\vrule$#$ 
&~$#$~\hfil 
&\vrule$#$
&~$#$~\hfil 
&\vrule$#$ 
&~$#$~\hfil 
&\vrule$#$
&~$#$~\hfil 
&\vrule$#$
&~$#$~\hfil 
&\vrule$#$\cr
\noalign{\hrule}
&  &&(\xi,\eta)  
&& \eta<-x && -x<\eta<1   &&  1<\eta<x&& x<\eta<\xi && \eta>\xi&& 
{\rm total\ phase} &\cr
\noalign{\hrule}
&(iv)&&\xi>x &&e^{i\pi(\hat\lambda_2+\hat\beta_2+\hat\gamma_2)} && 
e^{i\pi(\hat\lambda_2+\hat\gamma_2)} &&e^{i\pi\hat\gamma_2} && 1&& e^{i\pi\hat\kappa} &&\sigma_\lambda&\cr
\noalign{\hrule}}}$$
\vskip0pt
\centerline{\noindent{\bf Table 5:}
{\sl Phases $\Pi(x,\xi,\eta)$ along the integration region $(\xi,\eta)$ for $x>1$.}}
\vskip10pt}}}
\vskip-0.5cm
\br
Again, the different phase structures in the 
complex $\eta$--plane are shown in the next four figures.
\ifig\contii{The complex $\eta$--plane and the contour integrals for the two 
cases $(i)$ and~$(ii)$. }{\epsfxsize=0.5\hsize\epsfbox{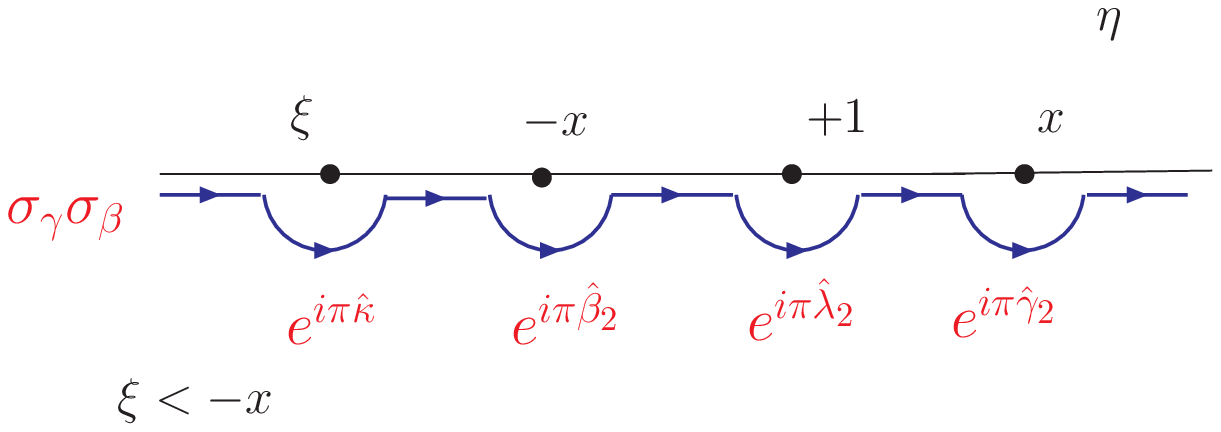}
\epsfxsize=0.5\hsize\epsfbox{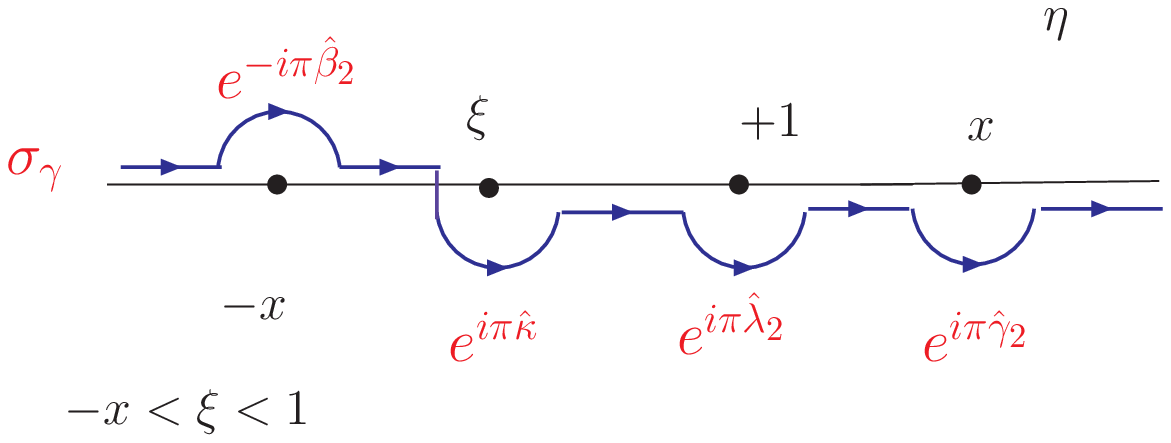}}
\ifig\contii{The complex $\eta$--plane and the contour integrals for the two 
cases~$(iii)$~and~$(iv)$.}{\epsfxsize=0.5\hsize\epsfbox{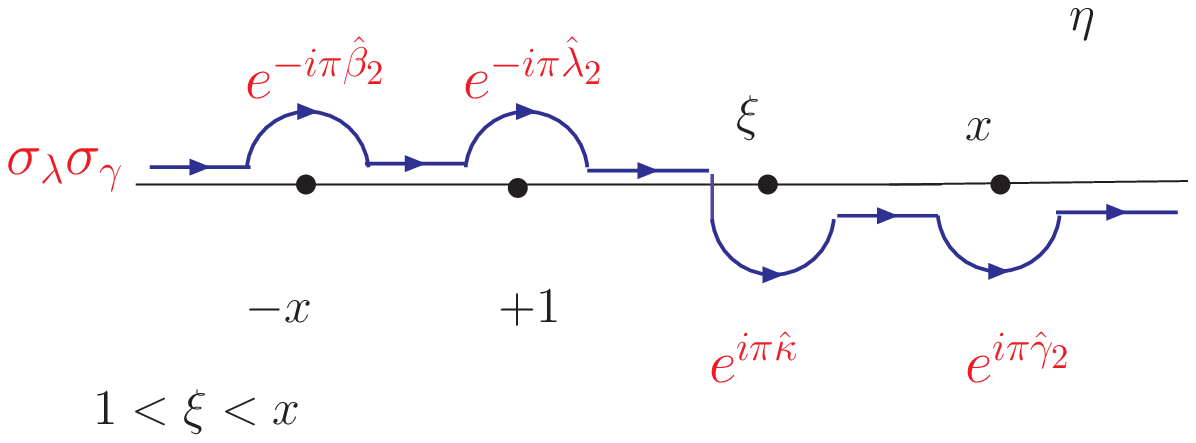}
\epsfxsize=0.5\hsize\epsfbox{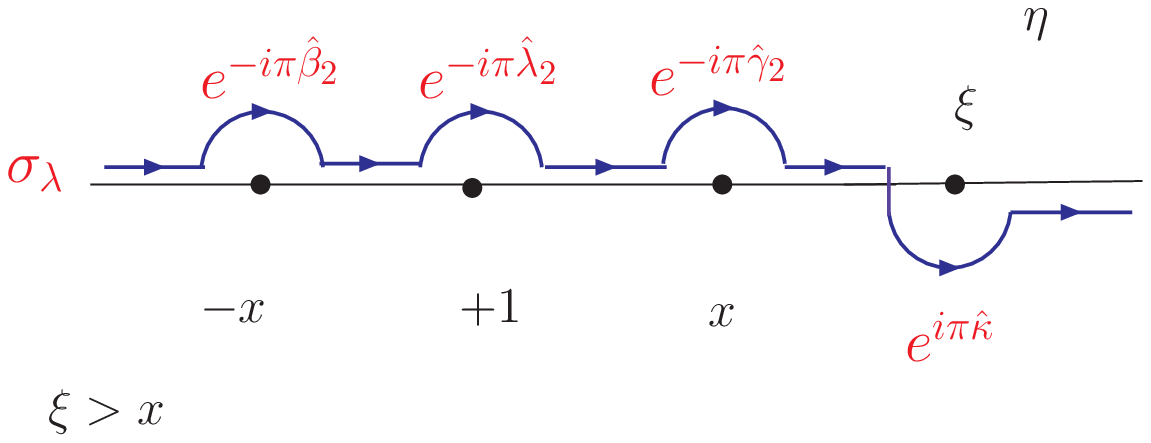}}
\noindent
After analyzing the structure of the contour integrals we find that the 
three cases $(ii),\ (iii)$ and $(iv)$ contribute to \ANGELLHU. 
Furthermore, in the two cases $(iii)$ and $(iv)$ there are contributions from  
a residuum at $\eta=\xi$.
For case $(ii)$ we may deform the contour in the complex $\eta$--plane and
integrate along the real $\eta$--axis from $-\infty$ to $-x$ and reverse taking into 
account the phase factors $e^{-i\pi \hatt\beta_2}$ and $e^{i\pi \hatt\beta_2}$, respectively.
On the other hand, in the case of $(iv)$ we may deform the contour and
integrate along the real $\eta$--axis from $\xi$ until $\infty$ and reverse
taking into account the corresponding phases
$e^{-i\pi \hatt\kappa}$ and $e^{i\pi \hatt\kappa}$, respectively, see the next Figure.
\ifig\Fontii{Deformed contours of the two 
cases~$(ii)$~and~$(iv)$.}{\epsfxsize=0.5\hsize\epsfbox{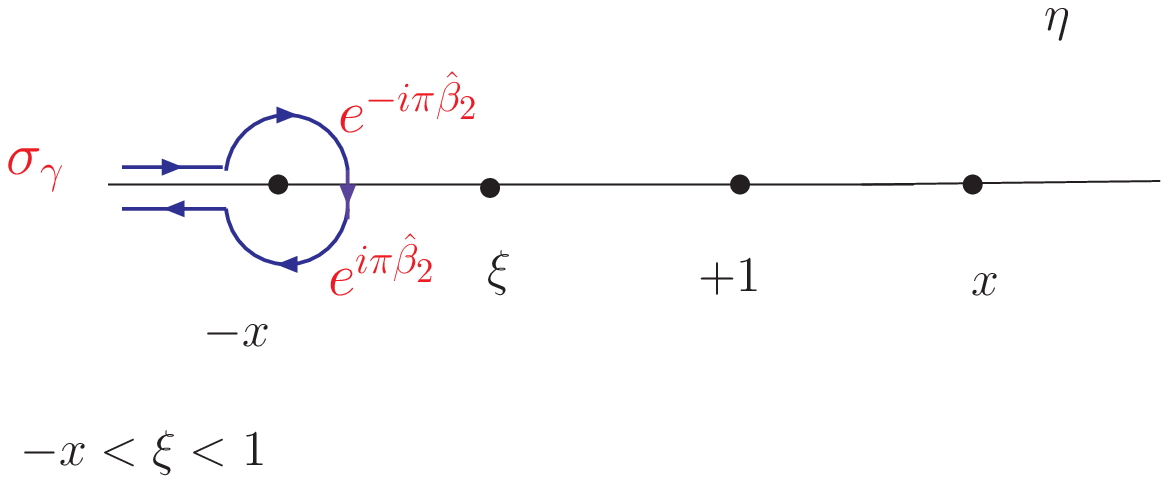}
\epsfxsize=0.5\hsize\epsfbox{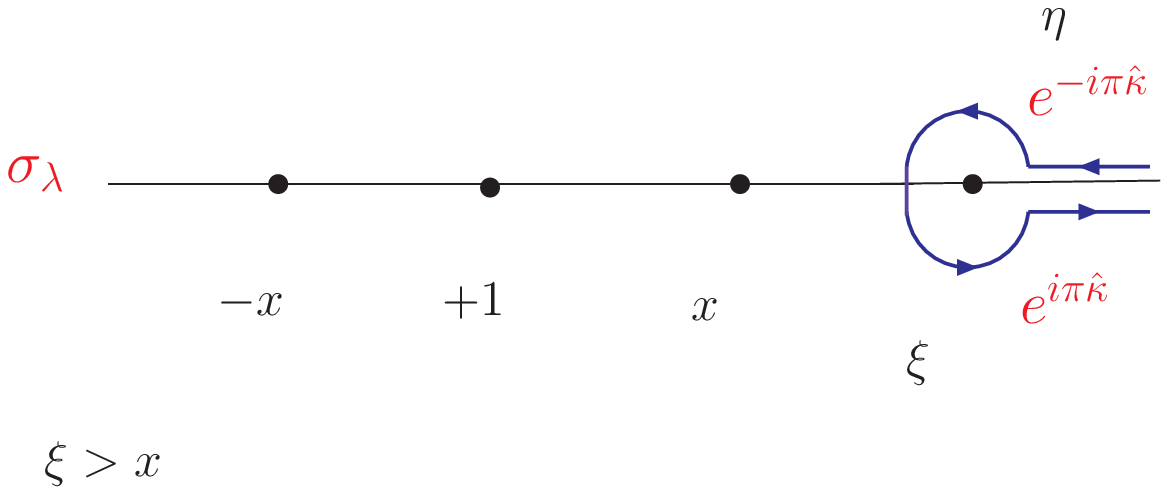}}
\noindent
Hence the contributions from $(ii)$ and $(iv)$ give:
\eqn\Iiia{\eqalign{
I^{(\al)}_{\si_3,ii}\lf[{\lambda_1,\gamma_1,\beta_1\atop\lambda_2,\gamma_2,\beta_2}\ri](x)&=
\sin(\pi\beta_2)\ \sigma_\gamma\ 
\int\limits_{-x}^1 d\xi\ \int\limits_{-\infty}^{-x} d\eta \ 
(1-\xi)^{\lambda_1}\ (x-\xi)^{\gamma_1}\ (\xi+x)^{\beta_1}\cr 
&\times (1-\eta)^{\lambda_2}\ (x-\eta)^{\gamma_2}\ (-\eta-x)^{\beta_2}\ 
(\xi-\eta)^\kappa\ ,\cr
I^{(\al)}_{\si_3,iv}\lf[{\lambda_1,\gamma_1,\beta_1\atop\lambda_2,\gamma_2,\beta_2}\ri](x)&=
\sin(\pi\kappa)\ \sigma_\lambda\ 
\int\limits_x^\infty d\xi\ \int\limits_\xi^\infty d\eta \ 
(\xi-1)^{\lambda_1}\ (\xi-x)^{\gamma_1}\ (\xi+x)^{\beta_1}\cr 
&\times (\eta-1)^{\lambda_2}\ (\eta-x)^{\gamma_2}\ (\eta+x)^{\beta_2}\ 
(\eta-\xi)^\kappa\ .}}
On the other hand, for case $(iii)$ we may first deform the whole contour in the 
complex $\eta$--plane and integrate along the real $\eta$--axis from $\xi$ until 
$\infty$ and reverse respecting 
the phase factors at $\eta=\xi$ and $\eta=x$, respectively, see the next Figure. 
\ifig\defConti{\ Deformed contour of the case ~$(iii)$.}
{\epsfxsize=0.5\hsize\epsfbox{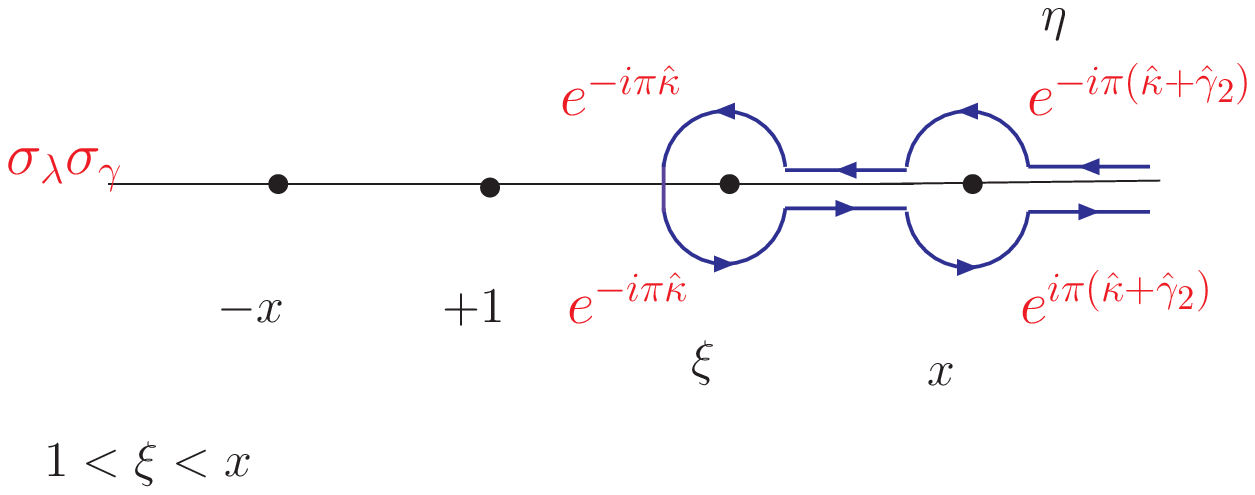}}
\noindent
Therefore, the contribution from $(iii)$ gives:
\eqn\Iiiia{\eqalign{
I^{(\al)}_{\si_3,iii}\lf[{\lambda_1,\gamma_1,\beta_1\atop\lambda_2,\gamma_2,\beta_2}\ri](x)
&=\sin[\pi(\kappa+\gamma_2)]\ \sigma_\lambda\ \sigma_\gamma\ 
\int\limits_1^x d\xi\ \int\limits_x^\infty d\eta \ 
(\xi-1)^{\lambda_1}\ (x-\xi)^{\gamma_1}\ (\xi+x)^{\beta_1}\cr 
&\times (\eta-1)^{\lambda_2}\ (\eta-x)^{\gamma_2}\ (\eta+x)^{\beta_2}\ 
(\eta-\xi)^\kappa\cr
&+\sin(\pi\kappa)\ \sigma_\lambda\ \sigma_\gamma\ 
\int\limits_1^x d\xi\ \int\limits_\xi^x d\eta\ (\xi-1)^{\lambda_1}\ 
(x-\xi)^{\gamma_1}\ (\xi+x)^{\beta_1}\cr 
&\times  (\eta-1)^{\lambda_2}\
(x-\eta)^{\gamma_2}\ (\eta+x)^{\beta_2}\ (\eta-\xi)^\kappa\ .}}

The case $x<-1$ gives the analogous expressions.
After taking into account the $x$--integration of \GENNHU\ the
integrals \Iiia\ and \Iiiia\ give rise to \Obtainvii.

\appendix\appD{Complex world--sheet integral 
$W^{(\kappa,\al_0,\al_3)}\lf[{\al_1,\lambda_1,\gamma_1,\beta_1,\eps_1\atop\al_2,\lambda_2,\gamma_2,\beta_2,\eps_2}\ri]$}

In this Appendix we undertake the computation of the complex integral \GENTO.
It has been argued in Subsection 2.4.5, that this amounts to consider the integral \GENNTO
\eqn\GENNTO{\eqalign{
W^{(\kappa,\al_0,\al_3)}\lf[{\alpha_1,\lambda_1,\gamma_1,\beta_1,\eps_1\atop
\alpha_2,\lambda_2,\gamma_2,\beta_2,\eps_2}\ri]&=2^{\al_0}\ i\  
 \int_{-1}^{1} dx\
|2x|^{\hatt\al_3}\ |1+x|^{\hatt\alpha_1}\ |1-x|^{\hatt\alpha_2}\cr 
&\times (2x)^{m_3}\ (1+x)^{1+m_1}\ (1-x)^{1+m_2}\ 
I^{(\kappa)}\lf[{\lambda_1,\gamma_1,\beta_1,\eps_1\atop\lambda_2,\gamma_2,\beta_2,\eps_2}\ri](x)\ ,}}
with:
\eqn\GENTO{\eqalign{
I^{(\kappa)}\lf[{\lambda_1,\gamma_1,\beta_1,\eps_1\atop\lambda_2,\gamma_2,\beta_2,\eps_2}\ri](x)&=
\int_{-\infty}^\infty d\xi\ \int_{-\infty}^\infty d\eta\  
|1-\xi|^{\hatt\lambda_1}\ |1+\xi|^{\hatt\gamma_1}\ |\xi-x|^{\hatt\beta_1}\ |\xi+x|^{\hatt\eps_1}\cr 
&\times |1-\eta|^{\hatt\lambda_2}\ |1+\eta|^{\hatt\gamma_2}\ |\eta-x|^{\hatt\beta_2}\ |\eta+x|^{\hatt\eps_2}\ 
|\xi+\eta|^{\hatt\kappa}\ \Pi(x,\xi,\eta)\cr
&\times (1-\xi)^{n_1}\ (1+\xi)^{n_3}\ (\xi-x)^{n_5}\ (\xi+x)^{n_7}\cr 
&\times (1-\eta)^{n_2}\ (1+\eta)^{n_4}\ (\eta-x)^{n_6}\ (\eta+x)^{n_8}\ (\xi+\eta)^{\tilde\kappa}\ .}}
Again, hatted quantities  refer to the non--integer part of the
parameter \parameterto. Furthermore we write $\kappa=\hatt\kappa+\tilde\kappa$
separating the non--integer part $\hatt\kappa$.
For a given $x\in\IR$ the phase factor $\Pi(x,\xi,\eta)$, which has been introduced in \GENTO, is determined by the variables $\xi$ and $\eta$.
Hence in the following for the range $-1<x<1$ we shall  analyze the phase factor
$\Pi(x,\xi,\eta)$
In what follows we have to discuss the two cases $(1)\ 0<x<1$, and
$(2)\ -1<x<0$, separately.

\subsec{Case\  $0<x< 1$}

For this case the values of the phase $\Pi(x,\xi,\eta)$ are displayed in the next Table.

\vskip0.1cm
{\vbox{\ninepoint{$$\centerline{
\vbox{\offinterlineskip\tabskip=0pt
\halign{\strut\vrule#
&~$#$~\hfil 
&\vrule$#$ 
&~$#$~\hfil 
&\vrule$#$ 
&~$#$~\hfil 
&\vrule$#$ 
&~$#$~\hfil 
&\vrule$#$
&~$#$~\hfil 
&\vrule$#$ 
&~$#$~\hfil 
&\vrule$#$
&~$#$~\hfil 
&\vrule$#$ 
&~$#$~\hfil 
&\vrule$#$
&~$#$~\hfil 
&\vrule$#$\cr
\noalign{\hrule}
&  &&\ss{(\xi,\eta)}  
&& \ss{\eta<-1} && \ss{-1<\eta<-x}  && \ss{-x<\eta<x}&& \ss{x<\eta<1} && 
\ss{1<\eta<-\xi}&& \ss{\eta>-\xi}&&\ss{{\rm total\ phase}} &\cr
\noalign{\hrule}
&\ss{(i)}&&\ss{\xi<-1} &&\ss{e^{i\pi\hat\kappa}} && \ss{e^{i\pi(\hat\gamma_2+\hat\kappa)}}  &&
\ss{e^{i\pi(\hat\gamma_2+\hat\eps_2+\hat\kappa)}}  && 
\ss{e^{i\pi(\hat\gamma_2+\hat\bet_2+\hat\eps_2+\hat\kappa)}}
&& \ss{e^{i\pi(\hat\lambda_2+\hat\gamma_2+\hat\bet_2+\hat\eps_2+\hat\kappa)}}
&& \ss{e^{i\pi(\hat\lambda_2+\hat\gamma_2+\hat\bet_2+\hat\eps_2)}}
&&\ss{\si_\gamma\si_\bet\si_\eps}&\cr
\noalign{\hrule}}}}$$
$$\centerline{\vbox{\offinterlineskip\tabskip=0pt
\halign{\strut\vrule#
&~$#$~\hfil 
&\vrule$#$ 
&~$#$~\hfil 
&\vrule$#$ 
&~$#$~\hfil 
&\vrule$#$ 
&~$#$~\hfil 
&\vrule$#$
&~$#$~\hfil 
&\vrule$#$ 
&~$#$~\hfil 
&\vrule$#$
&~$#$~\hfil 
&\vrule$#$
&~$#$~\hfil 
&\vrule$#$
&~$#$~\hfil 
&\vrule$#$\cr
\noalign{\hrule}
&  &&\ss{(\xi,\eta)}  
&&\ss{\eta<-1} && \ss{-1<\eta<-x}&& \ss{-x<\eta<x}&& \ss{x<\eta<-\xi} && 
\ss{-\xi<\eta<1}&&\ss{\eta>1}&& \ss{\rm total\ phase} &\cr
\noalign{\hrule}
&\ss{(ii)}&&\ss{-1<\xi<-x} &&\ss{e^{i\pi(\hat\gamma_2+\hat\kappa)}} && 
\ss{e^{i\pi\hat\kappa}}&&\ss{e^{i\pi(\hat\eps_2+\hat\kappa)}}
&&\ss{e^{i\pi(\hat\bet_2+\hat\eps_2+\hat\kappa)}}
&&\ss{e^{i\pi(\hat\bet_2+\hat\eps_2)}}&&\ss{e^{i\pi(\hat\lambda_2+\hat\bet_2+\hat\eps_2)}}  
&&\ss{\sigma_\bet\sigma_\eps}&\cr
\noalign{\hrule}}}}$$
$$\centerline{\vbox{\offinterlineskip\tabskip=0pt
\halign{\strut\vrule#
&~$#$~\hfil 
&\vrule$#$ 
&~$#$~\hfil 
&\vrule$#$ 
&~$#$~\hfil 
&\vrule$#$ 
&~$#$~\hfil 
&\vrule$#$
&~$#$~\hfil 
&\vrule$#$ 
&~$#$~\hfil 
&\vrule$#$
&~$#$~\hfil 
&\vrule$#$
&~$#$~\hfil 
&\vrule$#$
&~$#$~\hfil 
&\vrule$#$\cr
\noalign{\hrule}
&  &&\ss{(\xi,\eta)}  
&& \ss{\eta<-1} && \ss{-1<\eta<-x}   && \ss{-x<\eta<-\xi}&& \ss{-\xi<\eta<x} && 
\ss{x<\eta<1}&&\ss{\eta>1}&& \ss{\rm total\ phase} &\cr
\noalign{\hrule}
&\ss{(iii)}&&\ss{-x<\xi<x} &&\ss{e^{i\pi(\hat\gamma_2+\hat\eps_2+\hat\kappa)}}&&
\ss{e^{i\pi(\hat\eps_2+\hat\kappa)}}&&
\ss{e^{i\pi\hat\kappa}}&& \ss{1}  && 
\ss{e^{i\pi\hat\bet_2}} &&
\ss{e^{i\pi(\hat\lambda_2+\hat\bet_2)}}&&\ss{\si_\bet}&\cr
\noalign{\hrule}}}}$$
$$\centerline{\vbox{\offinterlineskip\tabskip=0pt
\halign{\strut\vrule#
&~$#$~\hfil 
&\vrule$#$ 
&~$#$~\hfil 
&\vrule$#$ 
&~$#$~\hfil 
&\vrule$#$ 
&~$#$~\hfil 
&\vrule$#$
&~$#$~\hfil 
&\vrule$#$ 
&~$#$~\hfil 
&\vrule$#$
&~$#$~\hfil 
&\vrule$#$
&~$#$~\hfil
&\vrule$#$
&~$#$~\hfil 
&\vrule$#$\cr
\noalign{\hrule}
&  &&\ss{(\xi,\eta)}  
&& \ss{\eta<-1} && \ss{-1<\eta<-\xi}   &&  \ss{-\xi<\eta<-x}&& \ss{-x<\eta<x} && 
\ss{x<\eta<1}&&\ss{\eta>1}&& \ss{\rm total\ phase} &\cr
\noalign{\hrule}
&\ss{(iv)}&&\ss{x<\xi<1} &&\ss{e^{i\pi(\hat\gamma_2+\hat\beta_2+\hat\eps_2+\hat\kappa)}} && 
\ss{e^{i\pi(\hat\beta_2+\hat\eps_2+\hat\kappa)}} &&\ss{e^{i\pi(\hat\beta_2+\hat\eps_2)}}
&& \ss{e^{i\pi\hat\beta_2}} && \ss{1}&& \ss{e^{i\pi\hat\lambda_2}} && \ss{1}  &\cr
\noalign{\hrule}}}}$$
$$\centerline{\vbox{\offinterlineskip\tabskip=0pt
\halign{\strut\vrule#
&~$#$~\hfil 
&\vrule$#$ 
&~$#$~\hfil 
&\vrule$#$ 
&~$#$~\hfil 
&\vrule$#$ 
&~$#$~\hfil 
&\vrule$#$
&~$#$~\hfil 
&\vrule$#$ 
&~$#$~\hfil 
&\vrule$#$
&~$#$~\hfil 
&\vrule$#$ 
&~$#$~\hfil 
&\vrule$#$
&~$#$~\hfil 
&\vrule$#$\cr
\noalign{\hrule}
&  &&\ss{(\xi,\eta)} 
&& \ss{\eta<-\xi} && \ss{-\xi<\eta<-1}   &&  \ss{-1<\eta<-x}&& \ss{-x<\eta<x} && 
\ss{x<\eta<1}&& \ss{\eta>1}&&\ss{\rm total\ phase} &\cr
\noalign{\hrule}
&\ss{(v)}&&\ss{\xi>1} &&
\ss{e^{i\pi(\hat\lambda_2+\hat\gamma_2+\hat\beta_2+\hat\eps_2+\hat\kappa)}} && 
\ss{e^{i\pi(\hat\lambda_2+\hat\gamma_2+\hat\bet_2+\hat\eps_2)}} &&
\ss{e^{i\pi(\hat\lambda_2+\hat\bet_2+\hat\eps_2)}}&& \ss{e^{i\pi(\hat\lambda_2+\hat\bet_2)}}
&& \ss{e^{i\pi\hat\lambda_2}}&&\ss{1} &&\ss{\sigma_\lambda}&\cr
\noalign{\hrule}}}}$$
\vskip0pt
\centerline{\noindent{\bf Table 6:}
{\sl Phases $\Pi(x,\xi,\eta)$ along the integration region $(\xi,\eta)$ for $0<x<1$.}}
\vskip10pt}}}
\vskip-0.5cm
\br
We have introduced the total phases
$\sigma_\lambda:=e^{i\pi(n_1+n_2)},\sigma_\gamma:=e^{i\pi(n_3+n_4)},
\sigma_\beta:=e^{i\pi(n_5+n_6)},$ and $\sigma_\eps:=e^{i\pi(n_7+n_8)}$.
The different phase structures in the 
complex $\eta$--plane are shown in the next five figures. More precisely, these figures
display the way, how to integrate in the 
complex $\eta$--plane to take into account the phases of Table 6.
\ifig\contii{The complex $\eta$--plane and the contour integrals for the two 
cases $(i)$ and~$(ii)$. }{\epsfxsize=0.55\hsize\epsfbox{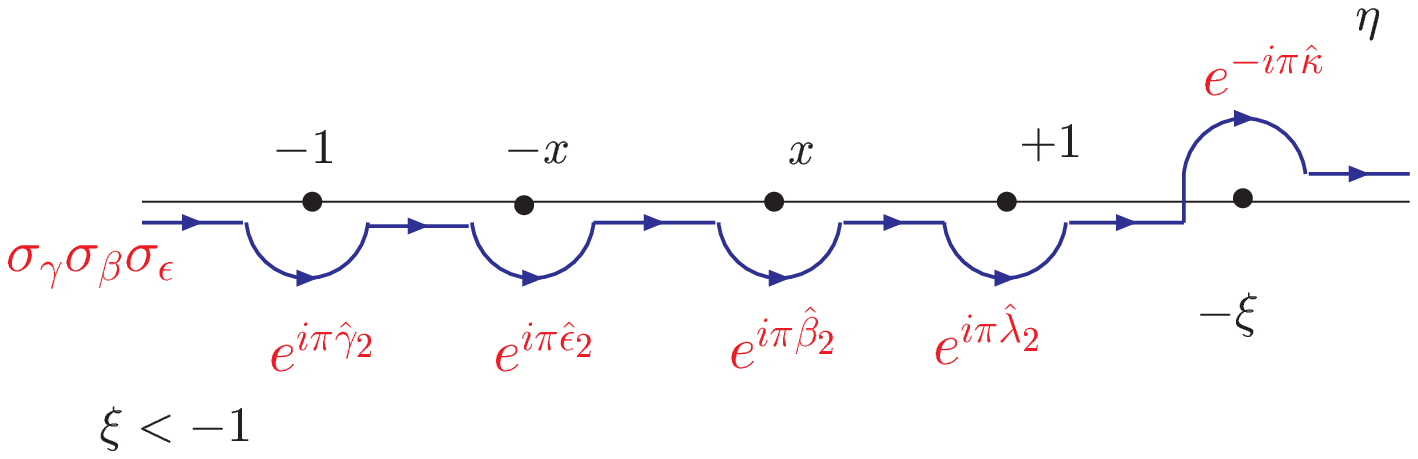}
\epsfxsize=0.55\hsize\epsfbox{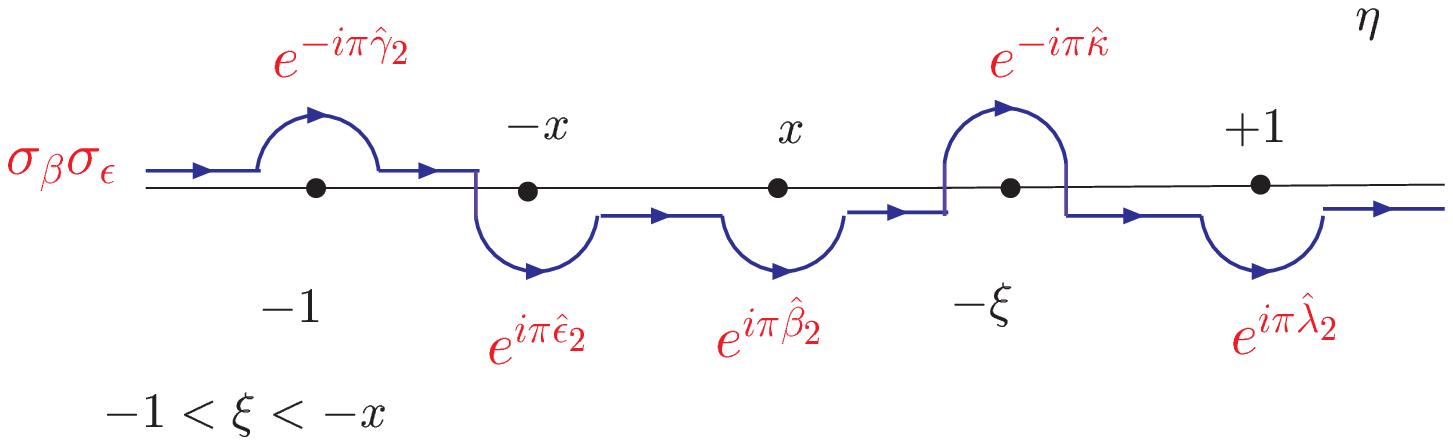}}
\ifig\contii{The complex $\eta$--plane and the contour integrals for the two 
cases~$(iii)$~and~$(iv)$.}{\epsfxsize=0.55\hsize\epsfbox{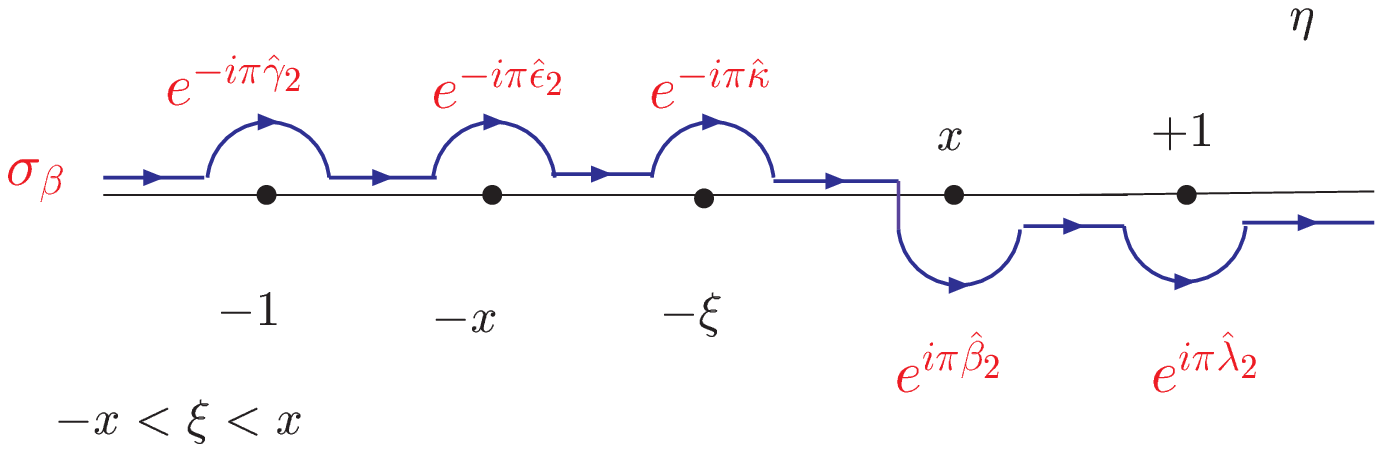}
\epsfxsize=0.55\hsize\epsfbox{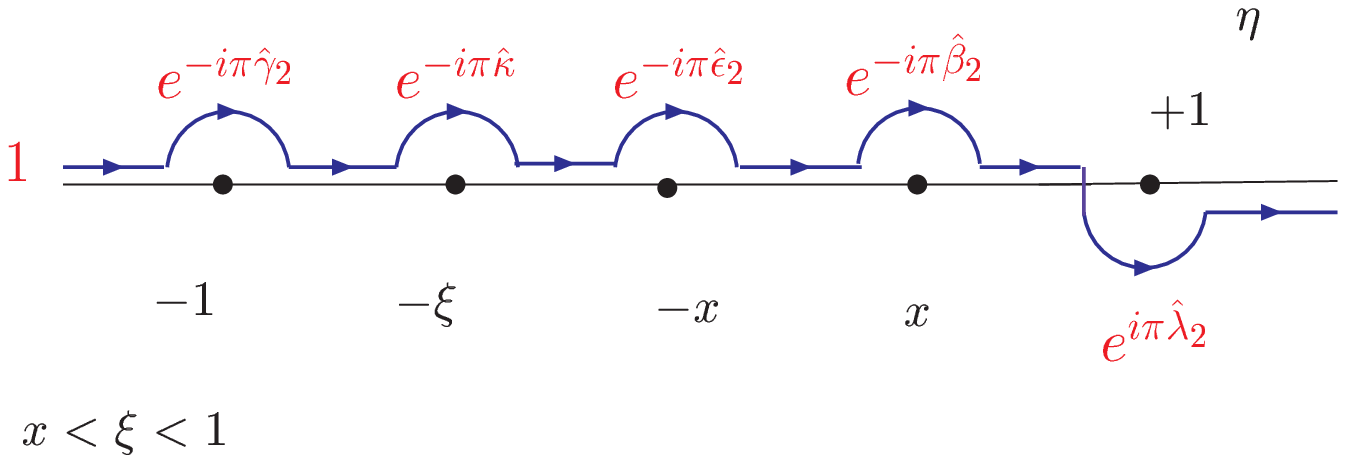}}
\ifig\defConti{\ The complex $\eta$--plane and the contour for the 
case~$(v)$.}
{\epsfxsize=0.55\hsize\epsfbox{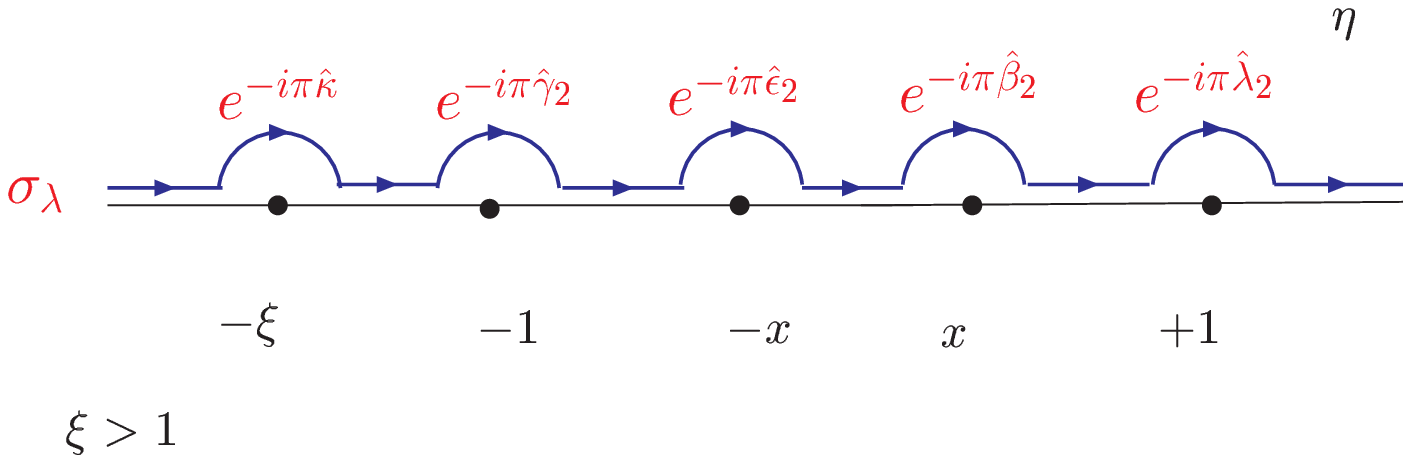}}
\noindent
After analyzing the structure of the contour integrals we find that the four cases
$(i)-(iv)$ contribute to \GENTO. For case $(i)$ we may deform the contour in the complex $\eta$--plane and integrate along the real $\eta$--axis from $-\xi$ to 
$\infty$ and reverse taking into account the phase factors $e^{-i\pi\hatt\kappa}$
and $e^{i\pi\hatt\kappa}$, respectively, \cf the left diagram of the next figure 
\ifig\Fontii{Deformed contours of the two 
cases~$(i)$~and~$(iv)$.}{\epsfxsize=0.5\hsize\epsfbox{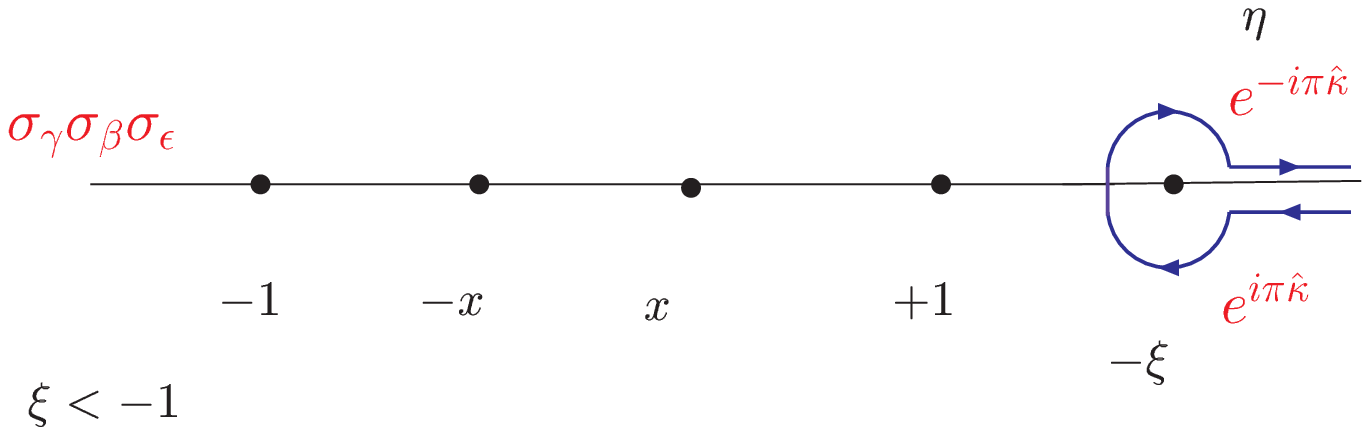}
\epsfxsize=0.5\hsize\epsfbox{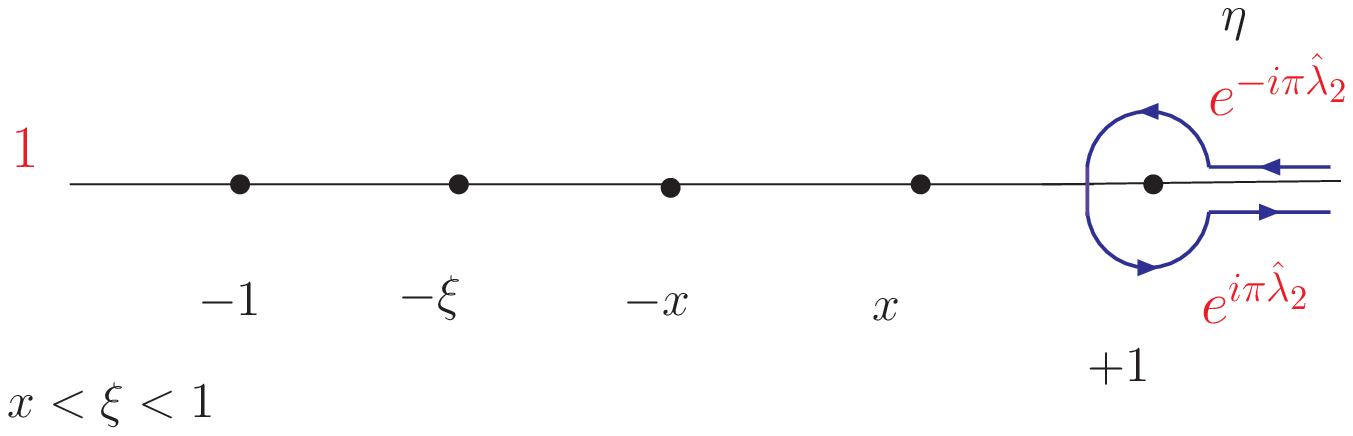}}
\noindent
On the other hand, in the case of $(iv)$ we may deform the contour and integrate along the real $\eta$--axis from $+1$ to $\infty$ and reverse taking into account the corresponding phases $e^{-i\pi\hatt\lambda_2}$
and $e^{i\pi\hatt\lambda_2}$.
Hence the contributions from $(i)$ and $(iv)$ give:
\eqn\Iiia{\eqalign{
I^{(\kappa)}_{a}\lf[{\lambda_1,\gamma_1,\beta_1,\eps_1\atop\lambda_2,\gamma_2,\beta_2,\eps_2}\ri](x)&=
\sin[\pi(\lambda_2+\gamma_2+\beta_2+\eps_2)]\ \si_\gamma\si_\beta\si_\eps\cr  
&\times\int\limits_{-\infty}^{-1} d\xi\ \int\limits_{-\xi}^\infty d\eta \ 
(1-\xi)^{\lambda_1}\ (-1-\xi)^{\gamma_1}\ (x-\xi)^{\bet_1}\ 
(-\xi-x)^{\eps_1}\cr 
&\times (\eta-1)^{\lambda_2}\ (1+\eta)^{\gamma_2}\ (\eta-x)^{\bet_2}\ (\eta+x)^{\eps_2}\ 
(\xi+\eta)^\kappa\ ,\cr
I^{(\kappa)}_{d}\lf[{\lambda_1,\gamma_1,\beta_1,\eps_1\atop\lambda_2,\gamma_2,\beta_2,\eps_2}\ri](x)&=
\sin(\pi\lambda_2)\ \int\limits_x^1 d\xi\ \int\limits_1^\infty d\eta \ 
(1-\xi)^{\lambda_1}\ (1+\xi)^{\gamma_1}\ (\xi-x)^{\bet_1}\ (\xi+x)^{\eps_1}\cr 
&\times (\eta-1)^{\lambda_2}\ (1+\eta)^{\gamma_2}\ (\eta-x)^{\bet_2}\ (\eta+x)^{\eps_2}\ (\xi+\eta)^\kappa .}}
On the other hand, for case $(iii)$ we may  deform the whole contour in the 
complex $\eta$--plane and integrate along the real $\eta$--axis from $x$ until 
$\infty$ and reverse respecting 
the phase factors $e^{i\pi \hatt\beta_2}$ at $\eta=x$ and $e^{i\pi (\hatt\lambda_2+\hatt\beta_2)}$, respectively, see the next Figure. 
\ifig\defConti{\ Deformed contour of the case ~$(iii)$.}
{\epsfxsize=0.55\hsize\epsfbox{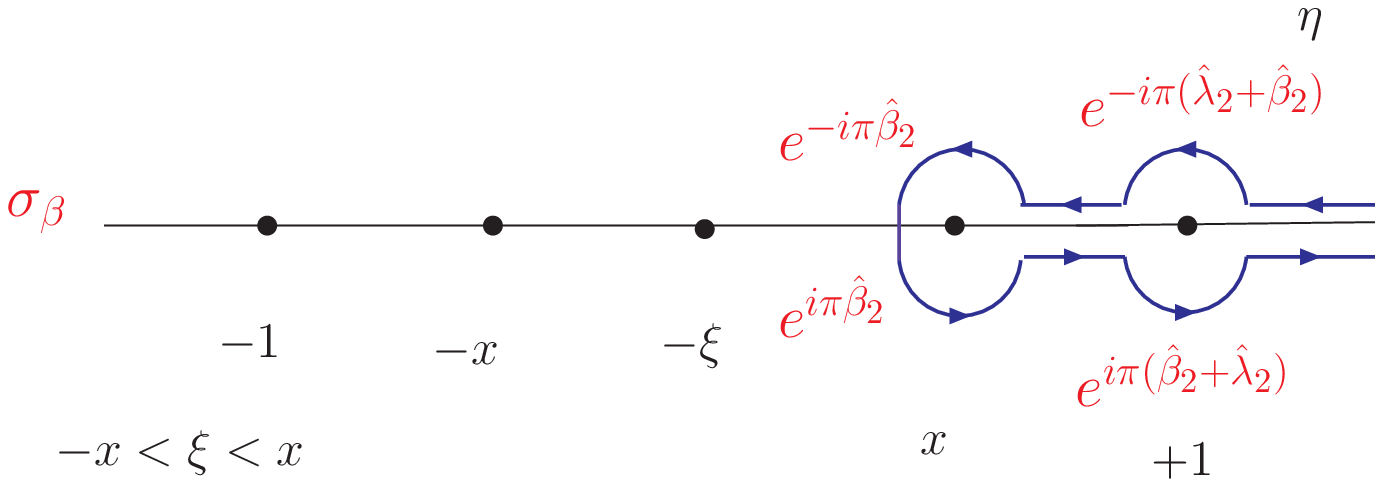}}
\noindent
Therefore, the two contributions from $(iii)$ give:
\eqn\Iiia{\eqalign{
I^{(\kappa)}_{c_1}\lf[{\lambda_1,\gamma_1,\beta_1,\eps_1\atop\lambda_2,\gamma_2,\beta_2,\eps_2}\ri](x)&=
\sin(\pi\beta_2)\ \si_\beta\ 
\int\limits_{-x}^{x} d\xi\ \int\limits_x^1 d\eta \ 
(1-\xi)^{\lambda_1}\ (1+\xi)^{\gamma_1}\ (x-\xi)^{\bet_1}\ (\xi+x)^{\eps_1}\cr 
&\times (1-\eta)^{\lambda_2}\ (1+\eta)^{\gamma_2}\ (\eta-x)^{\bet_2}\ 
(\eta+x)^{\eps_2}\ (\xi+\eta)^\kappa\ ,\cr
I^{(\kappa)}_{c_2}\lf[{\lambda_1,\gamma_1,\beta_1,\eps_1\atop\lambda_2,\gamma_2,\beta_2,\eps_2}\ri](x)&=
\sin[\pi(\lambda_2+\beta_2)]\ \si_\beta\ 
\int\limits_{-x}^{x} d\xi\ \int\limits_1^\infty d\eta \ 
(1-\xi)^{\lambda_1}\ (1+\xi)^{\gamma_1}\ (x-\xi)^{\bet_1}\cr 
&\times  (\xi+x)^{\eps_1}\ (\eta-1)^{\lambda_2}\ (1+\eta)^{\gamma_2}\ (\eta-x)^{\bet_2}\ (\eta+x)^{\eps_2}\ (\xi+\eta)^\kappa\ .}}
Finally, for case $(ii)$ we may  deform the whole contour in the 
complex $\eta$--plane to a contour from $\eta=-\xi$
until infinity, going back to $\eta=-\xi$ and encircling the latter point clockwise, see the next Figure. 
\ifig\defConti{\ Deformed contour of the case ~$(ii)$.}
{\epsfxsize=0.78\hsize\epsfbox{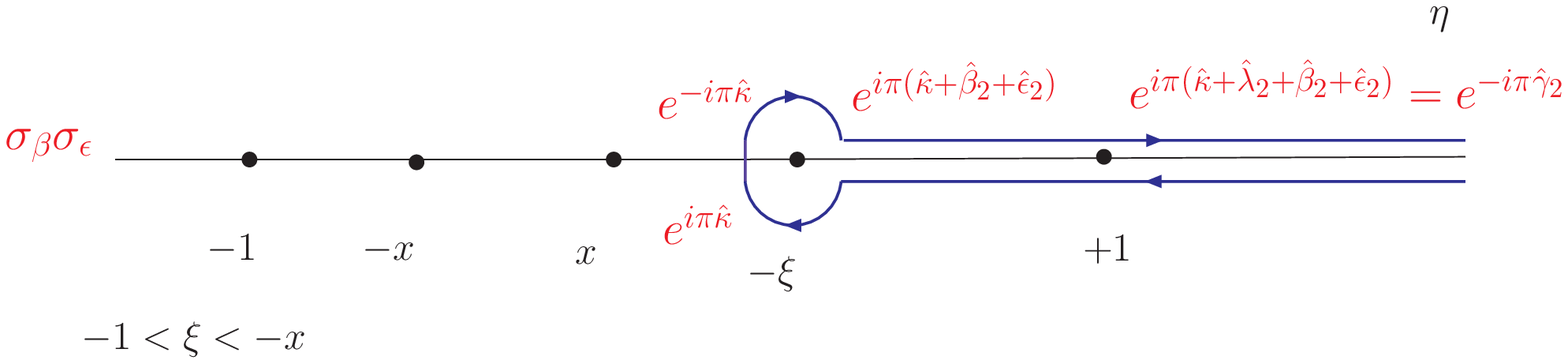}}
\noindent
In the region $-\xi<\eta<1$ we are left with the phase
$e^{i\pi(\hatt\kappa+\hatt\beta_2+\hatt\eps_2)}$. 
The latter receives an additional phase factor $e^{i\pi\hatt\lambda_2}$ 
after passing the point $\eta=1$ into the region $1<\eta<\infty$.
Therefore, the  total contribution to  $(ii)$ is
$$I^{(\kappa)}_{b_2}\lf[{\lambda_1,\gamma_1,\beta_1,\eps_1\atop\lambda_2,\gamma_2,\beta_2,\eps_2}\ri](x)+I^{(\kappa)}_{b_1}\lf[{\lambda_1,\gamma_1,\beta_1,\eps_1\atop\lambda_2,\gamma_2,\beta_2,\eps_2}\ri](x)\ ,$$
with:
\eqn\Iiia{\eqalign{
I^{(\kappa)}_{b_1}\lf[{\lambda_1,\gamma_1,\beta_1,\eps_1\atop\lambda_2,\gamma_2,\beta_2,\eps_2}\ri](x)&=
-\sin(\pi\kappa)\ \si_\beta\si_\eps\ e^{\pi i(\lambda_2+\beta_2+\eps_2+\kappa)}\cr 
&\times\int\limits_{-1}^{-x} d\xi\ \int\limits_{1}^{\infty} d\eta \ 
(1-\xi)^{\lambda_1}\ (1+\xi)^{\gamma_1}\ (x-\xi)^{\bet_1}\ 
(-\xi-x)^{\eps_1}\ \cr 
&\times (\eta-1)^{\lambda_2}\ (1+\eta)^{\gamma_2}\ (\eta-x)^{\bet_2}\ 
(\eta+x)^{\eps_2}\ (\xi+\eta)^\kappa\ ,\cr
I^{(\kappa)}_{b_2}\lf[{\lambda_1,\gamma_1,\beta_1,\eps_1\atop\lambda_2,\gamma_2,\beta_2,\eps_2}\ri](x)&=
-\sin(\pi\kappa)\ \si_\beta\si_\eps\ e^{\pi i(\beta_2+\eps_2+\kappa)}\cr 
&\times\int\limits_{-1}^{-x} d\xi\ \int\limits_{-\xi}^1 d\eta \ 
(1-\xi)^{\lambda_1}\ (1+\xi)^{\gamma_1}\ (x-\xi)^{\bet_1}\cr 
&\times (-\xi-x)^{\eps_1}\ (1-\eta)^{\lambda_2}\ (1+\eta)^{\gamma_2}\ (\eta-x)^{\bet_2}\ (\eta+x)^{\eps_2}\ (\xi+\eta)^\kappa\ .}}

\subsec{Case\  $-1<x<0$}

The case $-1<x<0$ can be infered from the previous case $0<x<1$ by using the relation
\follRel. As a result for the case under consideration the 
phase structure $\Pi(x,\xi,\eta)$ of \GENTO\ is obtained
from Table 6 by interchanging $-x$ and $x$ and permuting 
$\hatt\eps_i$ and $\hatt\bet_i$, \ie $\hatt\eps\leftrightarrow\hatt\bet$.

\listrefs
\end